\documentclass[longauth,traditabstract]{aa}
\usepackage{graphicx}
\usepackage{amsmath,amsfonts,amssymb}
\usepackage{txfonts}
\usepackage[table,usenames,dvipsnames]{xcolor}
\definecolor{linkcolor}{rgb}{0.6,0,0}
\definecolor{citecolor}{rgb}{0,0,0.75}
\definecolor{urlcolor}{rgb}{0.12,0.46,0.7}
\usepackage[breaklinks, colorlinks, urlcolor=urlcolor, linkcolor=linkcolor,citecolor=citecolor,pdfa=true]{hyperref}
\hypersetup{linktocpage}
\usepackage{natbib}
\usepackage{multirow}
\usepackage{epsf}
\usepackage{epsfig}
\usepackage{pgf,tikz}

\usepackage{ifthen}

\usepackage[T1]{fontenc}
\usepackage{lmodern}
\usepackage{ifxetex,ifluatex}

\usepackage{dblfloatfix}
\usepackage{morefloats}
\usepackage{siunitx}

\usepackage[switch]{lineno}

\bibpunct{(}{)}{;}{a}{}{,}

\def\setsymbol#1#2{\expandafter\def\csname #1\endcsname{#2}}
\def\getsymbol#1{\csname #1\endcsname}

\def\Planck{\textit{Planck}}

\newbox\tablebox    \newdimen\tablewidth
\def\leaderfil{\leaders\hbox to 5pt{\hss.\hss}\hfil}
\def\endPlancktable{\tablewidth=\columnwidth 
    $$\hss\copy\tablebox\hss$$
    \vskip-\lastskip\vskip -2pt}

\def\tablenote#1 #2\par{\begingroup \parindent=0.8em
    \abovedisplayshortskip=0pt\belowdisplayshortskip=0pt
    \noindent
    $$\hss\vbox{\hsize\tablewidth \hangindent=\parindent \hangafter=1 \noindent
    \hbox to \parindent{$^#1$\hss}\strut#2\strut\par}\hss$$
    \endgroup}
\def\doubleline{\vskip 3pt\hrule \vskip 1.5pt \hrule \vskip 5pt}

\def\L2{\ifmmode L_2\else $L_2$\fi}

\def\DeltaT{\ifmmode \Delta T\else $\Delta T$\fi}
\def\deltat{\ifmmode \Delta t\else $\Delta t$\fi}
\def\fknee{\ifmmode f_{\rm knee}\else $f_{\rm knee}$\fi}
\def\Fmax{\ifmmode F_{\rm max}\else $F_{\rm max}$\fi}
\def\solar{\ifmmode{\rm M}_{\mathord\odot}\else${\rm M}_{\mathord\odot}$\fi}
\def\Msolar{\ifmmode{\rm M}_{\mathord\odot}\else${\rm M}_{\mathord\odot}$\fi}
\def\Lsolar{\ifmmode{\rm L}_{\mathord\odot}\else${\rm L}_{\mathord\odot}$\fi}
\def\inv{\ifmmode^{-1}\else$^{-1}$\fi}
\def\mo{\ifmmode^{-1}\else$^{-1}$\fi}
\def\sup#1{\ifmmode ^{\rm #1}\else $^{\rm #1}$\fi}
\def\expo#1{\ifmmode \times 10^{#1}\else $\times 10^{#1}$\fi}
\def\,{\thinspace}
\def\lsim{\mathrel{\raise .4ex\hbox{\rlap{$<$}\lower 1.2ex\hbox{$\sim$}}}}
\def\gsim{\mathrel{\raise .4ex\hbox{\rlap{$>$}\lower 1.2ex\hbox{$\sim$}}}}

\def\simprop{\mathrel{\raise .4ex\hbox{\rlap{$\propto$}\lower 1.2ex\hbox{$\sim$}}}}
\def\deg{\ifmmode^\circ\else$^\circ$\fi}
\def\pdeg{\ifmmode $\setbox0=\hbox{$^{\circ}$}\rlap{\hskip.11\wd0 .}$^{\circ}
          \else \setbox0=\hbox{$^{\circ}$}\rlap{\hskip.11\wd0 .}$^{\circ}$\fi}
\def\arcs{\ifmmode {^{\scriptstyle\prime\prime}}
          \else $^{\scriptstyle\prime\prime}$\fi}
\def\arcm{\ifmmode {^{\scriptstyle\prime}}
          \else $^{\scriptstyle\prime}$\fi}
\newdimen\sa  \newdimen\sb
\def\parcs{\sa=.07em \sb=.03em
     \ifmmode \hbox{\rlap{.}}^{\scriptstyle\prime\kern -\sb\prime}\hbox{\kern -\sa}
     \else \rlap{.}$^{\scriptstyle\prime\kern -\sb\prime}$\kern -\sa\fi}
\def\parcm{\sa=.08em \sb=.03em
     \ifmmode \hbox{\rlap{.}\kern\sa}^{\scriptstyle\prime}\hbox{\kern-\sb}
     \else \rlap{.}\kern\sa$^{\scriptstyle\prime}$\kern-\sb\fi}
\def\ra[#1 #2 #3.#4]{#1\sup{h}#2\sup{m}#3\sup{s}\llap.#4}
\def\dec[#1 #2 #3.#4]{#1\deg#2\arcm#3\arcs\llap.#4}
\def\deco[#1 #2 #3]{#1\deg#2\arcm#3\arcs}
\def\rra[#1 #2]{#1\sup{h}#2\sup{m}}

\def\dots{\relax\ifmmode \ldots\else $\ldots$\fi}
\def\WHzsr{\ifmmode $W\,Hz\mo\,sr\mo$\else W\,Hz\mo\,sr\mo\fi}
\def\mHz{\ifmmode $\,mHz$\else \,mHz\fi}
\def\GHz{\ifmmode $\,GHz$\else \,GHz\fi}
\def\mKs{\ifmmode $\,mK\,s$^{1/2}\else \,mK\,s$^{1/2}$\fi}
\def\muKs{\ifmmode \,\mu$K\,s$^{1/2}\else \,$\mu$K\,s$^{1/2}$\fi}
\def\muKRJs{\ifmmode \,\mu$K$_{\rm RJ}$\,s$^{1/2}\else \,$\mu$K$_{\rm RJ}$\,s$^{1/2}$\fi}
\def\muKHz{\ifmmode \,\mu$K\,Hz$^{-1/2}\else \,$\mu$K\,Hz$^{-1/2}$\fi}
\def\MJysr{\ifmmode \,$MJy\,sr\mo$\else \,MJy\,sr\mo\fi}
\def\MJysrmK{\ifmmode \,$MJy\,sr\mo$\,mK$_{\rm CMB}\mo\else \,MJy\,sr\mo\,mK$_{\rm CMB}\mo$\fi}
\def\microns{\ifmmode \,\mu$m$\else \,$\mu$m\fi}

\def\muK{\ifmmode \,\mu$K$\else \,$\mu$\hbox{K}\fi}
\def\microK{\ifmmode \,\mu$K$\else \,$\mu$\hbox{K}\fi}
\def\muW{\ifmmode \,\mu$W$\else \,$\mu$\hbox{W}\fi}
\def\kms{\ifmmode $\,km\,s$^{-1}\else \,km\,s$^{-1}$\fi}
\def\kmsMpc{\ifmmode $\,\kms\,Mpc\mo$\else \,\kms\,Mpc\mo\fi}
\providecommand{\sorthelp}[1]{}

\def\eqref#1{(\ref{#1})}

\def\smica{{\tt SMICA}}
\def\nilc{{\tt NILC}}
\def\sevem{{\tt SEVEM}}
\def\commander{\texttt{Commander}}
\def\lowlcomm{\texttt{Lkl-Commander}}

\def\healpix{\texttt{HEALPix}}

\def\WMAP{\textrm{WMAP}}
\def\wmap{\textrm{WMAP}}

\def\LCDM{$\Lambda$CDM}

\def\nside{$N_{\mathrm{side}}$}

\def\eq{\begin{eqnarray}}
\def\qe{\end{eqnarray}}

\def\curl{\mathcal}
\def\({\left(}
\def\){\right)}

\def\and{\quad \mbox{and} \quad}
\def\barQ{\kern2pt\overline{\kern-2pt\curl{Q}}}

\def\barR{\kern2pt\overline{\kern-2pt\curl{R}}}

\def\bargamma{\kern2pt\overline{\kern-2pt\gamma}}

\def\leaderfil{\leaders\hbox to 5pt{\hss.\hss}\hfil}
\def\y{\vec{y}}

\newcommand{\cs}{Cold Spot}
\newcommand{\Qr}{$Q_\mathrm{r}$}
\newcommand{\Ur}{$U_\mathrm{r}$}
\newcommand{\be}{\begin{equation}}
\newcommand{\ee}{\end{equation}}
\newcommand{\bea}{\begin{eqnarray}}
\newcommand{\eea}{\end{eqnarray}}

\def\E{{\mathbf E}}

\renewcommand{\a}[0]{\mathbf{a}}

\newcommand{\F}[0]{\mathbf{F}}
\newcommand{\B}[0]{\mathbf{B}}

\renewcommand{\L}[0]{\mathbf{L}}

\newcommand{\M}[0]{\mathbf{M}}

\newcommand{\w}[0]{\mathbf{w}}

\renewcommand{\u}[0]{\mathbf{u}}

\renewcommand{\v}[0]{\mathbf{v}}

\def\inv{^{-1}}
\def\lm{{\ell m}}

\newcommand{\spinup}{\;\raise1.0pt\hbox{$'$}\hskip-6pt\partial\;}
\newcommand{\spindown}{\;\overline{\raise1.0pt\hbox{$'$}\hskip-6pt\partial}\;}

\newcommand{\pval}{{\it p}-value}

\defcitealias{planck2013-p09}{PCIS13}
\defcitealias{planck2014-a18}{PCIS15}

\title{\vglue -10mm\Planck\ 2018 results. VII. Isotropy and Statistics of the CMB}

\author{\small
Planck Collaboration: Y.~Akrami\inst{14, 49, 51}
\and
M.~Ashdown\inst{58, 5}
\and
J.~Aumont\inst{85}
\and
C.~Baccigalupi\inst{68}
\and
M.~Ballardini\inst{20, 36}
\and
A.~J.~Banday\inst{85, 8}~\thanks{Corresponding author: A.~J.~Banday \url{anthony.banday@irap.omp.eu}}
\and
R.~B.~Barreiro\inst{53}
\and
N.~Bartolo\inst{25, 54}
\and
S.~Basak\inst{75}
% ...
}
\institute{\small
AIM, CEA, CNRS, Universit\'{e} Paris-Saclay, Universit\'{e} Paris-Diderot, Sorbonne Paris Cit\'{e}, F-91191 Gif-sur-Yvette, France\goodbreak
\and
APC, AstroParticule et Cosmologie, Universit\'{e} Paris Diderot, CNRS/IN2P3, CEA/lrfu, Observatoire de Paris, Sorbonne Paris Cit\'{e}, 10, rue Alice Domon et L\'{e}onie Duquet, 75205 Paris Cedex 13, France\goodbreak
\and
African Institute for Mathematical Sciences, 6-8 Melrose Road, Muizenberg, Cape Town, South Africa\goodbreak
\and
Aix Marseille Univ, CNRS, CNES, LAM, Marseille, France\goodbreak
\and
Astrophysics Group, Cavendish Laboratory, University of Cambridge, J J Thomson Avenue, Cambridge CB3 0HE, U.K.\goodbreak
\and
Astrophysics \& Cosmology Research Unit, School of Mathematics, Statistics \& Computer Science, University of KwaZulu-Natal, Westville Campus, Private Bag X54001, Durban 4000, South Africa\goodbreak
\and
CITA, University of Toronto, 60 St. George St., Toronto, ON M5S 3H8, Canada\goodbreak
\and
CNRS, IRAP, 9 Av. colonel Roche, BP 44346, F-31028 Toulouse cedex 4, France\goodbreak
\and
Cahill Center for Astronomy and Astrophysics, California Institute of Technology, Pasadena CA,  91125, USA\goodbreak
\and
California Institute of Technology, Pasadena, California, U.S.A.\goodbreak
\and
Centre for Theoretical Cosmology, DAMTP, University of Cambridge, Wilberforce Road, Cambridge CB3 0WA, U.K.\goodbreak
\and
Computational Cosmology Center, Lawrence Berkeley National Laboratory, Berkeley, California, U.S.A.\goodbreak
\and
D\'{e}partement de Physique Th\'{e}orique, Universit\'{e} de Gen\`{e}ve, 24, Quai E. Ansermet,1211 Gen\`{e}ve 4, Switzerland\goodbreak
\and
D\'{e}partement de Physique, \'{E}cole normale sup\'{e}rieure, PSL Research University, CNRS, 24 rue Lhomond, 75005 Paris, France\goodbreak
\and
Departamento de Astrof\'{i}sica, Universidad de La Laguna (ULL), E-38206 La Laguna, Tenerife, Spain\goodbreak
\and
Departamento de F\'{\i}sica, Universidad de Oviedo, C/ Federico Garc\'{\i}a Lorca, 18 , Oviedo, Spain\goodbreak
\and
Department of Astrophysics/IMAPP, Radboud University, P.O. Box 9010, 6500 GL Nijmegen, The Netherlands\goodbreak
\and
Department of Mathematics, University of Stellenbosch, Stellenbosch 7602, South Africa\goodbreak
\and
Department of Physics \& Astronomy, University of British Columbia, 6224 Agricultural Road, Vancouver, British Columbia, Canada\goodbreak
\and
Department of Physics \& Astronomy, University of the Western Cape, Cape Town 7535, South Africa\goodbreak
\and
Department of Physics, Gustaf H\"{a}llstr\"{o}min katu 2a, University of Helsinki, Helsinki, Finland\goodbreak
\and
Department of Physics, Princeton University, Princeton, New Jersey, U.S.A.\goodbreak
\and
Department of Physics, University of California, Santa Barbara, California, U.S.A.\goodbreak
\and
Department of Physics, University of Illinois at Urbana-Champaign, 1110 West Green Street, Urbana, Illinois, U.S.A.\goodbreak
\and
Dipartimento di Fisica e Astronomia G. Galilei, Universit\`{a} degli Studi di Padova, via Marzolo 8, 35131 Padova, Italy\goodbreak
\and
Dipartimento di Fisica e Scienze della Terra, Universit\`{a} di Ferrara, Via Saragat 1, 44122 Ferrara, Italy\goodbreak
\and
Dipartimento di Fisica, Universit\`{a} La Sapienza, P. le A. Moro 2, Roma, Italy\goodbreak
\and
Dipartimento di Fisica, Universit\`{a} degli Studi di Milano, Via Celoria, 16, Milano, Italy\goodbreak
\and
Dipartimento di Fisica, Universit\`{a} degli Studi di Trieste, via A. Valerio 2, Trieste, Italy\goodbreak
\and
Dipartimento di Fisica, Universit\`{a} di Roma Tor Vergata, Via della Ricerca Scientifica, 1, Roma, Italy\goodbreak
\and
European Space Agency, ESAC, Planck Science Office, Camino bajo del Castillo, s/n, Urbanizaci\'{o}n Villafranca del Castillo, Villanueva de la Ca\~{n}ada, Madrid, Spain\goodbreak
\and
European Space Agency, ESTEC, Keplerlaan 1, 2201 AZ Noordwijk, The Netherlands\goodbreak
\and
Gran Sasso Science Institute, INFN, viale F. Crispi 7, 67100 L'Aquila, Italy\goodbreak
\and
Haverford College Astronomy Department, 370 Lancaster Avenue, Haverford, Pennsylvania, U.S.A.\goodbreak
\and
Helsinki Institute of Physics, Gustaf H\"{a}llstr\"{o}min katu 2, University of Helsinki, Helsinki, Finland\goodbreak
\and
INAF - OAS Bologna, Istituto Nazionale di Astrofisica - Osservatorio di Astrofisica e Scienza dello Spazio di Bologna, Area della Ricerca del CNR, Via Gobetti 101, 40129, Bologna, Italy\goodbreak
\and
INAF - Osservatorio Astronomico di Padova, Vicolo dell'Osservatorio 5, Padova, Italy\goodbreak
\and
INAF - Osservatorio Astronomico di Trieste, Via G.B. Tiepolo 11, Trieste, Italy\goodbreak
\and
INAF, Istituto di Radioastronomia, Via Piero Gobetti 101, I-40129 Bologna, Italy\goodbreak
\and
INAF/IASF Milano, Via E. Bassini 15, Milano, Italy\goodbreak
\and
INFN - CNAF, viale Berti Pichat 6/2, 40127 Bologna, Italy\goodbreak
\and
INFN, Sezione di Bologna, viale Berti Pichat 6/2, 40127 Bologna, Italy\goodbreak
\and
INFN, Sezione di Ferrara, Via Saragat 1, 44122 Ferrara, Italy\goodbreak
\and
INFN, Sezione di Milano, Via Celoria 16, Milano, Italy\goodbreak
\and
INFN, Sezione di Roma 2, Universit\`{a} di Roma Tor Vergata, Via della Ricerca Scientifica, 1, Roma, Italy\goodbreak
\and
Imperial College London, Astrophysics group, Blackett Laboratory, Prince Consort Road, London, SW7 2AZ, U.K.\goodbreak
\and
Institut d'Astrophysique Spatiale, CNRS, Univ. Paris-Sud, Universit\'{e} Paris-Saclay, B\^{a}t. 121, 91405 Orsay cedex, France\goodbreak
\and
Institut d'Astrophysique de Paris, CNRS (UMR7095), 98 bis Boulevard Arago, F-75014, Paris, France\goodbreak
\and
Institute Lorentz, Leiden University, PO Box 9506, Leiden 2300 RA, The Netherlands\goodbreak
\and
Institute of Astronomy, University of Cambridge, Madingley Road, Cambridge CB3 0HA, U.K.\goodbreak
\and
Institute of Theoretical Astrophysics, University of Oslo, Blindern, Oslo, Norway\goodbreak
\and
Instituto de Astrof\'{\i}sica de Canarias, C/V\'{\i}a L\'{a}ctea s/n, La Laguna, Tenerife, Spain\goodbreak
\and
Instituto de F\'{\i}sica de Cantabria (CSIC-Universidad de Cantabria), Avda. de los Castros s/n, Santander, Spain\goodbreak
\and
Istituto Nazionale di Fisica Nucleare, Sezione di Padova, via Marzolo 8, I-35131 Padova, Italy\goodbreak
\and
Jet Propulsion Laboratory, California Institute of Technology, 4800 Oak Grove Drive, Pasadena, California, U.S.A.\goodbreak
\and
Jodrell Bank Centre for Astrophysics, Alan Turing Building, School of Physics and Astronomy, The University of Manchester, Oxford Road, Manchester, M13 9PL, U.K.\goodbreak
\and
Kavli Institute for Cosmological Physics, University of Chicago, Chicago, IL 60637, USA\goodbreak
\and
Kavli Institute for Cosmology Cambridge, Madingley Road, Cambridge, CB3 0HA, U.K.\goodbreak
\and
Kavli Institute for the Physics and Mathematics of the Universe (Kavli IPMU, WPI), UTIAS, The University of Tokyo, Chiba, 277- 8583, Japan\goodbreak
\and
Laboratoire de Physique Subatomique et Cosmologie, Universit\'{e} Grenoble-Alpes, CNRS/IN2P3, 53, rue des Martyrs, 38026 Grenoble Cedex, France\goodbreak
\and
Laboratoire de Physique Th\'{e}orique, Universit\'{e} Paris-Sud 11 \& CNRS, B\^{a}timent 210, 91405 Orsay, France\goodbreak
\and
Low Temperature Laboratory, Department of Applied Physics, Aalto University, Espoo, FI-00076 AALTO, Finland\goodbreak
\and
Max-Planck-Institut f\"{u}r Astrophysik, Karl-Schwarzschild-Str. 1, 85741 Garching, Germany\goodbreak
\and
Mullard Space Science Laboratory, University College London, Surrey RH5 6NT, U.K.\goodbreak
\and
NAOC-UKZN Computational Astrophysics Centre (NUCAC), University of KwaZulu-Natal, Durban 4000, South Africa\goodbreak
\and
National Centre for Nuclear Research, ul. A. Soltana 7, 05-400 Otwock, Poland\goodbreak
\and
Nicolaus Copernicus Astronomical Center, Polish Academy of Sciences, Bartycka 18, 00-716 Warsaw, Poland\goodbreak
\and
SISSA, Astrophysics Sector, via Bonomea 265, 34136, Trieste, Italy\goodbreak
\and
San Diego Supercomputer Center, University of California, San Diego, 9500 Gilman Drive, La Jolla, CA 92093, USA\goodbreak
\and
School of Chemistry and Physics, University of KwaZulu-Natal, Westville Campus, Private Bag X54001, Durban, 4000, South Africa\goodbreak
\and
School of Physical Sciences, National Institute of Science Education and Research, HBNI, Jatni-752050, Odissa, India\goodbreak
\and
School of Physics and Astronomy, Cardiff University, Queens Buildings, The Parade, Cardiff, CF24 3AA, U.K.\goodbreak
\and
School of Physics and Astronomy, Sun Yat-sen University, 2 Daxue Rd, Tangjia, Zhuhai, China\goodbreak
\and
School of Physics and Astronomy, University of Nottingham, Nottingham NG7 2RD, U.K.\goodbreak
\and
School of Physics, Indian Institute of Science Education and Research Thiruvananthapuram, Maruthamala PO, Vithura, Thiruvananthapuram 695551, Kerala, India\goodbreak
\and
School of Physics, The University of New South Wales, Sydney NSW 2052, Australia\goodbreak
\and
Simon Fraser University, Department of Physics, 8888 University Drive, Burnaby BC, Canada\goodbreak
\and
Sorbonne Universit\'{e}, Observatoire de Paris, Universit\'{e} PSL, \'{E}cole normale sup\'{e}rieure, CNRS, LERMA, F-75005, Paris, France\goodbreak
\and
Sorbonne Universit\'{e}-UPMC, UMR7095, Institut d'Astrophysique de Paris, 98 bis Boulevard Arago, F-75014, Paris, France\goodbreak
\and
Space Research Institute (IKI), Russian Academy of Sciences, Profsoyuznaya Str, 84/32, Moscow, 117997, Russia\goodbreak
\and
Space Science Data Center - Agenzia Spaziale Italiana, Via del Politecnico snc, 00133, Roma, Italy\goodbreak
\and
Space Sciences Laboratory, University of California, Berkeley, California, U.S.A.\goodbreak
\and
The Oskar Klein Centre for Cosmoparticle Physics, Department of Physics, Stockholm University, AlbaNova, SE-106 91 Stockholm, Sweden\goodbreak
\and
UPMC Univ Paris 06, UMR7095, 98 bis Boulevard Arago, F-75014, Paris, France\goodbreak
\and
Universit\'{e} de Toulouse, UPS-OMP, IRAP, F-31028 Toulouse cedex 4, France\goodbreak
% ...
}

\authorrunning{Planck Collaboration}
\titlerunning{Isotropy and statistics of the CMB}

\begin{document}


\abstract{
  Analysis of the \Planck\ 2018 data set indicates that the
  statistical properties of the cosmic microwave background (CMB)
  temperature anisotropies are in excellent agreement with previous studies
  using the 2013 and 2015 data releases.  In particular, they are
  consistent with the Gaussian predictions of the $\Lambda$CDM
  cosmological model, yet also confirm the presence of several
  so-called ``anomalies'' on large angular scales. The novelty of the
  current study, however, lies in being a first attempt at a
  comprehensive analysis of the statistics of the polarization signal
  over all angular scales, using either maps of the Stokes parameters,
  $Q$ and $U$, or the $E$-mode signal derived from these using a new
  methodology (which we describe in an appendix).
  Although remarkable progress has been made in reducing the
  systematic effects that contaminated the 2015 polarization maps on
  large angular scales, it is still the case that residual systematics
  (and our ability to simulate them) can limit some tests of
  non-Gaussianity and isotropy.
  However, a detailed set of null tests applied to the maps
  indicates that these issues do not dominate the analysis on
  intermediate and large angular scales (i.e., $\ell \lesssim 400$).
  In this regime, no unambiguous detections of cosmological
  non-Gaussianity, or of anomalies corresponding to those seen in
  temperature, are claimed. Notably, the stacking of CMB polarization
  signals centred on the positions of temperature hot and cold spots
  exhibits excellent agreement with the $\Lambda$CDM cosmological
  model, and also gives a clear indication of how \Planck\ provides
  state-of-the-art measurements of CMB temperature and polarization on
  degree scales.  }

\maketitle



\section{Introduction}
\label{sec:introduction}

This paper, one of a set associated with the 2018 release of data from
the \Planck\footnote{\Planck\ (\url{http://www.esa.int/Planck}) is a
  project of the European Space Agency (ESA) with instruments provided
  by two scientific consortia funded by ESA member states and led by
  Principal Investigators from France and Italy, telescope reflectors
  provided through a collaboration between ESA and a scientific
  consortium led and funded by Denmark, and additional contributions
  from NASA (USA).}  mission \citep{planck2016-l01}, describes a compendium
of studies undertaken to determine the statistical properties of both
the temperature and polarization anisotropies of the cosmic microwave
background (CMB).

The \LCDM\ model explains the structure of the CMB in detail
\citep{planck2016-l06}, yet it
remains entirely appropriate to look for hints of departures from, or
tensions with, the standard cosmological model, by examining the
statistical properties of the observed radiation.
Indeed, in recent years, tantalizing evidence has emerged from the
WMAP and \Planck\ full-sky measurements of the CMB temperature
fluctuations of the presence of such ``anomalies,'' and 
indicating that a modest degree of deviation from global isotropy
exists. Such features appear to exert a statistically mild tension against the
mainstream cosmological models that themselves invoke the fundamental
assumptions of global statistical isotropy and Gaussianity.

A conservative explanation for the temperature anomalies is that they
are simply statistical flukes. This is particularly appealing given
the generally modest level of significance claimed, and the role of a
posteriori choices (also referred to as the ``look-elsewhere effect''),
i.e., whether interesting features in the data bias the choice
of statistical tests, or if arbitrary choices in the subsequent data
analysis enhance the significance of the features. However, determining
whether this is the case, or alternatively whether the anomalies are
due to real physical features of the cosmological model, cannot be
determined by further investigation of the temperature fluctuations on
the angular scales of interest, since those data are already
cosmic-variance limited.

Polarization fluctuations also have their origin in the primordial
gravitational potential, and have long been recognized as providing
the possibility to independently study the anomalies found in the
temperature data, given that they are largely sourced by different
modes.  The expectation, then, is that measurements of the full-sky
CMB polarization signal have the potential to provide an improvement
in significance of the detection of large-scale anomalies.
Specifically, it is important to determine in more detail whether any
anomalies are observed in the CMB polarization maps, and if so,
whether they are related to existing features in the CMB temperature
field.  Conversely, the absence of corresponding features in
polarization might imply that the temperature anomalies (if they are not
simply statistical excursions) could be due to a secondary effect such as the
integrated Sachs-Wolfe (ISW) effect
\citep{planck2013-p14,planck2014-a26}, or alternative scenarios in
which the anomalies arise from physical processes that do not
correlate with the temperature, e.g., texture or defect models.
Of course, there also remains the possibility that anomalies may be
found in the polarization data that are unrelated to existing features
in the temperature measurements.

In this paper, we present a first comprehensive attempt at assessing
the isotropy of the Universe via an analysis of the full-mission
\Planck\ full-sky polarization data. Analysis of the 2015 data set in
polarization \citep[hereafter
\citetalias{planck2014-a18}]{planck2014-a18} was limited on large
angular scales by the presence of significant residual systematic
artefacts in the High Frequency Instrument (HFI) data
\citep{planck2014-a08,planck2014-a09} that necessitated the high-pass
filtering of the component-separated maps.  This resulted in the
suppression of structure on angular scales larger than approximately
5\deg. However, the identification, modelling, and removal of
previously unexplained systematic effects in the polarization data, in
combination with new mapmaking and calibration procedures
\citep[][]{planck2014-a10,planck2016-l03}, means that such a procedure
is no longer necessary. Nevertheless, our studies remain limited both
by the relatively low signal-to-noise ratio of the polarization data,
and the presence of residual systematic artefacts that can be
significant with respect to detector sensitivity and comparable to the
cosmological signal. A detailed understanding of the latter, in
particular, have a significant impact on our ability to produce
simulations that are needed to allow a meaningful assessment of the
data.
These issues will be subsequently quantified and the impact on results
discussed.

The current work covers all relevant aspects related to the
phenomenological study of the statistical isotropy and Gaussian nature
of the CMB measured by the \Planck\ satellite.
Constraints on isotropy or non-Gaussianity, as might arise
from non-standard inflationary models, are provided in a companion
paper \citep{planck2016-l09}.  The current paper is organized as follows.
Section~\ref{sec:polarization_preamble} provides a brief introduction
to the study of polarized CMB data. 
Section~\ref{sec:data} summarizes the \Planck\ full-mission data used
for the analyses, and important limitations of the polarization maps
that are studied. Section~\ref{sec:simulations} describes the
characteristics of the simulations that constitute our reference set
of Gaussian sky maps representative of the null hypothesis. 
In Sect.~\ref{sec:non_gaussianity} the null hypothesis is tested with
a number of standard tests that probe different aspects of
non-Gaussianity. This includes tests of the statistical nature of the
polarization signal observed by \Planck\ using a local analysis of
stacked patches of the sky.  Several important anomalous features of
the CMB sky
are studied in Sect.~\ref{sec:anomalies}, using both temperature and
polarization data.  Aspects of the CMB fluctuations specifically
related to dipolar asymmetry are examined in Sect.~\ref{sec:dipmod}.
Section~\ref{sec:conclusions} provides the main conclusions of the
paper. Finally, in Appendix~\ref{sec:EB_maps} a detailed description
is provided of the novel method, called ``purified inpainting,'' used to generate
$E$- and $B$-mode maps from the Stokes $Q$ and $U$ data.

\section{Polarization analysis preamble}
\label{sec:polarization_preamble}

Traditionally, the Stokes parameters $Q$ and $U$ are used to describe
CMB polarization anisotropies \citep[e.g.,][]{Zaldarriaga1997}.  
However, unlike intensity, $Q$ and $U$ are not scalar quantities, but rather 
components of the rank-2 polarization tensor in a specific coordinate
basis associated with the map.
Such quantities are not rotationally invariant, thus in many analyses
it is convenient to consider alternate, but related,
polarization quantities. 

The polarization amplitude $P$ and polarization angle $\Psi$, defined
as follows,
\begin{linenomath*}
\begin{equation} \begin{array}{ccc}
       P & = & \sqrt{Q^2 + U^2}, \\[\smallskipamount]
  \Psi & = & \dfrac{1}{2}\arctan \dfrac{U}{Q}, \\
\end{array},
\label{eqn:Pol_amp_angle}
\end{equation}
\end{linenomath*}
are commonly used quantities in, for example, Galactic
astrophysics. However, completely unbiased estimators of these
quantities in the presence of anisotropic and/or correlated noise are
difficult to determine \citep{Plaszczynski2014}. Of course, it is
still possible to take the observed (noise-biased) quantity and directly
compare it to simulations analysed in the same manner.  As
an alternative, Sect.~\ref{sec:onepdf} works with the quantity $P^2$
and applies a correction for noise bias determined from simulations. A
cross-estimator based on polarization observations from two maps,
$P^2 = Q_1Q_2 + U_1U_2$ is also considered.

In addition, a local rotation of the Stokes parameters, resulting in
quantities denoted by $Q_{\mathrm{r}}$ and $U_{\mathrm{r}}$, is
employed in Sects.~\ref{sec:npoint_correlation} and \ref{sec:stacking}.
In this case, a local frame is defined with respect to a reference point
$\vec{\hat{n}}_\mathrm{ref}$ so that
\begin{linenomath*}
\begin{equation}
\begin{array}{ccc}
Q_\mathrm{r}\left(\vec{\hat{n}};\vec{\hat{n}}_\mathrm{ref}\right) & = & -Q\left(\vec{\hat{n}}\right)\cos{(2\phi)}-U\left(\vec{\hat{n}}\right)\sin{(2\phi)},  \\[\smallskipamount]
U_\mathrm{r}\left(\vec{\hat{n}};\vec{\hat{n}}_\mathrm{ref}\right) & = & \phantom{-}Q\left(\vec{\hat{n}}\right)\sin{(2\phi)}-U\left(\vec{\hat{n}}\right)\cos{(2\phi)},  \\
\end{array}
\label{eqn:local_Stokes}
\end{equation}
\end{linenomath*}
where $\phi$ denotes the angle between the axis aligned along a
meridian in the local coordinate system centred on the reference point 
and the great circle connecting this point to a position
$\vec{\hat{n}}$. 

Finally, the rotationally invariant quantities referred to as $E$ and
$B$ modes are commonly used for the global analysis of CMB
data. Since the quantities $Q\pm iU$, defined relative
to the direction vectors $\vec{\hat{n}}$, transform as spin-2 variables
under rotations around the $\vec{\hat{n}}$ axis, they can be expanded as
\begin{linenomath*}
\begin{equation}
(Q \pm iU) (\vec{\hat{n}}) = \sum_{\ell = 2}^{\infty} \sum_{m=-\ell}^{\ell}a_{\ell m}^{(\pm 2)} \,_{\pm 2}Y_{\ell m}(\vec{\hat{n}}),
\end{equation}
\end{linenomath*}
where $\,_{\pm 2}Y_{\ell m}(\vec{\hat{n}})$ denotes the spin-weighted
spherical harmonics and $a_{\ell m}^{(\pm 2)}$ are the corresponding
harmonic coefficients. If we define
\begin{linenomath*}
\begin{equation}
\begin{array}{ccc}
a^{E}_{\ell m} & = & -\dfrac{1}{2}\left(a_{\ell m}^{(2)} + a_{\ell m}^{(-2)}\right), \\[\medskipamount]
a^{B}_{\ell m} & = & \phantom{-}\dfrac{i}{2}\left(a_{\ell m}^{(2)} - a_{\ell m}^{(-2)}\right).
\end{array}
\end{equation}
\end{linenomath*}
then the invariant quantities are given by
\begin{linenomath*}
\begin{equation}\begin{array}{ccc}
E(\vec{\hat{n}})& = & \displaystyle \sum_{\ell = 2}^{\infty} \sum_{m=-\ell}^{\ell} a^{E}_{\ell m} Y_{\ell m}(\vec{\hat{n}}), \\
B(\vec{\hat{n}}) & = & \displaystyle \sum_{\ell = 2}^{\infty} \sum_{m=-\ell}^{\ell} a^{B}_{\ell m} Y_{\ell m}(\vec{\hat{n}}).
\end{array}
\end{equation}
\end{linenomath*}

In practice, the $Q$ and $U$ data sets that are analysed are the end
products of sophisticated component-separation
approaches. Nevertheless, the presence of residual foregrounds
mandates the use of a mask, the application of which during the generation
of $E$- and $B$-mode maps results in $E/B$ mixing
\citep{Lewis2002,Bunn2003}. 
In Appendix~\ref{sec:EB_maps}, we describe the method adopted 
in this paper to reduce such mixing.

\section{Data description}
\label{sec:data}

\begin{figure*}[hbtp!]
  \begin{center}
    \begin{tabular}{ccc}
      \includegraphics[width=0.29\linewidth]{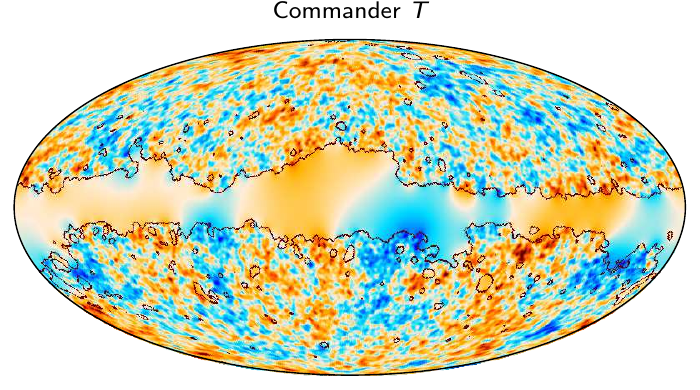}&
      \includegraphics[width=0.29\linewidth]{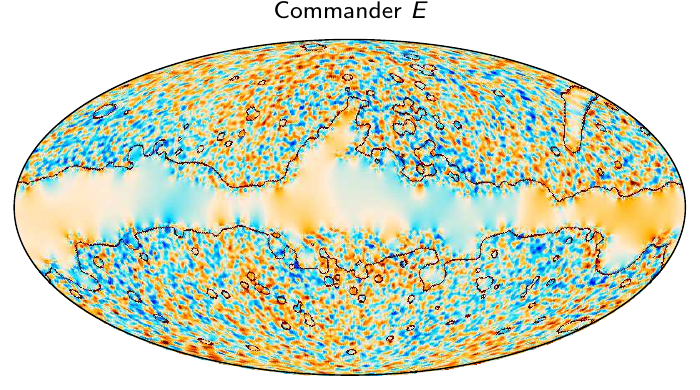}&
      \includegraphics[width=0.29\linewidth]{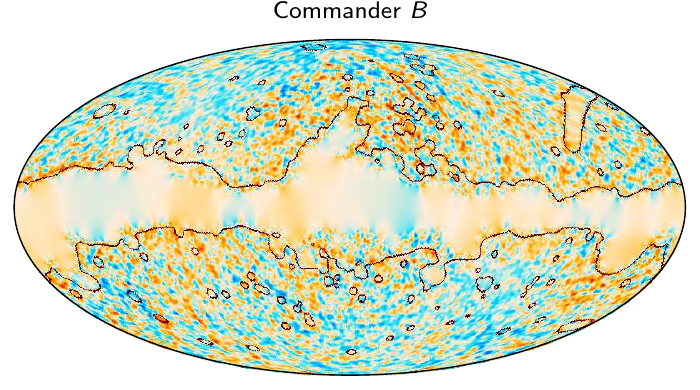}\\
      \includegraphics[width=0.29\linewidth]{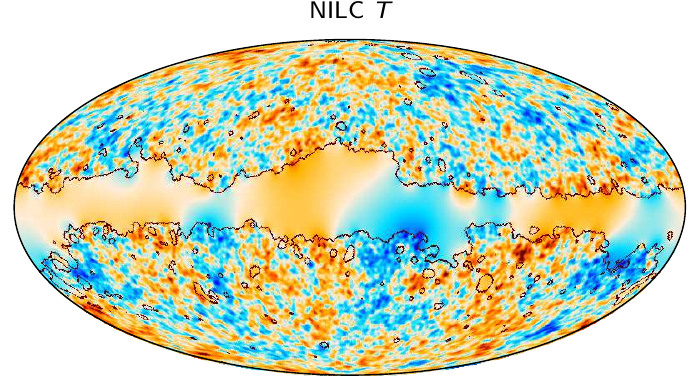}&
      \includegraphics[width=0.29\linewidth]{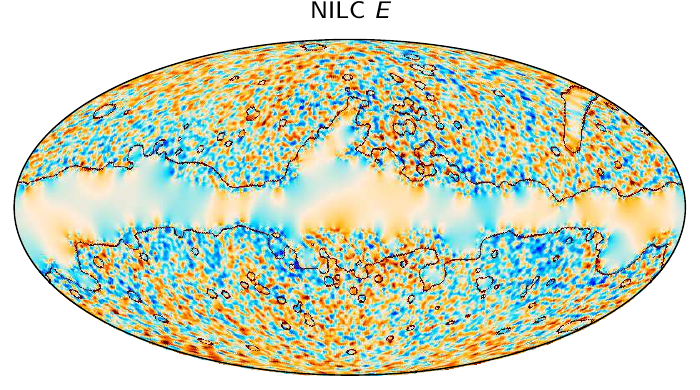}&
      \includegraphics[width=0.29\linewidth]{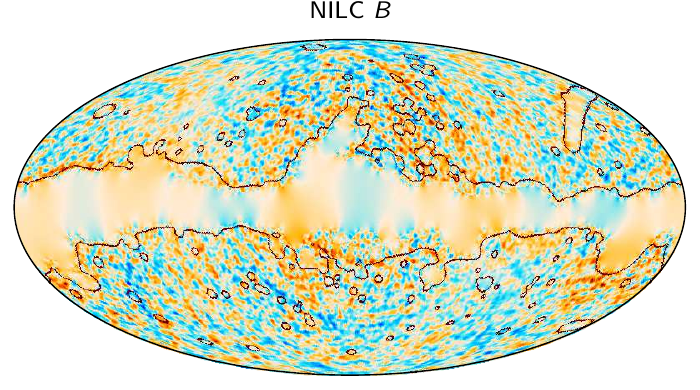}\\
      \includegraphics[width=0.29\linewidth]{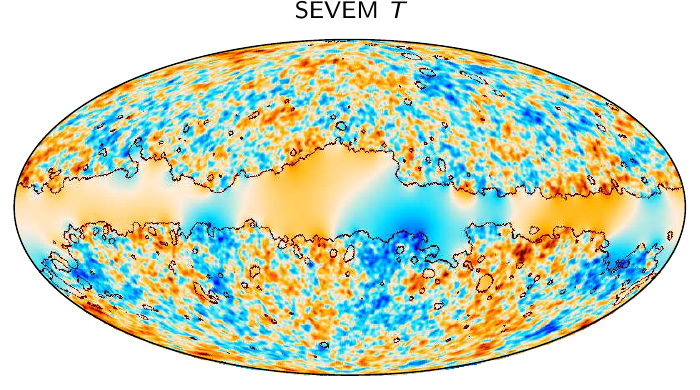}&
      \includegraphics[width=0.29\linewidth]{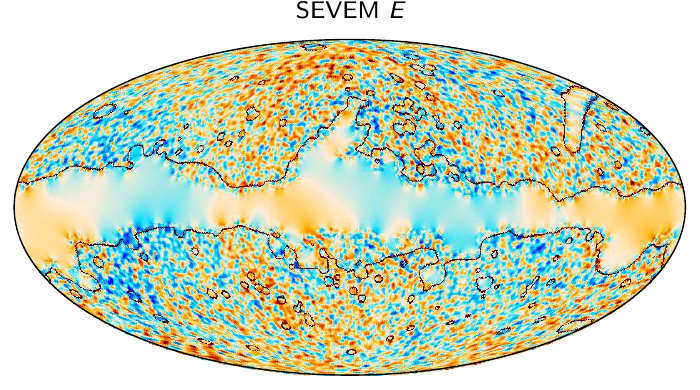}&
      \includegraphics[width=0.29\linewidth]{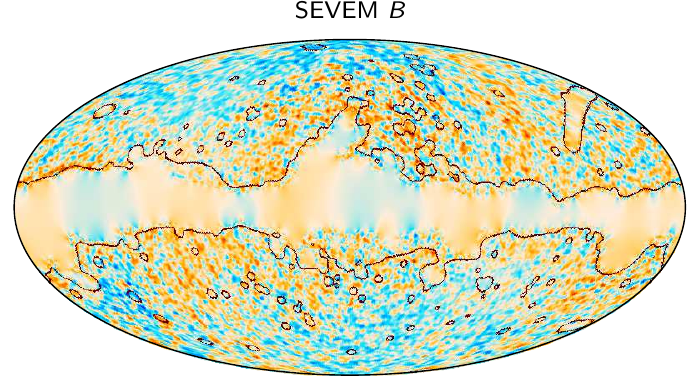}\\
      \includegraphics[width=0.29\linewidth]{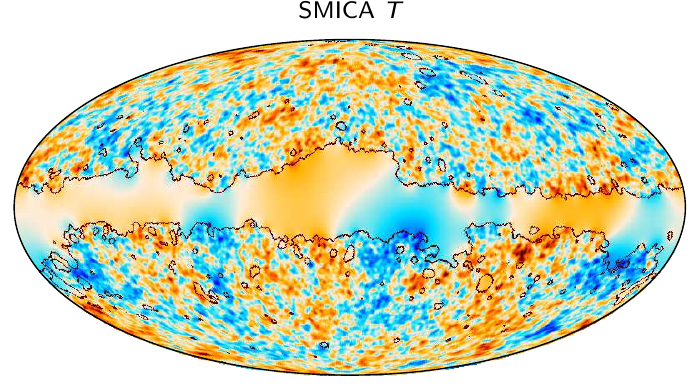}&
      \includegraphics[width=0.29\linewidth]{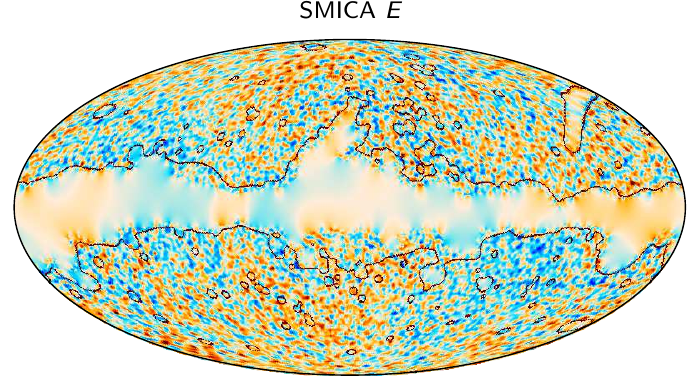}&
      \includegraphics[width=0.29\linewidth]{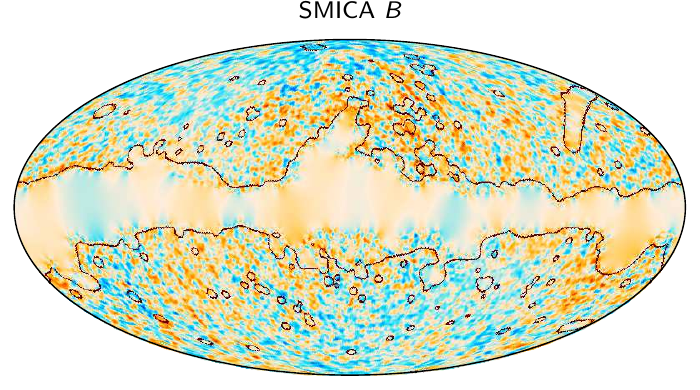}\\
      \includegraphics[width=0.25\linewidth]{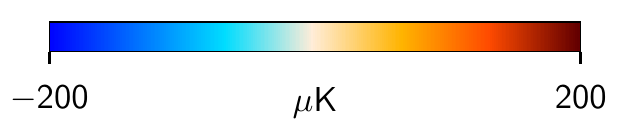}&
      \multicolumn{2}{c}{
        \includegraphics[width=0.25\linewidth]{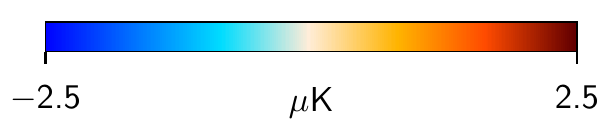}
      }
    \end{tabular}
  \end{center}
  \caption{Component-separated CMB maps at 80\arcmin\
    resolution. Columns show temperature $T$, and $E$- and $B$-mode
    maps, respectively, while rows show results derived with different
    component-separation methods. The temperature maps are inpainted
    within the common mask, but are otherwise identical to those
    described in \citet{planck2016-l04}. The $E$- and $B$-mode maps
    are derived from the Stokes $Q$ and $U$ maps following the method
    described in Appendix~\ref{sec:EB_maps}. The dark lines indicate the
    corresponding common masks used for analysis of the maps at this
    resolution. Monopoles and dipoles have been subtracted from the
    temperature maps, with parameters fitted to the data
    after applying the common  mask.}
\label{fig:cmb_maps}
\end{figure*}

\begin{figure*}[hbtp!]
  \begin{center}
    \begin{tabular}{ccc}
      \includegraphics[width=0.29\linewidth]{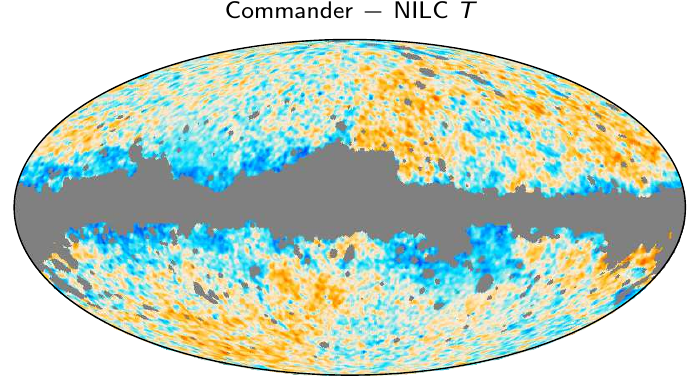}&
      \includegraphics[width=0.29\linewidth]{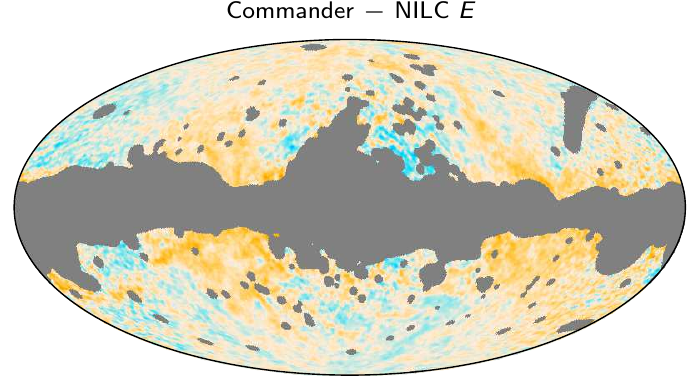}&
      \includegraphics[width=0.29\linewidth]{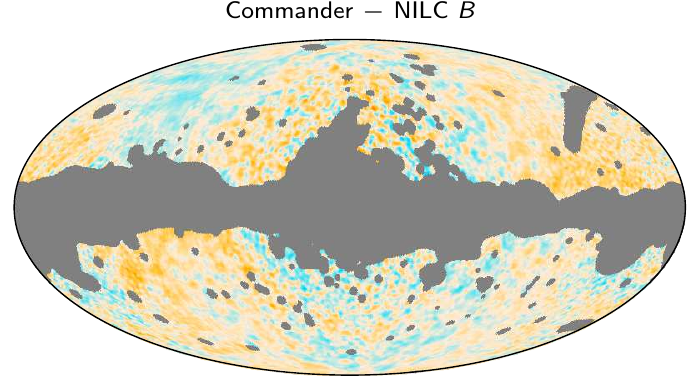}\\
      \includegraphics[width=0.29\linewidth]{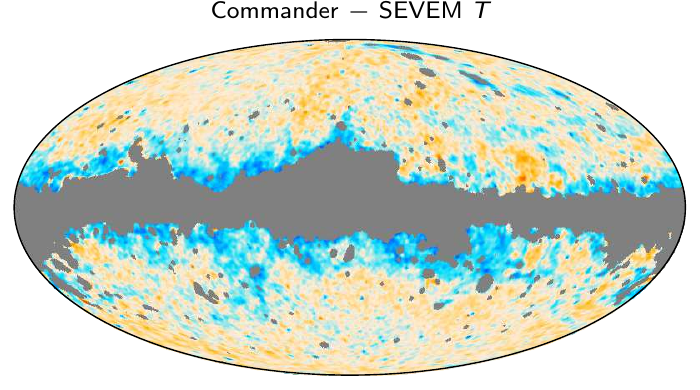}&
      \includegraphics[width=0.29\linewidth]{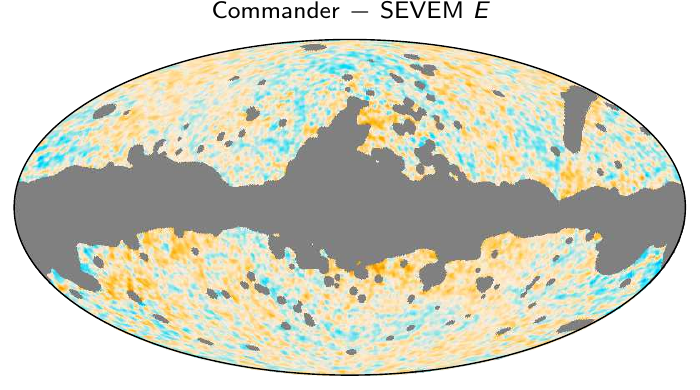}&
      \includegraphics[width=0.29\linewidth]{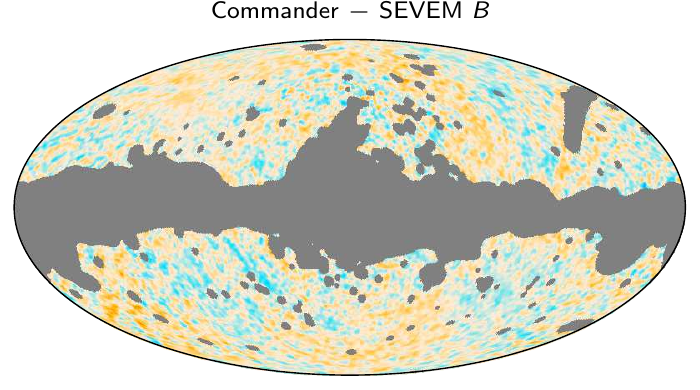}\\
      \includegraphics[width=0.29\linewidth]{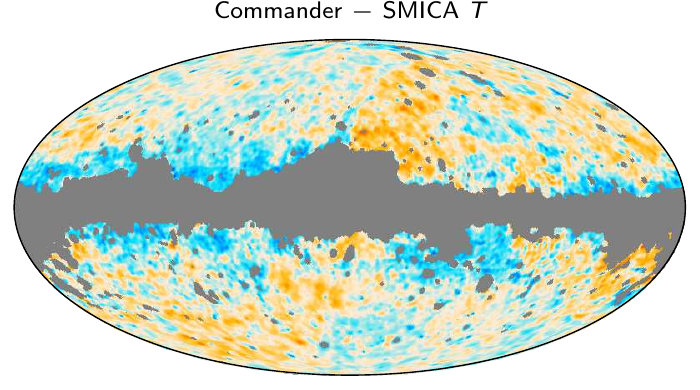}&
      \includegraphics[width=0.29\linewidth]{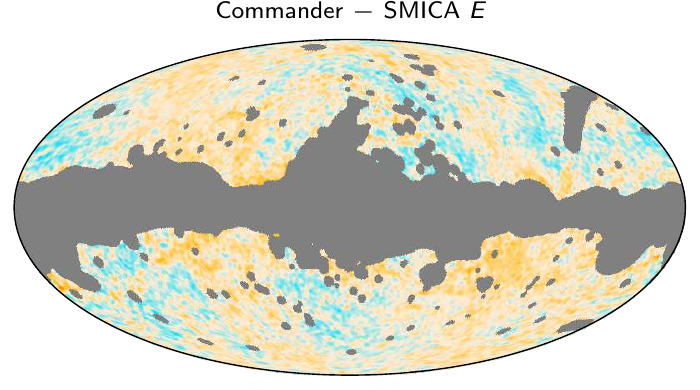}&
      \includegraphics[width=0.29\linewidth]{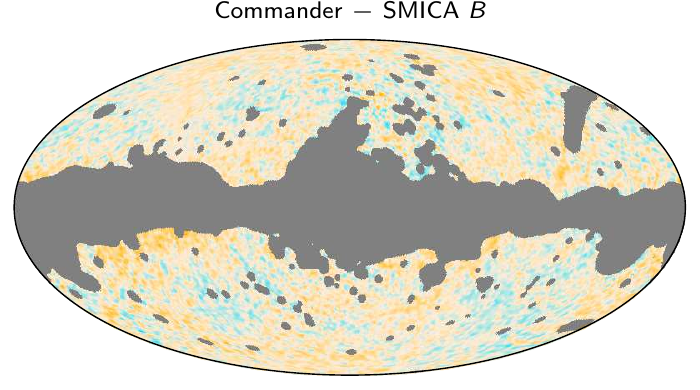}\\
      \includegraphics[width=0.29\linewidth]{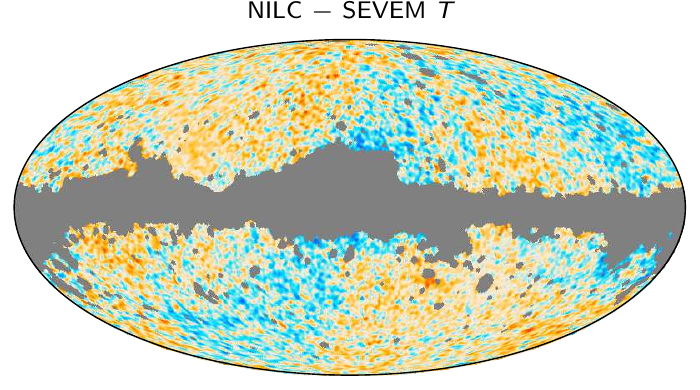}&
      \includegraphics[width=0.29\linewidth]{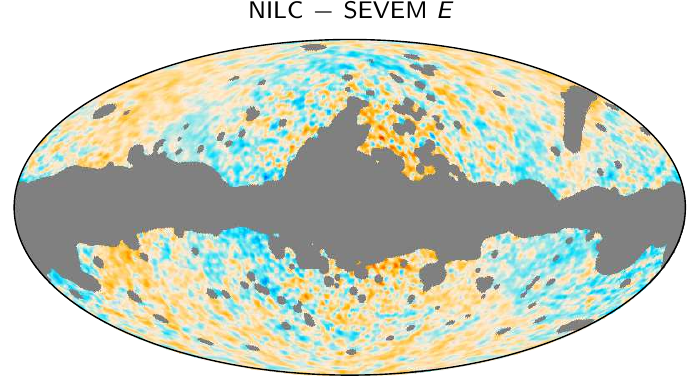}&
      \includegraphics[width=0.29\linewidth]{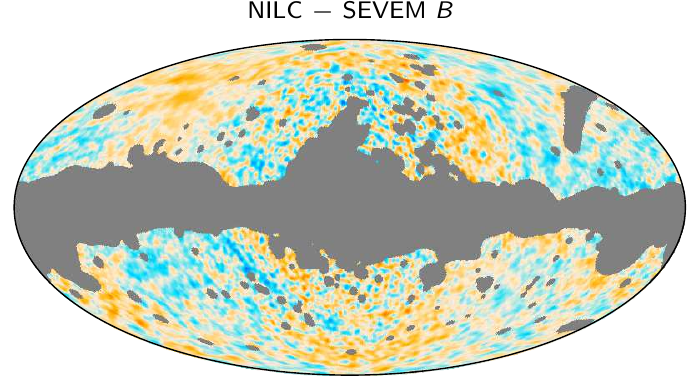}\\
      \includegraphics[width=0.29\linewidth]{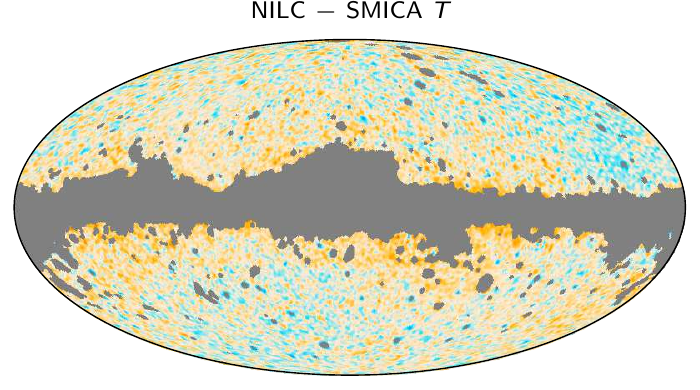}&
      \includegraphics[width=0.29\linewidth]{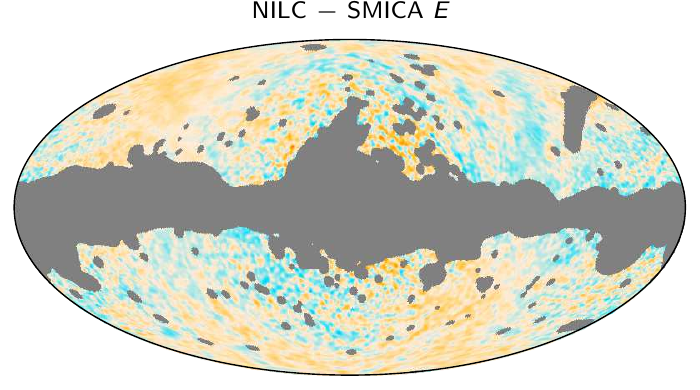}&
      \includegraphics[width=0.29\linewidth]{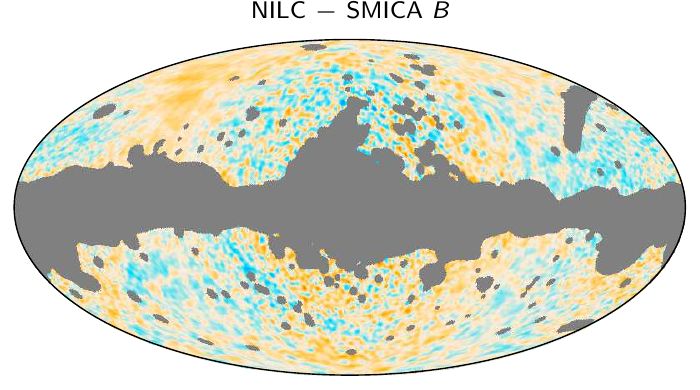}\\
      \includegraphics[width=0.29\linewidth]{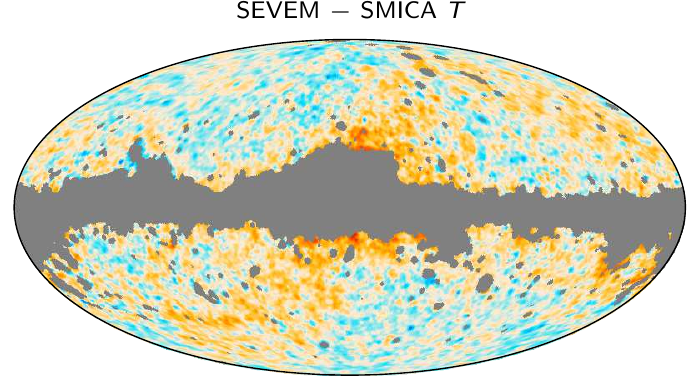}&
      \includegraphics[width=0.29\linewidth]{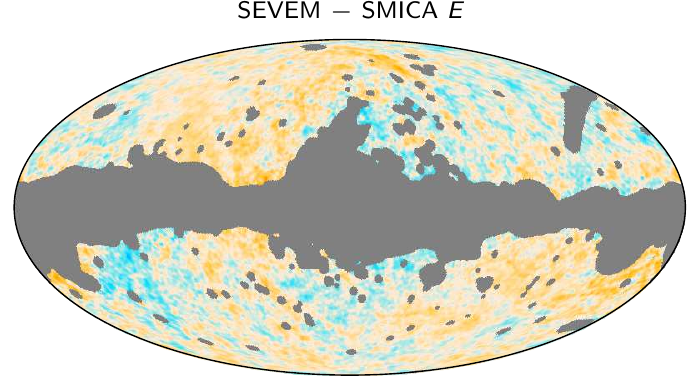}&
      \includegraphics[width=0.29\linewidth]{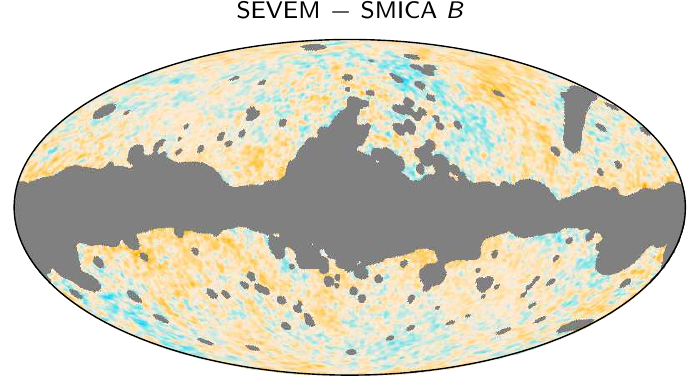}\\
      \includegraphics[width=0.25\linewidth]{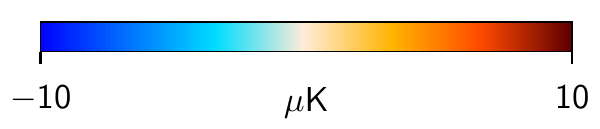}&
      \multicolumn{2}{c}{
        \includegraphics[width=0.25\linewidth]{colourbar_2p5uK}
      }
    \end{tabular}
  \end{center}
  \caption{ Pairwise differences between maps from the four CMB
    component-separation pipelines, smoothed to 80\arcmin\
    resolution. Columns show temperature, $T$, and $E$- and $B$-mode
    maps, respectively, while rows show results for different pipeline
    combinations.  The grey regions correspond to the appropriate
    common masks.  Monopoles and dipoles have been subtracted from the
    temperature difference maps, with parameters fitted to the data
    after applying the common  mask.}
\label{fig:cmb_diff_pipe_maps}
\end{figure*}

\begin{figure*}[hbtp!]
  \begin{center}
    \begin{tabular}{ccc}
      \includegraphics[width=0.29\linewidth]{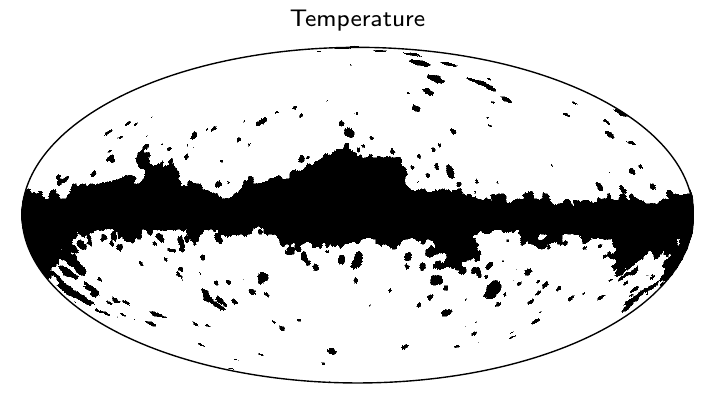}&
      \includegraphics[width=0.29\linewidth]{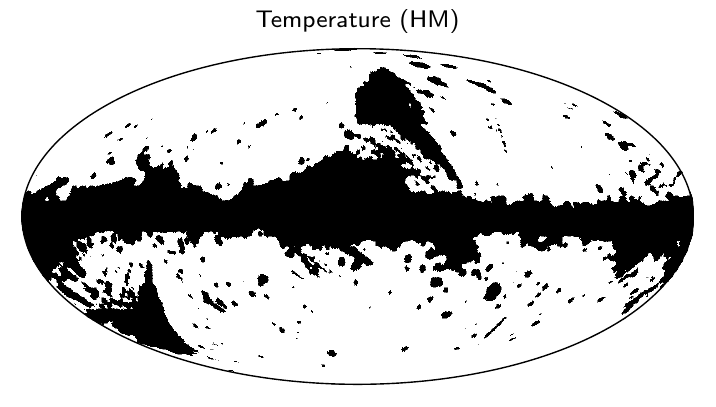}&
      \includegraphics[width=0.29\linewidth]{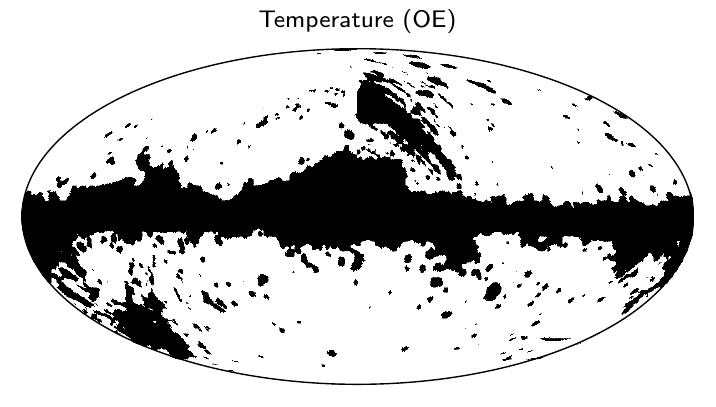}\\
      \includegraphics[width=0.29\linewidth]{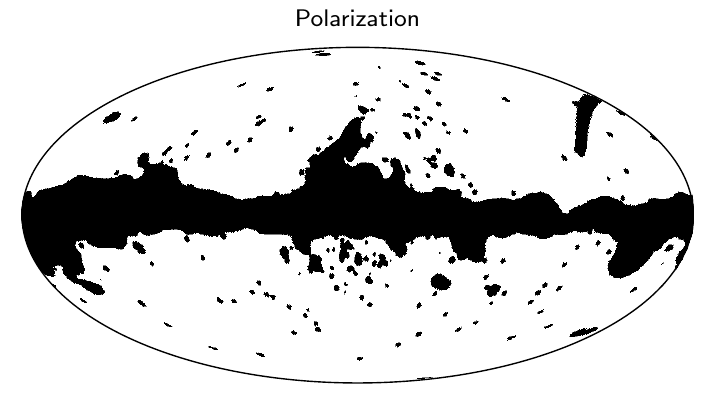}&
      \includegraphics[width=0.29\linewidth]{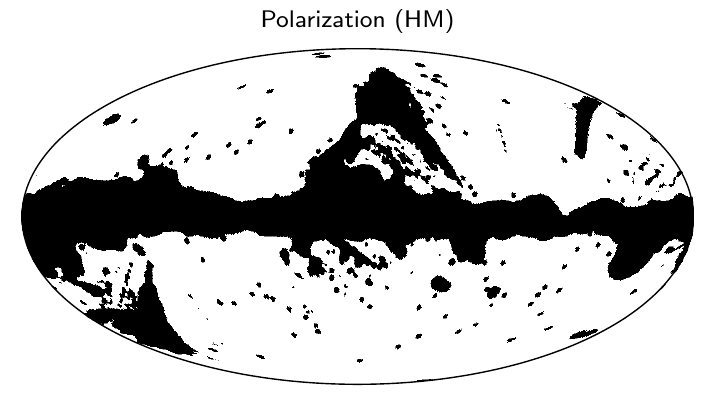}&
      \includegraphics[width=0.29\linewidth]{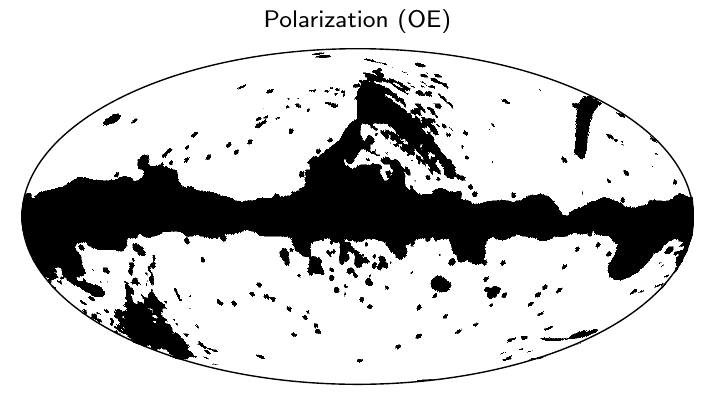}\\
      \includegraphics[width=0.29\linewidth]{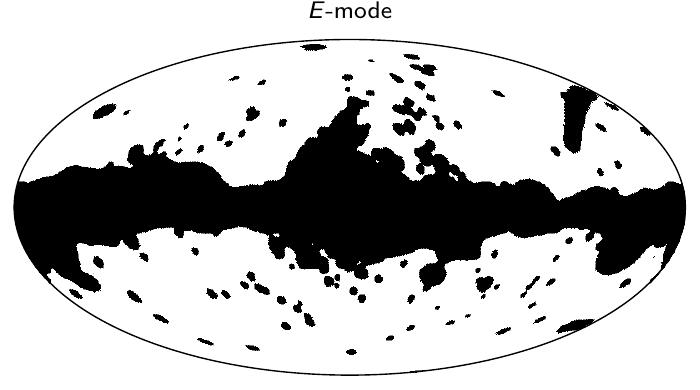}&
      \includegraphics[width=0.29\linewidth]{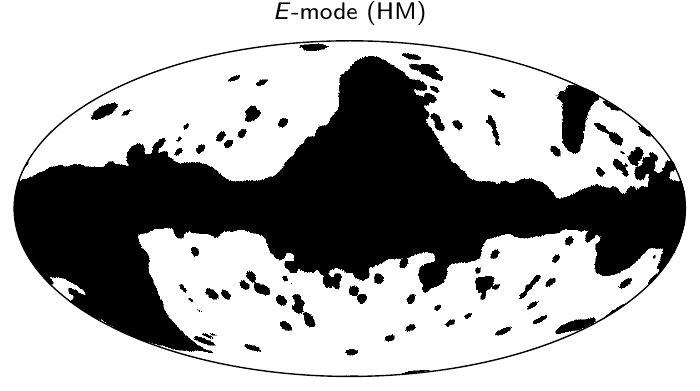}&
      \includegraphics[width=0.29\linewidth]{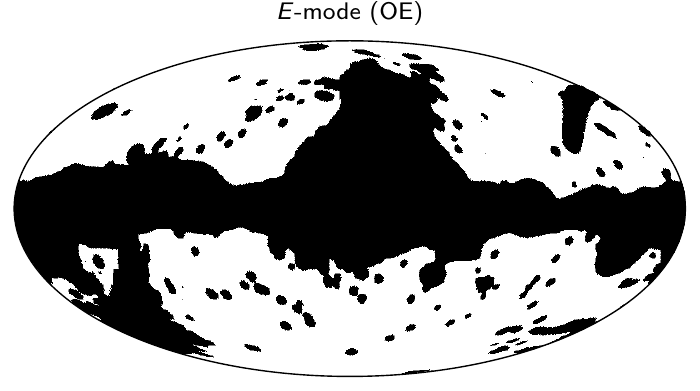}\\
    \end{tabular}
  \end{center}
  \caption{Examples of common masks. From top to bottom, the masks
    correspond to those used for analysing temperature maps,
    polarization represented by the Stokes $Q$ and $U$ parameters, and
    $E$-mode polarization data, at a resolution \nside\ = 128. From
    left to right, full-mission, HM, and OE masks are shown. Note that
    the masks for $E$- and $B$-mode analysis are extended relative to
    those derived for $Q$ and $U$ studies, in order to reduce the
    reconstruction residuals.  }
\label{fig:cmb_masks}
\end{figure*}

In this paper, we use data from the \Planck\ 2018 full-mission data
release (``PR3'') that are made available on the Planck Legacy Archive
(PLA\footnote{\url{http://pla.esac.esa.int}}).  The raw data are
identical to those used in 2015, except that the HFI omits 22 days of
observations from the final, thermally-unstable phase of the mission.
The release includes sky maps at nine frequencies in temperature, and
seven in polarization, provided in {\tt HEALPix} format
\citep{gorski2005},\footnote{\url{http://healpix.sourceforge.net}}
with a pixel size defined by the \nside\ parameter.\footnote{In {\tt
    HEALPix} the sphere is divided into $12\,N_{\rm side}^2$ pixels.
  At $N_{\rm side}=2048$, typical of \Planck\ maps, the mean pixel
  size is 1\farcm7.} For polarization studies, the 353-GHz maps are
based on polarization-sensitive bolometer (PSB) observations only
\citep[see][for details]{planck2016-l03}.

Estimates of the instrumental noise contribution and limits on
time-varying systematic artefacts can be inferred from maps that are
generated by splitting the full-mission data sets in various ways.
For LFI, half-ring maps are generated from the first and second half
of each stable pointing period, consistent with the approach in the
2013 and 2015 \Planck\ papers.  For HFI, odd-ring (O) and even-ring
(E) maps are constructed using alternate pointing periods, i.e.,
either odd or even numbered rings, to avoid the correlations observed
previously in the half-ring data sets.  However, for convenience and
consistency, we will refer to both of these ring-based splits as
``odd-even'' (OE), in part as recognition of the signal-to-noise
ratios of the LFI and HFI maps and their relative contributions to the
component-separated maps described below.  Half-mission (HM) maps are
generated from a combination of Years~1 and 3, and Years~2 and 4 for LFI, or
the first and second half of the full-mission data set in the case of
HFI. Note that important information on the level of noise and
systematic-effect residuals can be inferred from maps constructed from
half-differences of the half-mission (HMHD) and odd-even (OEHD)
combinations. In particular, the OE differences trace the instrumental
noise, but filter away any component fluctuating on timescales longer
than the pointing period, whereas the HM differences are sensitive to
the time evolution of instrumental effects.
A significant number of consistency checks are applied to this set of
maps.  Full details are provided in two companion papers
\citep{planck2016-l02,planck2016-l03}.

As in previous studies, we base our main results on estimates of the
CMB from four component-separation algorithms, --- \commander, \nilc,
\sevem, and \smica\ --- as described in \citet{planck2016-l04}.  These
provide data sets determined from combinations of the \Planck\ raw
frequency maps with minimal Galactic foreground residuals, although some
contributions from unresolved extragalactic sources are present in the
temperature solutions.  Foreground-cleaned versions of the 70-, 100-,
143-, and 217-GHz sky maps generated by the \sevem\ algorithm,
hereafter referred to as {\tt SEVEM-070}, {\tt SEVEM-100}, {\tt
  SEVEM-143}, and {\tt SEVEM-217}, respectively, allow us to test the
frequency dependence of the cosmological signal, either to verify its
cosmological origin, or to search for specific frequency-dependent
effects.  In all cases, possible residual emission is then mitigated
in the analyses by the use of sky-coverage masks.

The CMB temperature maps are derived using all channels, from 30 to
857\,GHz, and provided at a common angular resolution of 5\arcm\
full width at half maximum (FWHM)
and \nside\ = 2048. In contrast to the 2013 and 2015 releases, these do
not contain a contribution from the second order temperature
quadrupole \citep{kamionkowski2003}.  An additional window function,
applied in the harmonic domain, smoothly truncates power in the maps
over the range
$\ell_{\mathrm{min}}\ \leq \ell\ \leq \ell_{\mathrm{max}}$, such that
the window function is unity at $\ell_{\mathrm{max}}$ = 3400 and
zero at $\ell_{\mathrm{max}}$ = 4000.  The polarization solutions include
information from all channels sensitive to polarization, from 30 to
353\,GHz, at the same resolution as the temperature results, but only
including contributions from harmonic scales up to
$\ell_{\mathrm{max}}$ = 3000.  In the context of these CMB maps, we
refer to an ``odd-even'' data split that combines the LFI half-ring~1 with
the HFI odd-ring data, and the LFI half-ring 2 with the HFI even-ring
data.

Lower-resolution versions of these data sets are also used in the
analyses presented in this paper. The downgrading procedure is as
follows. The full-sky maps are decomposed into spherical harmonics at
the input \healpix\ resolution, these coefficients are then convolved
to the new resolution using the appropriate beam and pixel window
functions, then the modified coefficients are used to synthesize a map
directly at the output \healpix\ resolution.

Specific to this paper, we consider polarization maps determined via
the method of ``purified inpainting'' (described in
Appendix~\ref{sec:EB_maps}) from the component-separated $Q$ and $U$
data. Figure~\ref{fig:cmb_maps} presents the $E$- and $B$-mode maps
for the four component-separation methods at a resolution of \nside\ =
128, with the corresponding common masks overplotted.
\citet{planck2016-l04} notes that some broad large-scale features
aligned with the \Planck\ scanning strategy are observed in the
$Q$ and $U$ data. The detailed impact on $E$- and $B$-mode map generation is
unclear, thus some caution should be exercised in the interpretation
of the largest angular scales in the
data. Figure~\ref{fig:cmb_diff_pipe_maps} does indicate the presence of
large-scale residuals in the pairwise differences of the
component-separated maps.  Finally, we note that the $B$-mode
polarization is strongly noise dominated on all scales, therefore,
although shown here
for completeness, we do not present a comprehensive statistical
analysis of these maps.

\begin{table}[hbtp!]
\begingroup
\newdimen\tblskip \tblskip=5pt
\caption{ Standardized data sets used in this paper. The resolutions of
  the sky maps used are defined in terms of the \nside\ parameter and
  corresponding FWHM of the Gaussian beam with which they are
  convolved. The fraction of unmasked
  pixels in the corresponding common masks for the full-mission
  (Full), as well as the HM and OE data splits, are also specified. 
}
\label{tab:masks}
\nointerlineskip
\vskip -3mm
\footnotesize
\setbox\tablebox=\vbox{
   \newdimen\digitwidth
   \setbox0=\hbox{\rm 0}
   \digitwidth=\wd0
   \catcode`*=\active
   \def*{\kern\digitwidth}
   \newdimen\signwidth
   \setbox0=\hbox{+}
   \signwidth=\wd0
   \catcode`!=\active
   \def!{\kern\signwidth}
\halign{\hbox to 0.8in{#\leaderfil}\tabskip 5pt&
\hfil#\hfil\tabskip 5pt&
#\hfil\tabskip 5pt&
#\hfil\tabskip 5pt&
\hfil#\hfil\/\tabskip 0pt\cr                          
\noalign{\doubleline}
\omit& FWHM\hfil&\omit\hfil&  Fraction [\%]\hfil&\omit\hfil\cr
\omit\nside\ \hfil& [arcmin]\hfil& \hfil Full\hfil&\hfil HM&\hfil OE\hfil\cr
\noalign{\vskip 3pt\hrule\vskip 3pt}
\noalign{\vskip 3pt}
\multispan5\hfil Temperature\hfil\cr
\noalign{\vskip 3pt}
2048& **5& 77.9& \hfil 74.7& 76.3\cr
1024& *10& 76.9& \hfil 72.3& 74.0\cr
*512& *20& 75.6& \hfil 70.1& 71.6\cr
*256& *40& 74.7& \hfil 69.0& 70.2\cr
*128& *80& 73.6& \hfil 68.0& 69.0\cr
**64& 160& 71.3& \hfil 65.4& 66.6\cr
**32& 320& 68.8& \hfil 62.0& 63.6\cr
**16& 640& 64.5& \hfil 56.2& 58.2\cr
\noalign{\vskip 3pt\hrule\vskip 3pt}
\noalign{\vskip 3pt}
\multispan5\hfil $Q$ $U$ polarization\hfil\cr
\noalign{\vskip 3pt}
2048& **5& 78.1& \hfil 75.0& 76.5\cr
1024& *10& 77.7& \hfil 73.2& 74.8\cr
*512& *20& 77.0& \hfil 71.6& 73.0\cr
*256& *40& 76.1& \hfil 70.6& 71.7\cr
*128& *80& 74.5& \hfil 69.2& 70.0\cr
**64& 160& 72.4& \hfil 66.9& 67.9\cr
**32& 320& 69.5& \hfil 63.2& 64.7\cr
**16& 640& 63.6& \hfil 55.9& 57.9\cr
\noalign{\vskip 3pt\hrule\vskip 3pt}
\noalign{\vskip 3pt}
\multispan5\hfil $E$ $B$ polarization\hfil\cr
\noalign{\vskip 3pt}
2048& **5& \,\,\dots& \hfil \dots& \dots\cr
1024& *10& 64.8& \hfil \dots& \dots\cr
*512& *20& 64.9& \hfil \dots& \dots\cr
*256& *40& 64.9& \hfil 54.5& 54.7\cr
*128& *80& 64.9& \hfil 54.5& 54.8\cr
**64& 160& 64.0& \hfil 54.2& 54.2\cr
**32& 320& 62.6& \hfil 53.7& 54.4\cr
**16& 640& 55.4& \hfil 46.5& 48.6\cr
\noalign{\vskip 3pt\hrule\vskip 3pt}
\noalign{\vskip 3pt}}}
\endPlancktable 
\endgroup
\end{table}

In general, we make use of standardized masks made available for
temperature and polarization analysis, as described in detail in
\citet{planck2016-l04}.  These masks are then downgraded for
lower-resolution studies as follows.  The binary mask at the starting
resolution is first downgraded in the same manner as a temperature
map.  The resulting smooth downgraded mask is then thresholded by
setting pixels where the value is less than $0.9$ to zero and all
others to unity, in order to again generate a binary mask.

In the case of the data cuts, some additional care must be taken with
masking. Since the HFI HM and OE maps contain many unobserved
pixels\footnote{These are pixels that were either never seen by any of
  the bolometers present at a given frequency, or for which the
  polarization angle coverage is too poor to support a reliable
  decomposition into the three Stokes parameters. Note that the number
  of unobserved pixels has increased significantly between the 2015
  and 2018 data sets, due to a change in the condition number
  threshold at the map-making stage.}  at a given frequency, some
pre-processing is applied to them before the application of the
component-separation algorithms. Specifically, the value of any
unobserved pixel is replaced by the value of the corresponding
\nside\ = 64 parent pixel. Analysis of the component-separated maps
derived from the data cuts then requires masking of these
pixels. However, a simple merge of the unobserved pixel masks at each
frequency for a given data cut is likely to be insufficient, since the
various convolution and deconvolution processes applied by the
component-separation algorithms will cause leakage of the inpainted values
into neighbouring pixels. The masks are therefore extended as follows.
Starting with the initial merge of the unobserved pixels over all
frequencies, the unobserved pixels are selected and their neighbouring
pixels are also masked. This is repeated three times. Lower resolution
versions are generated by degrading the binary mask to the target
resolution, then setting all pixels with values less than a threshold
of 0.95 to zero, while all other pixels have their values set to unity.
Masks appropriate for the analysis of the HM and OE maps are generated
by combining the unobserved pixel masks with the full-mission
standardized masks. 

The masks for $E$- and $B$-mode analysis are extensions of the those
applied to the $Q$ and $U$ maps before executing the purified
inpainting technique. Specifically, an optimal confidence mask is
defined by performing reconstructions on simulated CMB-plus-noise
realizations, as propagated through all four \Planck\
component-separation pipelines, then evaluating the residuals with
respect to the input full-sky maps. The final mask is specified by requiring
that the maximal rms level of the residuals observed in the simulations is
less than $0.5\,\mu$K, significantly below the cosmological $E$-mode
signal. Examples of common masks are shown in Fig.~\ref{fig:cmb_masks}.

In what follows, we will undertake analyses of the data at a given
resolution denoted by a specific \nside\ value. Unless otherwise
stated, this implies that the data have been smoothed to a
corresponding FWHM as described above, and a standardized mask
employed. Often, we will simply refer to such a mask as the ``common
mask,'' irrespective of the resolution or data split in
question. However, in the latter case, we will refer to full-mission,
HM or OE common masks, where appropriate, to avoid
confusion. Table~\ref{tab:masks} lists the \nside\ and FWHM values
defining the resolution of these maps, together with the different
masks and their sky coverage fractions that accompany the signal maps.

\section{Simulations}
\label{sec:simulations}

The results presented in this paper are derived using Monte Carlo (MC)
simulations. These provide both the reference set of sky maps used for
the null tests employed here, and form the basis of any debiasing in
the analysis of the real data, as required by certain statistical
methods.  The simulations include Gaussian CMB signals and
instrumental noise realizations that capture important characteristics
of the \Planck\ scanning strategy, telescope, detector responses, and
data-reduction pipeline over the full-mission period.  These are
extensions of the ``full focal-plane'' simulations described in
\citet{planck2014-a14}, with the latest set being known as ``FFP10.''

The fiducial CMB power spectrum 
corresponds the cosmology described by the parameters in
Table~\ref{tab:params}.
Note that the preferred value of $\tau$ in \citet{planck2016-l06} is
slightly lower, at 
$\tau = 0.054\pm 0.007$.
1000 realizations of the CMB sky are generated including lensing,
Rayleigh scattering, and Doppler boosting\footnote{Doppler boosting, due
to our motion with respect to the CMB rest frame, induces both a
dipolar modulation of the temperature anisotropies and an aberration
that corresponds to a change in the apparent arrival directions of the
CMB photons, where both effects are aligned with the CMB dipole
\citep{challinor2002,planck2013-pipaberration}.
Both contributions are present in the FFP10 simulations. 
} effects, the latter two of
which are frequency-dependent.  The signal realizations include the
frequency-specific beam properties of the LFI and HFI data sets
implemented by the \texttt{FEBeCoP} \citep{mitra2010} beam-convolution
approach.

\begin{table*}
\begingroup
\newdimen\tblskip \tblskip=5pt
\caption{Cosmological parameters for the FFP10 simulations, used to make
the simulated maps in this paper, and throughout the
\Planck\ 2018 papers.}
\label{tab:params}
\nointerlineskip
\vskip -3mm
\footnotesize
\setbox\tablebox=\vbox{
   \newdimen\digitwidth
   \setbox0=\hbox{\rm 0}
   \digitwidth=\wd0
   \catcode`*=\active
   \def*{\kern\digitwidth}
   \newdimen\signwidth
   \setbox0=\hbox{+}
   \signwidth=\wd0
   \catcode`!=\active
   \def!{\kern\signwidth}
\halign{ \hbox to 2.75in{#\leaderfil}\tabskip 0.5em&
         #\hfil\tabskip 1.5em&
         \hfil#\hfil\tabskip 0pt\cr                           
\noalign{\doubleline}
\multispan2\hfil Parameter\hfil& Value\cr   
\noalign{\vskip 4pt\hrule\vskip 6pt}
Baryon density& $\omega_{\rm b} = \Omega_{\rm b}h^2$& 0.022166\cr
Cold dark matter density& $\omega_{\rm c} =\Omega_{\rm c}h^2$& 0.12029\cr
Neutrino energy density& $\omega_{\nu} = \Omega_{\nu}h^2$& 0.000645\cr
Density parameter for cosmological constant&
$\Omega_{\Lambda}$& 0.68139\cr
Hubble parameter& $h=H_0/100\,\mathrm{km}\,\mathrm{s}^{-1}\,\mathrm{Mpc}^{-1}$&
 0.67019\cr
Spectral index of power spectrum&
$n_{\rm s}$& 0.96369\cr
Amplitude of power
(at $k=0.05\,\mathrm{Mpc}^{-1}$)& $A_{\rm s}$& $2.1196\times 10^{-9}$\cr
Thomson optical depth&
$\tau$& 0.06018\cr
\noalign{\vskip 3pt\hrule\vskip 4pt}}}
\endPlancktable                    
\endgroup
\end{table*}

Given that the instrumental noise properties of the \Planck\ data are
complex, we make use of a set of so-called ``end-to-end'' simulations.
For HFI, residual systematics must be accounted for in the scientific
analysis of the polarized sky signal, thus the simulations include
models of all systematic effects, together with noise and sky signal (a fixed CMB
plus foregrounds fiducial sky). Realistic time-ordered information for
all HFI frequencies are then generated and subsequently propagated
through the map-making algorithm to produce frequency maps. Finally,
the sky signal is removed and the resulting maps of noise and residual
systematics can be added to the set of CMB realizations. More details
can be found in \citet{planck2014-a10} and \citet{planck2016-l03}.
A similar approach is followed by LFI to generate noise MCs that
capture important characteristics of the scanning strategy, detector
response, and data-reduction pipeline over the full-mission period
\citep{planck2016-l02}.  
A total of 300 realizations are generated at each \Planck\ frequency, for the
full-mission, HM and OE data splits.  In what follows, we will often
refer to simulations of the noise plus systematic effects simply as
``noise simulations.''  The noise and CMB realizations are then considered
to form the FFP10 full-focal plane simulations.

Finally, the CMB signal and noise simulations are propagated through
the various component-separation pipelines using the same weights as
derived from the \Planck\ full-mission data analysis
\citep{planck2016-l04}. The signal and noise realizations are then
permuted to generate 999 simulations\footnote{During analysis it was
  determined that CMB realization 970 was corrrupted and thereafter was
  omitted from the MC data set.} for each component-separation method
to be compared to the data.

In the analyses presented in this paper, we often quantify the significance
of a test statistic in terms of the \pval. This is the probability of
obtaining a test statistic at least as extreme as the observed one,
under the assumption that the null hypothesis (i.e., primordial
Gaussianity and isotropy of the CMB) as represented by the simulations
is true. However, this also requires that the simulated reference data
set adopts a cosmological model that is sufficiently consistent with
that preferred by the data. We have noted above that the $\tau$-value
used in the FFP10 simulations is high relative to that preferred by
the latest cosmological analysis.  As a preliminary assessment, we
have considered the predicted variance of the CMB signal for two
values of $\tau$, specifically $0.060$ (as adopted by the FFP10
simulations), and $0.052$ \citep[which is representative of the value determined
by an analysis of the HFI data in][]{planck2016-l06} with
$A_{\rm s}$ set to an appropriate value. We find that the latter
reduces the polarization variance by approximately $20\,\%$ at 
\nside\ = 16 and 32. It may be necessary to take this effect into account when
interpreting the polarization results in what follows.

Similar considerations apply to the simulated noise and residual
systematic effects, particularly given the signal-to-noise regime of the
polarized data.  In order to quantify the agreement of the noise
properties and systematic effects in the data and simulations, we use
differences computed from various subsets of the full-mission data
set.  Note that detailed comparisons have been undertaken using the
power spectra of the individual frequency maps. Figure~18 of
\citet{planck2016-l02} compares half-ring half-difference (HRHD)
spectra for the LFI 30-, 44- and 70-GHz data with simulations, finding
good agreement over most angular scales. Figure~17 of
\citet{planck2016-l03} makes a similar comparison for the HFI 100-,
143-, 217- and 353-GHz half-mission HMHD and OEHD data.

Of more importance to this paper, however, is the consistency of the
data and simulations after various component-separation methods have
been applied.  As established in \citet{planck2016-l04}, the
corresponding end-to-end simulations exhibit biases at the level of
several percent with respect to the observations on intermediate and
small scales, with reasonable agreement on larger scales. These
discrepancies in part originate from the individual frequency
bands. For example, the power in the 100--217\,GHz HFI simulations
underestimates the noise in the data \citep{planck2016-l03}.
Alternatively, biases can arise due to the lack of foreground
residuals in the simulations. On small angular scales, the power
observed in the temperature data exceeds that of the simulations due
to a point-source residual contribution not included in FFP10.  It
should, therefore, be apparent that systematic shifts over some ranges
of angular scale could contribute to \pval\ uncertainties in
subsequent studies.  

We attempt to verify that the analyses presented in this paper are not
sensitive to the differences between the simulations and data.  In
particular, the comparison of the HMHD and OEHD maps for each
component-separation method with those computed from the ensemble of
FFP10 simulations allows us to define the angular scales over which
the various statistical tests applied to the data can be considered
reliable. These may vary depending on the analysis being undertaken.

\section{Tests of non-Gaussianity}
\label{sec:non_gaussianity}

A key prediction of the standard cosmological model is that an early
phase of accelerated expansion, or inflation, gave rise to
fluctuations that correspond to a homogeneous and isotropic Gaussian
field, and that the corresponding statistical properties were
imprinted directly on the primordial CMB 
\citep{planck2013-p17,planck2014-a24,planck2016-l10}. 
Searching for departures
from this scenario is crucial for its validation, yet there is no
unique signature of non-Gaussianity. Nevertheless, the application of
a variety of tests\footnote{One of the more important
  tests in the context of inflationary cosmology is related to the
  analysis of the bispectrum. This is explored thoroughly in
  \citet{planck2016-l09}, and is therefore not discussed further in
  this paper. } over a range of angular scales allows us to probe the
data for inconsistencies with the theoretically motivated Gaussian
statistics.

In previous work \citepalias{planck2013-p09,planck2014-a18}, we
demonstrated that the \Planck\ temperature anisotropies are indeed
consistent with Gaussianity, except for a few apparent anomalies discussed
further in the following section. 
Here, we again apply a non-exhaustive set of tests to the
temperature fluctuations in order to confirm previous results, then
extend the studies to the polarization data.
Of course, significant evidence of deviation from Gaussianity in the
statistics of the measured CMB anisotropies is usually considered to
be an indicator of the presence of residual foregrounds or systematic
artefacts in the data. It is important to be able to mitigate against
such possibilities, particularly in the case of polarization
anisotropies, where the signal-to-noise remains relatively low.  The
analyses are therefore applied to all four component-separation
products (\commander, \nilc, \sevem, and \smica) at a given resolution
with the accompanying common mask, and significance levels are
determined by comparison with the corresponding results derived from
the FFP10 simulations. The consistency of the results derived from the various
component-separation techniques then provides a strong argument against
significant contamination of the data. However, the fidelity of the
simulations is limited by the accuracy with which the systematic
effects can be modelled; therefore we use HMHD and OEHD null tests to
evaluate the agreement of the data and simulations over the scales of
interest. It is plausible that the simulations of the polarized signal show
evidence of a small level of non-Gaussianity depending on the statistical test
applied,  given the significant level of the systematic effects modelled therein.

\subsection{One-dimensional moments}
\label{sec:onepdf}

In this section we consider simple tests of Gaussianity based on
moments of the CMB temperature and polarization anisotropy maps.

For the temperature analysis, we repeat the study performed in
\citetalias{planck2014-a18} and measure the variance, skewness, and
kurtosis of the \Planck\ 2018 component-separated maps using the
unit-variance estimator \citep{Cruz2011}. This method requires a normalized
variance sky map, $u^X$ defined as:
\begin{linenomath*}
\begin{equation}\label{normalizedCMBmapT}
u^X_i(\sigma^2_{X,0}) = \frac{X_i}{\sqrt{\sigma^2_{X,0}+\sigma^2_{i,\mathrm{N}}}} \, ,
\end{equation}
\end{linenomath*}
where $X_i$ is the observed temperature at pixel $i$, $\sigma^2_{X,0}$
is the variance of the CMB signal, and $\sigma^2_{i,\mathrm{N}}$ is
the variance of the noise for that pixel, estimated using the FFP10 MC
simulations. The CMB variance is then determined by finding the
$\hat{\sigma}^2_{X,0}$ value for which the variance of the normalized map
$u^X$ is unity. The skewness and kurtosis are then subsequently
computed from the appropriately normalized map.

\begin{figure}
\includegraphics[scale=0.44]{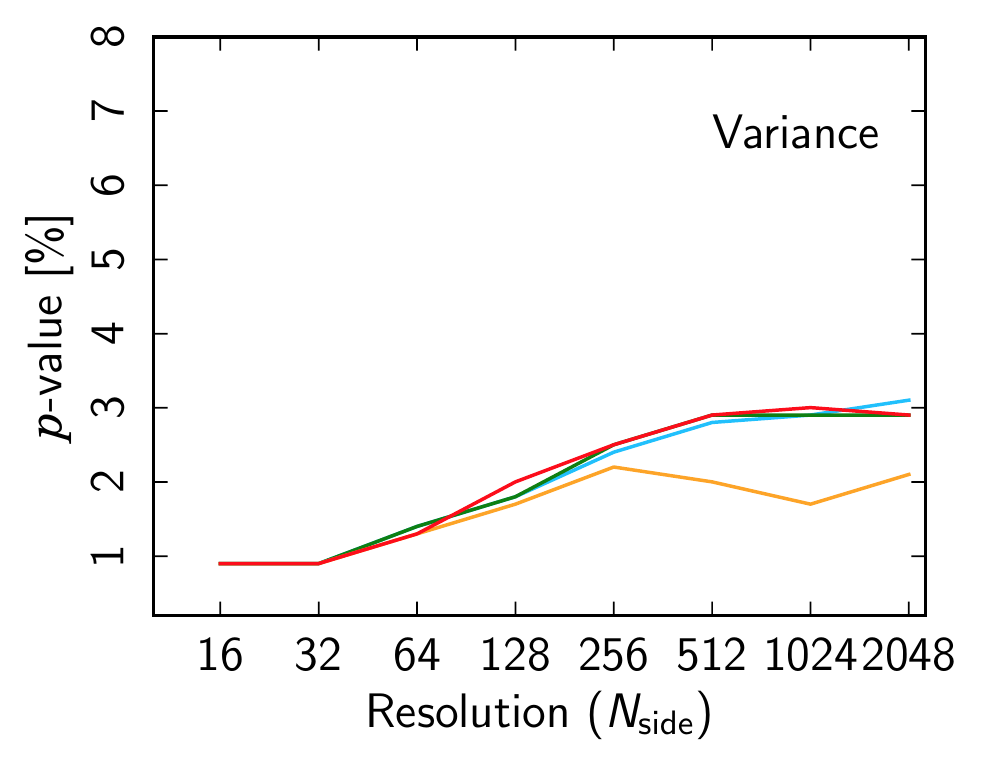}
\includegraphics[scale=0.44]{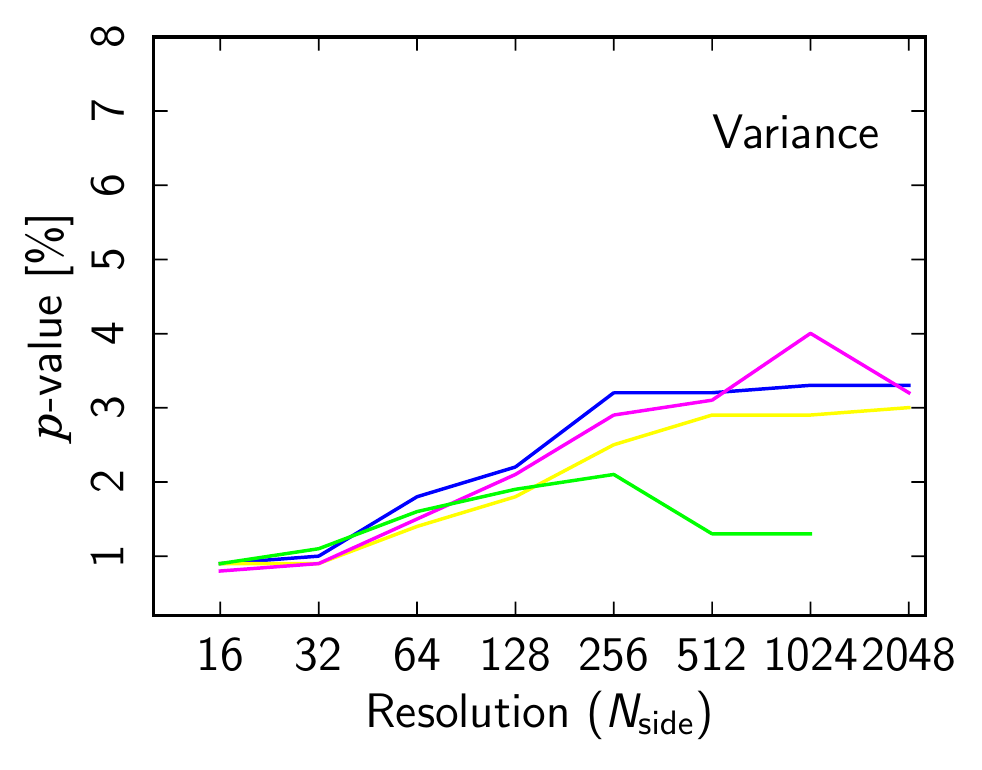}\\
\includegraphics[scale=0.44]{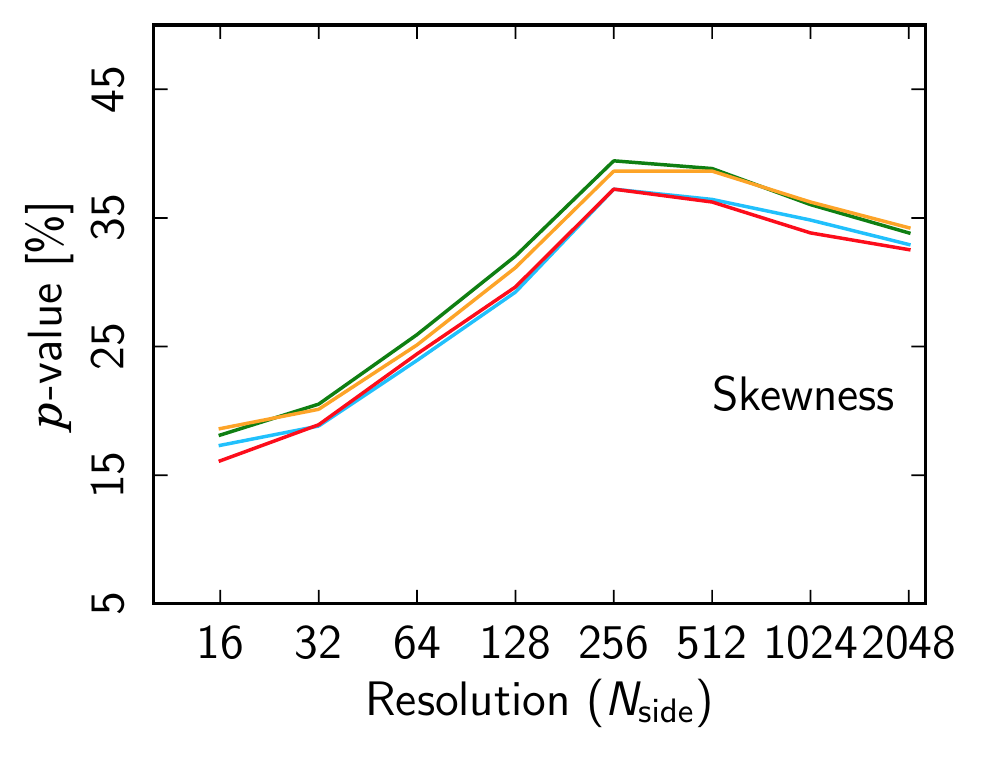}
\includegraphics[scale=0.44]{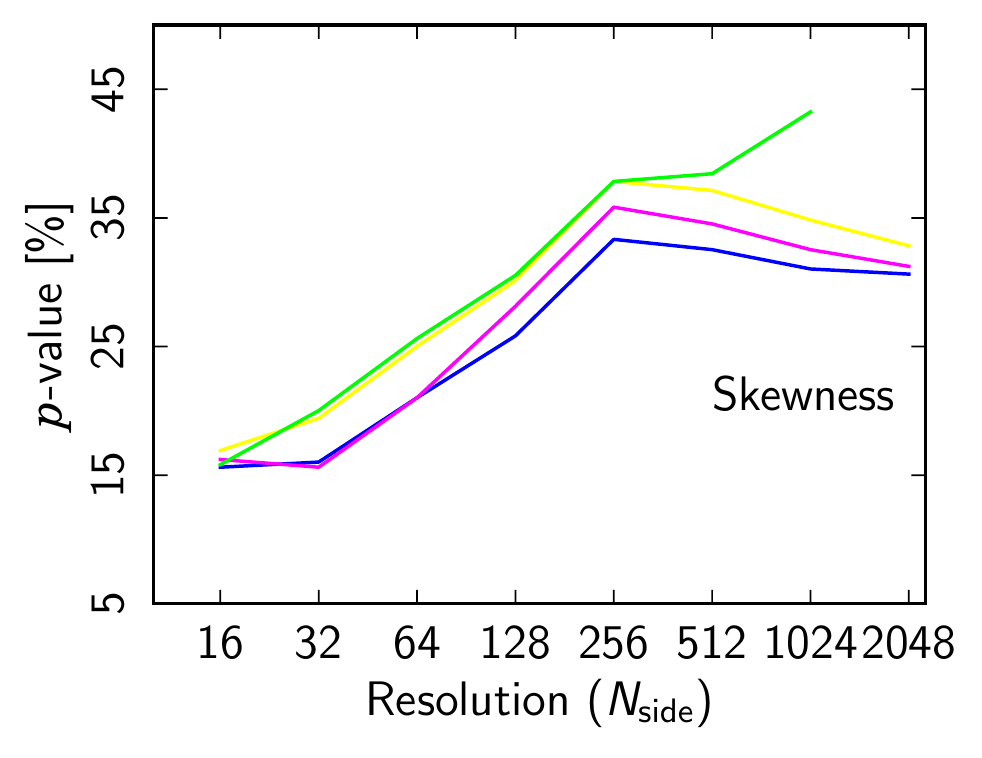}\\
\includegraphics[scale=0.44]{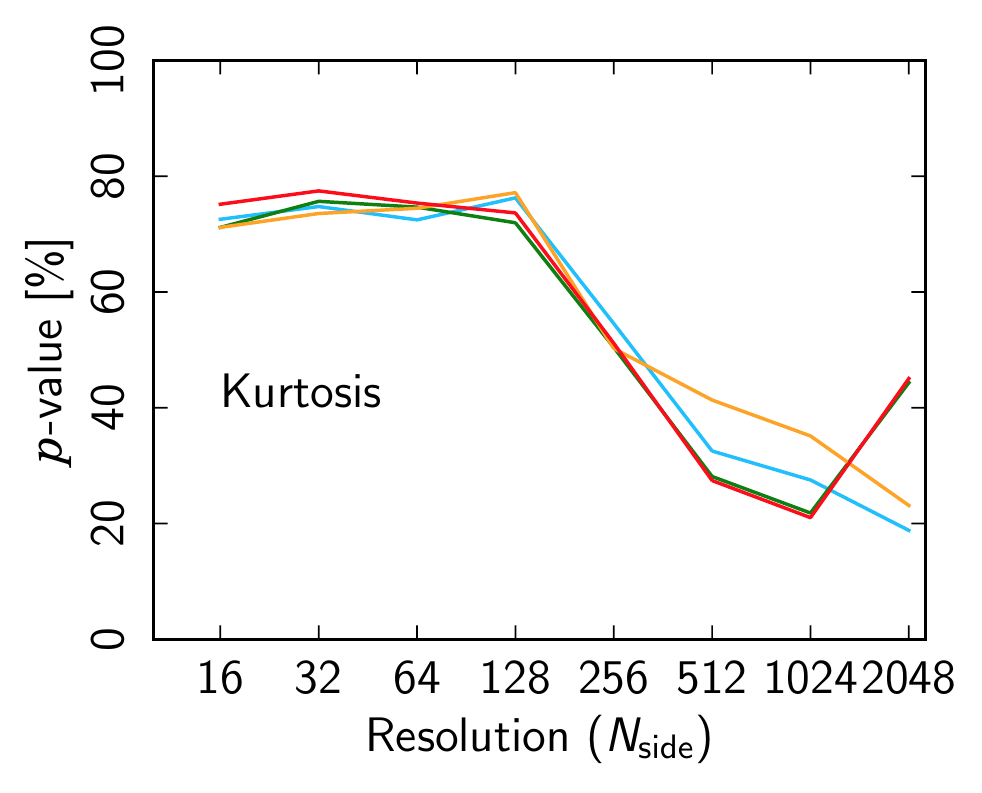}
\includegraphics[scale=0.44]{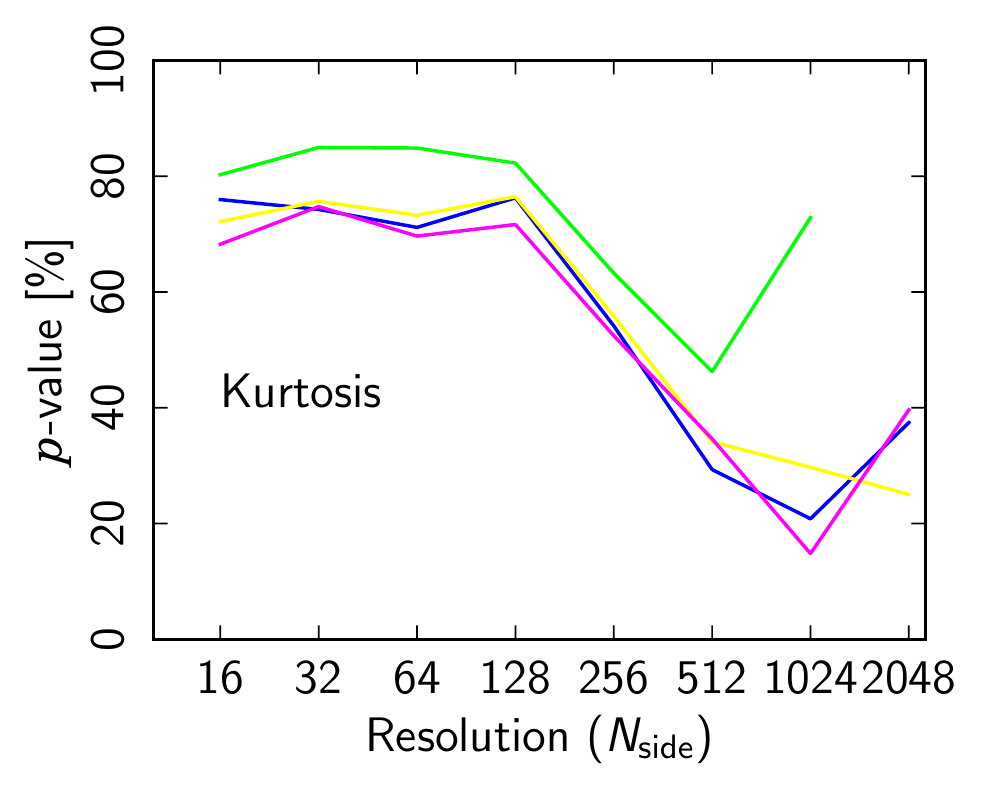}
\caption{Lower-tail probabilities of the variance (top), skewness
  (centre), and kurtosis (bottom), determined from the \commander\
  (red), \nilc\ (orange), \sevem\ (green), and \smica\ (blue)
  component-separated temperature maps (left) and the {\tt SEVEM-070}
  (light green), {\tt SEVEM-100} (dark blue), {\tt SEVEM-143}
  (yellow), and {\tt SEVEM-217} (magenta) frequency-cleaned maps
  (right) at different resolutions.}
\label{onepdf_T_PTE}
\end{figure}

In Fig.~\ref{onepdf_T_PTE} we show the lower-tail probability of the
variance, skewness, and kurtosis determined at different resolutions
from the four component-separated maps (left columns) and from the
\sevem\ frequency-cleaned maps (right columns), after applying the
appropriate common mask.  There is good agreement between the maps,
although the \nilc\ results indicate a slightly lower {\it p}-value
for the variance at intermediate and high resolutions. This may
be related to the small relative power deficit observed between \nilc\ and the
other component separation methods over the multipole range
$\ell\,{=}\,100$--300, as shown in figure~15 of
\citet{planck2016-l04}.  We note that \citet{planck2016-l04} has
demonstrated the presence of a noise mismatch between the observed
data and simulations, as traced by the HMHD and OEHD maps. However,
this is not relevant for analysis of the temperature data, given its
very high signal-to-noise ratio.  The results for 1D moments presented here are in
very good agreement with the \Planck\ 2015 analysis
\citepalias{planck2014-a18}, showing a decreasing lower-tail probability
with decreasing resolution. This lower-tail probability is related to
the presence of the well known lack of power on large angular
scales. However, in the previous analysis we found a minimum value for
the probability of 0.5\,\% at $N_{\mathrm{side}}\,{=}\,16$ for all the maps
considered, compared to a probability of roughly 1\,\% here.  The
difference can be explained by the fact that the 2018 common mask
rejects less of the sky than the 2016 common mask, and previous work
(\citetalias{planck2014-a18}; \citealt{Gruppuso2013})
has shown that the low variance
anomaly becomes less significant with increasing sky coverage. Indeed,
when we apply the 2016 common mask to the current data set, the
probability decreases to 0.7--0.8\,\%, in better agreement with the
previous results. The skewness and kurtosis results do not show
any anomalous behaviour, in agreement with earlier analyses.

In polarization we follow a different approach, as a consequence of 
the lower signal-to-noise ratio. Specifically, 
we subtract the noise contribution to the total variance of the
polarization maps and define the estimator

\begin{linenomath*}
\begin{equation}\label{VarianceP}
\hat{\sigma}^2_\mathrm{CMB} = \langle Q^2+U^2 \rangle - \langle
Q_{\rm N}^2+U_{\rm N}^2 \rangle_\mathrm{MC} \, ,\\
\end{equation}
\end{linenomath*}
where $Q$ and $U$ are the Stokes parameters of the observed polarization maps,
and $\langle Q_{\rm N}^2+U_{\rm N}^2 \rangle_\mathrm{MC}$ are
noise estimates determined from MC simulations.
\citet{planck2016-l04} indicates a mismatch between the noise in the
data and that in simulations for map resolutions above \nside\,{=}\,256. 
This corresponds to a few percent of the theoretical CMB variance
up to \nside\,{=}\,1024, while it is much larger
at the highest resolution. Since the noise mismatch is likely to affect
the less signal-dominated polarization results, 
we also define a cross-variance estimator that determines the variance from the
two maps available for each data split, HM or OE, respectively:

\begin{linenomath*}
\begin{equation}\label{VarianceP_cross}
\hat{\sigma}^2_\mathrm{CMB} = \langle Q_1Q_2+U_1U_2 \rangle - \langle
Q_1^{\rm N}Q_2^{\rm N}+U_1^{\rm N}U_2^{\rm N} \rangle_\mathrm{MC} \, ,
\end{equation}
\end{linenomath*}
where $Q_1$, $Q_2$, $U_1$, and $U_2$ are the Stokes parameters of the two
maps from either the HM- or OE-cleaned data split, and
$\langle Q_1^{\rm N}Q_2^{\rm N}+U_1^{\rm N}U_2^{\rm N} \rangle_\mathrm{MC}$
is the corresponding noise contribution to the total variance in polarization
estimated from the corresponding simulations. Note that a
cross-estimator should be less affected by noise mismatch,
although correlated noise remains an issue. However, it is impossible
to assess if the latter is well described by the simulations.

\begin{figure}
\includegraphics[width=\columnwidth]{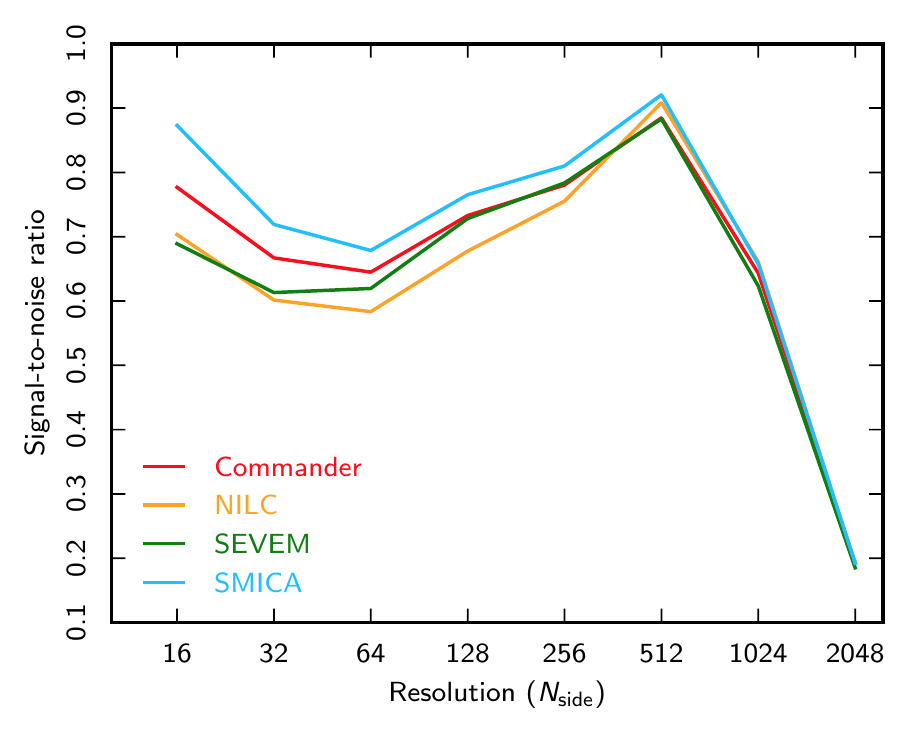}
\caption{Signal-to-noise ratio for the variance
    estimator in polarization for \commander\ (red), \nilc\ (orange),
    \sevem\ (green), and \smica\ (blue), obtained by comparing
    the theoretical variance from the \Planck\ FFP10 fiducial model with
    an MC noise estimate (right-hand term of Eq.~\eqref{VarianceP}).
    Note that the same colour scheme for distinguishing the four
    component-separation maps is used throughout this paper.}
\label{Variance_S2Nratio}
\end{figure}

In Fig.~\ref{Variance_S2Nratio} we show the expected signal-to-noise
ratio
of the polarization variance for the component-separated maps,
determined by comparing the theoretical variance of the signal at
different resolutions (as evaluated from the \Planck\ FFP10 fiducial
model, including beam and pixel window function effects) to the
corresponding MC estimate of the noise,
$\langle Q_{\rm N}^2+U_{\rm N}^2 \rangle_\mathrm{MC}$ . All of the methods show
similar behaviour, with a maximum signal-to-noise ratio of about 0.8 on
intermediate scales, \nside\,{=}\,512. The minimum ratio is observed at
\nside\,{=}\,64, as explained by the fact that the $EE$ angular power
spectrum exhibits a low amplitude over the multipole range $\ell\,{=}\,10$--100.
At very large scales, \nside\,{=}\,16, the signal-to-noise ratio
increases again, but with an amplitude that depends noticeably on the
component-separation method considered .  At very high resolutions the
signal-to-noise ratio drops, as expected.

\begin{figure}
\includegraphics[scale=0.44]{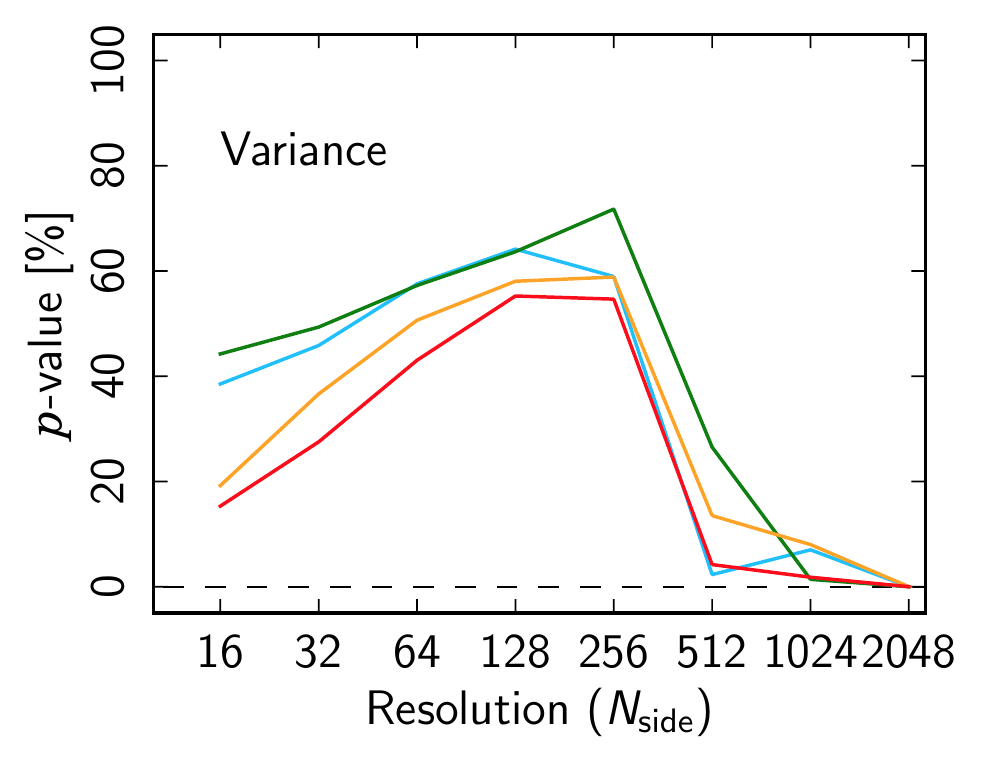}
\includegraphics[scale=0.44]{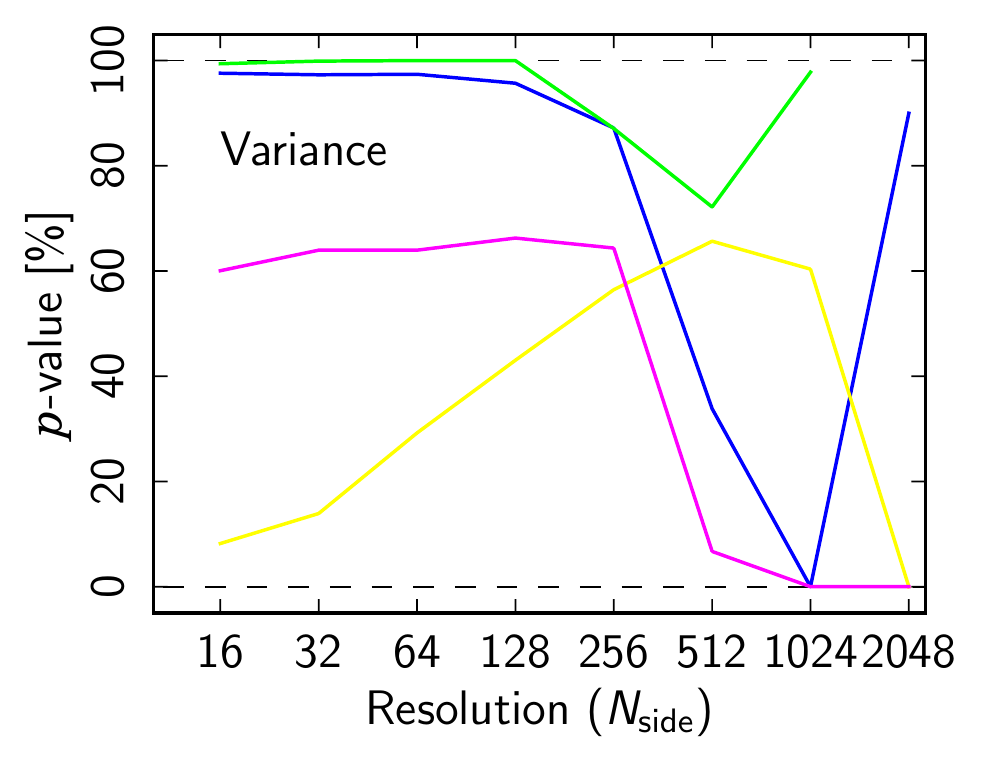}\\
\includegraphics[scale=0.44]{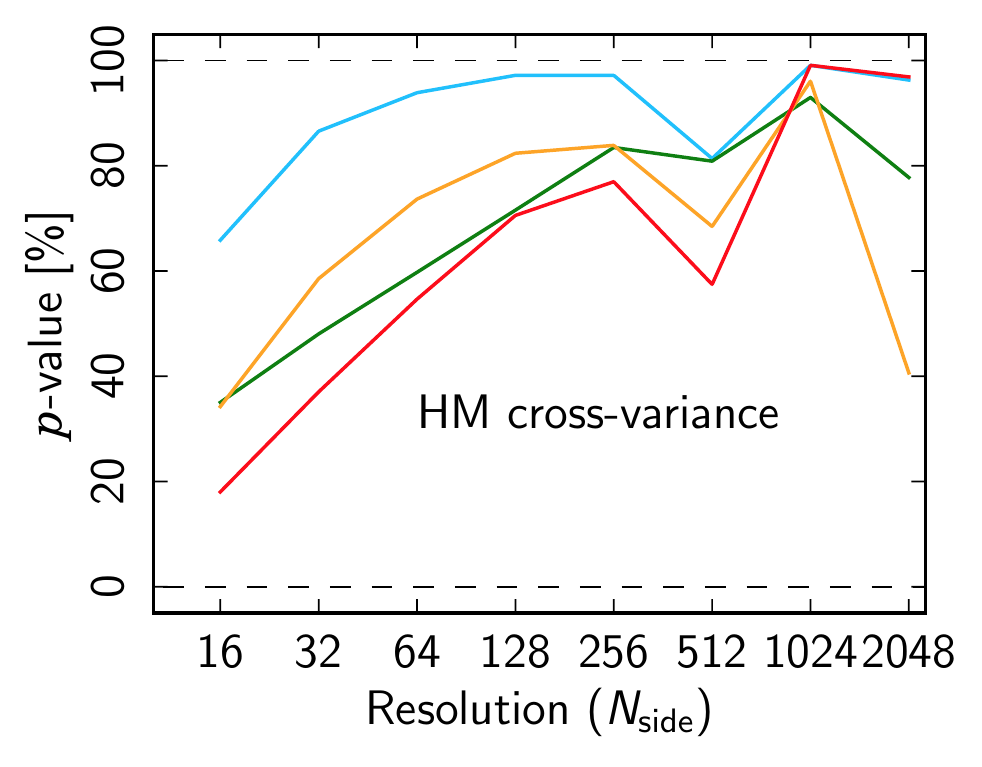}
\includegraphics[scale=0.44]{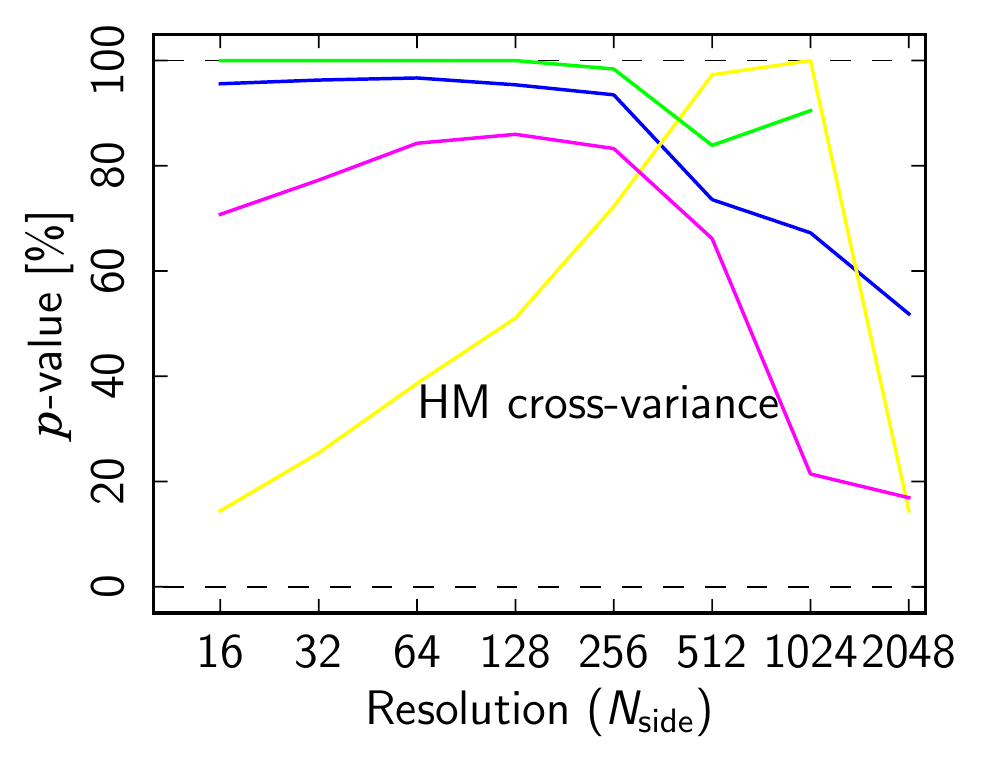}\\
\includegraphics[scale=0.44]{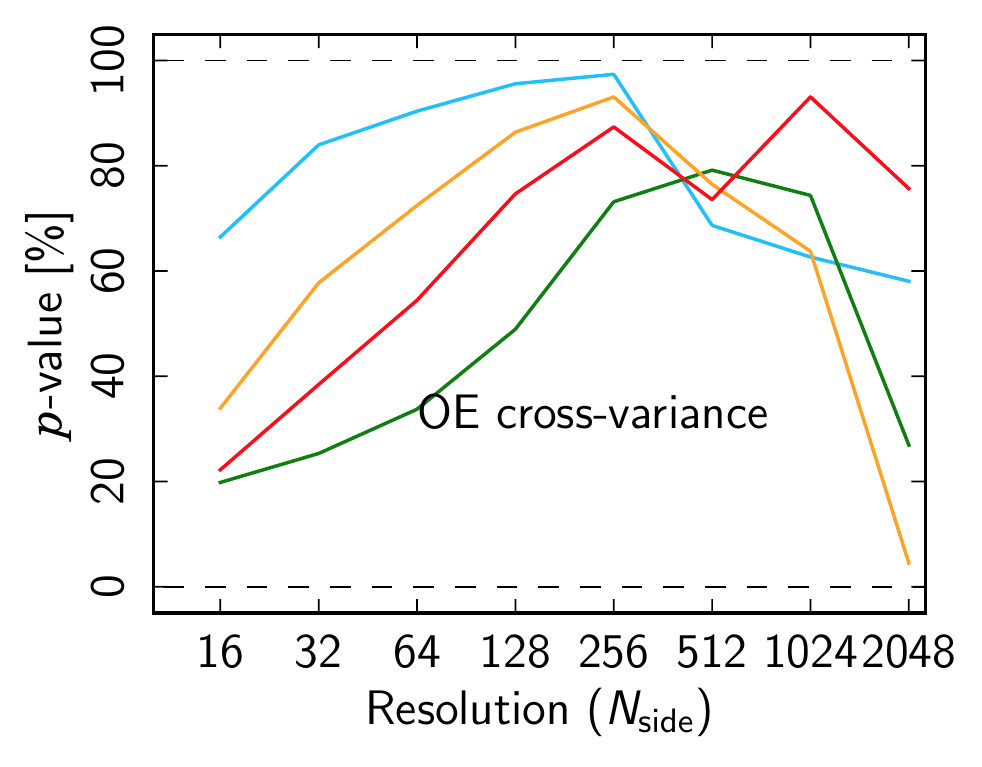}
\includegraphics[scale=0.44]{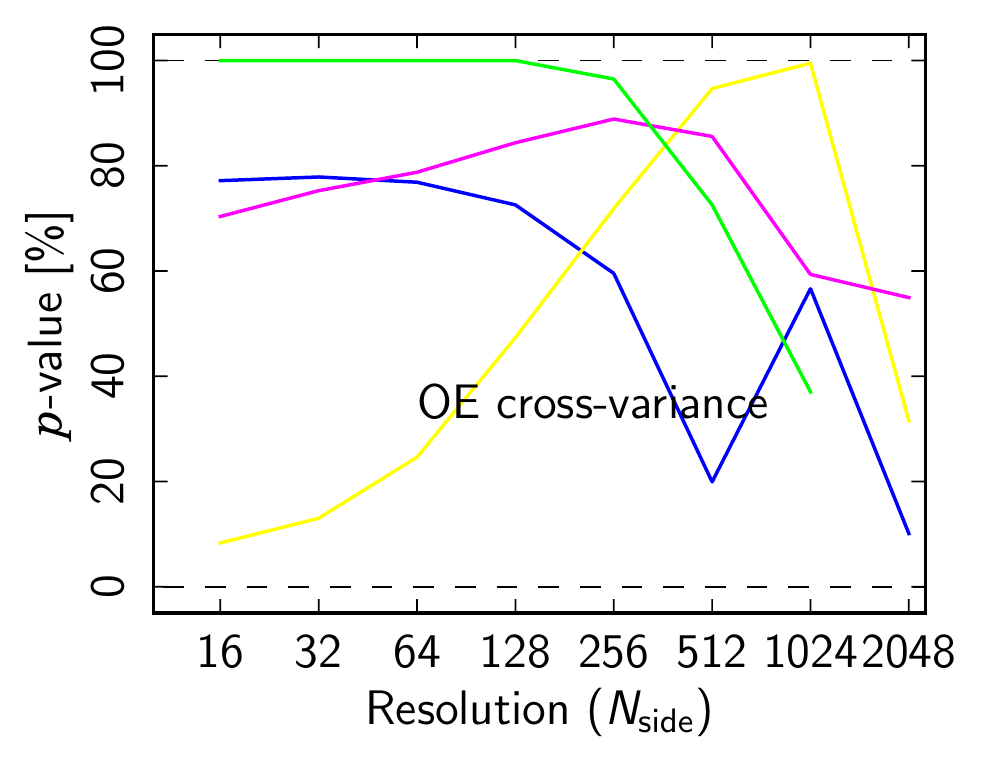}
\caption{Lower-tail probabilities of the variance determined from the
  \commander\ (red), \nilc\ (orange), \sevem\ (green), and \smica\
  (blue) component-separated polarization maps (left) and the {\tt
    SEVEM-070} (light green), {\tt SEVEM-100} (dark blue), {\tt
    SEVEM-143} (yellow), and {\tt SEVEM-217} (magenta)
  frequency-cleaned maps (right) at different resolutions. The top,
  middle and bottom rows correspond to results evaluated with the
  full-mission, HM- and OE-cross-variance estimates, respectively.
  In this figure, small $p$-values would correspond to anomalously
  low variance.}
\label{onepdf_P_PTE}
\end{figure}

In Fig.~\ref{onepdf_P_PTE} we show the lower-tail probabilities of the
variance determined from the full-mission and the HM and OE data
splits using the appropriate common mask, compared to the corresponding
results from MC simulations.
At high resolutions, the lower-tail probability determined from the
variance of the full-mission data approaches zero. As previously
noted, this is due to the poor agreement between the noise properties
of the data and the MC simulations, in particular at high
resolution. This explanation is further supported by the fact that the
lower-tail probability becomes more compatible with the MC simulations
when we consider the cross-variance analyses. However, given the
uncertainties in the properties of the correlated noise in the
simulations, we prefer to focus on the intermediate and large angular scales,
$N_{\rm side}\,{\leq}\,256$.
We note that there is a trend towards lower probabilities as the
resolution decreases from \nside\,{=}\,256, similar to what is observed with
the temperature data. This behaviour is common to all of the component-separated methods and also to the {\tt SEVEM-143} frequency-cleaned
data, although with different probabilities at a given resolution.
The {\tt SEVEM-070} frequency-cleaned map is not compatible with the
MC simulations for resolutions lower than \nside\,{=}\,128. This may be due
to the presence of either residual foregrounds in the data or
systematic effects that are not sufficiently well represented by the
MC simulations. Although the compatibility of the data with the MC
simulations is generally adequate, the large variation of
probabilities seen for different component-separation methods and
resolutions, even when a cross-estimator is considered, suggests that
correlated noise and residual systematics in the data may not be
sufficiently well described by the current set of simulations.

\begin{figure}
\includegraphics[scale=0.45]{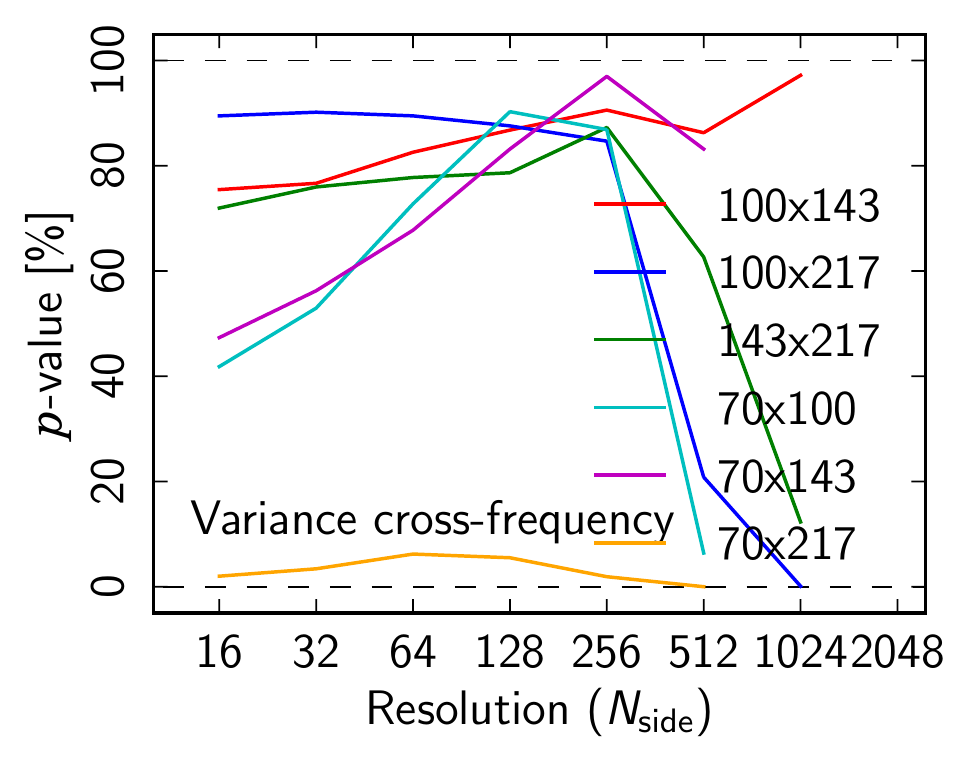}\\
\includegraphics[scale=0.45]{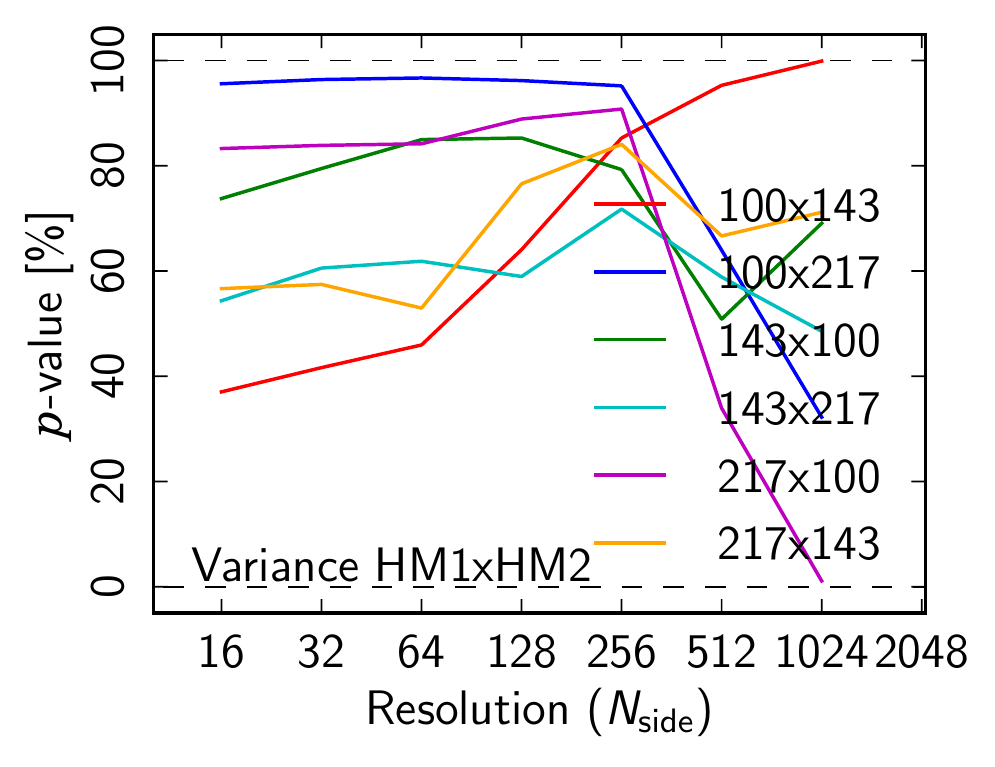}
\includegraphics[scale=0.45]{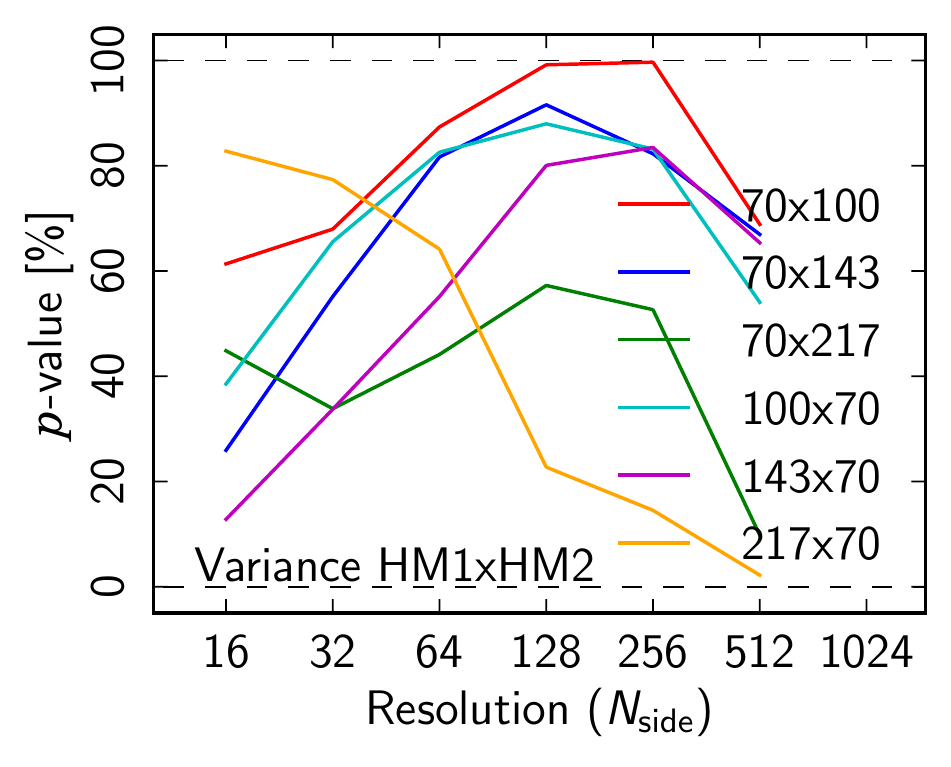}\\
\includegraphics[scale=0.45]{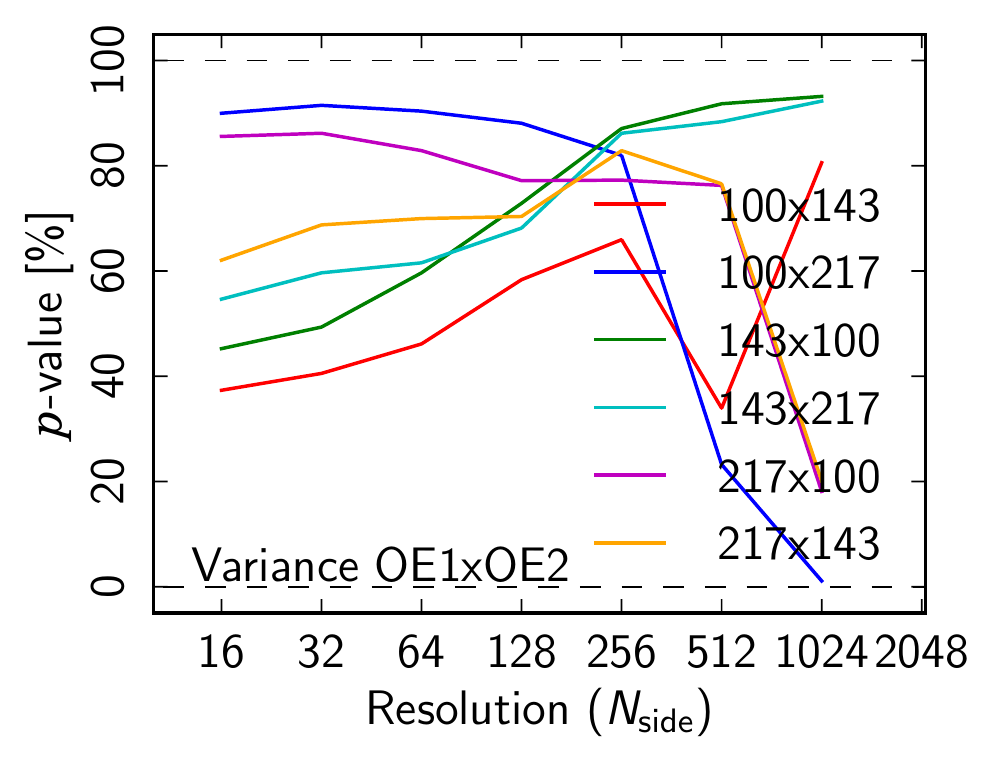}
\includegraphics[scale=0.45]{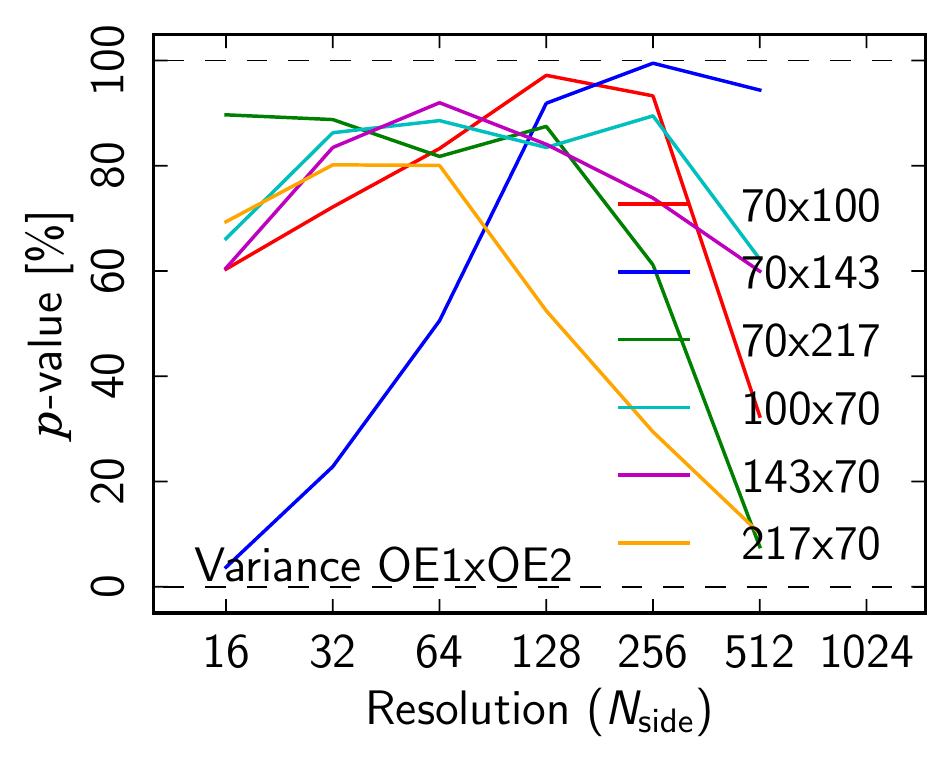}
\caption{Lower-tail probabilities of the cross-variance determined
  between the \sevem\ frequency-cleaned polarization maps (top) or
  from the HM (centre) or OE (bottom) \sevem\ frequency-cleaned maps
  at different resolutions.  In this figure, small $p$-values would
  correspond to anomalously low variance.}
\label{onepdf_P_freq_PTE}
\end{figure}

In an attempt to minimize the impact of correlated noise, we consider
the cross-variance estimated between pairs of frequencies from the
\sevem\ frequency-cleaned maps, using full-mission, HM and OE data
sets. The results are shown in Fig.~\ref{onepdf_P_freq_PTE}.
Note that the combination of the {\tt SEVEM-070} data with higher
frequency maps is consistent with the MC simulations, supporting the
idea that residual systematic effects in the former, which are not well
described by the corresponding simulations, bias results computed
only with the 70-GHz cleaned data.  In addition, the cross-variance
determined between the {\tt SEVEM-070} and {\tt SEVEM-217} maps yields
a particular low probability. Since the 217-GHz data are used
to clean the 70-GHz map, it is probable that this particular
combination is more affected by correlated residuals than elsewhere.
However, the effect disappears when considering the HM or OE cross-variance
data.

In summary, we confirm previous results based on the analysis of the
temperature anisotropy \citepalias{planck2013-p09, planck2014-a18},
indicating that the data are consistent with Gaussianity, although
exhibiting low variance on large angular scales, with a probability of
about 1\,\% as compared to our fiducial cosmological model.  In
polarization we find reasonable consistency with MC simulations on
intermediate and large angular scales, but there is a considerable
range of {\it p}-values found, depending on the specific combinations
of data considered.  This indicates that the lower signal-to-noise
ratio of the \Planck\ data in polarization, and, more specifically,
the uncertainties in our detailed understanding of the noise
characterization (both in terms of amplitude and correlations between
angular scales) limits our ability to pursue further investigate
the possible presence of anomalies in the 1D moments.

\subsection{N-point correlation functions}
\label{sec:npoint_correlation}

In this section, we present tests of the non-Gaussianity of the
\Planck\ 2018 temperature and polarization CMB data using real-space
$N$-point correlation functions.

An $N$-point correlation function is defined as the average product of
$N$ observables, measured in a fixed relative orientation on the sky,
\begin{linenomath*}
\begin{equation}
C_{N}(\theta_{1}, \ldots, \theta_{2N-3}) = \left\langle X(\vec{\hat{n}}_{1})\cdots
X(\vec{\hat{n}}_{N}) \right\rangle,
\label{eqn:npoint_def}
\end{equation}
\end{linenomath*}
where the unit vectors $\vec{\hat{n}}_1, \ldots,\vec{\hat{n}}_{N}$
span an $N$-point polygon. If statistical isotropy is assumed, these
functions do not depend on the specific position or orientation of the
$N$-point polygon on the sky, but only on its shape and size.  In the
case of the CMB, the fields, $X$, correspond to the temperature, $T$, and
the two Stokes parameters, $Q$ and $U$, which describe the linearly
polarized radiation in direction $\vec{\hat{n}}$. 
Following the standard CMB
convention, $Q$ and $U$ are defined with respect to the local meridian
of the spherical coordinate system of choice.
To obtain coordinate-system-independent $N$-point correlation
functions, we define Stokes parameters in a
radial system, denoted by $Q_\mathrm{r}$ and $U_\mathrm{r}$, according to
Eq.~\eqref{eqn:local_Stokes},
where the reference point, $\vec{\hat{n}}_\mathrm{ref}$, is specified 
by the centre of mass of the polygon \citep{gjerlow:2010}. 
In the case of the 2-point function, this corresponds to defining a local
coordinate system in which the local meridian passes through the two points
of interest \citep[see][]{kamionkowski:1997}.

The correlation functions are estimated by simple product averages
over all sets of $N$ pixels fulfilling the geometric requirements set
by the $2N-3$ parameters $\theta_1, \ldots, \theta_{2N-3}$ characterizing
the shape and size of the polygon,
\begin{linenomath*}
\begin{equation}
\hat{C}_{N}(\theta_{1}, \ldots, \theta_{2N-3}) 
= \frac{\sum_i \left(w_1^i \cdots w_N^i \right) \left( X_1^i \cdots
X_N^i \right) }{\sum_i w_1^i \cdots w_N^i} \ .
\label{eqn:npoint_estim}
\end{equation}
\end{linenomath*}
Here, pixel weights $w_1^i,\cdots,w_N^i$ 
represent masking and are set to 1 or 0 for included or excluded
pixels, respectively.

The shapes of the polygons selected for the analysis are not more
optimal for testing Gaussianity than other configurations, but are
chosen because of ease of implementation and for comparison of the
results with those for the 2013 and 2015 \Planck\ data sets. In
particular, we consider the 2-point function, as well as the
pseudo-collapsed and equilateral configurations for the 3-point
function.
Following \citet{eriksen2005}, the pseudo-collapsed configuration
corresponds to an (approximately) isosceles triangle, where the length
of the baseline falls within the second bin of the separation
angles and the length of the longer edge of the triangle, $\theta$,
parametrizes its size. Analogously, in the case of the equilateral
triangle, the size of the polygon is parametrized by the
length of the edge, $\theta$. 

We use a simple $\chi^2$ statistic to quantify the agreement between
the observed data and simulations. This is defined by
\begin{linenomath*}
\begin{equation}
\chi^2 = \sum_{i,j=1}^{N_{\rm bin}} \Delta_N (\theta_i)
\tens{M}_{ij}^{-1} \Delta_N(\theta_j)
\ .
\label{eqn:chisqr_stat}
\end{equation}
\end{linenomath*}
Here,
$\Delta_N(\theta_i) \equiv \left( \hat{C}_N(\theta_i) - \left< C_N
    (\theta_i) \right> \right) / \sigma_N(\theta_i)$ is the difference
between the observed, $\hat{C}_N(\theta_i)$, and the corresponding average
from the MC simulation ensemble, $\left< C_N(\theta_i) \right>$, of
the $N$-point correlation function for the bin with separation angle
$\theta_i$, normalized by the standard deviation of the difference,
$\sigma_N(\theta_i)$, and $N_\mathrm{bin}$ is the number of bins used
for the analysis. If $\Delta^{(k)}_N(\theta_i)$ is the $k$th simulated
$N$-point correlation function difference and $N_\mathrm{sim}$ is the
number of simulations, then the covariance matrix (normalized to unit
variance) $\tens{M}_{ij}$ is estimated by
\begin{linenomath*}
\begin{equation}
 \tens{M}_{ij} = \frac{1}{N'_\mathrm{sim}}\sum_{k=1}^{N_\mathrm{sim}}
\Delta^{(k)}_N(\theta_i) \, \Delta^{(k)}_N(\theta_j) ,
\label{eqn:cov_mat}
\end{equation}
\end{linenomath*}
where $N'_\mathrm{sim} = N_\mathrm{sim} - 1$. However, due to
degeneracies in the covariance matrix resulting from an overdetermined
system and a precision in estimation of the matrix elements of order
\mbox{$\Delta \tens{M}_{ij} \sim \sqrt{2 / N_\mathrm{sim}}$}, the
inversion of the matrix is unstable. To avoid this, a singular-value
decomposition (SVD) of the matrix is performed, and only those modes
that have singular values larger than $\sqrt{2 / N_\mathrm{sim}}$ are
used in the computation of the $\chi^2$ statistic
\citep{gaztanaga2005}. We note that this
is a modification of the procedure used in previous \Planck\ analyses
\citepalias{planck2013-p09,planck2014-a18}.
Finally, we also correct for bias in the inverse
covariance matrix by multiplying it by a factor
$(N'_\mathrm{sim}-N_\mathrm{bin}-1)/N'_\mathrm{sim}$
\citep{hartlap2007}.

\begin{figure*}[htp!]
\begin{center}
\includegraphics[width=\textwidth]{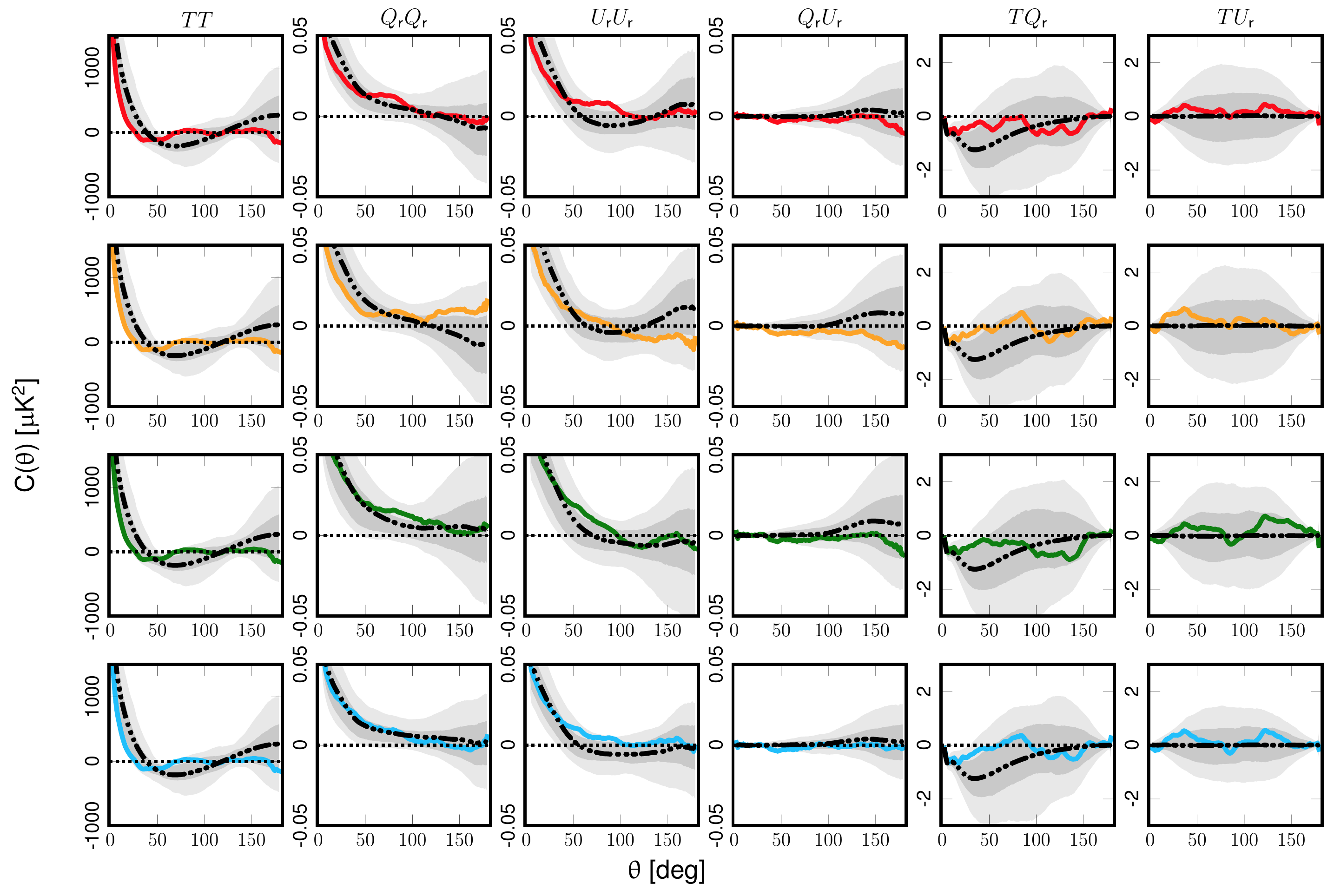}
\caption{2-point correlation functions determined from the \nside\ = 64
  \Planck\ CMB 2018 temperature and polarization maps.
  Results are shown for the \commander, \nilc, \sevem, and \smica\
  maps (first, second, third, and fourth rows, respectively). The solid
  lines correspond to the data, while the black three dots-dashed lines
  indicate the mean determined from the corresponding FFP10 simulations,
  and the shaded dark and light grey areas indicate the
  corresponding 68\,\% and 95\,\% confidence regions,
  respectively.}
\label{fig:2pt_data}
\end{center}
\end{figure*}

\begin{figure*}[htp!]
\begin{center}
\includegraphics[width=0.8\textwidth]{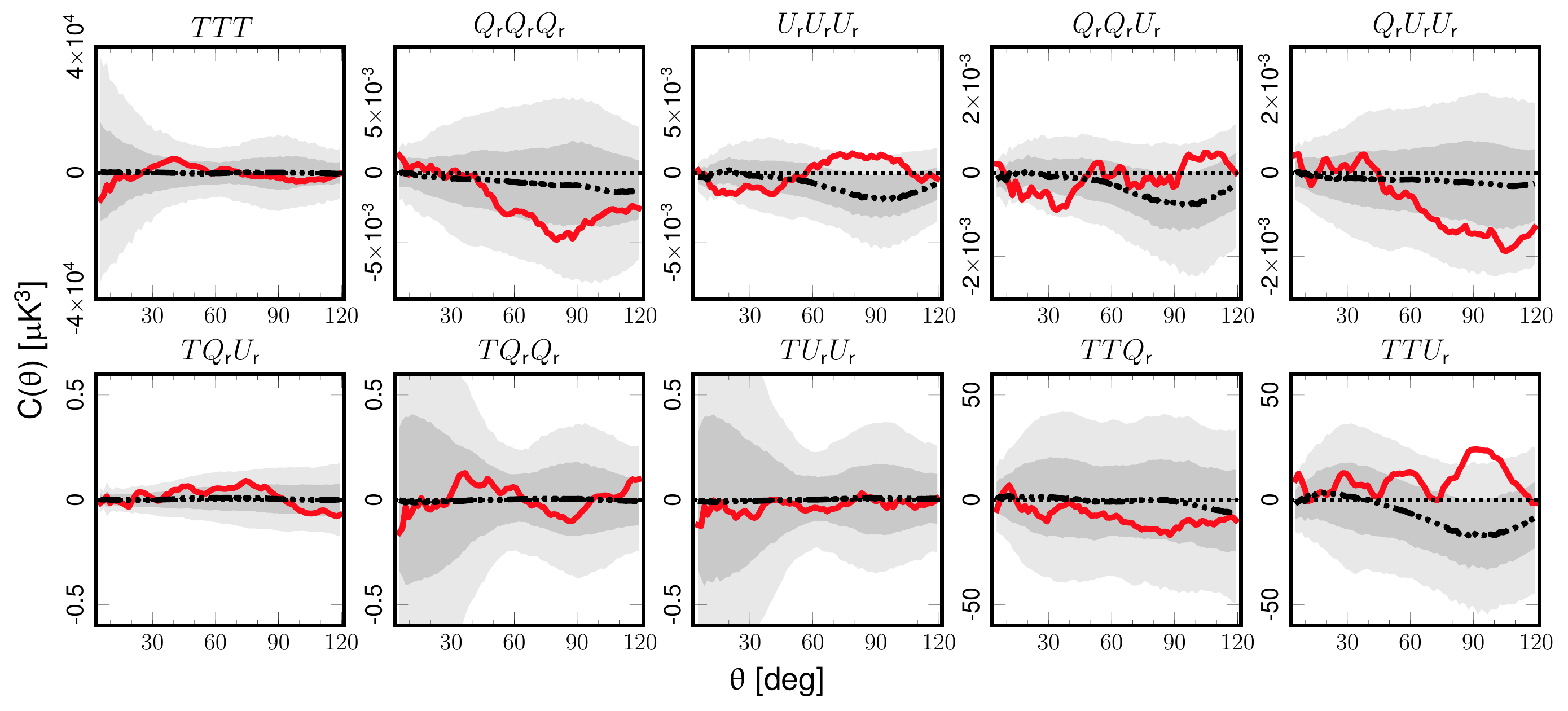}\\
\includegraphics[width=0.8\textwidth]{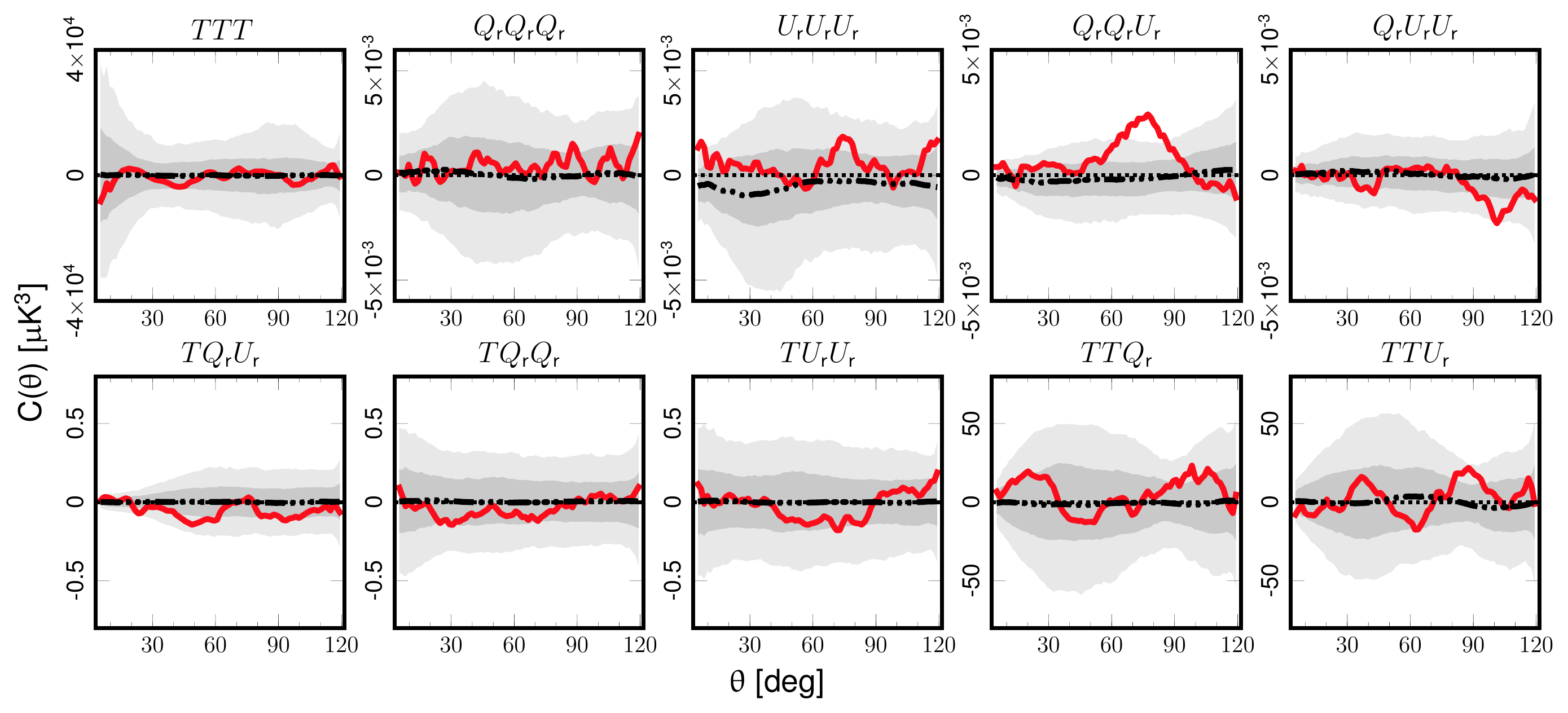}
\caption{
3-point correlation functions determined from the \nside\ = 64
  \Planck\ CMB \commander\ 2018 temperature and polarization maps.
  Results are shown for the pseudo-collapsed 3-point
  (upper panel) and equilateral 3-point (lower panel) functions. The
  red solid line corresponds to the data, while
  the black three dots-dashed line indicates the mean determined from the FFP10 \commander\
  simulations, and the shaded dark and light grey regions indicate the
  corresponding 68\,\% and 95\,\% confidence areas,
  respectively. See Sect.~\ref{sec:npoint_correlation} for the
  definition of the separation angle $\theta$.
}
\label{fig:npt_data_commander}
\end{center}
\end{figure*}

We analyse the CMB estimates at a resolution of \nside\ = 64 due to
computational limitations.  The results for the 2-point correlation
functions of the CMB maps are presented in Fig.~\ref{fig:2pt_data},
while in Fig.~\ref{fig:npt_data_commander} the 3-point functions for
the \commander\ maps are shown.
In the figures, the $N$-point functions for the data are compared with
the mean values estimated from the FFP10 MC simulations.
Note that the mean behaviour of the 3-point functions derived from the
simulations indicates the presence of small non-Gaussian contributions,
presumably associated with modelled systematic effects that are
included in the simulations.  Furthermore, both the mean and
associated confidence regions vary between component-separation
methods, which reflects the different weightings given to the
individual frequency maps that contribute to the CMB estimates, and
the systematic residuals contained therein.
Some evidence for this behaviour can also be found in the analysis of
HMHD and OEHD maps in the companion paper \citet{planck2016-l04}.
To avoid biases, it is essential to compare the statistical properties
of a given map with the associated simulations. Comparing with simulations
without systematic effects could lead to incorrect conclusions.

\begin{table}[tp] 
\begingroup
\newdimen\tblskip \tblskip=5pt
\caption{Probabilities of obtaining values for the $\chi^2$ statistic
  of the $N$-point functions determined
  from the \Planck\ fiducial $\Lambda$CDM model at least as
  large as those obtained from the {\tt Commander}, {\tt NILC}, {\tt SEVEM},
  and {\tt SMICA} temperature and polarization ($Q$ and $U$) maps at
  \nside\ = 64 resolution.  In this table, large $p$-values would
  correspond to anomalously low values of $\chi^2$. }
\label{tab:prob_npt}
\nointerlineskip
\vskip -3mm
\footnotesize
\setbox\tablebox=\vbox{
   \newdimen\digitwidth 
   \setbox0=\hbox{\rm 0} 
   \digitwidth=\wd0 
   \catcode`*=\active 
   \def*{\kern\digitwidth}
   \newdimen\signwidth 
   \setbox0=\hbox{+} 
   \signwidth=\wd0 
   \catcode`!=\active 
   \def!{\kern\signwidth}
\halign{\tabskip 0pt\hbox to 1.6in{#\leaderfil}\tabskip 4pt&
\hfil#\hfil&
\hfil#\hfil&
\hfil#\hfil&
\hfil#\hfil\tabskip 0pt\cr
\noalign{\doubleline\vskip -1pt}
\omit&\multispan4 \hfil Probability [\%]\hfil\cr
\noalign{\vskip -4pt}
\omit&\multispan4\hrulefill\cr
\omit\hfil Function\hfil&{\tt Comm.}&{\tt NILC}&{\tt SEVEM}&{\tt SMICA}\cr
\noalign{\vskip 3pt\hrule\vskip 6pt}
\multispan5\hfil 2-point functions\hfil\cr
\noalign{\vskip 3pt}
$TT$& 74.1& 75.5& 75.1& 76.7\cr
$Q_\mathrm{r} Q_\mathrm{r}$& 68.3& 25.7& 51.5& 29.1\cr
$U_\mathrm{r} U_\mathrm{r}$& 71.4& 52.5& 40.2& 35.0\cr
$T Q_\mathrm{r}$& 55.5& 77.7& 80.0& 59.8\cr
$T U_\mathrm{r}$& 67.2& 55.4& 60.6& 16.7\cr
$Q_\mathrm{r} U_\mathrm{r}$& 29.8& 22.2& 14.7& 23.3\cr
\noalign{\vskip 6pt\hrule\vskip 6pt}
\multispan5\hfil Pseudo-collapsed 3-point functions\hfil\cr
\noalign{\vskip 3pt}
$TTT$& 91.3& 89.7& 90.6& 90.2\cr
$Q_\mathrm{r} Q_\mathrm{r} Q_\mathrm{r}$& 23.5& 53.9& 22.3& 40.8\cr
$U_\mathrm{r} U_\mathrm{r} U_\mathrm{r}$& 27.2& 30.2& 18.3& 13.6\cr
$TTQ_\mathrm{r}$& 98.8& 97.7& 97.2& 99.2\cr
$TTU_\mathrm{r}$& 32.1& 29.5& 39.8& 46.7\cr
$TQ_\mathrm{r} Q_\mathrm{r}$& 73.5& 81.1& 85.2& 59.4\cr
$TU_\mathrm{r} U_\mathrm{r}$& 93.9& 96.4& 88.5& 90.8\cr
$TQ_\mathrm{r} U_\mathrm{r}$& 51.8& 46.9& 52.9& 18.6\cr
$Q_\mathrm{r} Q_\mathrm{r} U_\mathrm{r}$& *7.6& *8.3& 17.4& *9.0\cr
$Q_\mathrm{r} U_\mathrm{r} U_\mathrm{r}$& 59.1& 92.4& 23.9& 52.9\cr
\noalign{\vskip 6pt\hrule\vskip 6pt}
\multispan5\hfil Equilateral 3-point functions\hfil\cr
\noalign{\vskip 3pt}
$TTT$& 95.6& 95.3& 94.8& 95.3\cr
$Q_\mathrm{r} Q_\mathrm{r} Q_\mathrm{r}$& 15.8& 24.3& 26.2& 12.2\cr
$U_\mathrm{r} U_\mathrm{r} U_\mathrm{r}$& 34.7& 75.5& *9.3& 37.6\cr
$TTQ_\mathrm{r}$& 76.9& 91.2& 63.8& 91.2\cr
$TTU_\mathrm{r}$& 58.5& 83.2& 88.7& 73.9\cr
$TQ_\mathrm{r} Q_\mathrm{r}$& 99.5& 85.8& 99.5& 90.7\cr
$TU_\mathrm{r} U_\mathrm{r}$& 79.0& 84.6& 84.8& 90.3\cr
$TQ_\mathrm{r} U_\mathrm{r}$& 66.7& 55.4& 82.4& 52.0\cr
$Q_\mathrm{r} Q_\mathrm{r} U_\mathrm{r}$& 46.0& 91.6& 18.7& 33.5\cr
$Q_\mathrm{r} U_\mathrm{r} U_\mathrm{r}$& 41.1& 16.0& 28.7& 50.8\cr
 \noalign{\vskip 6pt\hrule\vskip 3pt}}}
\endPlancktable                    
\endgroup
\end{table}

\begin{table}[tp] 
\begingroup
\newdimen\tblskip \tblskip=5pt
\caption{Probabilities of obtaining values for the $\chi^2$
    statistic of the $N$-point functions  determined
    from the \Planck\ fiducial $\Lambda$CDM model at least as
    large as the those obtained from the {\tt Commander}, {\tt NILC},
    {\tt SEVEM}, and {\tt SMICA} temperature and polarization ($E$-mode) 
    maps at \nside\ = 64 resolution.  In this table, large $p$-values
    would correspond to anomalously low values of $\chi^2$. }
\label{tab:prob_npt_teb}
\nointerlineskip
\vskip -3mm
\footnotesize
\setbox\tablebox=\vbox{
   \newdimen\digitwidth 
   \setbox0=\hbox{\rm 0} 
   \digitwidth=\wd0 
   \catcode`*=\active 
   \def*{\kern\digitwidth}
   \newdimen\signwidth 
   \setbox0=\hbox{+} 
   \signwidth=\wd0 
   \catcode`!=\active 
   \def!{\kern\signwidth}
\halign{\tabskip 0pt\hbox to 1.6in{#\leaderfil}\tabskip 4pt&
\hfil#\hfil&
\hfil#\hfil&
\hfil#\hfil&
\hfil#\hfil\tabskip 0pt\cr
\noalign{\doubleline\vskip -1pt}
\omit&\multispan4 \hfil Probability [\%]\hfil\cr
\noalign{\vskip -4pt}
\omit&\multispan4\hrulefill\cr
\omit\hfil Function\hfil&{\tt Comm.}&{\tt NILC}&{\tt SEVEM}&{\tt SMICA}\cr
\noalign{\vskip 3pt\hrule\vskip 6pt}
\multispan5\hfil 2-point functions\hfil\cr
\noalign{\vskip 3pt}
$EE$& 90.1& 69.0& 59.9& 89.2\cr
$TE$& 60.2& 45.4& 72.5& 42.6\cr
\noalign{\vskip 6pt\hrule\vskip 6pt}
\multispan5\hfil Pseudo-collapsed 3-point functions\hfil\cr
\noalign{\vskip 3pt}
$EEE$& 65.3& 57.5& 49.7& 64.6\cr
$TTE$& 99.7& 97.6& 97.0& 98.4\cr
$TEE$& 98.2& 87.7& 86.2& 87.7\cr
\noalign{\vskip 6pt\hrule\vskip 6pt}
\multispan5\hfil Equilateral 3-point functions\hfil\cr
\noalign{\vskip 3pt}
$EEE$& 76.1& 46.3& 89.1& 49.2\cr
$TTE$& 98.2& 98.0& 98.9& 95.6\cr
$TEE$& 94.8& 87.4& 95.0& 85.5\cr
 \noalign{\vskip 6pt\hrule\vskip 3pt}}}
\endPlancktable                    
\endgroup
\end{table}

The probabilities of obtaining values of the $\chi^2$ statistic for
the \Planck\ fiducial $\Lambda$CDM model at least as large as the
observed temperature and polarization values are provided in
Table~\ref{tab:prob_npt}.
It is worth noting that the values of the $N$-point functions and
their associated errors are strongly correlated between different
angular separations. The estimated probabilities, which take into
account such correlations, therefore provide more reliable information
on the goodness-of-fit between the data and the simulations than a
simple inspection of the figures can reveal.

The $N$-point function results show excellent consistency between the
CMB temperature maps estimated using the different
component-separation methods. Some differences between results for the
2015 and 2018 temperature data sets are caused by the use of different
masks in the analysis, and the adoption of the pseudo-inverse matrix
in the computation of the $\chi^2$ statistic, as described by
Eq.~\eqref{eqn:chisqr_stat}.
In the case of polarization, some scatter is 
observed between the functions computed for different methods, which
is a consequence of the relatively low signal-to-noise ratio of the
polarized data on large angular scales.
Interestingly, a tendency towards very high probability values is
observed for the pseudo-collapsed $TTQ_\mathrm{r}$ 3-point functions
for all methods, and for the equilateral $TQ_\mathrm{r}Q_\mathrm{r}$
functions in the case of \commander\ and \sevem.

As an alternative to the Stokes parameters, we also consider
$N$-point functions computed from the temperature and $E$-mode
polarizations maps.
The probabilities of obtaining values of the $\chi^2$ statistic for
the \Planck\ fiducial $\Lambda$CDM model at least as large as the
observed temperature and polarization values are provided in
Table~\ref{tab:prob_npt_teb}. 
Here, we see that the most significant deviations between the data and
the simulations occur for the $TTE$ 3-point functions for 
all component-separation methods. 

Nevertheless, we conclude that no strong evidence is found for statistically
significant deviations from Gaussianity of the CMB temperature and
polarization maps using $N$-point correlation functions.

Finally, we note that the results for the $TT$ correlation function
confirm the lack of structure at large separation angles, noted in the
WMAP first-year data by \citet{bennett2003a} and in previous
\Planck\ analyses \citepalias{planck2013-p09,planck2014-a18}.  We will
discuss this issue further in Sect.~\ref{sec:n_point_anomalies}, where
we also consider the behaviour of the $TQ_\mathrm{r}$ correlation
function.

\subsection{Minkowski functionals}
\label{sec:Minkowski_functionals}

\begin{table}[hbtp!]  
\begingroup     
\newdimen\tblskip \tblskip=5pt
\caption{Probability $P\left( \chi^2>\chi^2_{\rm Planck}\right)$ as a
  function of resolution determined using normalized real-space MFs for the
  temperature and polarization $E$-mode data.  In this table,
  large probabilities correspond to anomalously low $\chi^2$. }
\label{tab:mfs_smoothscales_te}
\nointerlineskip 
\vskip -3mm 
\footnotesize  
\setbox\tablebox=\vbox{ 
   \newdimen\digitwidth 
   \setbox0=\hbox{\rm 0} 
   \digitwidth=\wd0 
   \catcode`*=\active 
   \def*{\kern\digitwidth}
   \newdimen\signwidth 
   \setbox0=\hbox{+} 
   \signwidth=\wd0 
   \catcode`!=\active 
   \def!{\kern\signwidth}
\halign{\hbox to 1.2in{#\leaderfil}\tabskip 4pt&
\hfil#\hfil&
\hfil#\hfil&
\hfil#\hfil&
\hfil#\hfil\tabskip 0pt\cr
\noalign{\doubleline}
\noalign{\vskip -1pt}
\omit&\multispan4 \hfil Probability [\%] \hfil\cr
\noalign{\vskip -4pt}
\omit&\multispan4\hrulefill\cr
\omit\hfil $N_{\rm side}$\hfil& {\tt Comm.}& {\tt NILC}& {\tt SEVEM}& {\tt SMICA}\cr
\noalign{\vskip 3pt\hrule\vskip 6pt}
\multispan5\hfil Temperature\hfil\cr
\noalign{\vskip 3pt}
2048& 98.8& 99.8& 66.0& 98.0\cr 
1024& 69.2& 81.9& 63.2& 76.6\cr 
*512& 78.0& 87.2& 81.7& 80.7\cr 
*256& 95.4& 93.8& 81.5& 93.9\cr 
*128& 85.7& 97.2& 79.7& 85.4\cr 
**64& 99.0& 98.2& 63.2& 90.0\cr 
**32& 94.4& 86.9& 86.1& 96.9\cr 
**16& 71.5& 57.5& 52.5& 24.0\cr 
\noalign{\vskip 6pt\hrule\vskip 6pt}
\multispan5 \hfil $E$ polarization\hfil\cr
\noalign{\vskip 3pt}
1024& 90.2& 67.2& 58.0& 43.0\cr 
*512& 17.8& 84.8& *9.1& 75.8\cr 
*256& 94.3& *5.3& 46.2& 10.4\cr 
*128& 49.1& 15.0& 50.4& 26.2\cr 
**64& 14.7& 18.8& 80.4& 34.9\cr 
**32& 39.7& 14.1& 82.5& 29.2\cr 
**16& 93.9& *1.0& *7.5& 63.1\cr 
\noalign{\vskip 6pt\hrule\vskip 3pt}}} %
\endPlancktable 
\endgroup %
\end{table}

\begin{table}[hbtp!]  
\begingroup     
\newdimen\tblskip \tblskip=5pt 
\caption{Probability $P\left( \chi^2>\chi^2_{\rm Planck}\right)$ as a
  function of needlet scale.  In this table, large probabilities
  correspond to anomalously low $\chi^2$.}
\label{tab:mfs_needlet_te}
\nointerlineskip 
\vskip -3mm 
\footnotesize  
\setbox\tablebox=\vbox{ 
   \newdimen\digitwidth 
   \setbox0=\hbox{\rm 0} 
   \digitwidth=\wd0 
   \catcode`*=\active 
   \def*{\kern\digitwidth}
   \newdimen\signwidth 
   \setbox0=\hbox{+} 
   \signwidth=\wd0 
   \catcode`!=\active 
   \def!{\kern\signwidth}
\halign{\hbox to 1.2in{#\leaderfil}\tabskip 4pt&
\hfil#\hfil&
\hfil#\hfil&
\hfil#\hfil&
\hfil#\hfil\tabskip 0pt\cr
\noalign{\doubleline}
\noalign{\vskip -1pt}
\omit&\multispan4 \hfil Probability [\%] \hfil\cr
\noalign{\vskip -4pt}
\omit&\multispan4\hrulefill\cr
\omit\hfil Needlet scale ($\ell$ range)\hspace{3pt}\hfil& {\tt Comm.}& {\tt NILC}& {\tt SEVEM}& {\tt SMICA}\cr
\noalign{\vskip 3pt\hrule\vskip 6pt}
\multispan5\hfil Temperature\hfil\cr
\noalign{\vskip 3pt}
1 (1--4)&      *4.1& *7.1& *8.7& 20.6\cr 
2 (2--8)&      *4.5& *3.9& *8.7& *5.7\cr 
3 (4--16)&     *9.0& 11.5& 30.1& *4.9\cr 
4 (8--32)&     *7.0& *4.6& *2.8& *2.4\cr 
5 (16--64)&    87.8& 94.1& 90.9& 89.1\cr 
6 (32--128)&   46.1& 47.2& 29.6& 53.6\cr 
7 (64--256)&   23.4& *8.9& 25.7& 48.5\cr 
8 (128--512)&  47.5& 91.1& 62.5& 55.6\cr 
9 (256--1024)& 52.4& 91.5& 91.1& 40.9\cr
\noalign{\vskip 3pt\hrule\vskip 6pt}
\multispan5\hfil $E$ polarization\hfil\cr
\noalign{\vskip 3pt}
1 (1--4)&      37.5& 22.8& 44.8& 27.4\cr 
2 (2--8)&      22.9& 52.5& 79.4& 67.4\cr 
3 (4--16)&     64.1& 85.1& 92.1& 29.5\cr 
4 (8--32)&     97.3& 76.5& 72.4& 54.7\cr 
5 (16--64)&    45.2& 58.9& *3.7& 61.4\cr 
6 (32--128)&   64.7& 93.0& 74.7& 92.3\cr 
7 (64--256)&   40.3& 52.8& 84.0& 97.9\cr 
8 (128--512)&  23.5& 83.8& 64.6& 59.3\cr 
9 (256--1024)& 53.0& 63.7& *1.6& *5.9\cr 
\noalign{\vskip 3pt\hrule\vskip 3pt}}} %
\endPlancktable 
\endgroup %
\end{table}

In this section, we present a morphological analysis of the 
\Planck\ 2018 temperature and polarization CMB maps using
Minkowski functionals. 
The Minkowski functionals (hereafter MFs) describe the morphology of
fields in any dimension and have long been used to investigate
non-Gaussianity and anisotropy in the CMB 
\citep[see][and references therein]{planck2013-p09}.
They are additive for disjoint regions of the sky and invariant under
rotations and translations. 
For the polarization data, we analyse the scalar $E$-mode 
representation, since the MFs computed from the
spin-2 $Q$ and $U$ Stokes parameters are no longer invariant under
rotation after the application of a mask
\citep{Chingangbam2017}. 

We compute MFs for the regions colder and hotter than a given
threshold $\nu$, usually defined in units of
the sky rms amplitude, $\sigma_0$. 
The three MFs, namely the area $V_0(\nu)=A(\nu)$, the
perimeter $V_1(\nu)=C(\nu)$, and the genus $V_2(\nu)=G(\nu)$, are
defined respectively as
\begin{linenomath*}
\begin{align} 
V_0(\nu) &\equiv \frac{N_\nu}{N_{\rm pix}}, \\
V_1(\nu) &\equiv \frac{1}{4A_{\rm tot}}\sum_{i}S_i, \\
V_2(\nu) &\equiv \frac{1}{2\pi A_{\rm tot}}\big( N_{\rm hot} - N_{\rm
  cold}\big), 
\end{align}
\end{linenomath*}
where $N_{\nu}$ is the number of pixels with $|\Delta T| / \sigma_0 >
|\nu|$, $N_{\rm pix}$ is the total number of available pixels, $A_{\rm
  tot}$ is the total area of the available sky, $N_{\rm hot} (\nu)$ is the
number of compact hot spots, $N_{\rm cold}(\nu)$ is the number of compact
cold spots, and $S_i(\nu)$ is the contour length of each hot or cold spot.  
There are two approaches to the calculation of $\sigma_{0}$.
The first possibility is to use a population rms, which can be inferred from the
average variance of the simulations. Using this estimator provides robust
results for low resolutions. An alternative is to use the sample rms,
estimated directly from the map in question. \citet{cm1603} have shown
that this approach increases the sensitivity of MF-based tests, and thus
we adopt this definition of $\sigma_{0}$ in our analysis.

Furthermore, 
the MFs can be written as a product of a function $A_k$ ($k=0,1,2$),
which depends only on the Gaussian power spectrum, and $v_k$, which is
a function only of the threshold $\nu$ \citep[see
e.g.,][]{vanmarcke1983,pogosyan2009a,gay2012a,matsubara2010,fantaye2014a}. 
This factorization is valid in the weakly non-Gaussian case.  In this
paper, we use the normalized MFs, $v_k$, to focus on deviations from
Gaussianity, with reduced sensitivity to the cosmic variance of the
Gaussian power spectrum. However, we have verified that the results
derived using both normalized or unnormalized MFs are consistent in
every 
configuration.\footnote{However, we note that, for the unnormalized
  MFs, the \nside\ = 16 map is sensitive to the 
  $\sigma_0$ definition. Using the population rms yields more
  consistent results with the normalized MFs, and between the data and
  simulations, than when using the sample rms.}
The analytical expressions are
\begin{linenomath*}
\begin{align}
V_{k}(\nu) &=A_{k}v_{k}(\nu),\\
v_{k}(\nu) &= {\rm e}^{-\nu^{2}/2} H_{k-1}(\nu), \quad k \leq 2, \label{eq:nuk1}
\end{align}
\end{linenomath*}
with $H_n$, the Hermite function,
\begin{linenomath*}
\begin{equation}
H_{n}(\nu)={\rm e}^{\nu^{2}/2}\left(-\frac{\rm d}{{\rm d}\nu}\right)^{n} {\rm e}^{-\nu^{2}/2}.
\end{equation}
\end{linenomath*}

The amplitude $A_k$ depends only on the shape of the power spectrum
$C_{\ell}$ through the parameters $\sigma_{0}$ and $\sigma_{1}$,
the rms of the field and its first derivative, respectively:
\begin{linenomath*}
\begin{equation} 
A_{k} = \frac{1}{(2\pi)^{(k+1)/2}}\frac{\omega_{2}}{\omega_{2-k} \;\omega_{k}}\left( \frac{\sigma_{1}}{\sqrt{2}\sigma_{0}}\right)^{k},\quad k \leq 2,
 \end{equation}
\end{linenomath*}
with $\omega_{k}\equiv\pi^{k/2}/\Gamma(k/2+1)$.

In order to characterize the MFs, we consider two approaches for the
scale-dependent analysis of the temperature and polarization sky maps:
in real space via a standard Gaussian smoothing and degradation of the
maps; and in harmonic space by using needlets. Such a complete
investigation should provide insight regarding the harmonic and spatial
nature of possible non-Gaussian features detected with the MFs.

First, we undertake a real-space analysis by computing the three
normalized functionals described above at different resolutions and
smoothing scales for each of the four component-separation
methods. The appropriate common mask is applied for a given scale.
The MFs are evaluated for 12 thresholds ranging between $-3$ and $3$
in $\sigma_0$ units, providing a total of 36 different statistics
$\y=\lbrace v_0, v_1, v_2\rbrace$. A $\chi^2$ value is then
computed by combining these, assuming a Gaussian likelihood for
  the MFs at every threshold, taking into account their correlations
\citep{ducout2012} using a covariance matrix computed from the FFP10
simulations: 
\begin{align}
\chi^2(\y) &\equiv [\y - \bar{\y}^{\:\rm sim}]^T\tens{C}^{-1}[\y-\bar{\y}^{\:\rm sim}]\\
            &= \sum_{i,j=1}^{36}C_{ij}^{-1}[y_i-\bar{y}_i^{\:\rm sim}][y_j-\bar{y}_j^{\:\rm sim}],
\end{align}
where $\bar{\y}^{\:\rm sim} \equiv \langle \y^{\:\rm sim }\rangle$ is
the mean of the statistics $\y$ computed on the simulations, 
$i,j$ are the threshold indices from the combined MFs, and
\begin{equation}
C_{ij} \equiv \langle (y_i^{\:\rm sim} - \bar{y}_i^{\:\rm sim}) (y_j^{\:\rm sim} - \bar{y}_j^{\:\rm sim}) \rangle
\end{equation}
is the covariance matrix estimated from the FFP10 simulations. 
The covariance matrix is well converged for this low number of
statistics (i.e., 36).

\begin{figure*} [hbtp!]
\begin{center}
  \includegraphics[width=\textwidth]{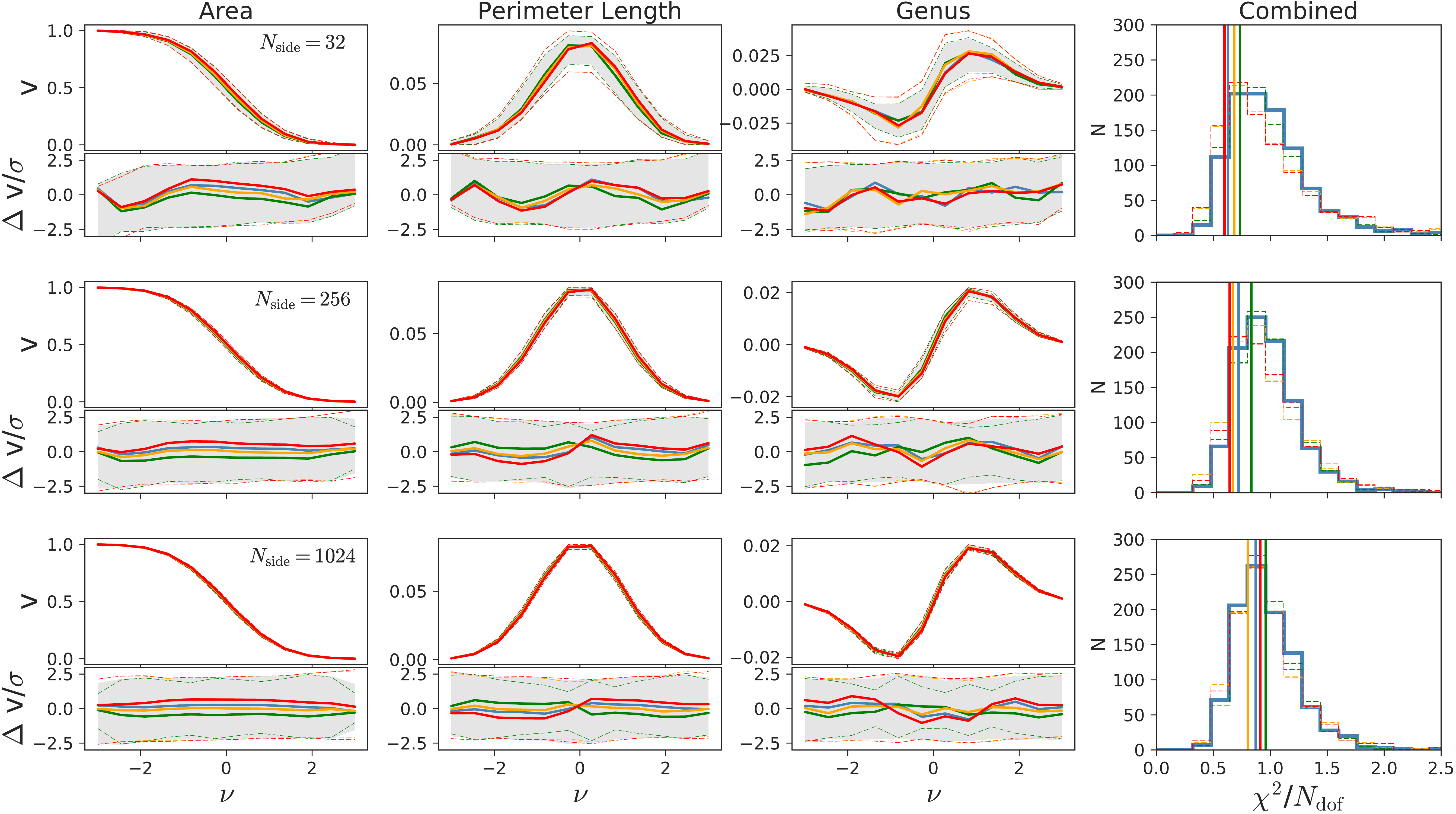}
  \caption{Real-space normalized MFs determined from the \Planck\ 2018 temperature
    data using the four component-separated maps, \commander\ (red),
    \nilc\ (orange), \sevem\ (green), and \smica\ (blue). The grey
    region corresponds to the 99th percentile area, estimated from the
    FFP10 simulations processed by the \smica\ method, while the
    dashed curves with matching colours outline the same interval for the
    other component-separation methods. Results are shown for analyses
    at \nside\ = 32, 256, and 1024.
    The right-most column shows the $\chi^2$
    obtained by combining the three MFs in real space with an
    appropriate covariance matrix derived from FFP10 simulations. The
    vertical lines correspond to values from the \Planck\ data.}
\label{fig:mfs_smoothscales_t}
\end{center}
\end{figure*}

\begin{figure*} [hbtp!]
\begin{center}
  \includegraphics[width=\textwidth]{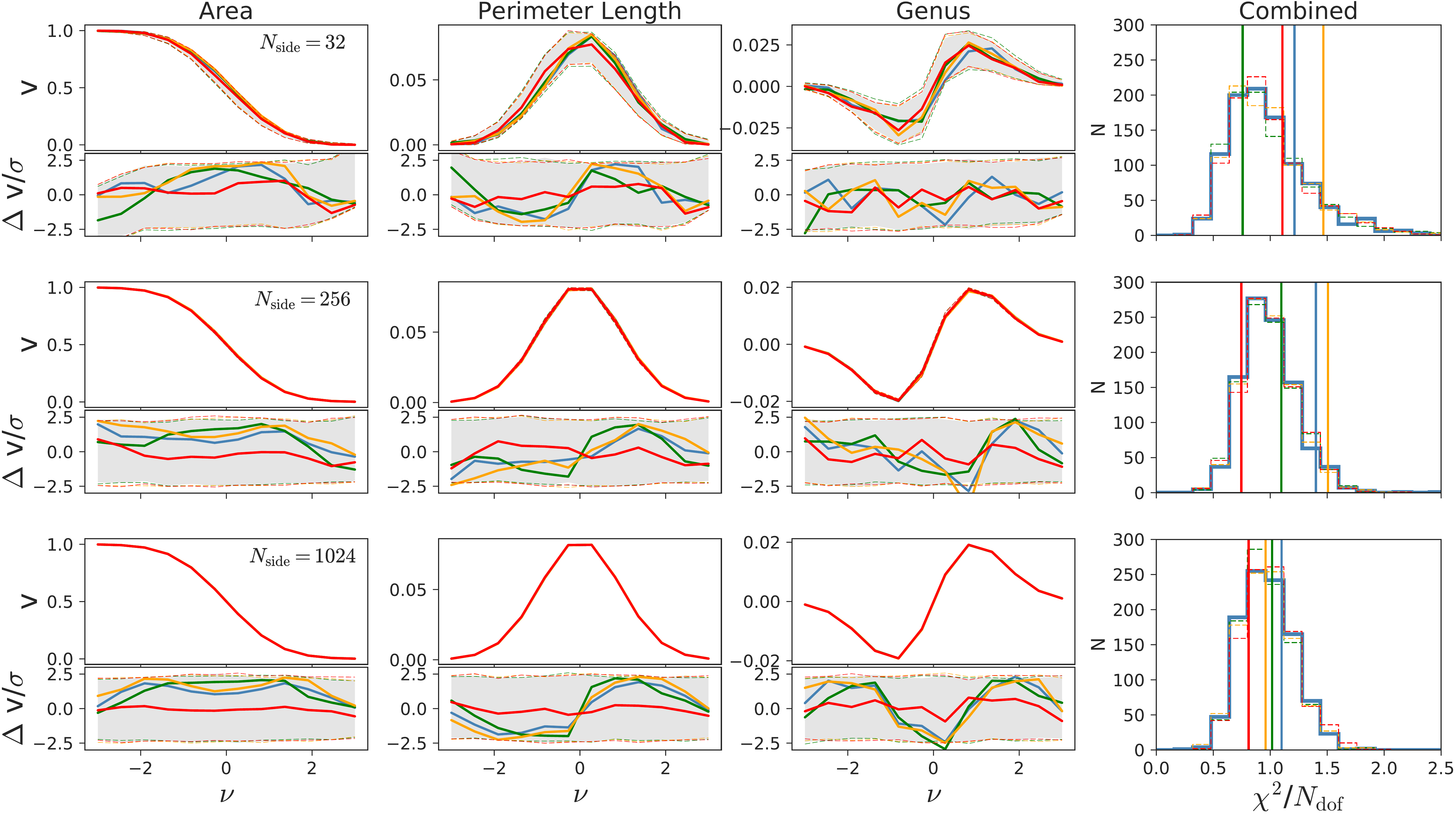}
\caption{As in Fig.~\ref{fig:mfs_smoothscales_t}, but for the \Planck\ 2018
  $E$-mode
  polarization data. The \Planck\ data are consistent with the Gaussian
  FFP10 simulations, but variations between the
  different component-separation methods are evident.}
\label{fig:mfs_smoothscales_e}
\end{center}
\end{figure*}

The results for temperature and $E$-mode polarization data are shown in
Figs.~\ref{fig:mfs_smoothscales_t} and
\ref{fig:mfs_smoothscales_e}, respectively. 
The first three columns of panels in these figures show the normalized
MFs together with their variance-weighted difference with respect to
the mean of the simulations for the three MFs.  The right-most column 
of panels in Figs.~\ref{fig:mfs_smoothscales_t} and 
\ref{fig:mfs_smoothscales_e} presents the $\chi^2$
obtained when the three MFs are combined with an appropriate
covariance matrix derived using the FFP10 simulations.  The vertical lines in 
these figures represent the data, with different colours for the
different component-separation methods. The grey shaded regions in the
MFs plot and the histogram in the $\chi^2$ plot are determined from
the FFP10 simulations.  
Table~\ref{tab:mfs_smoothscales_te} presents the corresponding {\it
  p}-values determined for the different component-separation
techniques and map resolutions, between \nside\ = 16 and \nside\ =
2048 for temperature, and between \nside\ = 16 and \nside\ = 1024 for
the polarization $E$-mode.

For the temperature results, the $\chi^2$ values computed for the
different component-separation methods are more consistent than was
the case for the \Planck\ 2015 analysis,
for all scales.  
In the case of the $E$-mode results, we find no significant
discrepancy between the \Planck\ data and the FFP10 simulations. The
striking variation in the {\it p}-values for the four
component-separation methods, is also observed when considering
individual realizations in the set of simulations.

As a complement to the pixel-based analysis, we also determine the MFs of
needlet coefficient maps on various scales
(see Table~\ref{tab:mfs_needlet_te}).  Measuring the MFs in
needlet space, as compared to the usual pixel-space case,
has two clear advantages: the needlet maps are minimally
affected by masked regions due to the localization of the needlet
filter in pixel space, especially at high-frequency; and the
double-localization properties of needlets (in real and harmonic
space) allow a much more precise, scale-by-scale, interpretation of
any possible anomalies. While the behaviour of standard all-scale
(pixel-based) MFs is contaminated by the large cosmic variance of the
low multipoles, this is no longer the case for MFs evaluated at the
highest needlet scales; in such circumstances, the variance of
normalized components may be shown to decrease steadily, entailing a
much greater detection power in the presence of anomalies. Finally,
and most importantly, the needlet MFs are more sensitive to the shape
of the power spectrum than the corresponding all-scale MFs. This is
because if one changes the shape of the power spectrum while still
keeping $\sum_{\ell}[2\ell+1]C_{\ell}$ constant, the pixel-space MFs
will not change but the needlet MFs are affected. This sensitivity to
the shape of the power spectrum can be used to understand in more
detail the nature of any possible non-Gaussianity detected by the
MF analysis.

The needlet components of a scalar CMB field are defined by
\citet{marinucci08a} and \citet{baldi09a}, and are given by
\begin{linenomath*}
\begin{align}
\beta _{j}(\vec{\hat n}) & =\sum_{\ell ={\mathbb{B}}^{j-1}}^{{\mathbb{B}}^{j+1}}b^{2}\(\frac{\ell }{{\mathbb{B}}^{j}}%
\)a_{\ell m}Y_{\ell m}(\vec{\hat n}) \\
 & =\sum_{\ell ={\mathbb{B}}^{j-1}}^{{\mathbb{B}}^{j+1}}b^{2}\(\frac{\ell }{%
{\mathbb{B}}^{j}}\)X_{\ell }(\vec{\hat n})\text{ ,}
\label{needfield} 
\end{align}
\end{linenomath*}
where $j$ on the left-hand side is the needlet index and $j$ on the
right-hand side is a power.
Here, $X_{\ell }(\vec{\hat n})$ denotes the component at multipole
$\ell$ of the CMB map $X(\vec{\hat n})$ (corresponding to temperature or the
polarization $E$ or $B$ modes), i.e.,
\begin{linenomath*}
\begin{equation}
X(\vec{\hat n})=\sum_{\ell }X_{\ell }(\vec{\hat n})\text{ ,}
\end{equation}
\end{linenomath*}
where $\vec{\hat n} \in S^{2}$ denotes the pointing direction, $\mathbb{B}$ is
a fixed parameter that controls the needlet's band width 
(usually taken to be between $1$ and $3$), and $b(.)$ is a smooth
function such that $\sum_{j}b^{2}(\ell/{\mathbb{B}}^{j})=1$  for all
$\ell$.  In \citet{fantaye2014a}, it is shown that a general analytical
expression for MFs at a given needlet scale $j$ can be written as
\begin{linenomath*}
\begin{equation} 
V_k^j = \sum_{i=0}^k  t_{(2-i)} A_i^j v_i, 
\end{equation}
\end{linenomath*}       
where $t_0=2$, $t_1=0$, and $t_2=4\pi$, are the Euler-Poincar\'{e}
characteristic, boundary length, and area of the full sphere,
respectively. The quantities $v_k$ are the normalized MFs given in 
Eq.~\eqref{eq:nuk1}, while the needlet-scale amplitudes $A_k^j$ have a similar
form to $A_k$, but with the variances of the map and its first
derivative given by
\begin{linenomath*}
\begin{align} 
\sigma_0^2 &= \sum_{\ell }b^{4}\(\frac{\ell }{{\mathbb{B}}^{j}}\)C_{\ell }%
\frac{2\ell +1}{4\pi } , \\
\sigma_1^2 &= \sum_{\ell }b^{4}\(\frac{\ell }{{\mathbb{B}}^{j}}\)C_{\ell }%
\frac{2\ell +1}{4\pi }\frac{\ell (\ell +1)}{2} .
 \end{align}
\end{linenomath*}

We adopted the needlet
parameters ${\mathbb{B}}=2$, $j=1$, 2, 3, 4, 5, 6, 7, 8, 9, and 10
for this analysis. Note that the $j$th
needlet scale has compact support over the multipole range
$[2^{j-1}, 2^{j+1}]$. For clarity in all the figures, we refer to the
different needlet scales by their central multipole $\ell_\mathrm{c} = 2^j$.

To obtain the needlet maps at different scales, we initially decompose
into spherical harmonics the temperature and polarization maps, inpainted
using diffusive and purified inpainting (see Appendix~\ref{sec:EB_maps}),
respectively, at \nside=1024.  In all cases, we set the maximum multipole
to $\ell_{\rm max}=2048$, which is twice the maximum resolution
considered for the needlet MF analysis. We then obtain the $j$th
needlet-scale map, by computing Eq.~\eqref{needfield} using the
\healpix\ \texttt{map2alm} routine at the appropriate
\nside. Specifically, we use \nside=16 for $j=1,2,3,4$, and
\nside=$2^{j}$ for the remaining needlet scales. These choices allow
us to adopt the same masks used for the pixel-space analysis without
alteration.

\begin{figure*} [hbtp!]
\begin{center}
\includegraphics[width=1\textwidth]{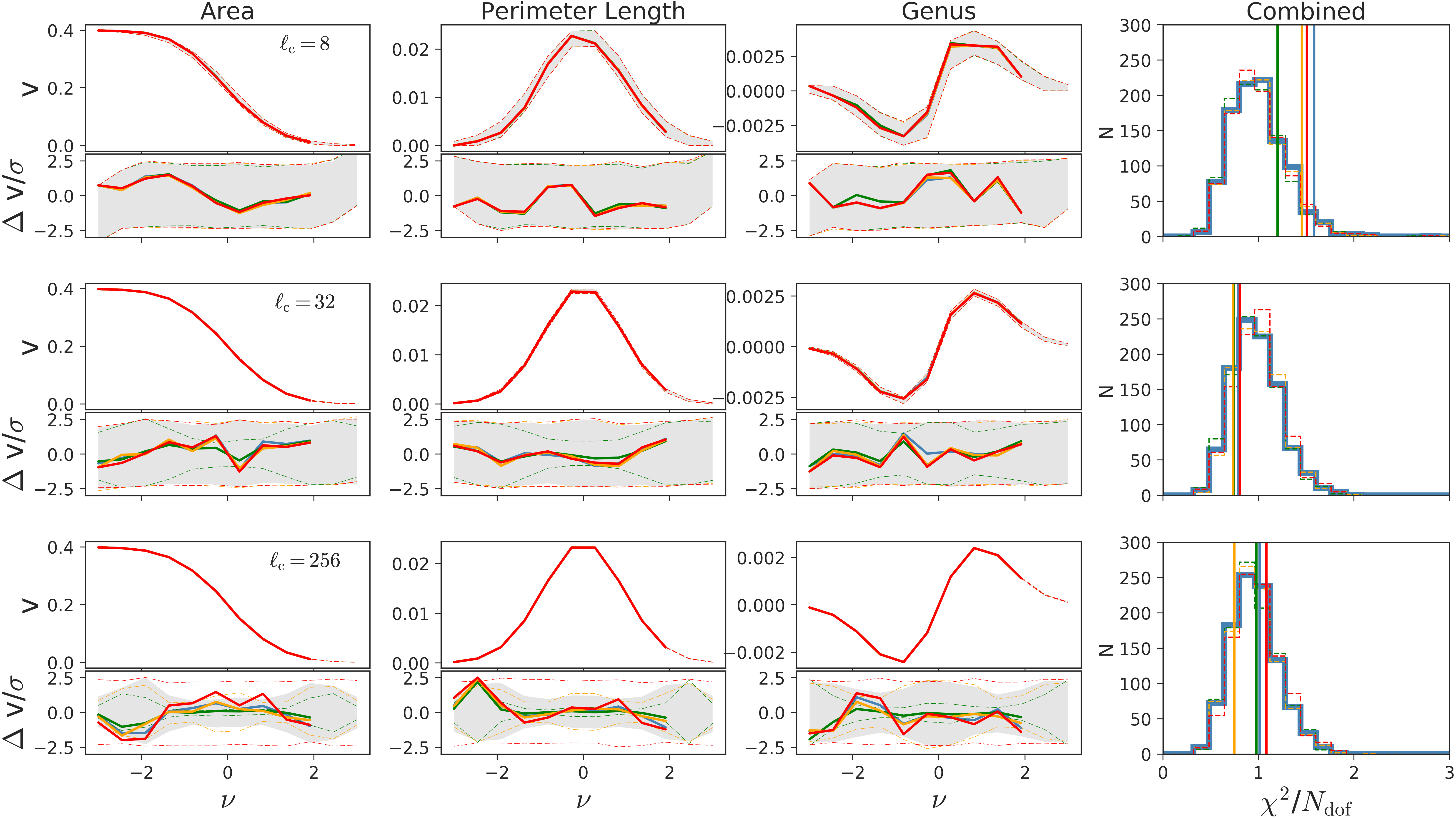}%
\caption{Needlet-space normalized MFs of the
 \Planck\ 2018 temperature data using the four
 component-separated maps, \commander\ (red), \nilc\ (orange),
 \sevem\ (green), and \smica\ (blue). The grey region corresponds to
 the 99th percentile area, estimated from the FFP10 simulations processed by
 the \smica\ method, while the dashed curves with matching colours 
 outline the same interval for
 the other component-separation methods. 
 The needlet MFs are denoted by the central
 multipole of the needlet filter $\ell_\mathrm{c} = 2^j$; the $j$th needlet
 parameter has compact support over the multipole range $[2^{j-1},
 2^{j+1}]$. The right-most column shows the $\chi^2$ obtained
 by combining the three MFs in needlet space
 with an appropriate covariance matrix derived from FFP10 simulations. The
 vertical lines correspond to values from the \Planck\ data.}
\label{fig:mfneedle1}
\end{center}
\end{figure*}

Once the needlet maps are obtained, we follow the identical procedure as in
the pixel-space case to compute the three MFs.  The results derived
from the \commander, \nilc, \sevem, and \smica\ component-separated
temperature maps for needlet scales ${\mathbb{B}}=2$, $j=2, 6, 9$ are shown in
Fig.~\ref{fig:mfneedle1}.  
As can be seen from both the MF and $\chi^2$ plots, the \Planck\ 2018
temperature data are consistent with the Gaussian FFP10
simulations. This is true for all the needlet scales considered.
Similar to the pixel-space MFs case, the results indicate a high
degree of consistency among the four component-separation methods.

\begin{figure*} [hbtp!]
\begin{center}
\includegraphics[width=1\textwidth]{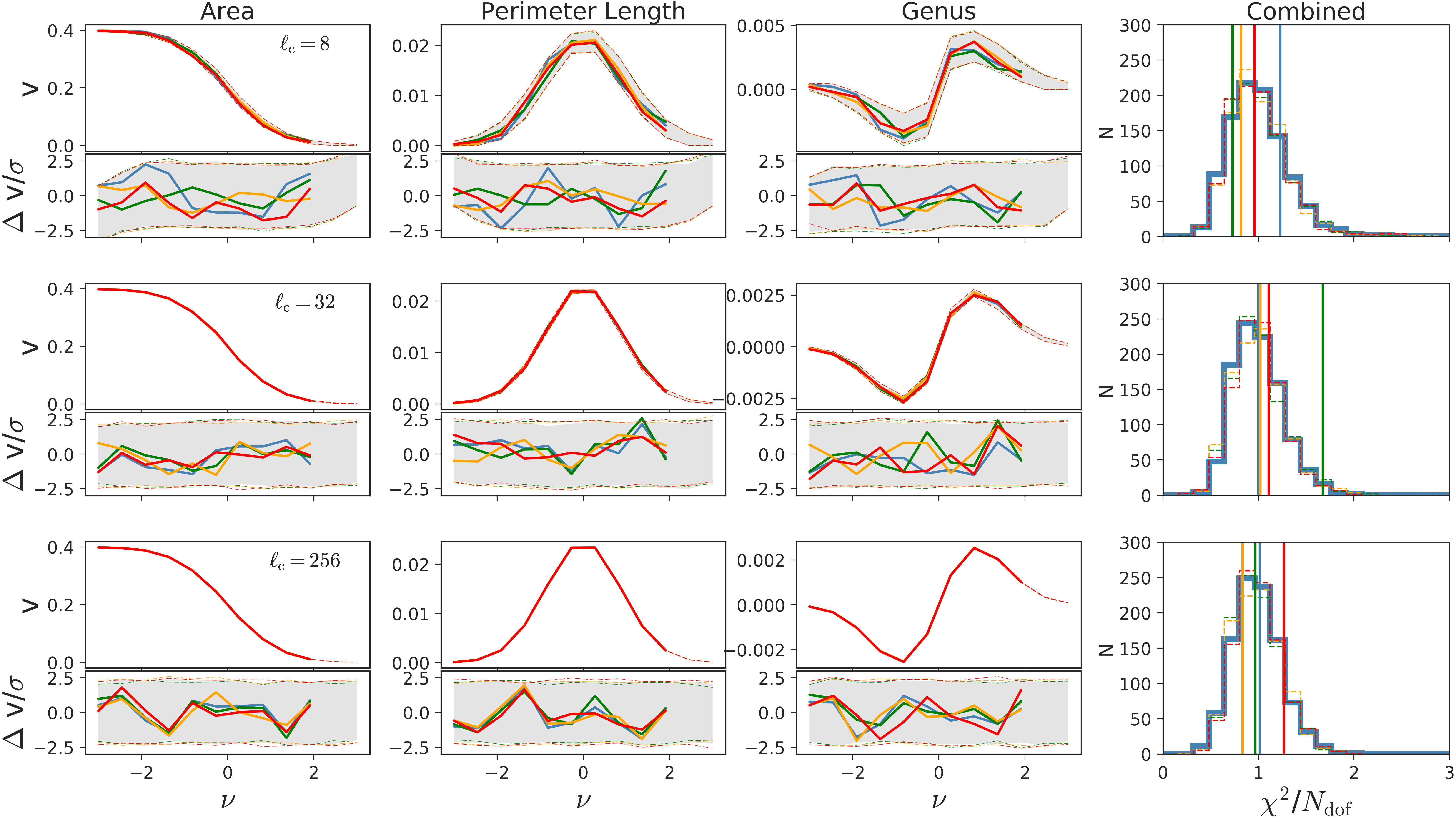}
\caption{As in Fig.~\ref{fig:mfneedle1}, but for the \Planck\ 2018 $E$-mode
  polarization data. The \Planck\ data are consistent with the Gaussian
  FFP10 simulations, but some variations between the
  different component-separation methods are evident. }
\label{fig:mfneedle2}
\end{center}
\end{figure*}

The results for the $E$-mode polarization data are shown in
Fig.~\ref{fig:mfneedle2}. Compared to the temperature results, the
different component-separation methods show greater variation,
although no significant deviations between data and simulations are
observed. A similar degree of scatter in the results determined for the
OEHD and HMHD component-separated maps suggests that 
noise can play a significant role in explaining
this variation.

In summary, both the pixel- and needlet-space MF analyses show that
the 2018 \Planck\ temperature and polarization maps are consistent with
the Gaussian simulations over angular scales corresponding to
$\ell_{\rm max}=2048$.

\subsection{Peak statistics}
\label{sec:peakstat}

\begin{figure*}
  \centering
  \begin{tabular}{cc}
  \includegraphics[width=88mm]{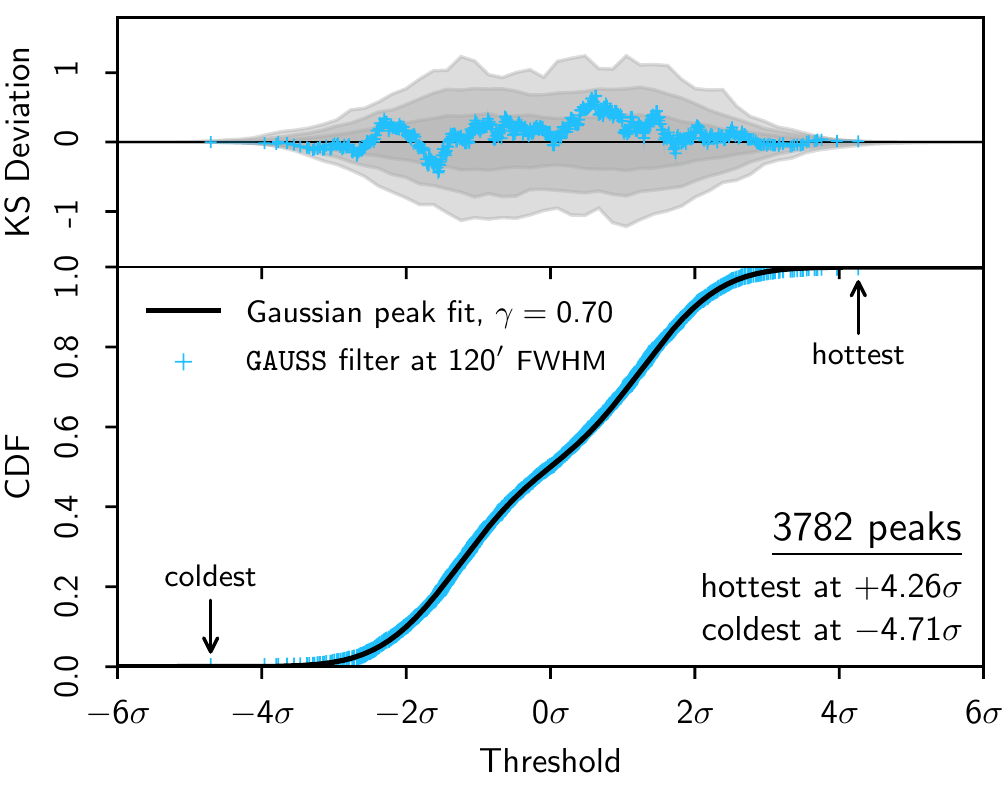} &
  \includegraphics[width=88mm]{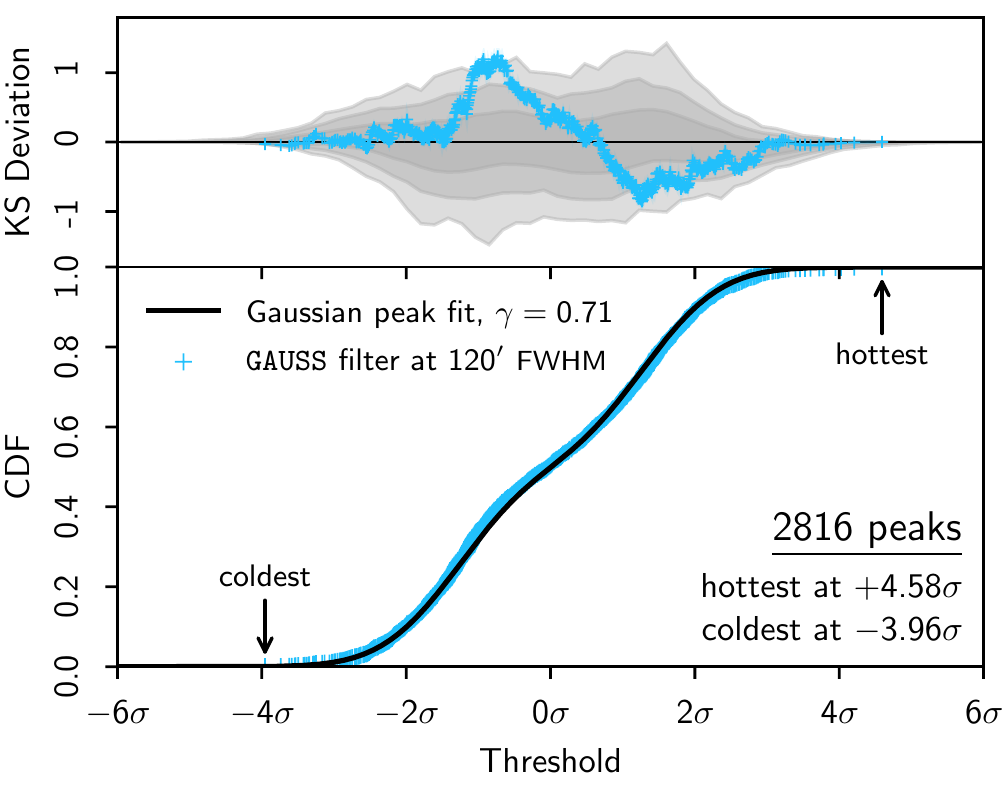} \\
  \includegraphics[width=88mm]{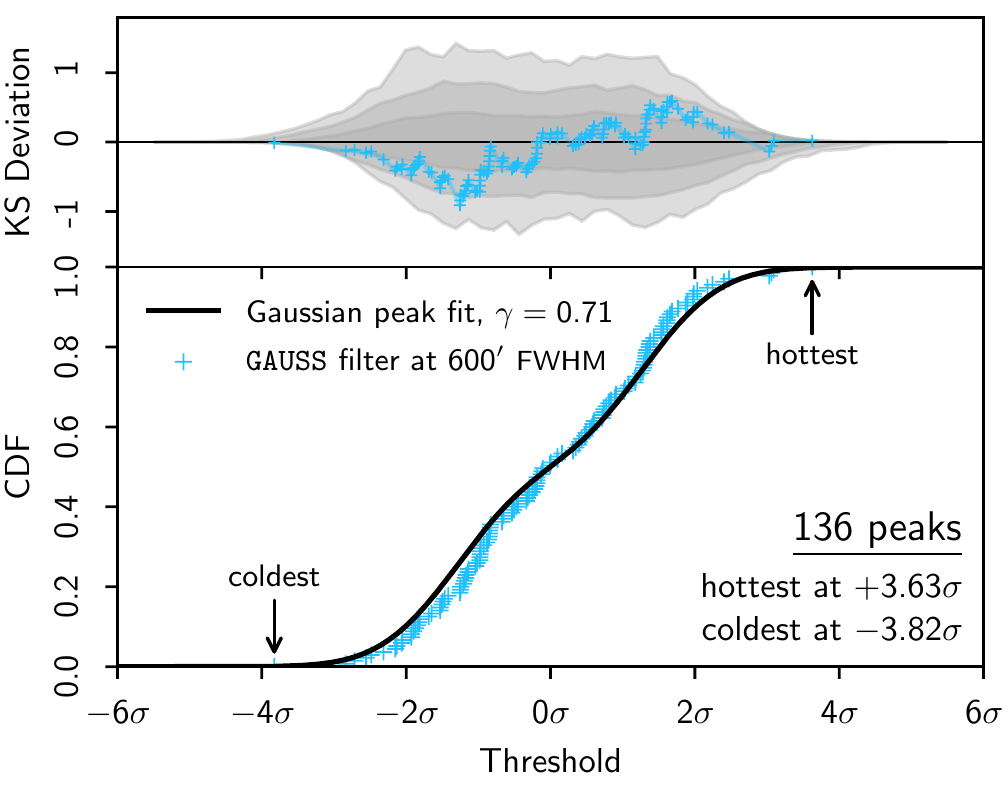} &
  \includegraphics[width=88mm]{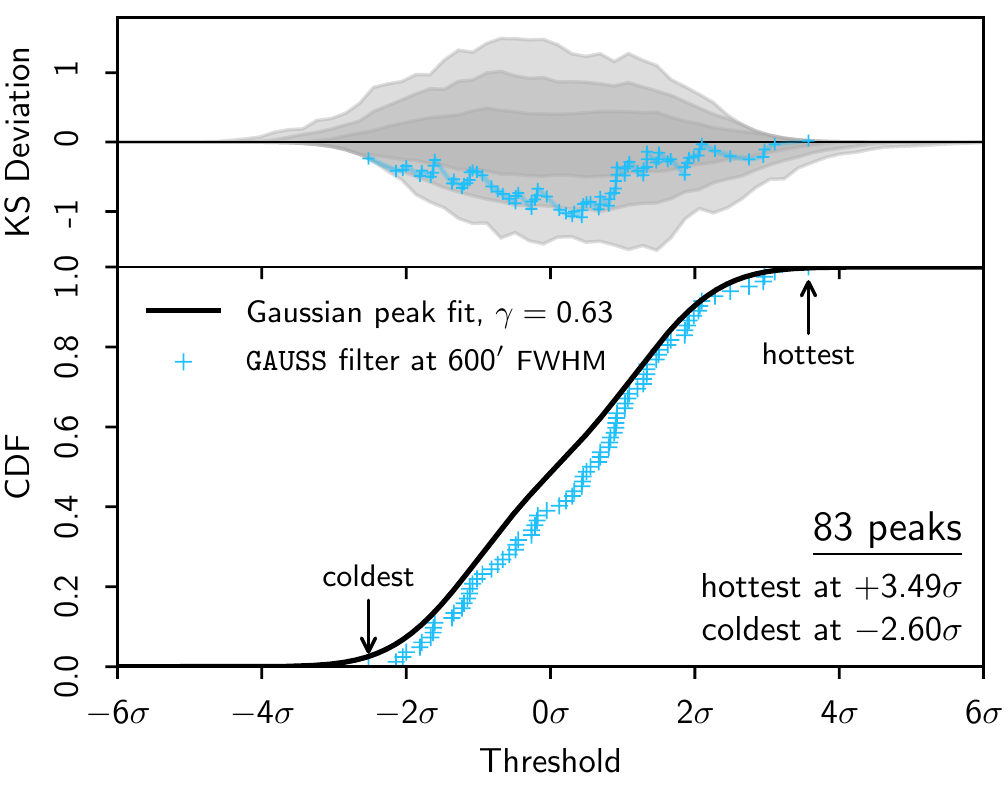}
  \end{tabular}
  \caption{ Cumulative density-function of the peak distributions for
    the \smica\ temperature $T$ (left) and reconstructed $E$-mode
    polarization (right) maps. The top row shows the peak CDF filtered
    with a GAUSS kernel of 120\arcmin~FWHM, the bottom row
    shows the peak CDF filtered with the same kernel of
    600\arcmin~FWHM. The spectral shape parameter $\gamma$ (see
    Eq.~\ref{eq:peaks:be}) is the best-fit value for the simulated
    ensemble, as indicated by the cyan circle in
    Fig.~\ref{fig:peaks:params}. No significant deviations from
    Gaussian expectations are observed. Similar results are obtained
    for other component-separation methods.}
  \label{fig:peaks:stat}
\end{figure*}

\begin{figure*}
  \centering
  \begin{tabular}{cc}
  \includegraphics[width=88mm]{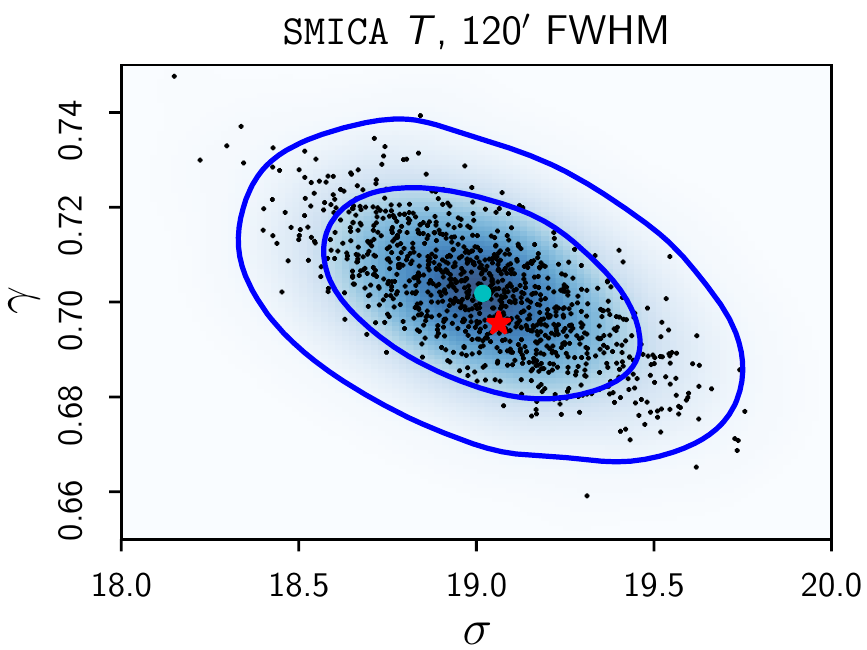} &
  \includegraphics[width=88mm]{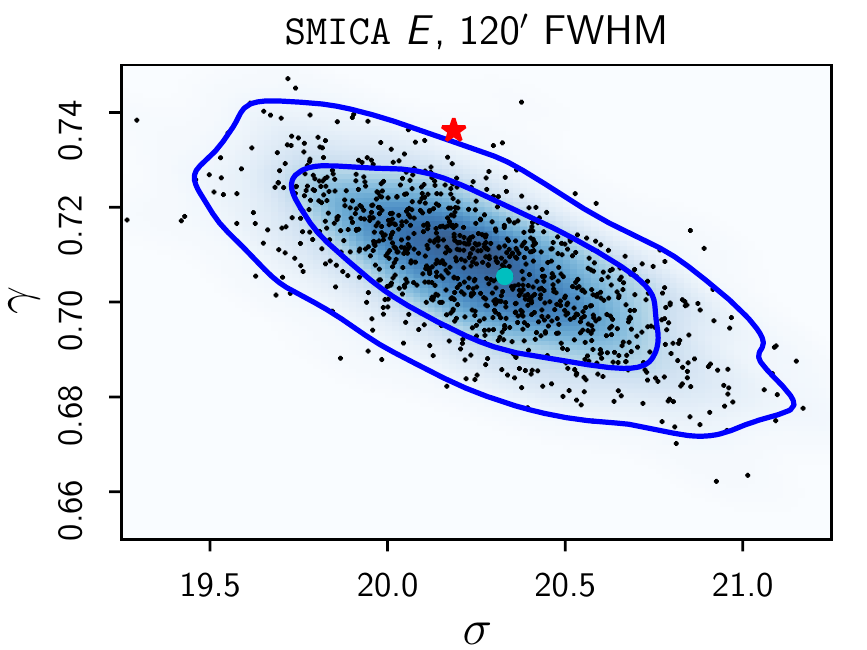} \\
  \includegraphics[width=88mm]{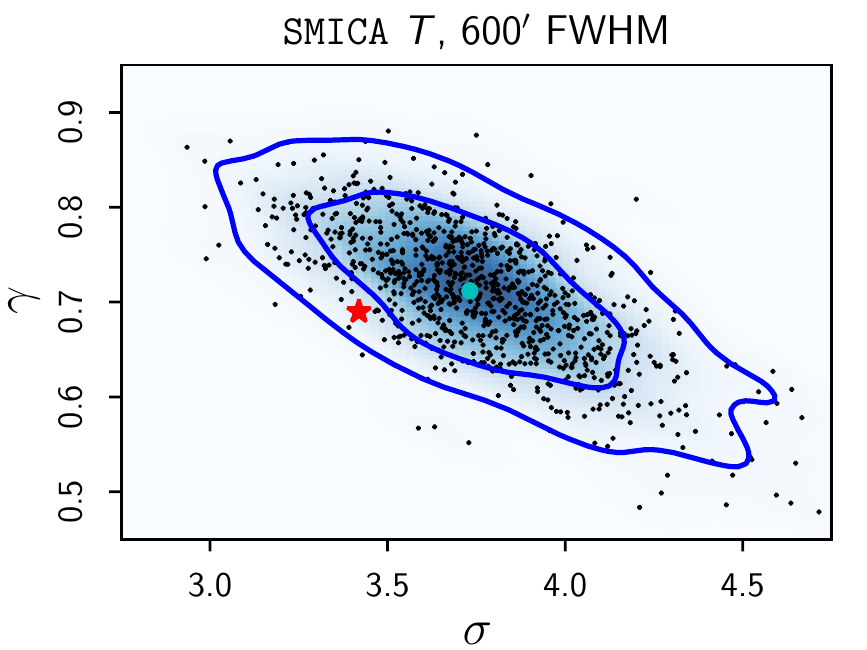} &
  \includegraphics[width=88mm]{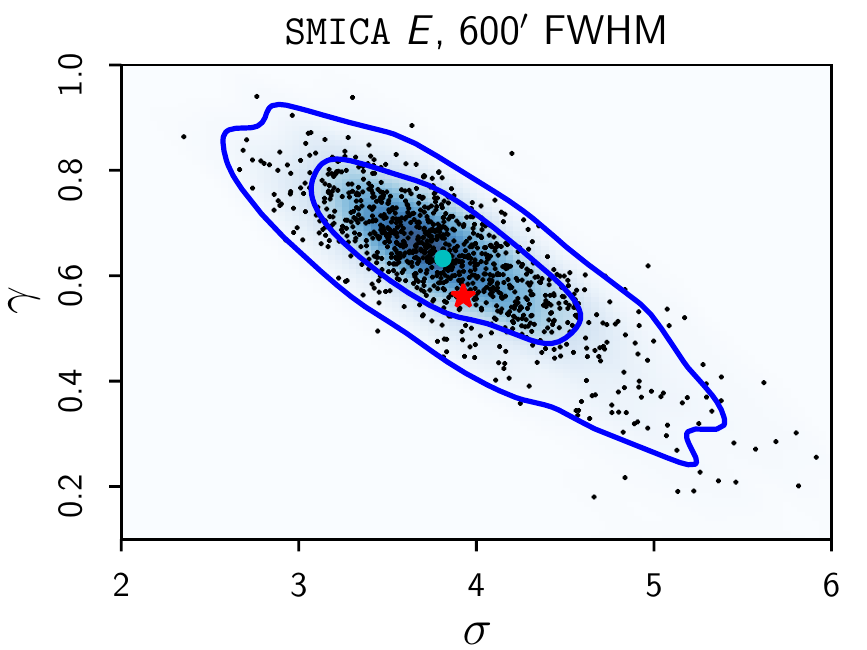}
  \end{tabular}
  \caption{ Distribution of best-fit Gaussian peak CDF spectral shape
    parameters, $\sigma$ and $\gamma$ (as defined in Eq.~\ref{eq:peaks:be}),
    recovered from FFP10 simulations, as indicated by the black dots and the
    smoothed density map, and compared to those derived for the observed sky
    (shown by the red star) for the \smica\ temperature $T$ (left) and the
    reconstructed $E$-mode polarization (right) maps. The blue contours enclose
    68\,\% and 95\,\% of the parameter distribution, and the cyan circle
    represents the best-fit parameters for the median peak CDF determined
    from simulations. The upper panel shows the peak CDF parameters for
    the \smica\ map filtered with a GAUSS kernel of $120\arcm$ FWHM, while
    the lower panel shows the corresponding peak CDF using the same kernel
    with $600\arcm$ FWHM. Similar results are obtained for the other
    component-separation methods.
  }
  \label{fig:peaks:params}
\end{figure*}

\begin{figure*}
  \centering
  \begin{tabular}{cc}
  \includegraphics[width=88mm]{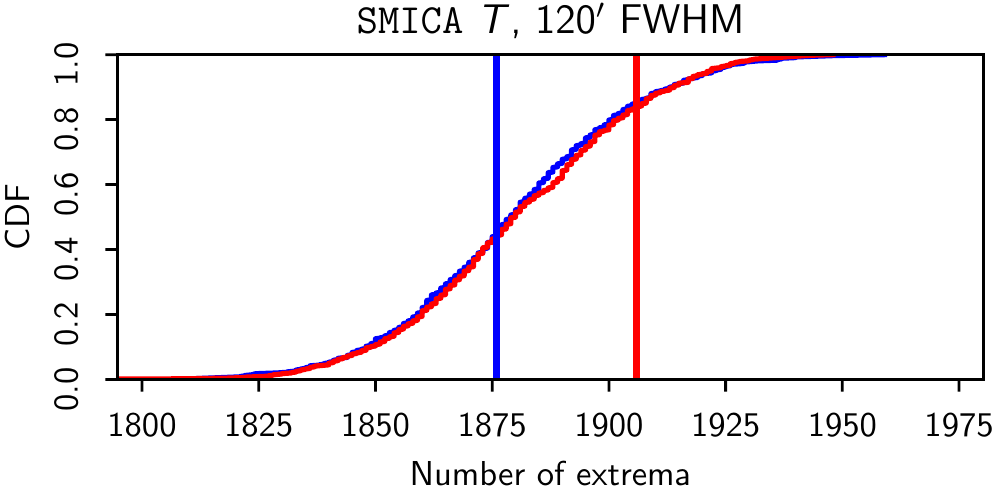} &
  \includegraphics[width=88mm]{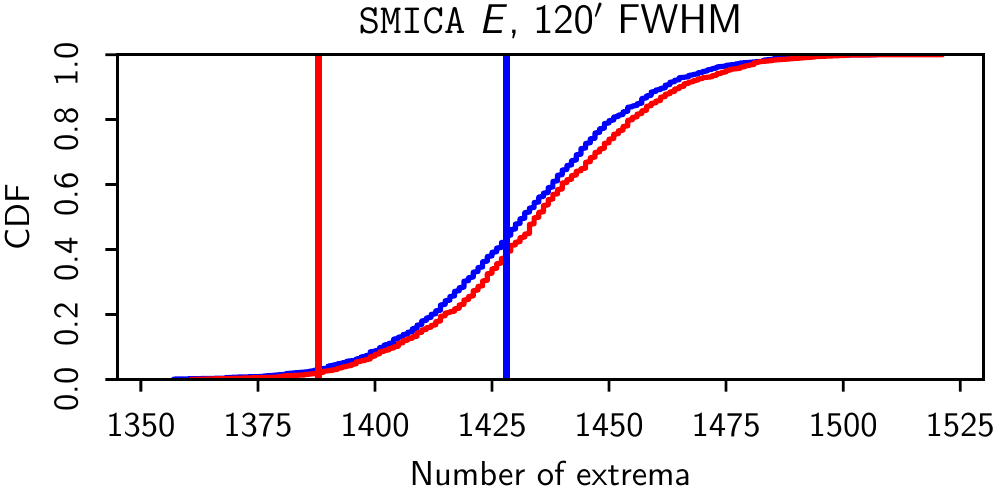} \\
  \includegraphics[width=88mm]{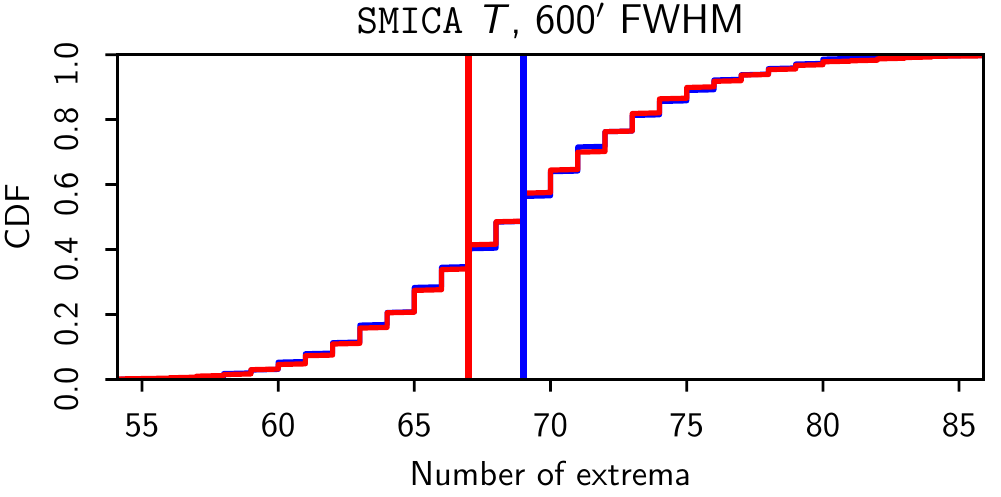} &
  \includegraphics[width=88mm]{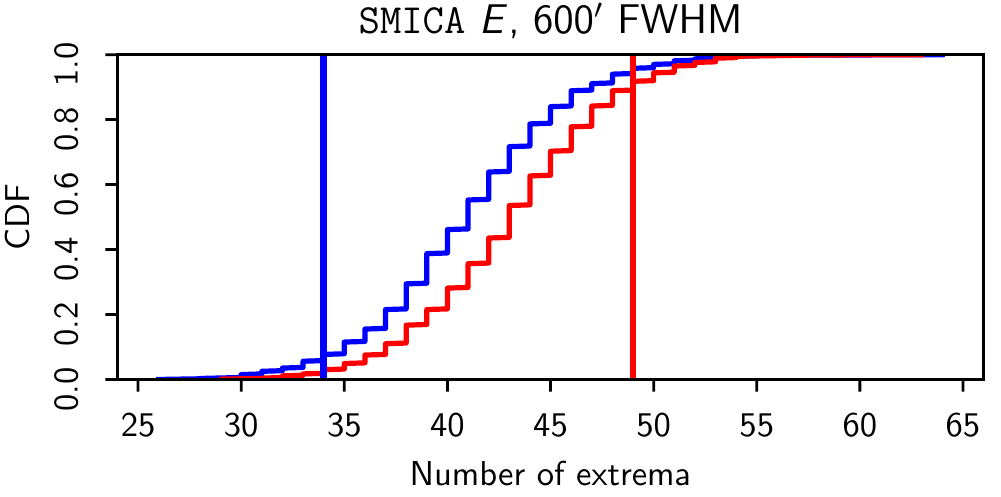}
  \end{tabular}
  \caption{ Cumulative density-function of the number of all extrema, 
    maxima (red) and minima (blue), derived from simulations, compared to
    the equivalent values observed for the \smica\ temperature $T$ (left) and the
    reconstructed $E$-mode polarization (right). The upper panel shows the peak
    counts for maps filtered with a GAUSS kernel of $120\arcm$ FWHM.
    The lower panel shows the corresponding peak count CDF for the same
    kernel of $600\arcm$ FWHM. Similar results are obtained for the other
    component-separation methods.
  }
  \label{fig:peaks:counts}
\end{figure*}

\begin{figure*}
  \centering
  \begin{tabular}{cc}
  \includegraphics[width=88mm]{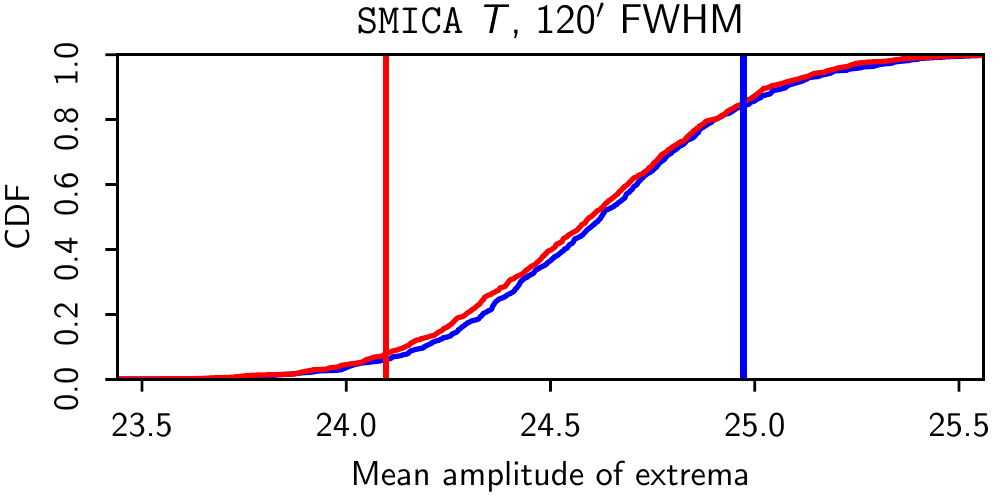} &
  \includegraphics[width=88mm]{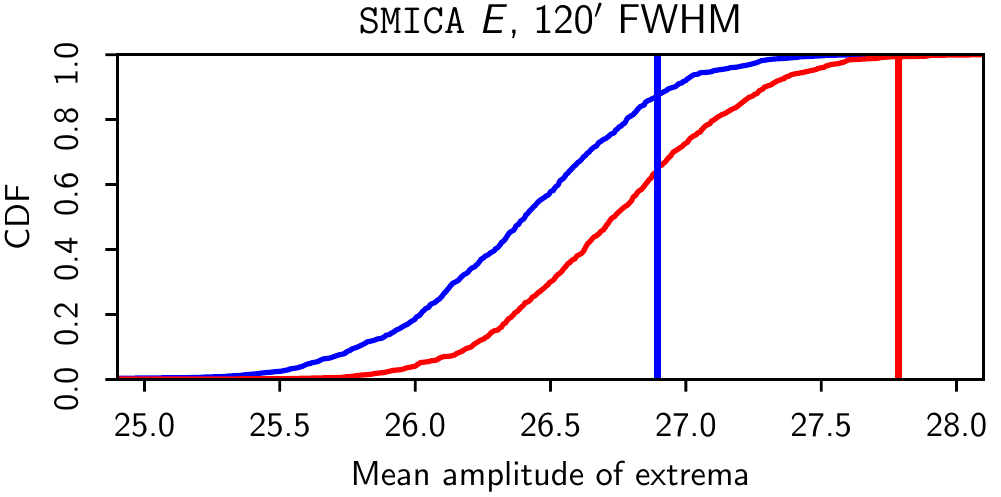} \\
  \includegraphics[width=88mm]{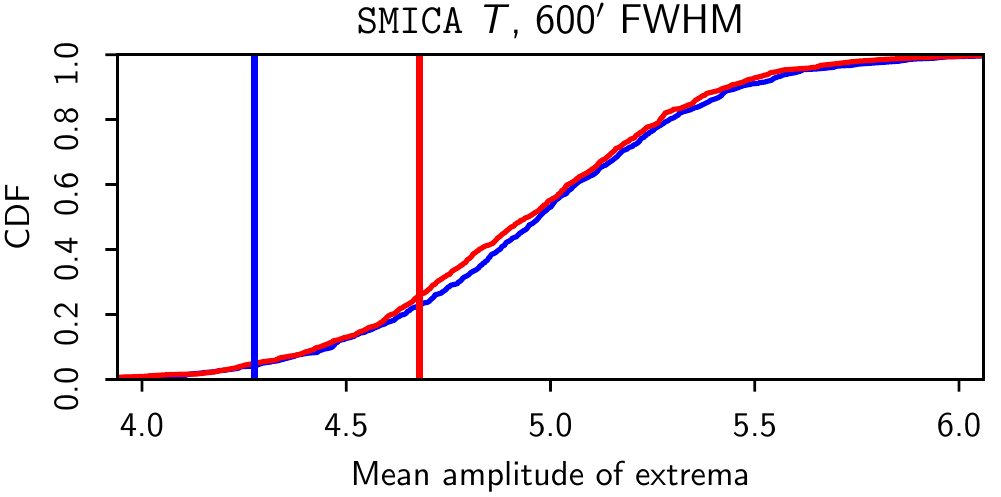} &
  \includegraphics[width=88mm]{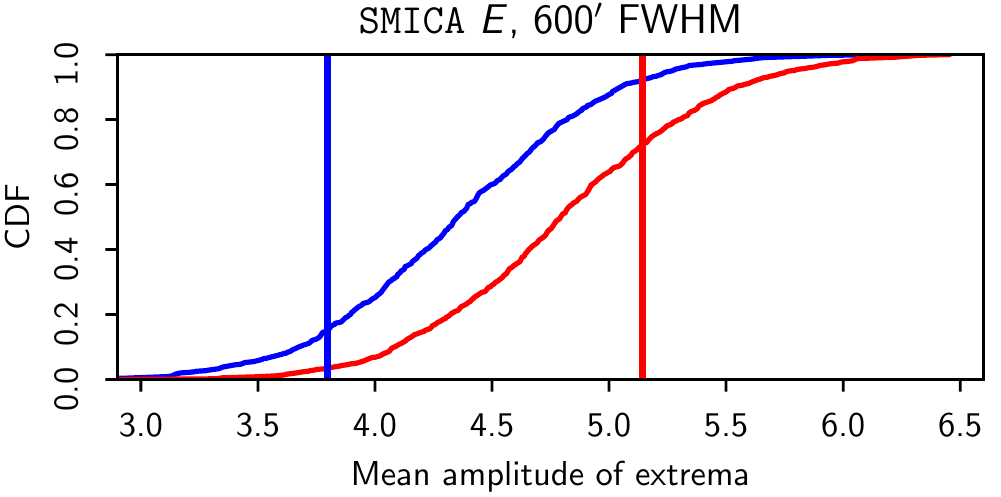}
  \end{tabular}
  \caption{ Cumulative density-function of the mean amplitude of all extrema, 
    maxima (red) and minima (blue), derived from simulations, compared to
    the equivalent values observed for the \smica\ temperature $T$ (left) and the
    reconstructed $E$-mode polarization (right). The upper panel shows the peak
    mean amplitudes for maps filtered with a GAUSS kernel of $120\arcm$ FWHM.
    The lower panel shows the corresponding peak CDF for the same
    kernel of $600\arcm$ FWHM. Similar results are obtained for the other
    component separation methods. The amplitude values are shown
    in arbitrary (dimensionless) units determined by map pre-whitening.
  }
  \label{fig:peaks:means}
\end{figure*}

In this section, we present non-Gaussianity tests of the \Planck\ 2018
temperature and polarization data using the statistical properties of
local extrema (both minima and maxima, to be referred to collectively
as ``peaks'') determined from the component-separated CMB maps. The
peaks, defined as pixels whose amplitudes are either higher or lower
than the corresponding values for all of their nearest neighbours,
compress the information contained in the map and provide tests
complementary to spherical-harmonics-based methods. Peak statistics
are particularly sensitive to non-Gaussian features localized in real
space.

The statistical properties of peaks for an isotropic Gaussian random field
were derived in \citet{Bond:1987ub}. In particular, the fraction of peaks
with amplitudes $x$ above a certain threshold $x/\sigma>\nu$ is given by
\begin{equation}\label{eq:peaks:be}
  F(\nu) = 
    \sqrt{\frac{3}{2\pi}}\, \gamma^2\, \nu \exp\left(-\frac{\nu^2}{2}\right) +
    \frac{1}{2}\, \text{\rm erfc} \left[ \frac{\nu}{\sqrt{2-\frac{4}{3}\,\gamma^2}} \right],
\end{equation}
where $\sigma$ is the rms random field amplitude, and $\gamma$ is
the shape parameter, dependent on the spectrum of the Gaussian random field.
Peak locations and amplitudes, and various derived quantities, such as
their correlation functions, have previously been used to characterize
the WMAP maps in \citet{Larson:2004vm, Larson:2005vb} and \citet{Hou:2009au},
and \Planck\ data in \citetalias{planck2014-a18}.

We consider peak statistics from the \Planck\ component-separated
temperature and polarization maps at \nside\ = 1024, with $E$-mode
maps reconstructed by the purified inpainting method described in
Appendix~\ref{sec:EB_maps}. The maps are pre-whitened by convolving
them with an isotropic function derived from the isotropic
best-fit CMB power spectrum, combined with a diagonal approximation to
the instrumental noise covariance. Then, a confidence
mask is applied, and weighted convolution is performed with a
2D-Gaussian smoothing kernel (that we label as ``GAUSS''),
as described in Appendix~A
of \citetalias{planck2014-a18}. The mask is further extended by rejecting
pixels with an effective convolution weight that differs from unity by
more than 12\,\%, and peaks within it are extracted and analysed.  The
empirical cumulative density-function (CDF) of peak values $x$,
defined for a set of $n$ peaks at values $\{X_i\}$,
\begin{equation}\label{eq:peaks:cdf}
F_n(x) = \frac{1}{n} \sum\limits_{i=1}^{n} I_{X_i \le x}, \hspace{1em}
I_{X_i \le x} \equiv \left\{ \begin{array}{cl}1,& \text{if }X_i \le x\\ 0,& \text{otherwise}\end{array} \right.
\end{equation}
is generated by sorting the peak values $\{X_i\}$ extracted from the
map in ascending order, and comparison to the median CDF $\bar{F}(x)$
derived from an identical analysis of the simulations. A statistical
measure of the difference of the two distributions is provided by the
Kolmogorov-Smirnov deviation
\begin{equation}\label{eq:peaks:ks}
K_n \equiv \sqrt{n}\, \mathop{\text{sup}}\limits_{x} \left|F_n(x) - \bar{F}(x)\right|.
\end{equation}
Although the Kolmogorov-Smirnov deviation has a known limiting
distribution, to evaluate {\it p}-values we derive its CDF directly
from the simulations.

The peak distributions for $T$ and $E$-mode peaks of the \smica\ CMB map filtered
at two different scales ($120\arcm$ and $600\arcm$ FWHM) are shown in
Fig.~\ref{fig:peaks:stat}. The lower panels show empirical peak CDFs
and total peak counts, compared to the Gaussian random field peak CDFs
derived from Eq.~\eqref{eq:peaks:be} by fitting parameters $\sigma$
and $\gamma$ to the median CDF $\bar{F}(x)$ from simulations. The upper
panels show the difference between the observed and median simulated
CDF values, $\sqrt{n}\, [F_n(x) - \bar{F}(x)]$, with the grey bands
representing the 68.3\,\%, 95.4\,\%, and 99.7\,\% regions of the
simulated CDF distributions. Other component separation methods produce
similar results, and no significant deviations from Gaussian expectations
are observed in the polarization peak statistics.

A further statistical test of isotropic Gaussian random field
expectations is to check if the best-fit parameters, $\sigma$ and
$\gamma$, to the observed empirical peak CDF, agree with those derived
from individual simulated realizations. The distribution of best-fit
values of $\sigma$ and $\gamma$ from simulations is compared to the
observed value in Fig.~\ref{fig:peaks:params} for the same data as
presented in Fig.~\ref{fig:peaks:stat}. Once again, the polarization
results are consistent with Gaussian expectations.

Extending the analysis of \citet{Larson:2004vm} to polarization data, we
also evaluate  whether the distributions of maxima and minima are separately
consistent with simulations.
Counts of maxima and minima in the filtered maps are compared to
the distributions determined from simulations in Fig.~\ref{fig:peaks:counts}.
The mean of all maxima, and the negative of the mean of all minima, are
calculated for the filtered map, and the observed values are compared to
the simulated distributions in Fig.~\ref{fig:peaks:means}. The observed
minima/maxima counts and means are not significantly different from
the fiducial model.

To summarize, the temperature and $E$-mode peak statistics determined
from the \Planck\ component-separated maps show no significant
anomalies, except perhaps for the previously known Cold Spot discussed in
Section~\ref{sec:large_scale_peaks}. They are consistent with the
predictions of the $\Lambda$CDM model, and our understanding of the
instrument and noise properties of the 2018 \Planck\ data.

\subsection{Stacking of CMB peaks}
\label{sec:stacking}

\subsubsection{Non-oriented stacking}

The stacking of CMB anisotropies in both temperature and polarization
around the locations of extrema generates characteristic patterns that
connect to the physics of recombination and anisotropy power spectra,
as discussed in detail in~\citetalias{planck2014-a18}
and~\citet{marcos-caballero2016}.
Comparison of the results from the \Planck\ CMB maps with the
predictions of the \Planck\ fiducial \LCDM\ model acts as both a test
of their consistency, and an assessment of the quality of the data at
the map level. Furthermore, the stacking procedure is expected to
mitigate the impact of small-scale noise and residual systematic
effects, thus minimizing the impact of inconsistencies in these
properties between the data and simulations.

Hot (or cold) peaks are selected in the CMB intensity map as local
maxima (or minima) by comparison with their nearest neighbour pixels,
and grouped into different ranges above (below) a given threshold
$\nu$ (in rms units of the intensity map). In order to facilitate the
comparison of the polarization signal between different
peaks, 
we use transformed Stokes parameters, $Q_\mathrm{r}$ and
$U_\mathrm{r}$, defined by Eq.~\eqref{eqn:local_Stokes}, where the
reference point, $\vec{\hat{n}}_\mathrm{ref}$, is specified by the
centre of each extremum $\vec{\hat{n}}_0$.
The $Q_\mathrm{r}$ component then traces the linear polarization in
terms of radial ($Q_\mathrm{r} > 0$) and tangential
($Q_\mathrm{r} < 0$) contributions with respect to the centre of the
peak.  This stacking method is referred to as ``non-oriented,''
because the orientation is defined relative to the local meridian
rather than any property of the data themselves.

\begin{figure*}[hbtp!]
\begin{center}
\includegraphics[scale=0.5]{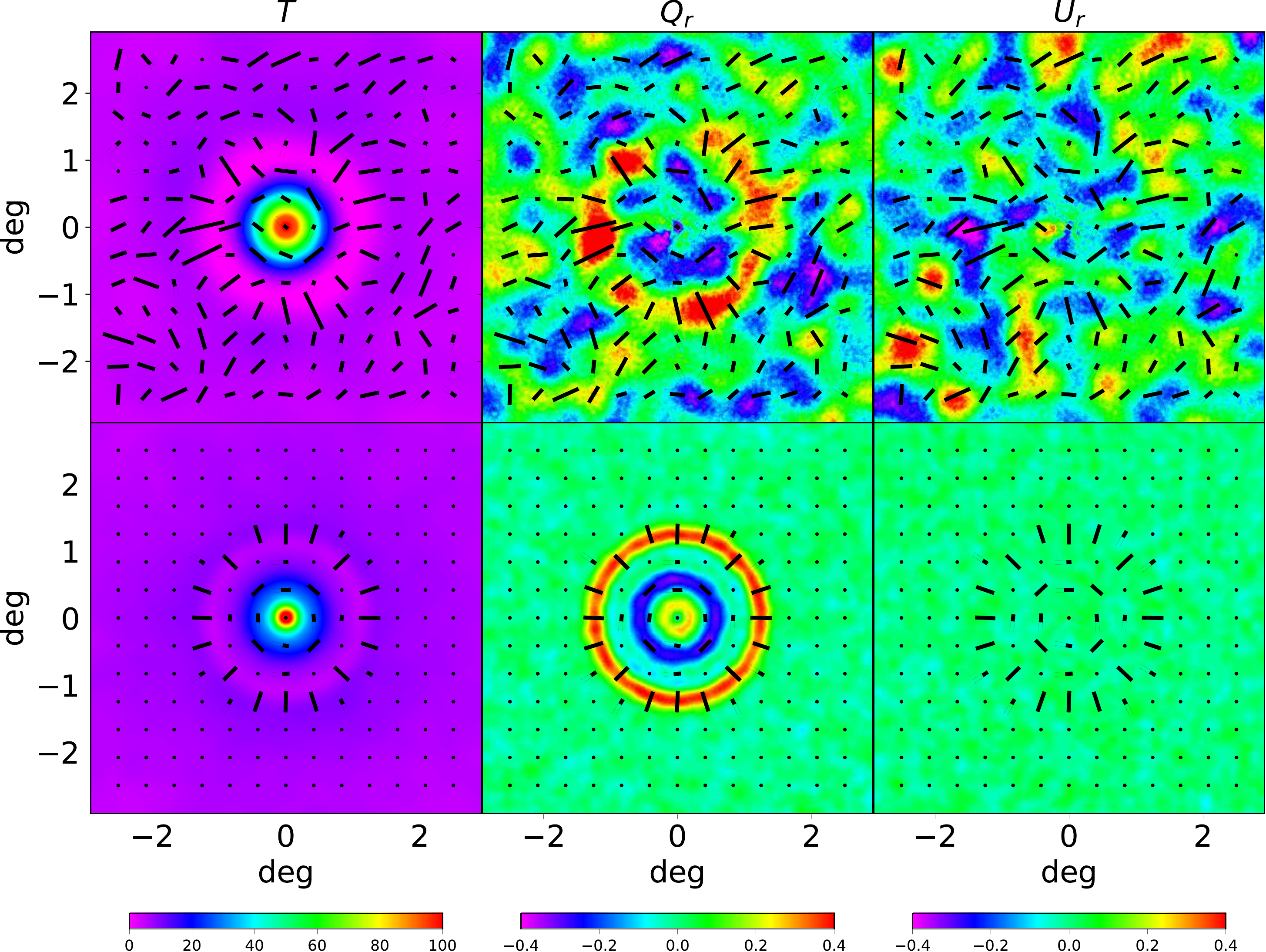} 

\ \\

\includegraphics[scale= 0.5]{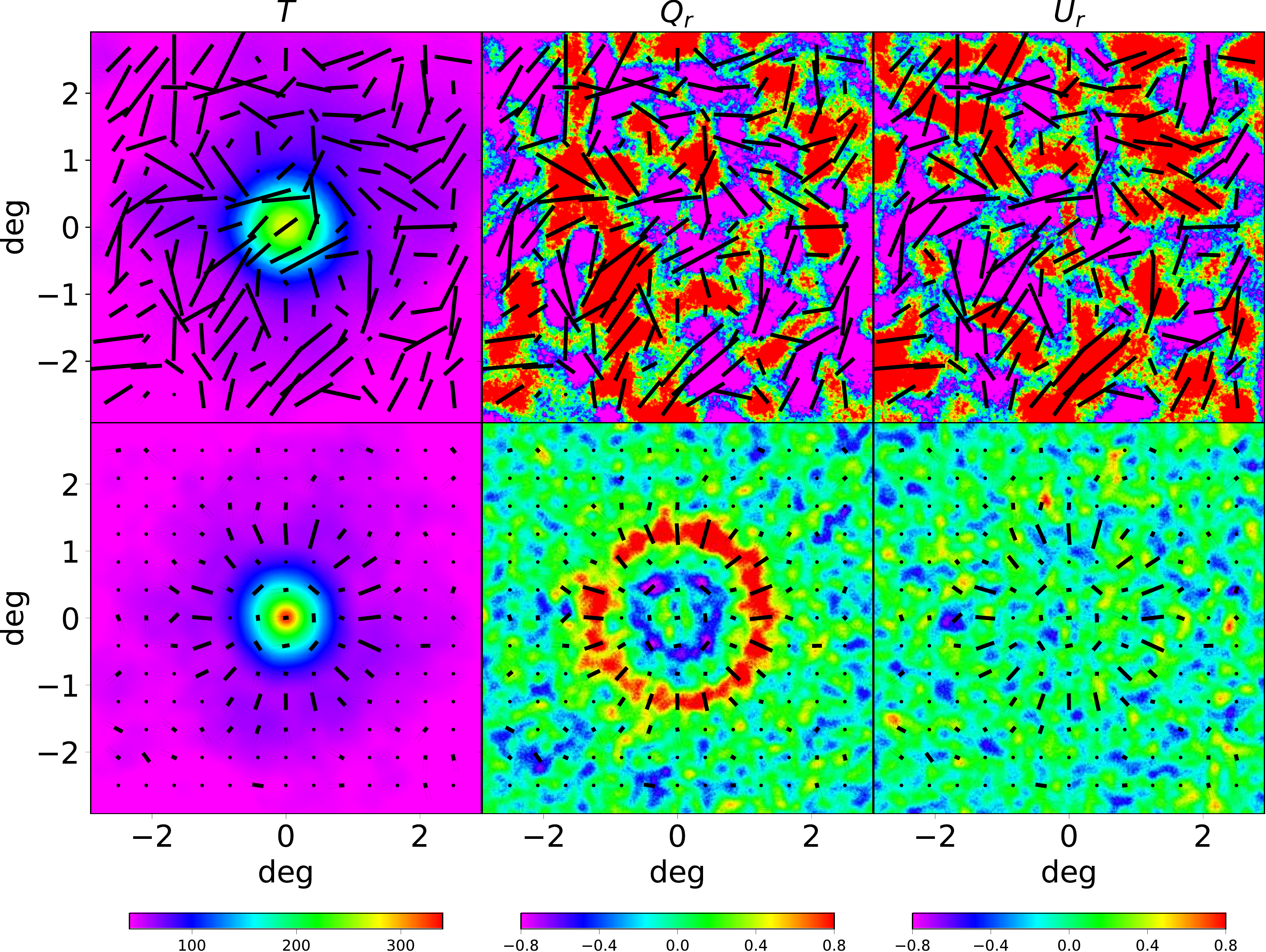}
\end{center}
\caption{{\it Upper panels}: Patches of $T$, $Q_\mathrm{r}$, and
  $U_\mathrm{r}$ in micro-Kelvin for maxima above $\nu=0$ from the WMAP V+W data at
  a {\tt HEALPix} resolution of \nside\ = 512 convolved with a Gaussian beam
  of 30\arcm\ (first row), and from the \sevem\ data at a {\tt HEALPix}
  resolution of \nside\ = 1024 convolved with a Gaussian beam of
  10\arcm\ (second row). {\it Lower panels}: the same thing, but only for maxima above
  $\nu=3$. The polarization direction is represented as a headless
  vector. Note that, in the bottom panel, the length of the headless vectors is divided by 3
 with respect to the scale used in the upper panel.}
\label{fig:stacking_nonoriented_patch}
\end{figure*}

Figure~\ref{fig:stacking_nonoriented_patch} compares the patterns 
seen in the \Planck\ and WMAP data when averaging over patches
centred on CMB intensity maxima above $\nu$ = 0 and 3. 
To enhance the visualization, a random rotation of the patch is
performed around each maximum before stacking (in particular,
this allows a residual pattern in $U_\mathrm{r}$ due to pixelization
effects to be removed).
Specifically, we consider the \sevem\
foreground-cleaned map at a {\tt HEALPix} pixel resolution of
$N_\mathrm{side} = 1024$ convolved with a Gaussian beam of
10\arcm, and a noise-weighted combination of the WMAP V and W bands
at a {\tt HEALPix} pixel resolution of $N_\mathrm{side} = 512$
convolved with a Gaussian beam of 30\arcm. Note that when the
signal-to-noise is sufficient, the stacking
procedure tends to provide an image with azimuthal symmetry about its
centre, due to the almost uncorrelated orientations of the
temperature peaks. Indeed, the \Planck\ analysis for maxima above
$\nu=3$ clearly reveals two rings, while the WMAP data are noise dominated
at the same resolution.  Furthermore, when maxima are
selected above $\nu=0$, the enhanced resolution of the \Planck\ data
allows additional inner rings to be observed compared to WMAP.

We now consider the consistency of the \Planck\ non-oriented results
with the predictions of \LCDM\ by focussing on the mean value of the
angular profiles $\mu(\theta)$ estimated as the average of the
profiles around all hot (cold) peaks above (below) a certain threshold
$\nu$. Although the analysis is performed on data
at a resolution \nside\ = 1024,
the profiles are only sampled with $16$ bins to ensure that the
covariance matrix can be well estimated from the simulations.

A $\chi^2$ estimator is used to quantify the differences between the
$\mu(\theta)$ profiles obtained from the data and the expected values
estimated with simulations:
\begin{linenomath*}
\begin{equation}
\chi^2 = \left[\mathbf{\mu} - \mathbf{\bar{\mu}}\right]^{\sf T} \tens{C}^{-1}\left[\mathbf{\mu} - \mathbf{\bar{\mu}}\right], 
\label{eq:profile_chi2}
\end{equation}
\end{linenomath*}
with the covariance matrix defined as
\begin{linenomath*}
\begin{equation}
\tens{C}(\theta_i,\theta_j) = \dfrac{1}{N-1}\sum_{k=1}^N{\left[\mu_k(\theta_i)- \bar{\mu}(\theta_i)\right]\left[\mu_k(\theta_j)- \bar{\mu}(\theta_j)\right]},
\end{equation}
\end{linenomath*}
where the index $k$ denotes a simulation, $N$ is the total number of
simulations used to estimate this matrix, and $\mathbf{\bar{\mu}}$ is
the ensemble average.

\begin{figure*}[hbtp!]
\begin{center}
\includegraphics[width=0.48\textwidth]{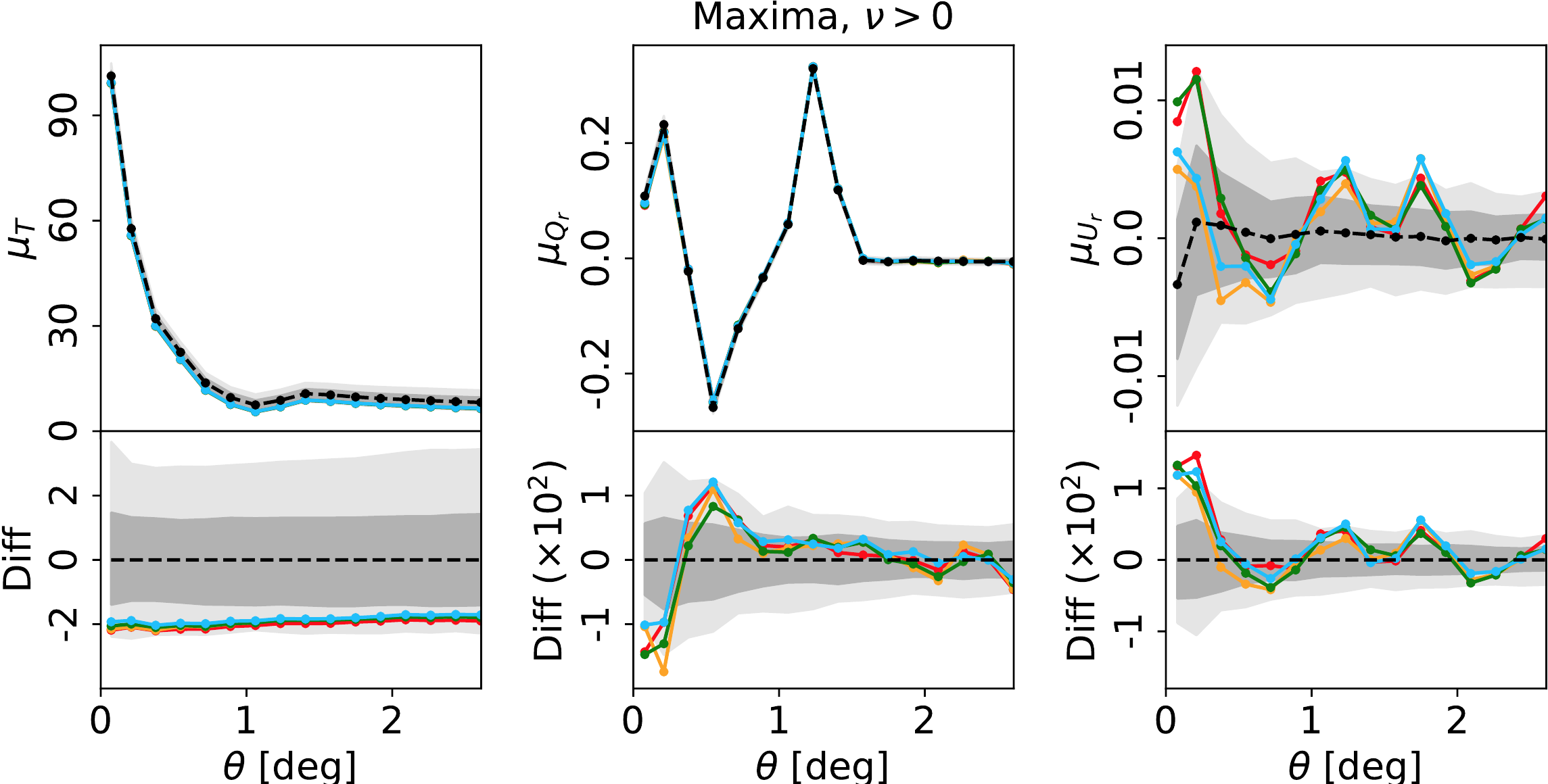}
\includegraphics[width=0.48\textwidth]{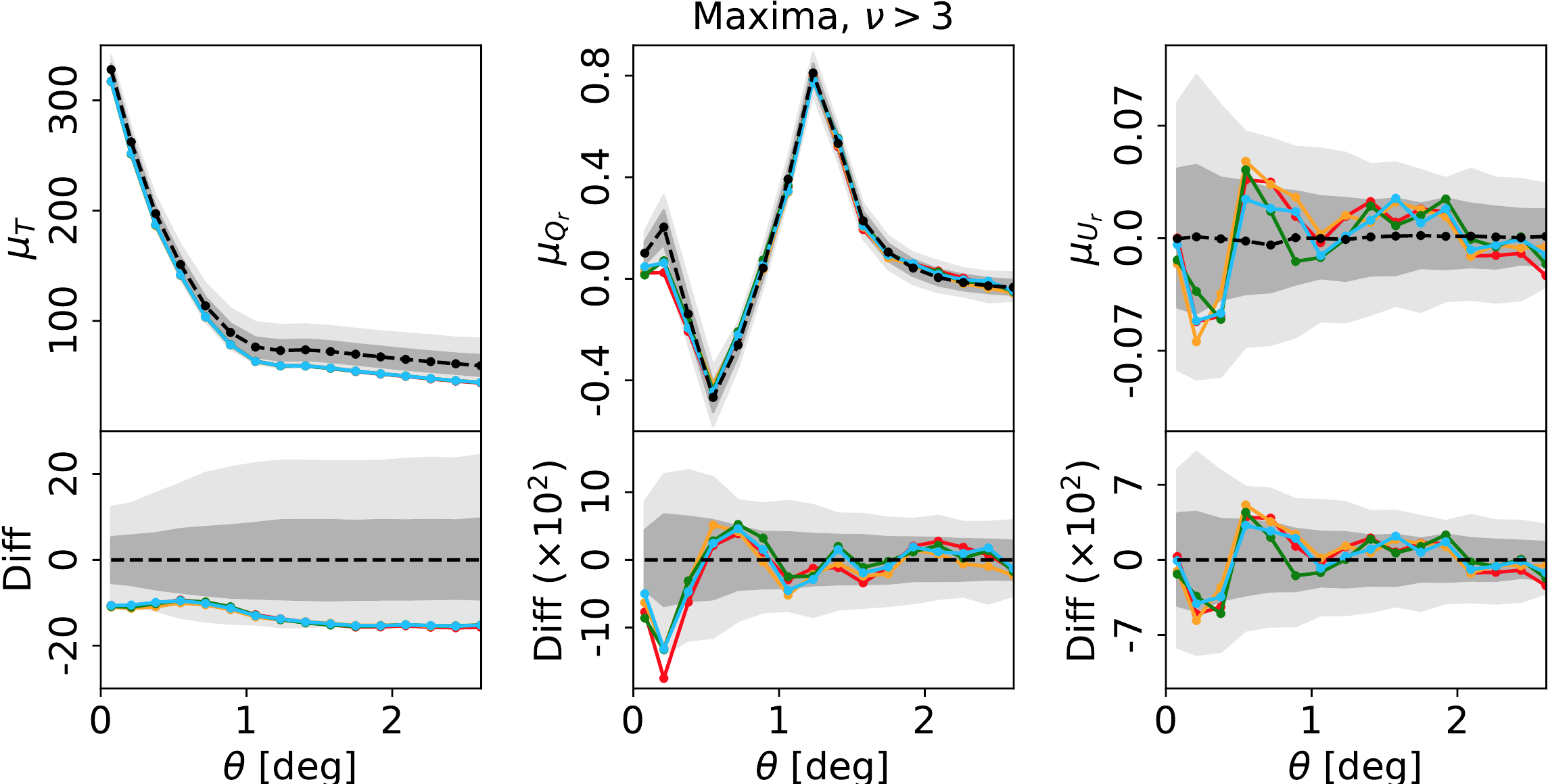}
\\
\includegraphics[width=0.48\textwidth]{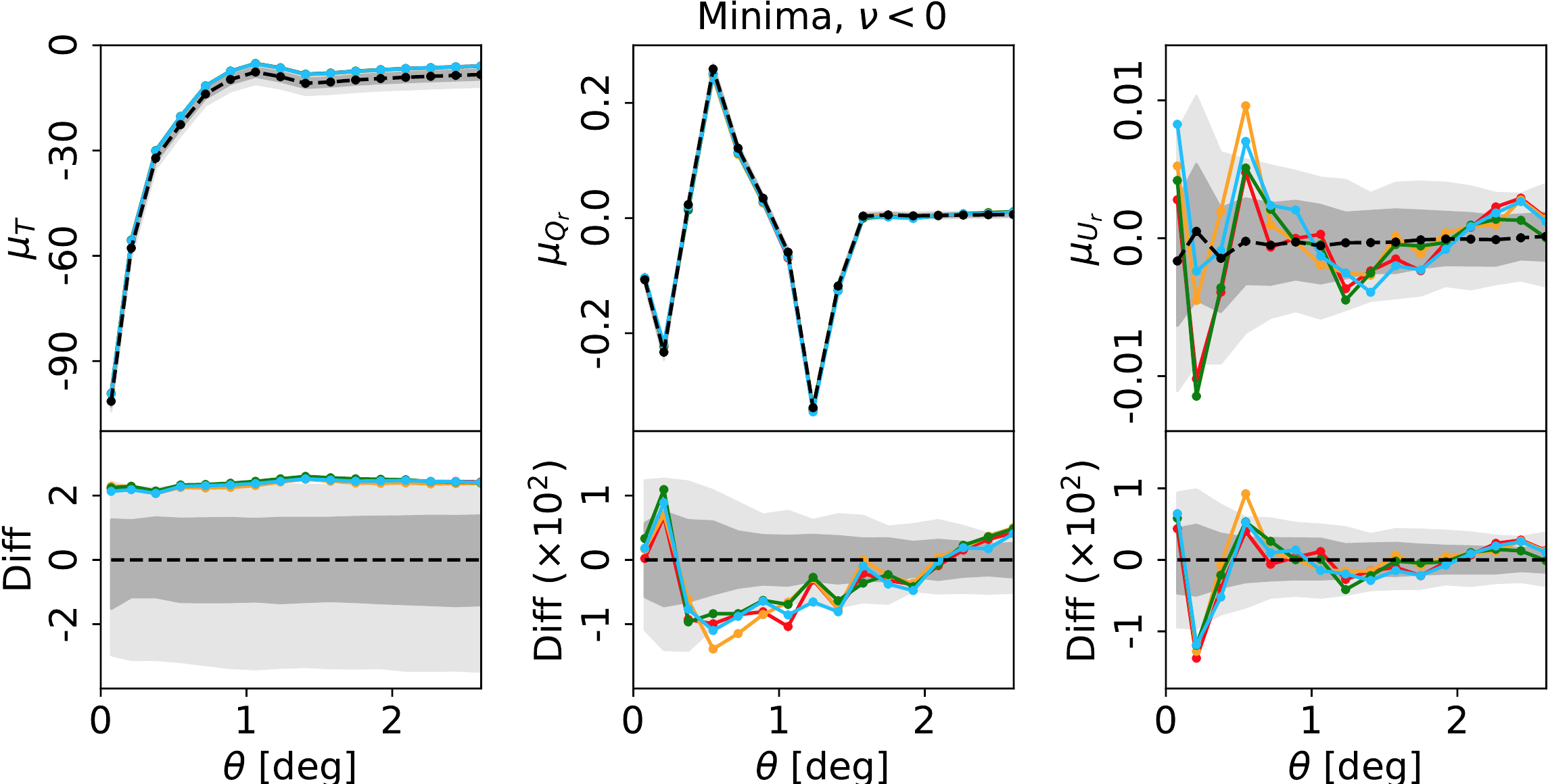}
\includegraphics[width=0.48\textwidth]{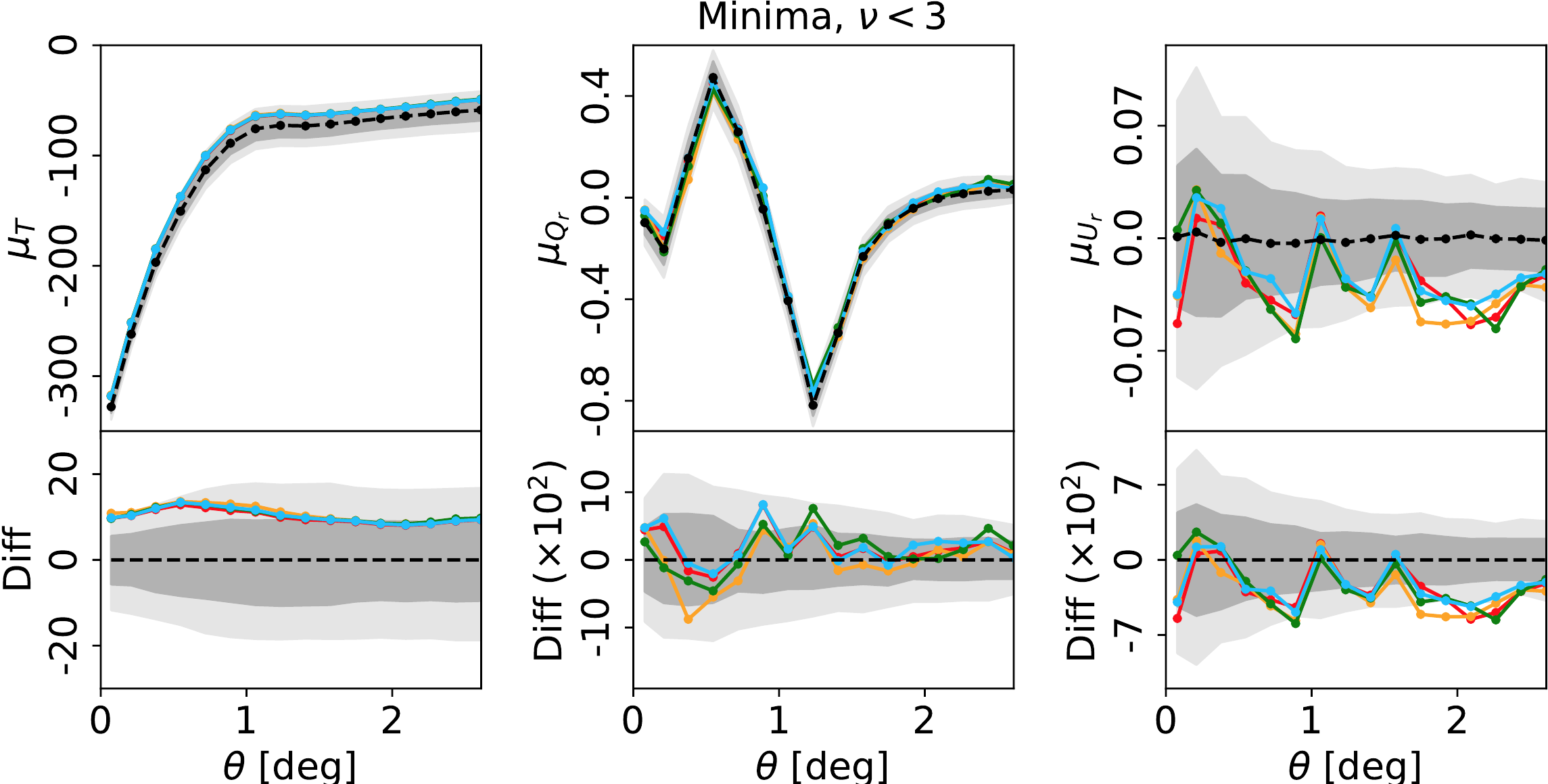}
\end{center}
\caption{Mean radial profiles of $T$, $Q_\mathrm{r}$, and
  $U_\mathrm{r}$ in micro-kelvin obtained for \commander\ (red), \nilc\
  (orange), \sevem\ (green), and \smica\ (blue) at \nside\ = 1024.  Each
  individual panel contains the mean radial profiles (top part) and the
  differences between the mean profiles of the data and those computed
  from the ensemble mean of the simulations (denoted ``Diff,'' lower part).
  Results based on stacks around temperature maxima and minima are
  shown in the upper and lower rows, respectively.  The left three columns 
  present results for peaks selected above the null
  threshold, while the right three columns show the equivalent results for
  peak amplitudes above (maxima) or below (minima) 3 times the
  dispersion of the temperature map.  The black dots (connected by
  dashed lines) show the mean value from simulations and the shaded
  regions correspond to the $\pm1\,\sigma$ ($68\,\%$) and $\pm 2\,\sigma$ ($95\,\%$)
  error bars estimated from \sevem\ simulations. Note that the
  ``Diff'' curves for each component-separation method are computed using
  the corresponding set of ensemble averages, although only the ensemble
  average from \sevem\ is shown here.}
\label{fig:stacking_nonoriented_prof}
\end{figure*}

Figure~\ref{fig:stacking_nonoriented_prof} presents the results for
maxima and minima selected at thresholds of $\nu$ = 0 and 3 for the
CMB maps provided by the four component-separation pipelines, compared
to the simulations. The corresponding {\it p}-values for
the comparison are presented in Table~\ref{table:chi2_stacking}.  We
see that all component-separation methods yield consistent results. No
significant differences for the intensity profiles $\mu_T(\theta)$ are
observed with respect to the results found in the \Planck-2015
analysis. As in the latter case, a systematic deviation between the
data and the mean value of simulations is present. This was previously
interpreted in \citetalias{planck2014-a18} as an effect connected with
the deficit in the observed power spectrum at low multipoles.

\citetalias{planck2014-a18} also demonstrated that 
the $\chi^2$ statistic is suboptimal when considering the
systematic shift between data and simulations, as seen in the intensity
profiles $\mu_T$, and may lead to misleading {\it p}-values.
We therefore consider an alternative quantity, the
integrated profile deviation $\Delta\mu_T(W)$, to
evaluate the consistency between the data and the model.
This is defined as
\begin{linenomath*}
\begin{equation}
\Delta\mu_T(W) =  \int_{0}^{R} \left[\mu_T(\theta) - \bar{\mu}_T(\theta)\right] W(\theta)\, {\rm d}\theta,  \label{eq:profile_difference}
\end{equation}
\end{linenomath*}
where $R$ represents the size of stacking patches ($3^{\circ}$ in this
case), and the weighting function $W$ is chosen to be proportional to
the expected profile. The {\it p}-values obtained in this case are
given in Table~\ref{table:deltamu_stacking}, and are consistent with
the deviations shown in Fig.~\ref{fig:stacking_nonoriented_prof}. 

Turning to polarization, while Table~\ref{table:chi2_stacking}
generally shows consistent results between the data and simulations,
somewhat low {\it p}-values for the minima below $\nu = 0$ are observed.
However, as is apparent from
Fig.~\ref{fig:stacking_nonoriented_prof}, the deviation between
the data and the mean value of simulations at this threshold is less
significant for the maxima.
We note that no evidence of asymmetry is found between the number of
maxima and minima in the data when compared to simulations.  In
addition, the sum of the profiles from maxima and minima is consistent
with zero for the data with respect to the simulations.
Since similar {\it p}-values are found when comparing the mean
angular profiles from the HMHD and OEHD data splits to corresponding
simulations, we consider that it is unlikely that this discrepancy for
the minima is cosmological in origin.

We find that the noise level traced by the $U_\mathrm{r}$ profile
seems to be slightly higher than was seen
in~\citetalias{planck2014-a18}.
In addition, for $\nu$ = 0 and low values of $\theta$
($<0\pdeg5$), the mean determined from the simulations does
not tend to a null value, especially for the \nilc\ and \smica\ maps
(although this is not observed in the HMHD and OEHD analyses).  Both
effects may be related to noise correlations introduced by changes in
the raw data processing, and variations in the weights ascribed to the
frequency maps by the different component-separation methods. In
addition, it was shown in \citet{planck2014-a23} that there is a
systematic effect in the $U_\mathrm{r}$ component due to uncertainty
in the orientation of the polarization sensitive detectors.

In summary, given our understanding of the \Planck\ data and
simulations, the results from non-oriented stacking are consistent
with the predictions of the \LCDM\ model.

\begin{table}[tp] 
\begingroup
\newdimen\tblskip \tblskip=5pt
\caption{Fraction of simulations with higher values of $\chi^2$ than the
  observed ones for the $T$, \Qr, and \Ur\ angular profiles, computed
  from the stacking of hot and cold extrema selected above or below the
  $\nu=0$ and $\nu=3$ thresholds.
  In this table, small $p$-values would be considered anomalous.}
\label{table:chi2_stacking}
\nointerlineskip
\vskip -3mm
\footnotesize
\setbox\tablebox=\vbox{
   \newdimen\digitwidth 
   \setbox0=\hbox{\rm 0} 
   \digitwidth=\wd0 
   \catcode`*=\active 
   \def*{\kern\digitwidth}
   \newdimen\signwidth 
   \setbox0=\hbox{+} 
   \signwidth=\wd0 
   \catcode`!=\active 
   \def!{\kern\signwidth}
\halign{\hbox to 1.0in{#\leaderfil}\tabskip 3pt&
\hfil#\hfil\tabskip 10pt&
\hfil#\hfil\tabskip 8pt&
\hfil#\hfil\tabskip 8pt&
\hfil#\hfil\/\tabskip 0pt\cr
\noalign{\doubleline}
\noalign{\vskip -1pt}
\omit&\multispan4 \hfil Probability [\%]\hfil\cr
\noalign{\vskip -4pt}
\omit&\multispan4\hrulefill\cr
\omit& {\tt Comm.}& {\tt NILC}& {\tt SEVEM}& {\tt SMICA}\cr
\noalign{\vskip 3pt\hrule\vskip 3pt}
\multispan5\hfil $\nu = 0$ (hot spots)\hfil\cr
 $T$ & 83.0 & 47.5 & 79.1 & 72.4 \cr
 $Q_{\mathrm{r}}$ & 24.3 & 14.0 & 25.5 & 48.7 \cr
 $U_{\mathrm{r}}$ & *0.7 & *4.1 & *7.3 & *3.2 \cr
\noalign{\vskip 3pt\hrule\vskip 3pt}
\multispan5\hfil $\nu = 3$ (hot spots)\hfil\cr
 $T$ & 33.1 & 35.9 & 34.8 & 45.8 \cr
 $Q_{\mathrm{r}}$ & 62.3 & 78.6 & 82.1 & 70.1 \cr
 $U_{\mathrm{r}}$ & 86.8 & 95.3 & 86.6 & 97.6 \cr
\noalign{\vskip 3pt\hrule\vskip 3pt}
\multispan5\hfil $\nu = 0$ (cold spots)\hfil\cr
 $T$ & 34.8 & 32.9 & 54.0 & 51.3 \cr
 $Q_{\mathrm{r}}$ & *0.3 & *0.3 & *1.0 & *0.5 \cr
 $U_{\mathrm{r}}$ & 11.4 & *3.8 & 26.9 & *4.7 \cr
\noalign{\vskip 3pt\hrule\vskip 3pt}
\multispan5\hfil $\nu = 3$ (cold spots)\hfil\cr
 $T$ & 43.2 & 36.4 & 51.2 & 40.7 \cr
 $Q_{\mathrm{r}}$ & 80.2 & 70.0 & 55.2 & 60.2 \cr
 $U_{\mathrm{r}}$ & 11.9 & *8.0 & 10.3 & 26.3 \cr
\noalign{\vskip 3pt\hrule\vskip 3pt}
}}
\endPlancktable                    
\endgroup
\end{table}

\begin{table}[tp] 
\begingroup
\newdimen\tblskip \tblskip=5pt
\caption{Fraction of simulations with higher values of $\chi^2$ than the
  observed ones for $\Delta \mu_T$, computed from the stacking of hot and
  cold spots selected above the $\nu=0$ and $\nu=3$ thresholds.
  In this table, small $p$-values would be considered anomalous.}
\label{table:deltamu_stacking}
\nointerlineskip
\vskip -3mm
\footnotesize
\setbox\tablebox=\vbox{
\halign{\hbox to 1.0in{#\leaderfil}\tabskip 3pt&
\hfil#\hfil\tabskip 10pt&
\hfil#\hfil\tabskip 8pt&
\hfil#\hfil\tabskip 8pt&
\hfil#\hfil\/\tabskip 0pt\cr
\noalign{\doubleline}
\noalign{\vskip -1pt}
\omit&\multispan4 \hfil Probability [\%]\hfil\cr
\noalign{\vskip -4pt}
\omit&\multispan4\hrulefill\cr
\omit& {\tt Comm.}& {\tt NILC}& {\tt SEVEM}& {\tt SMICA}\cr
\noalign{\vskip 3pt\hrule\vskip 3pt}
\multispan5\hfil Hot spots\hfil\cr
 $T$ ($\nu = 0$) & 96.0 & 95.0 & 95.0 & 94.3 \cr
 $T$ ($\nu = 3$) & 97.3 & 97.7 & 97.0 & 97.0 \cr
\noalign{\vskip 3pt\hrule\vskip 3pt}
\multispan5\hfil Cold spots\hfil\cr
 $T$ ($\nu = 0$) & 97.3 & 97.0 & 97.7 & 97.0 \cr
 $T$ ($\nu = 3$) & 93.7 & 96.0 & 94.7 & 94.3 \cr
\noalign{\vskip 3pt\hrule\vskip 3pt}
}}
\endPlancktable                    
\endgroup
\end{table}

\subsubsection{Oriented stacking}

\begin{figure*}[htbp!]
  \centering
  \begin{tabular}{ccc}
        $T$ & $E$ & $B$ \\
        \includegraphics[height=4.8cm]{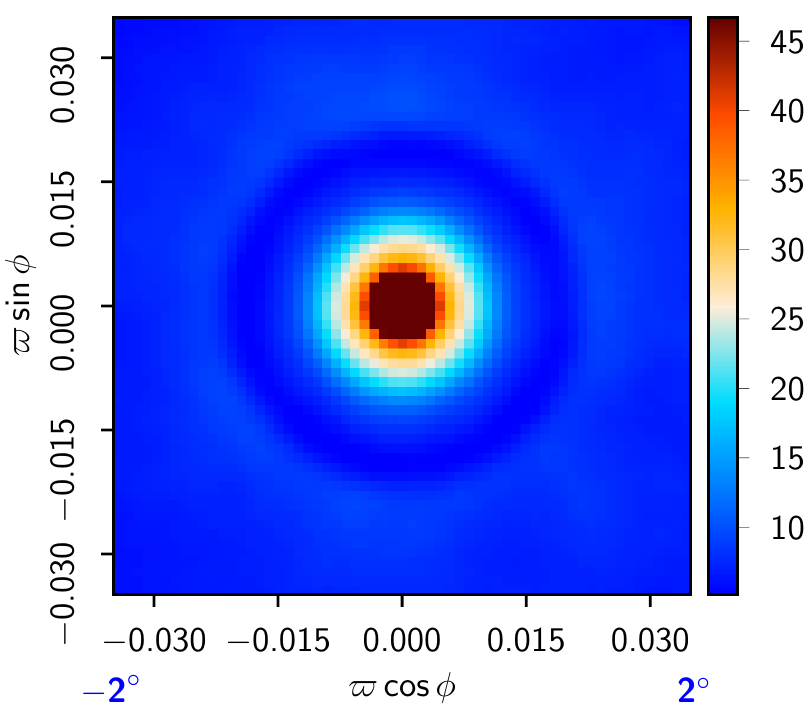} &
        \includegraphics[height=4.8cm]{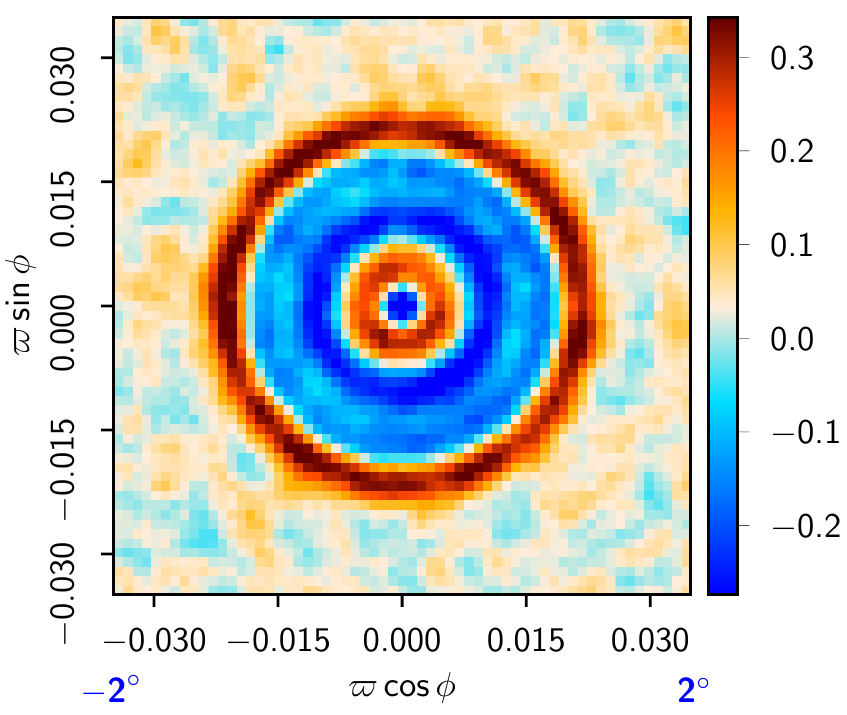} &
        \includegraphics[height=4.8cm]{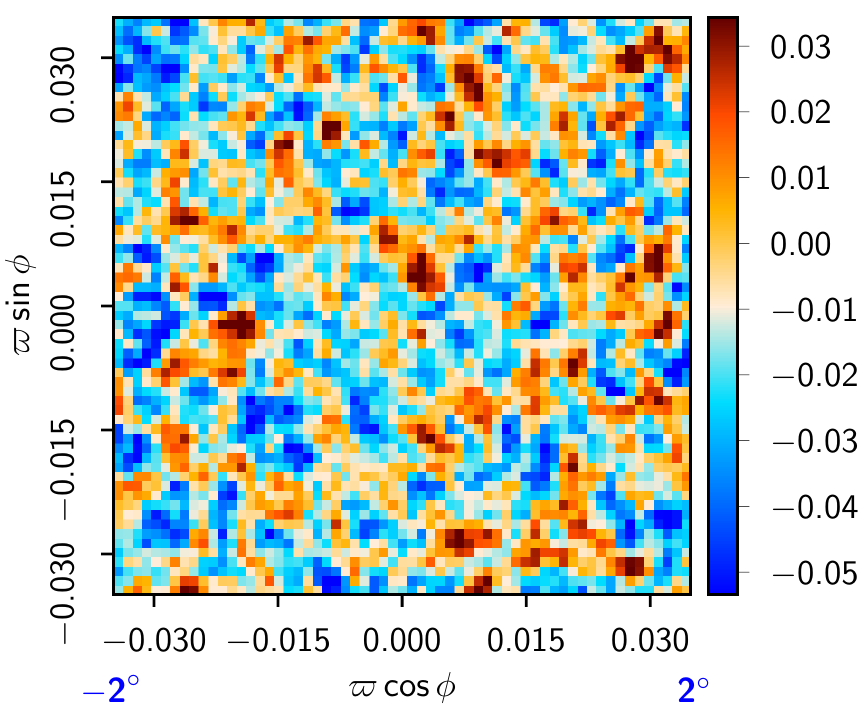} \\
        \includegraphics[height=4.8cm]{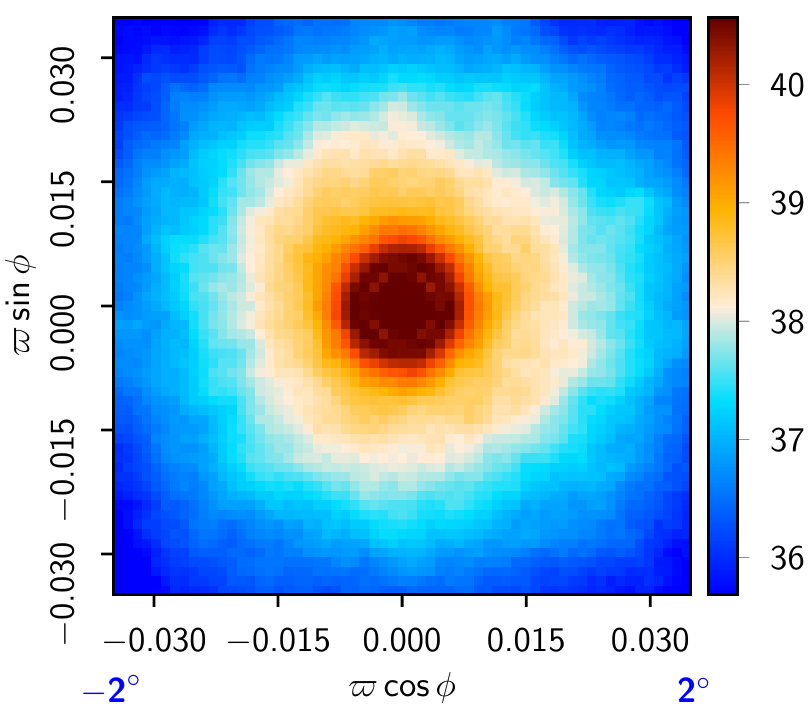} &
        \includegraphics[height=4.8cm]{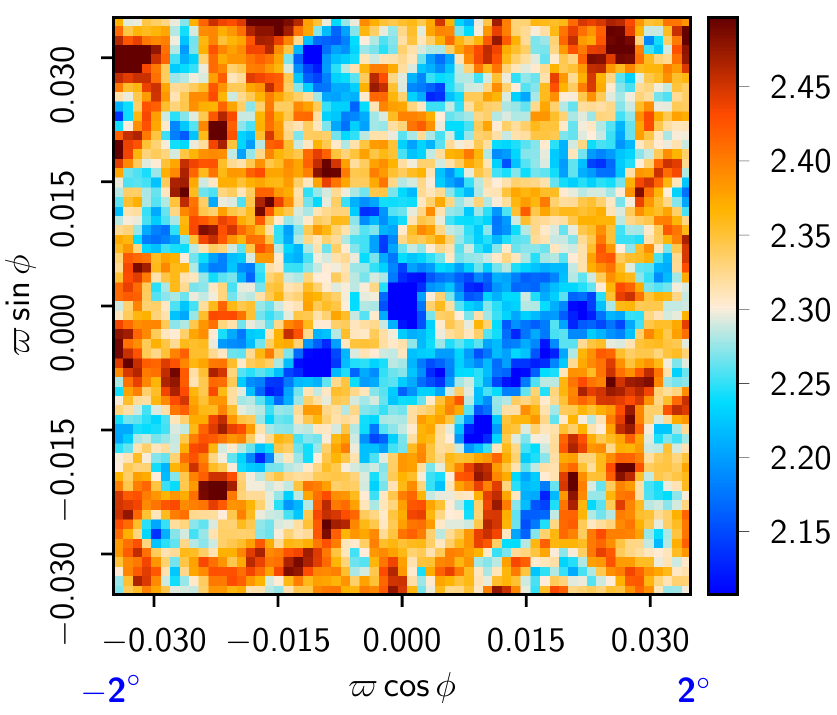} &
        \includegraphics[height=4.8cm]{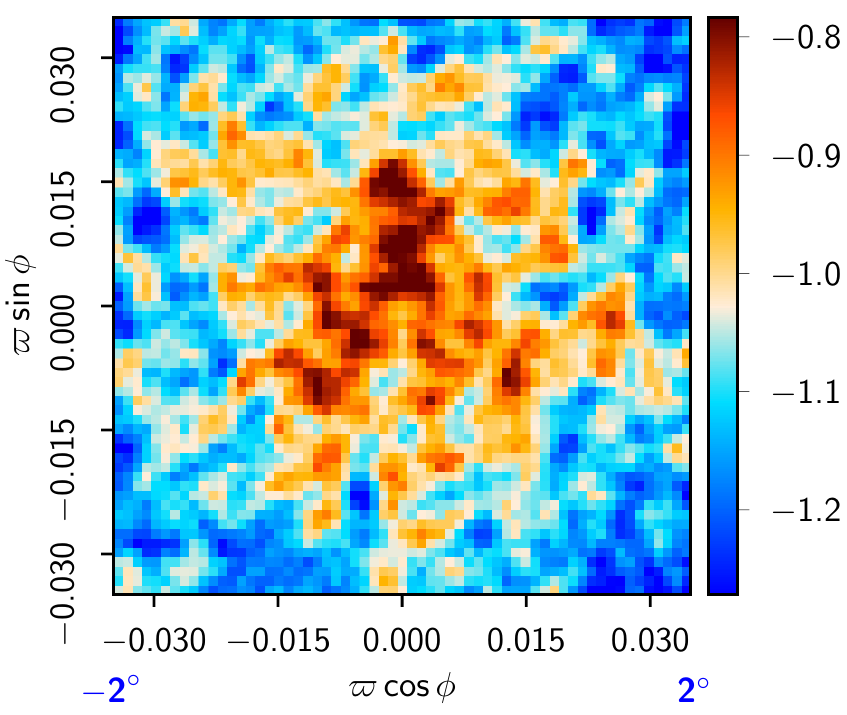} \\
        \includegraphics[height=4.8cm]{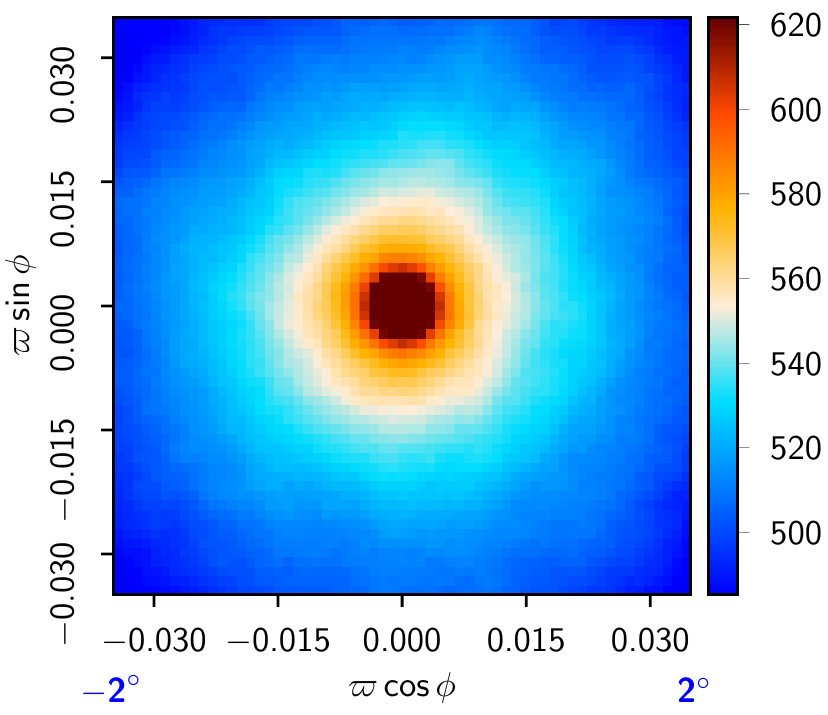} &
        \includegraphics[height=4.8cm]{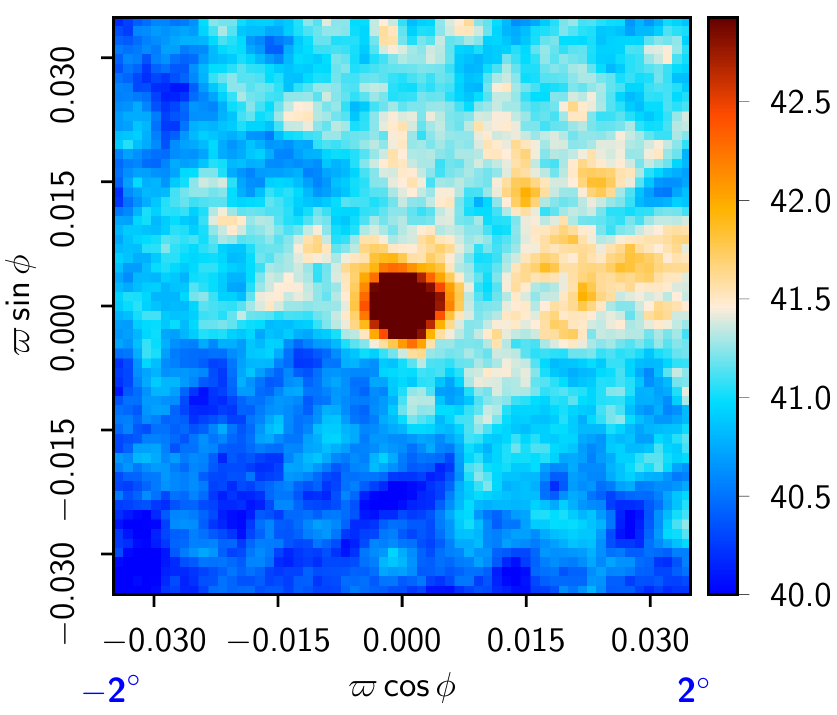} &
        \includegraphics[height=4.8cm]{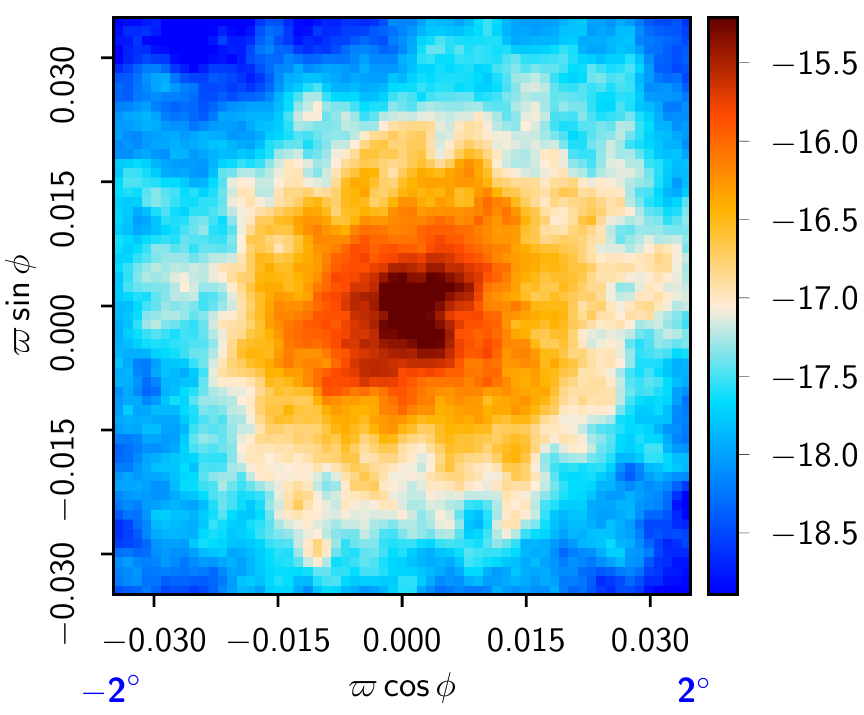}
  \end{tabular}
  \caption{Non-oriented stacks of intensity $T$ (left), reconstructed $E$-mode (centre)
    and $B$-mode (right) polarization stacked on intensity maxima for the CMB (top row, \smica\ map at \nside=1024 and
    10\arcm\ FWHM beam), low-frequency foregrounds (middle row, LFI 30-GHz map smoothed using a 10\arcm\ FWHM beam,
    with subtraction of the \smica\ CMB map at 10\arcm\ FWHM resolution smoothed by the 30-GHz LFI beam), and high-frequency
foregrounds (bottom row, 353-GHz HFI polarization-sensitive bolometer map smoothed using a
    10\arcm\ FWHM beam, with subtraction of the \smica\ CMB map at 10\arcm\ FWHM resolution smoothed by the 353-GHz HFI beam).
    The common temperature mask was used for temperature peak selection, and the polarization confidence mask of Appendix~\ref{sec:EB_maps} was used for $E$- and $B$-mode stacks.}
    \label{fig:stack:unoriented}
\end{figure*}

\begin{figure*}[htbp!]
  \centering
  \begin{tabular}{ccc}
        $T$ & $E$ & $B$ \\
        \includegraphics[height=4.8cm]{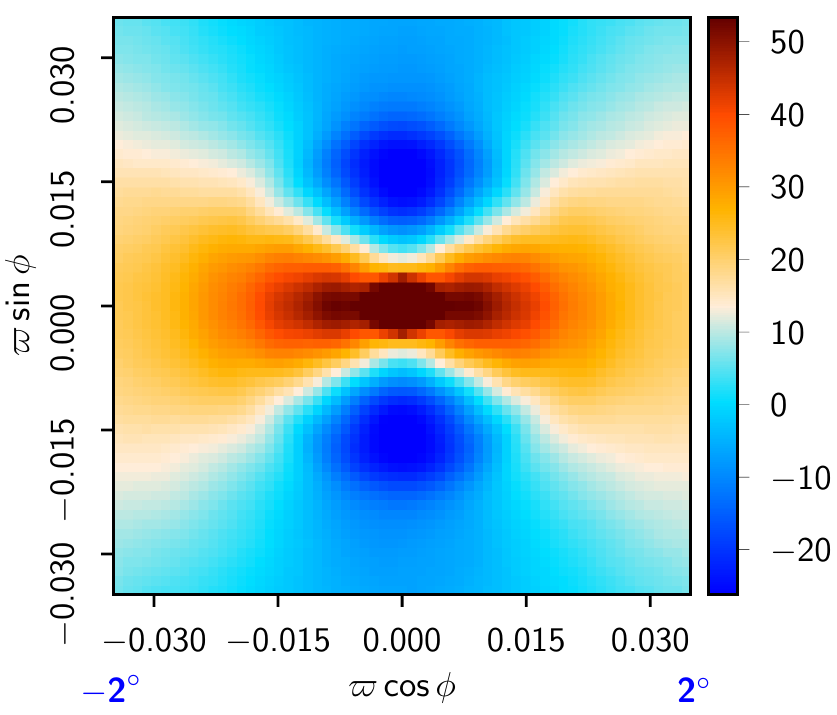} &
        \includegraphics[height=4.8cm]{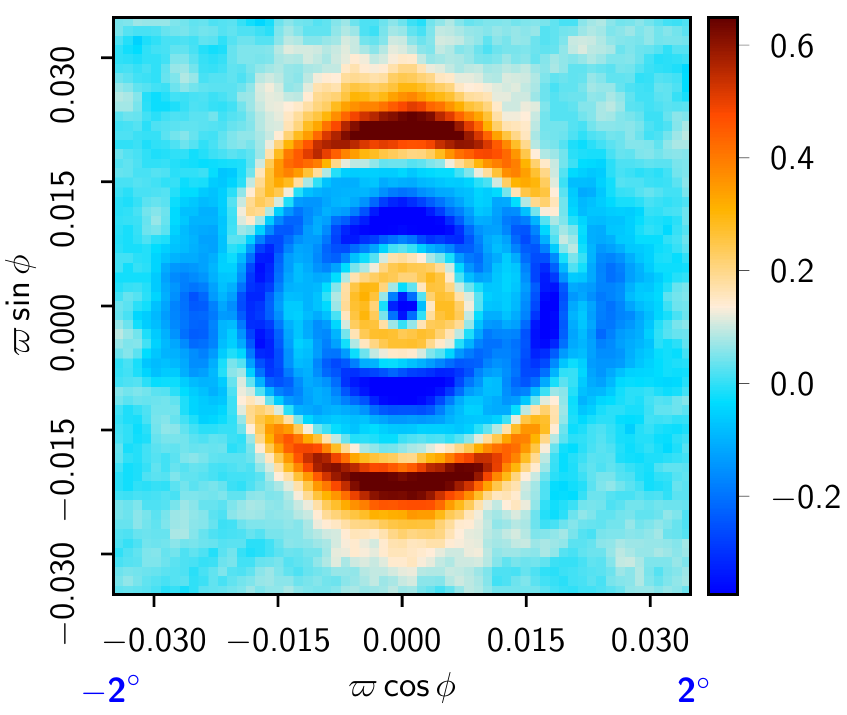} &
        \includegraphics[height=4.8cm]{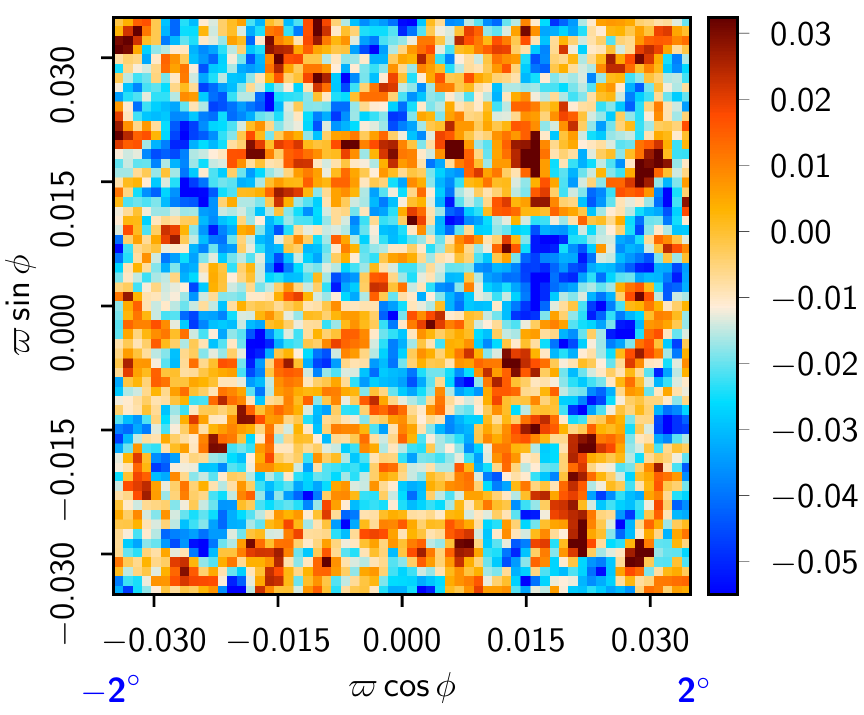} \\
        \includegraphics[height=4.8cm]{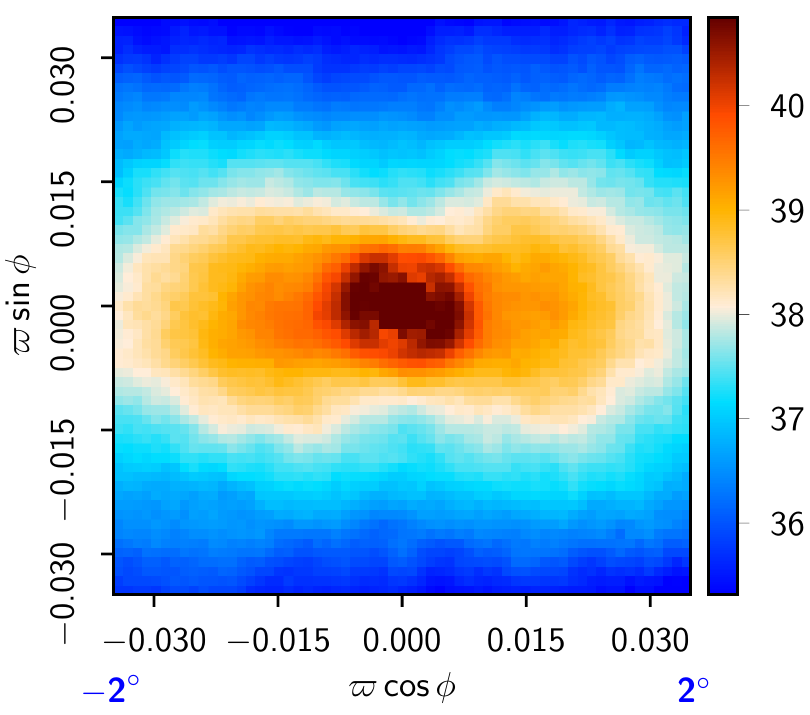} &
        \includegraphics[height=4.8cm]{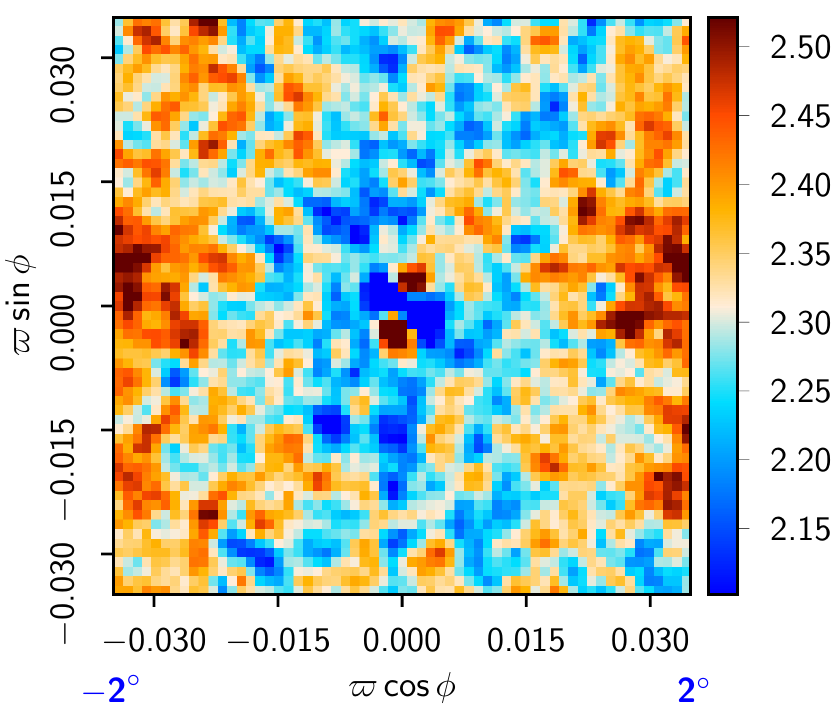} &
        \includegraphics[height=4.8cm]{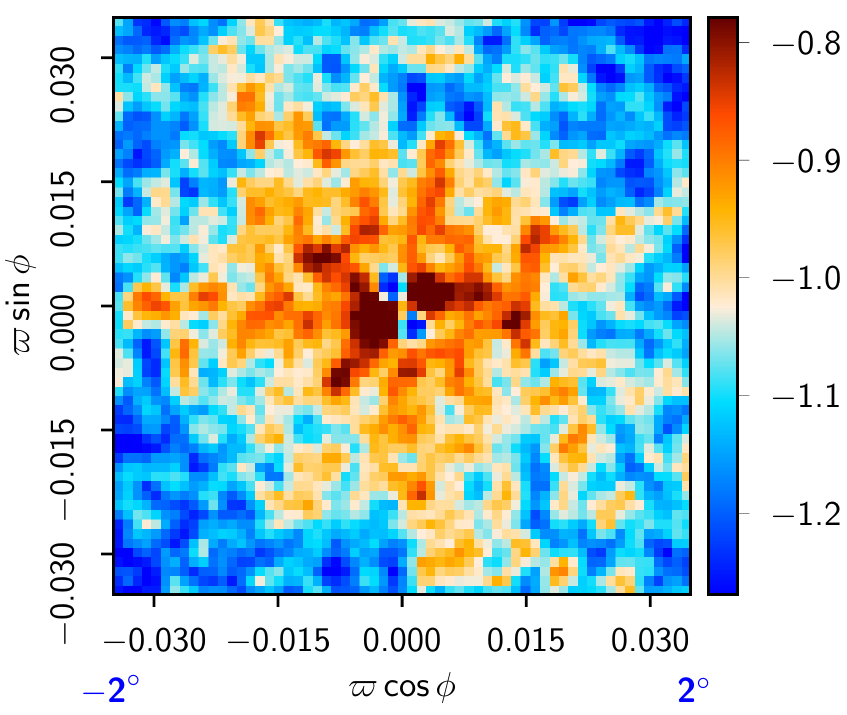} \\
        \includegraphics[height=4.8cm]{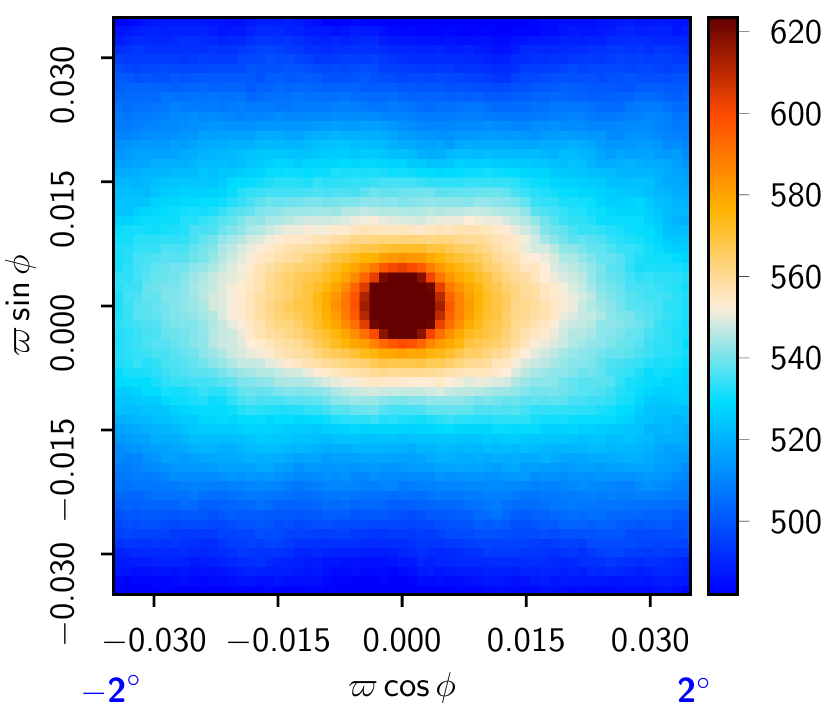} &
        \includegraphics[height=4.8cm]{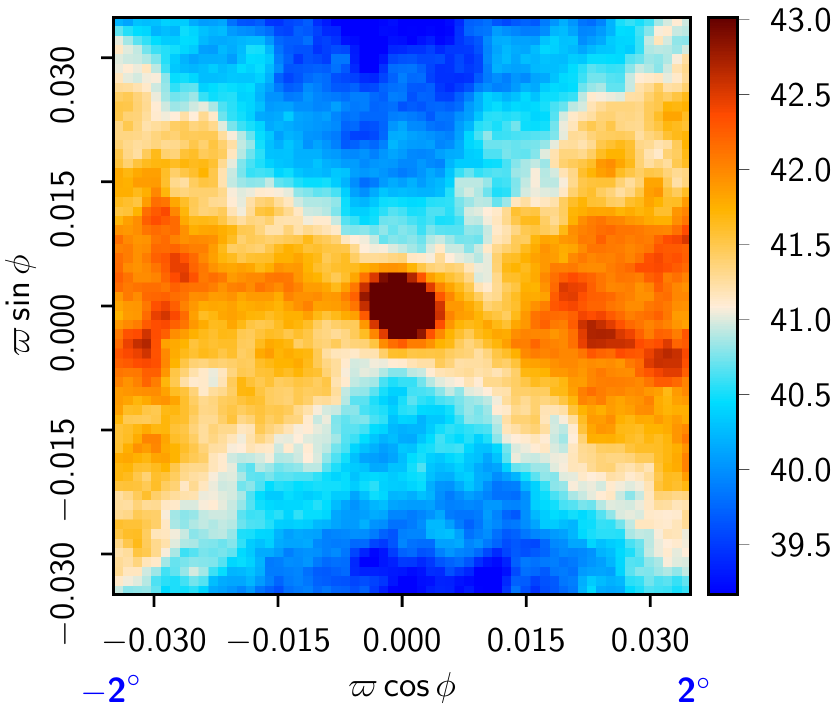} &
        \includegraphics[height=4.8cm]{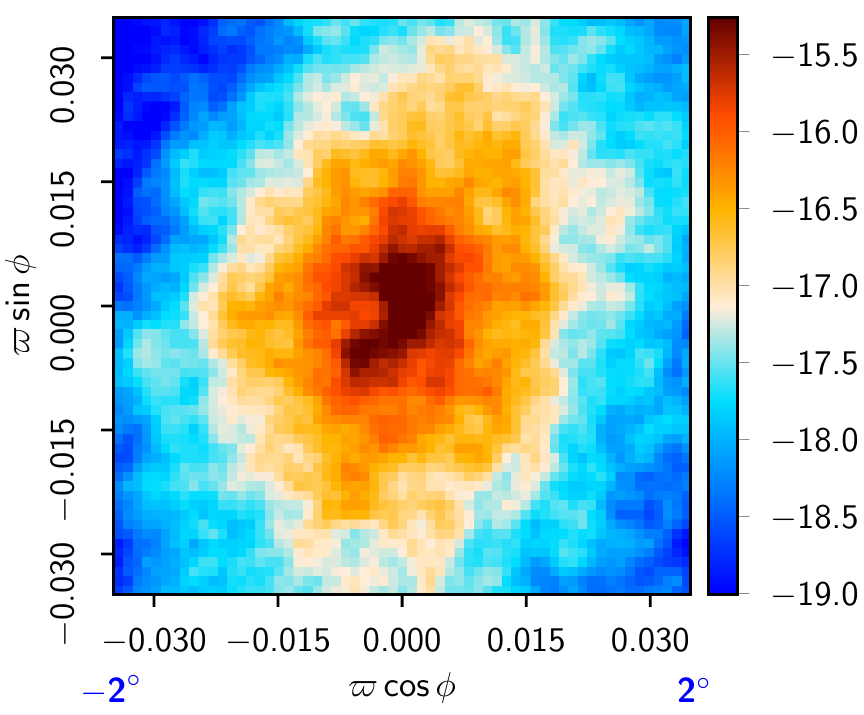}
  \end{tabular}
  \caption{Oriented stacks of intensity $T$ (left), reconstructed $E$-mode (centre)
    and $B$-mode (right) polarization stacked on intensity maxima for the CMB (top row, \smica\ map at \nside=1024 and
    10\arcm\ FWHM beam), low-frequency foregrounds (middle row, LFI 30-GHz map smoothed using a 10\arcm\ FWHM beam,
    with subtraction of the \smica\ CMB map at 10\arcm\ FWHM resolution smoothed by the 30-GHz LFI beam), and high-frequency
foregrounds (bottom row, 353-GHz HFI polarization-sensitive bolometer map smoothed using a
    10\arcm\ FWHM beam, with subtraction of the \smica\ CMB map at 10\arcm\ FWHM resolution smoothed by the 353-GHz HFI beam).
    The orientation of the stacked patch is chosen so that that the principal directions of the tensor
    $\nabla_i\nabla_j\nabla^{-2} T$ are aligned with the horizontal and vertical axes of the stack.
    The common temperature mask was used for temperature peak selection, and the polarization confidence
    mask of Appendix~\ref{sec:EB_maps} was used for the $E$- and $B$-mode stacks.}
    \label{fig:stack:oriented}
\end{figure*}

\begin{figure*}[htbp!]
  \centering
  \begin{tabular}{cccc}
    ~~~~$T_0(\varpi)$ & ~~~~$T_2(\varpi)$ &
    ~~~~$E_0(\varpi)$ & ~~~~$E_2(\varpi)$ \\
    \includegraphics[scale=0.82]{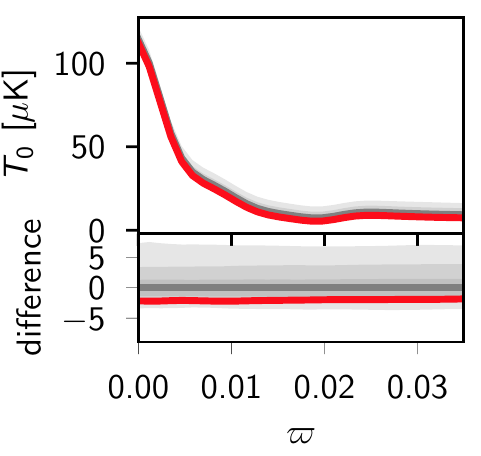} &
    \includegraphics[scale=0.82]{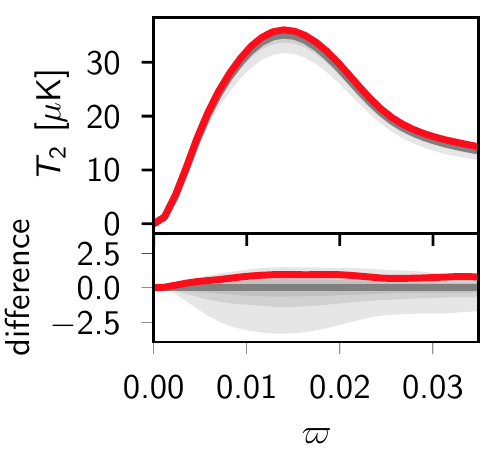} &
    \includegraphics[scale=0.82]{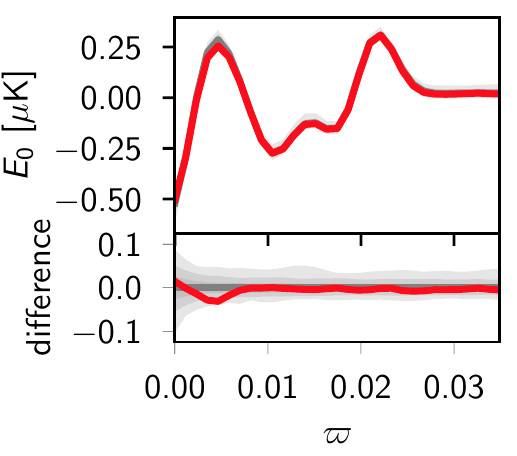} &
    \includegraphics[scale=0.82]{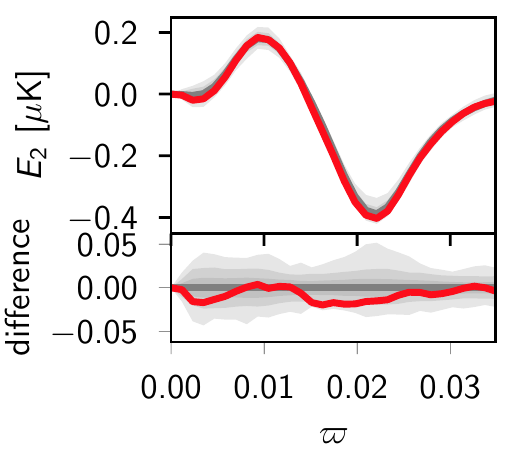} \\
    \includegraphics[scale=0.82]{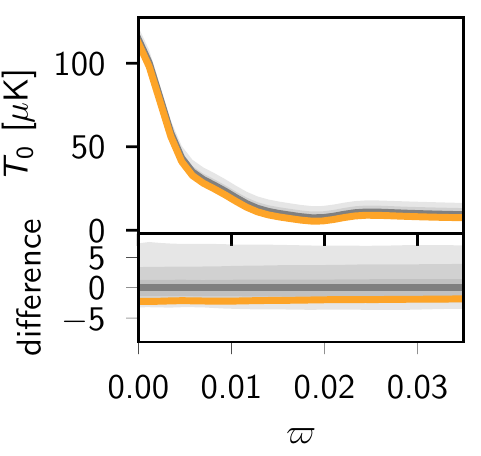} &
    \includegraphics[scale=0.82]{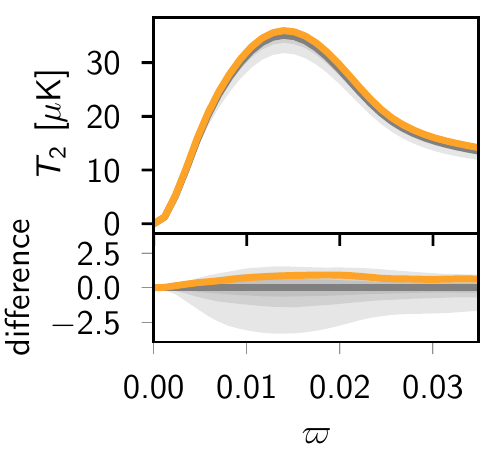} &
    \includegraphics[scale=0.82]{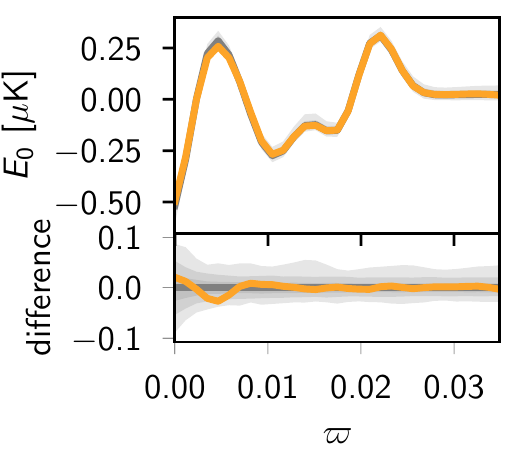} &
    \includegraphics[scale=0.82]{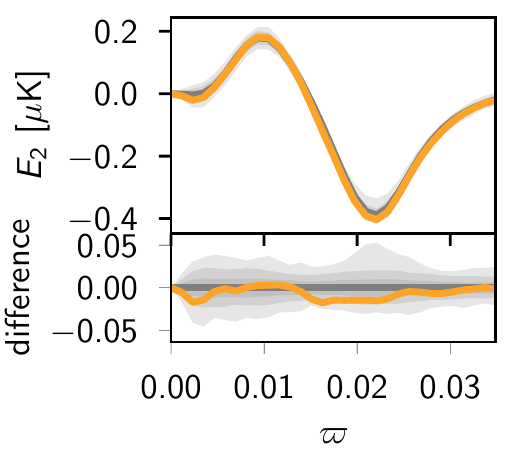} \\
    \includegraphics[scale=0.82]{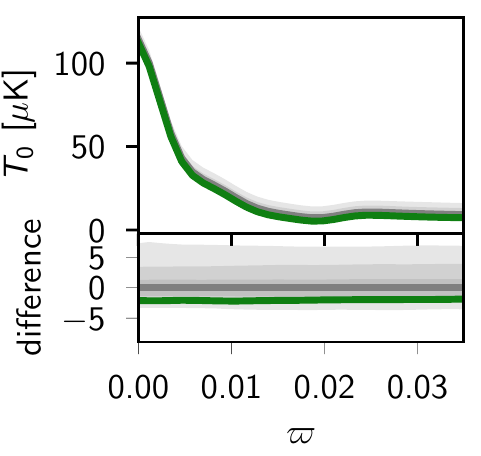} &
    \includegraphics[scale=0.82]{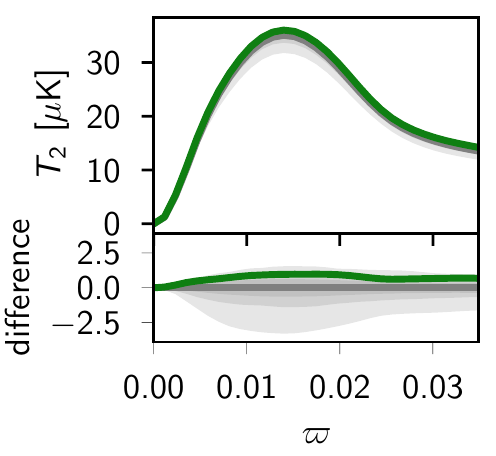} &
    \includegraphics[scale=0.82]{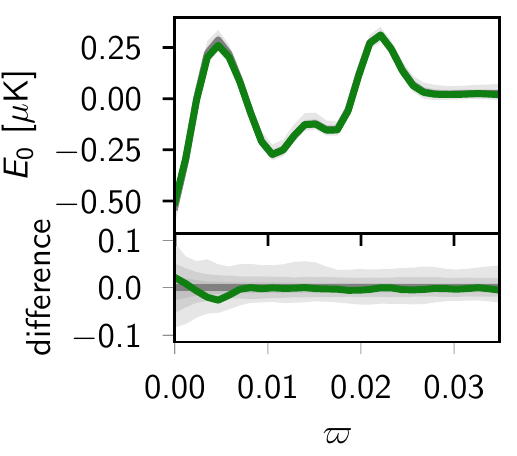} &
    \includegraphics[scale=0.82]{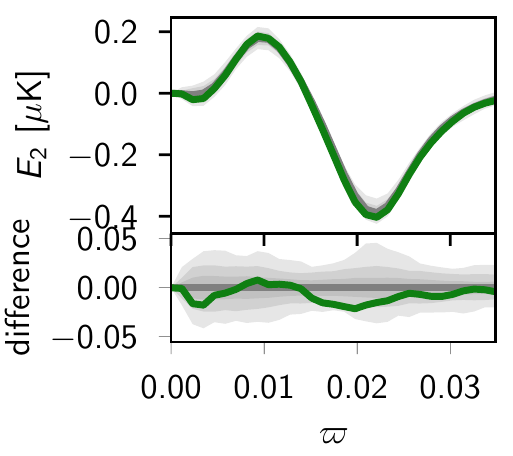} \\
    \includegraphics[scale=0.82]{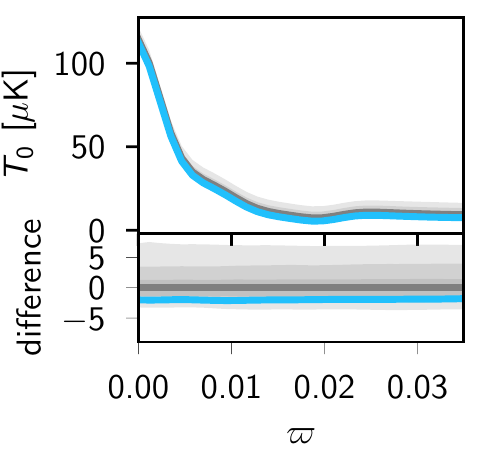} &
    \includegraphics[scale=0.82]{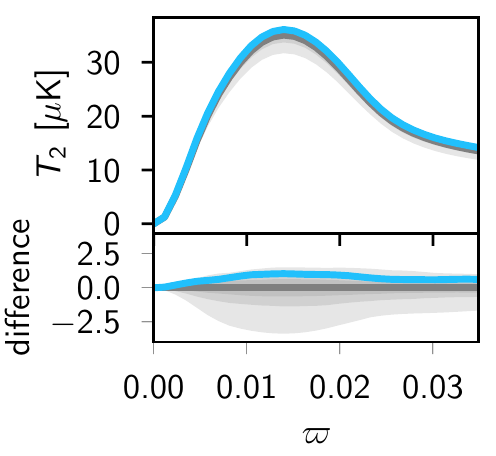} &
    \includegraphics[scale=0.82]{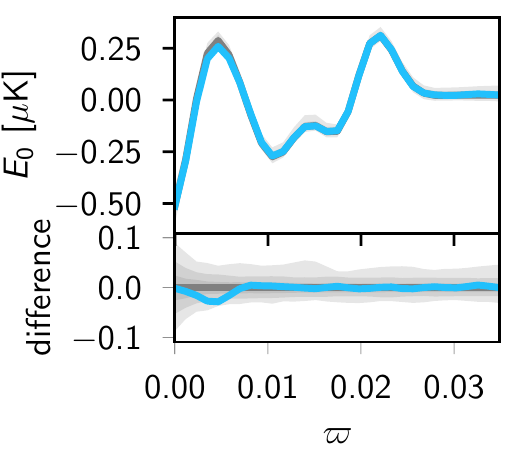} &
    \includegraphics[scale=0.82]{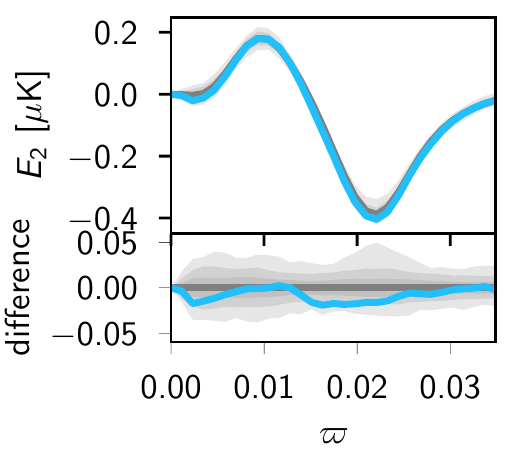} \\
  \end{tabular}
  \caption{Radial profiles of oriented stacks of intensity $T$ (left two columns)
    and reconstructed $E$ mode (right two columns)
    for \commander, \nilc, \sevem, and \smica\ pipelines. The top panel in each plot
    shows the radial profile compared to the simulation average (grey line), while the lower
    panel shows the difference of the radial profile and simulation average on smaller
    scales. The shaded grey regions represent 68\,\%, 95\,\%, and minimum-to-maximum bounds for individual
    realizations.}
    \label{fig:stack:profiles}
\end{figure*}

The stacking method of the previous subsection can be generalized by orienting the
local coordinate frame of the patch to be stacked in a way that is correlated
with the map being stacked. This approach, which we call ``oriented stacking,''
first introduced in \citetalias{planck2014-a18}, allows extra information to be extracted 
from the stacked data. For unoriented stacking, the ensemble average cannot result
in any intrinsic angular dependence, since it would be averaged by the uncorrelated
orientation choices. For example, the ensemble average of the combination of polarization Stokes
parameters $Q+iU$ around unoriented temperature peaks has overall angular
dependence $e^{2i\phi}$, which can be removed by a local rotation
(Eq.~\ref{eqn:local_Stokes}), as was carried out in the previous section.
However, for oriented stacking,
the angular dependence is a linear combination of a few Fourier modes, $e^{im\phi}$,
with the exact mode content determined by the spin of the field being stacked and the spin
of the orientation operator. For scalar fields stacked using orientations determined by the
spin-2 operator, only $m=0$ and $m=2$ modes are present.

Choices of what to stack, where to centre the patch, and how to orient it, provide
a multitude of statistical tests that are complementary to the auto- and cross-correlation
power spectra; these can be used to characterize non-Gaussian data (such as polarized
foreground emission) and have the advantage of being easy to visualize. Here we focus on patch
positions and orientations determined by the highest signal-to-noise channel, namely
temperature $T$, and present stacks of temperature and polarization data on
temperature peaks (either all peaks, or just those selected above a certain threshold $\nu$). The orientation
of the patch can be random, or determined by the second derivative of the temperature
(since gradients vanish at the peak). While the most straightforward way to orient
a patch centred on a temperature peak is to align the axes with principal directions
defined by a local quadratic expansion of the temperature field around the peak, this would be 
susceptible to noise, and a better choice is to use the principal directions of
a local quadratic expansion of the inverse Laplacian of the temperature $\nabla^{-2}T$,
namely the tensor $\nabla_i\nabla_j\nabla^{-2} T$.

The inverse Laplacian of the temperature $\nabla^{-2}T$ is readily computed using
the spherical harmonic transform
\begin{equation}
  \nabla^{-2} T(\hat{\vec{n}}) = - \sum\limits_{\ell=1}^{\ell_{\text{max}}} \sum\limits_{m=-\ell}^{\ell} \frac{a_{\ell m}^T}{\ell(\ell+1)}\, Y_{\ell m}(\hat{\vec{n}}),
\end{equation}
and its covariant derivatives can be computed using standard \healpix\ routines.
Perhaps a simpler interpretation of the tensor $\nabla_i\nabla_j\nabla^{-2} T$
is suggested by its projection onto the Stokes parameters using spin-2 spherical harmonics:
\begin{equation}
  (Q_T \pm i U_T)(\hat{\vec{n}}) = \sum\limits_{\ell=2}^{\ell_{\text{max}}} \sum\limits_{m=-\ell}^{\ell} a_{\ell m}^T\ _{\pm 2} Y_{\ell m}(\hat{\vec{n}}).
\end{equation}
This highlights the fact that the direction defined by the tensor relates to the
temperature $T$ in the same way as the direction of polarization relates to the $E$ mode.
In the flat-sky approximation, the temperature-derived Stokes parameters $Q_T$
and $U_T$ are
\begin{equation}
  Q_T \approx (\partial_x^2 - \partial_y^2)(\nabla^{-2}T), \hspace{1em}
  U_T \approx -2 \partial_x \partial_y (\nabla^{-2}T).
\end{equation}
We orient the patch so that $U_T$ vanishes and $Q_T$ is positive for the central
peak. We use an equal-area azimuthal Lambert projection to represent the stacks,
which introduces the radial variable
\begin{equation}
  \varpi = 2 \sin \frac{\theta}{2} \approx \theta,
\end{equation}
which is almost identical to the usual angular variable $\theta$ in the flat-sky limit,
but allows for less deformation if large angular sizes are considered. Rectangular
coordinates on the patch are defined as usual by $x=\varpi\cos\phi$ and $y=\varpi\sin\phi$.
Given the stacked field $X_{\text{stack}}(\varpi,\phi)$, information can be further
compressed by extracting radial profiles of the non-trivial angular modes:
\begin{equation}\label{eq:stack:radial}
  X_m(\varpi) = \frac{1}{(1+\delta_{m0})\pi} \int\limits_0^{2\pi} X_{\text{stack}}(\varpi,\phi) \cos m\phi\, {\rm d}\phi,
\end{equation}
where $\delta_{m0}$ is the Kronecker delta and $X$ stands for $T$ or $E$.
Statistical isotropy of the stacked field $X$ would imply that ensemble-averaged
radial functions for odd $m$s vanish. Similar to non-oriented stacking (Eq.~\ref{eq:profile_difference}),
the integrated profile deviation,
\begin{equation}\label{eq:stack:deltaXm}
\Delta X_m(W) =  \int_{0}^{R} \left[X_m(\varpi) - \bar{X}_m(\varpi)\right] W(\varpi)\, {\rm d}\varpi,
\end{equation}
can be used to evaluate the consistency between the data and the model. Once
again, the weighting function $W$ is chosen to be proportional to the expected
profile $\bar{X}_m(\varpi)$.

To assess foreground contamination in component-separated CMB
temperature and polarization data, we present unoriented and oriented
stacks of temperature $T$ and full-sky reconstructions of $E$ and
$B$ modes for \Planck\ CMB maps compared to low- and high-frequency
foregrounds in Figs.~\ref{fig:stack:unoriented} and
\ref{fig:stack:oriented}, respectively.  The stacking procedure is
identical, except for the data sets used.
The top rows of these figures show stacks of \smica\
component-separated CMB maps at $N_{\text{side}}=1024$, with a
10\arcmin\ FWHM beam.
The middle rows show stacks of the LFI 30-GHz maps at
$N_{\text{side}}=1024$ (with leakage correction applied), further
convolved with a 10\arcmin\ FWHM Gaussian beam, and corrected for the
CMB contribution by subtracting the \smica\ CMB map (at 10\arcmin\
FWHM resolution) itself beam-convolved with the LFI 30-effective beam,
thus representing the total low-frequency foreground emission.
The bottom rows show the HFI 353-GHz polarization-sensitive bolometer
maps convolved with a 10\arcmin\ FWHM Gaussian beam and degraded to
$N_{\text{side}}=1024$, then CMB-corrected using the \smica\ map
further beam-convolved with the HFI effective beam, thereby
representing the total high-frequency emission.
The polarization maps are transformed to $E$ and $B$ modes via the
full-sky spherical harmonic transform, while the $Q_T$ and $U_T$
signals used to define patch orientations are derived from full-sky
temperature maps. Temperature peaks are selected within the common
intensity mask above the $\nu = 0$ threshold, and patches either
rotated randomly, as shown in Fig.~\ref{fig:stack:unoriented}, or oriented so
that $U_T=0$ is at the peak, as in Fig.~\ref{fig:stack:oriented}. The
maps to be stacked are masked with the common intensity mask for the
temperature channel, and with the confidence mask of
Appendix~\ref{sec:EB_maps} for the polarization channels; monopole and
dipole patterns in unmasked regions are fitted and subtracted out from
all maps (a procedure that is more applicable to CMB maps, but is
also applied to foregrounds as well for consistency), and the result
stacked on temperature peaks selected as described above. All stacks
are presented in units of $\mu$K$_{\text{CMB}}$, and cover square
patches of size 4\deg\ by 4\deg.

The CMB data show a characteristic ringing pattern in the temperature
stacks, associated with the first acoustic peak, a high
signal-to-noise pattern in the $E$-mode stack (as expected from the CMB
$TE$ correlation), and no evidence of detectable patterns in the
$B$-mode stack (reaching a noise floor of about $0.05\,\mu$K), which
serves as a null test for foreground contamination (given the fact
that both low- and high-frequency foregrounds display $TB$
correlations, as described below). Polarized dust foregrounds in
particular were previously investigated in
\citet{planck2016-l11A}. Both low- and high- frequency foreground
stacks show no acoustic peak ringing, with oriented temperature stacks
being notably different from the CMB. The high-frequency stacks show
strong $TE$ and pronounced $TB$ correlations, likely driven by the
properties of the polarized dust emission that is dominant at
353\,GHz. Low-frequency foreground stacks show weaker, but still
noticeable, $TE$ and $TB$ correlations, with no prominent pattern in
the $E$ mode, as is the case for high-frequency foregrounds. All
foreground stacks are quite distinct from CMB ones, and foreground
leakage should result in extraneous correlations in CMB
component-separated stacks, which are not observed. Our results
indicate that the residual foreground contamination in \Planck\
component-separated maps
is below levels that are measurable by this method.

\begin{table}[htbp!]
\begingroup
\newdimen\tblskip \tblskip=5pt
\caption{{\it p}-values of $\chi^2$ computed from the oriented stacking
  of hot and cold spots selected above the $\nu=0$ threshold.
  In this table, small $p$-values would correspond to anomalously
  large values of $\chi^2$.}
\label{table:stack:oriented}
\nointerlineskip
\vskip -3mm
\footnotesize
\setbox\tablebox=\vbox{
   \newdimen\digitwidth 
   \setbox0=\hbox{\rm 0} 
   \digitwidth=\wd0 
   \catcode`*=\active 
   \def*{\kern\digitwidth}
   \newdimen\signwidth 
   \setbox0=\hbox{+} 
   \signwidth=\wd0 
   \catcode`!=\active 
   \def!{\kern\signwidth}
\halign{\hbox to 1.0in{#\leaderfil}\tabskip 3pt&
\hfil#\hfil\tabskip 10pt&
\hfil#\hfil\tabskip 8pt&
\hfil#\hfil\tabskip 8pt&
\hfil#\hfil\/\tabskip 0pt\cr
\noalign{\doubleline}
\noalign{\vskip -1pt}
\omit&\multispan4 \hfil Probability [\%]\hfil\cr
\noalign{\vskip -4pt}
\omit&\multispan4\hrulefill\cr
\omit& {\tt Comm.}& {\tt NILC}& {\tt SEVEM}& {\tt SMICA}\cr
\noalign{\vskip 3pt\hrule\vskip 3pt}
 $T$ ($m = 0$) & 60.3 & 96.4 & 80.7 & 88.4 \cr
 $T$ ($m = 2$) & 56.6 & 88.6 & 83.1 & 48.1 \cr
 $E$ ($m = 0$) & 17.8 & *8.9 & 58.7 & 36.7 \cr
 $E$ ($m = 2$) & 67.7 & 88.0 & 85.1 & 87.6 \cr
\noalign{\vskip 3pt\hrule\vskip 3pt}
}}
\endPlancktable                    
\endgroup
\end{table}

\begin{table}[htbp!]
\begingroup
\newdimen\tblskip \tblskip=5pt
\caption{{\it p}-values of $\Delta X_m$ computed from the oriented stacking
  of hot and cold spots selected above the $\nu=0$ threshold.
  In this table, small $p$-values would correspond to anomalously
  large {\it or\/} small values of $\Delta X_m$.}
\label{table:stack:deltaXm}
\nointerlineskip
\vskip -3mm
\footnotesize
\setbox\tablebox=\vbox{
   \newdimen\digitwidth 
   \setbox0=\hbox{\rm 0} 
   \digitwidth=\wd0 
   \catcode`*=\active 
   \def*{\kern\digitwidth}
   \newdimen\signwidth 
   \setbox0=\hbox{+} 
   \signwidth=\wd0 
   \catcode`!=\active 
   \def!{\kern\signwidth}
\halign{\hbox to 1.0in{#\leaderfil}\tabskip 3pt&
\hfil#\hfil\tabskip 10pt&
\hfil#\hfil\tabskip 8pt&
\hfil#\hfil\tabskip 8pt&
\hfil#\hfil\/\tabskip 0pt\cr
\noalign{\doubleline}
\noalign{\vskip -1pt}
\omit&\multispan4 \hfil Probability [\%]\hfil\cr
\noalign{\vskip -4pt}
\omit&\multispan4\hrulefill\cr
\omit& {\tt Comm.}& {\tt NILC}& {\tt SEVEM}& {\tt SMICA}\cr
\noalign{\vskip 3pt\hrule\vskip 3pt}
 $T$ ($m = 0$) & 12.2 & 11.9 & 12.5 & 14.3 \cr
 $T$ ($m = 2$) & *7.7 & 11.5 & *7.9 & *7.7 \cr
 $E$ ($m = 0$) & *6.0 & *3.3 & *7.9 & 16.1 \cr
 $E$ ($m = 2$) & 12.2 & 15.4 & *8.2 & 12.4 \cr
\noalign{\vskip 3pt\hrule\vskip 3pt}
}}
\endPlancktable                    
\endgroup
\end{table}

To investigate the quantitative agreement of the \Planck\
component-separated maps with fiducial cosmological and noise models,
we compare the radial profiles (Eq.~\ref{eq:stack:radial}) of oriented
temperature and polarization stacks for all component-separation
methods with identically processed FFP10 simulations.  We inpaint
temperature $T$ within the common intensity mask, and reconstruct $E$
and $B$ modes using purified inpainting within the common polarization
mask, as described in Appendix~\ref{sec:EB_maps}. Monopoles and
dipoles are fitted and subtracted from the resulting maps, and once
again we stack on temperature peaks above a threshold $\nu=0$ within
the common intensity mask oriented so that $U_T=0$ at the peak. The
confidence mask of Appendix~\ref{sec:EB_maps} is used for the $E$- and
$B$-mode stacks, and the common intensity mask for $T$ stacks.  Radial
profiles (Eq.~\ref{eq:stack:radial}) for $m=0$, 2, and 4 are extracted
from the stacked data, and are compared to simulations. The results
for $m=0$ and $2$ are shown in Fig.~\ref{fig:stack:profiles}; the
results for $m=4$ are consistent with zero signal, as should be the
case for scalar maps stacked with the spin-2 orientation operator.

The top panels of Fig.~\ref{fig:stack:profiles} show the observed
radial profiles of oriented stacks compared to the ensemble average of
the FFP10 simulations, while the lower panels show the differences on
a magnified scale. The shaded grey regions represent 68\,\% and 95\,\%
confidence, and minimum-to-maximum bounds using profiles obtained from
individual realizations. All observed radial profiles in oriented
stacking agree with simulations within the variance
bounds. Table~\ref{table:stack:oriented} summarizes the upper-tailed
$p$-values of the $\chi^2$ deviation (Eq.~\ref{eq:profile_chi2}) of
oriented radial profiles, while Table~\ref{table:stack:deltaXm}
summarizes the two-tailed $p$-values of the integrated
profile deviations (Eq.~\ref{eq:stack:deltaXm}). None of the
$p$-values appear anomalous.

To summarize this section, oriented stacking results reinforce the conclusion
of the previous section that the stacked \Planck\ CMB data are consistent
with the predictions of the standard $\Lambda$CDM model, given our
understanding of the instrument and simulations.

\section{Anomalies in the microwave sky}
\label{sec:anomalies}

The previous section established the lack of evidence for significant
non-Gaussianity in the \Planck\ temperature and polarization data.
Here we reconsider several noteworthy features detected both in the
WMAP temperature sky maps, and later confirmed in the \Planck\
analyses described in \citetalias{planck2013-p09} and \citetalias{planck2014-a18}.
These include a lack of large-angle correlations, a hemispherical
power asymmetry (either a simple excess of power in one hemisphere or
a continuous dipolar modulation of the CMB anistropy over the sky), a
preference for odd-parity modes in the angular power spectrum, and an
unexpectedly large temperature decrement in the southern hemisphere.
Tests that involve dipolar power asymmetry, either directly or via
measures of directionality, are collected together in
Sect.~\ref{sec:dipmod}, but in this section we consider tests of other
kinds of anomaly.

The existence of these features is uncontested, but, given the modest
significances at which they deviate from the standard \LCDM\
cosmological model, and the a posteriori nature of their detection,
the extent to which they provide evidence for a violation of isotropy
in the CMB remains unclear. It is plausible that they are indeed
simply statistical fluctuations.  Nevertheless, if any one of them
has a physical origin, it would be extremely important, and hence
further investigation is certainly worthwhile.  However, given that
the \Planck\ temperature data are cosmic variance limited on both large
and intermediate angular scales, new information is required to
determine if a real physical effect on the primordial fluctuations
is indicated, or otherwise. This can be achieved by the analysis of
the \Planck\ polarization data.

Of course, the $E$-mode polarization is partially correlated with the
temperature anisotropy, so that it is not a fully statistically
independent probe of the anomalies. Indeed, this correlation could
result in a polarized feature due to the presence of a chance
fluctuation in the temperature map, or could modify any intrinsic
polarization anomaly. In principle, it is possible to split the
polarization (temperature) signal into two parts, one that is
correlated with the temperature (polarization), and one that is
uncorrelated. Such an approach has already been applied to the WMAP
data by \citet{Frommert2010}. However, the methodology requires a
mathematical description of the noise plus residuals of the systematic
and component-separation effects that does not exist for the
component-separated data that we analyse here. Therefore our initial
approach is simply to test for evidence of anomalies in the
polarization data, in addition to verifying once again those seen
previously in the temperature maps.

\subsection{Lack of large-angle correlations}
\label{sec:n_point_anomalies}

We assess the lack of correlation in the 2-point angular correlation
function at large angular separations, as previously noted for both the
WMAP and \Planck\ temperature maps
(\citealt{bennett2003a,copi2013a}; \citetalias{planck2014-a18}). 
In particular, we extend the analysis to polarization data, which were previously 
too noisy and/or contaminated by residual systematic artefacts on
such scales to use them for verification of the temperature anomaly.
We consider the statistic proposed by \citet{copi2013b} in an analysis of the
WMAP data:
\begin{linenomath*}
\begin{equation}
S^{XY}(\theta_1, \theta_2) =  \int_{\cos \theta_2}^{\cos \theta_1} \left[ \hat{C}^{XY}_2(\theta) \right]^2 {\rm d}(\cos \theta)\ ,
\label{eqn:s_stat}
\end{equation}
\end{linenomath*}
where $X,Y$ can denote the temperature anisotropy $T$ or the two Stokes
parameters $Q_\mathrm{r}$ and $U_\mathrm{r}$
(as defined in Sect.~\ref{sec:npoint_correlation}),
and $\hat{C}^{XY}_2(\theta)$ is an
estimate of the corresponding 2-point correlation function. 
This is a generalization of the $S_{1/2}$ statistic
\citep{spergel2003} computed from the temperature auto-correlation
function where the lower, $\theta_1$, and upper, $\theta_2$, limits of the
separation angle range considered are 60\deg\ and 180\deg,
respectively. 

To check the consistency of the current data set with previous
analyses, we again adopt this range of separation angles and present
the results in Tables~\ref{tab:stat_2pt} and \ref{tab:prob_stat_2pt}.
We find that the temperature data show a lack of correlation on large
angular scales, with a significance consistent with that found by
\citet{copi2013a} (although note that the sense of the {\it p}-values
differs between the papers).  The {\it p}-values for the temperature
maps are slightly larger than those determined from the 2015 data set
\citepalias{planck2014-a18}. This could be caused either by changes in
the mask, or the inclusion of systematic effects in the FFP10
simulations. However, in temperature, the latter are relatively small
compared to the cosmological signal on large angular scales.

\citet{copi2013b} demonstrated that the signal-to-noise ratio in the
\WMAP\ polarization maps was insufficient to allow meaningful
estimates of $S^{XY}$ to be made.  For the \Planck\ polarization
data, we have estimated errors on the statistics from the standard
deviation of these values computed from the corresponding FFP10 HMHD
maps. 
We find that the errors are $\sigma_{S_{1/2}} \la 5\times 10^{-5}\,
\mu {\rm K}^4$ for the $S^{TT}_{1/2}$ statistic,
$\sigma_{S_{1/2}} \la 7\times 10^{-6}
\,\mu {\rm K}^4$ for the remaining statistics with estimated values at least
of order $10^{-5}\,\mu {\rm K}^4$, and $\sigma_{S_{1/2}} \la 5\times 10^{-7}
\,\mu {\rm K}^4$ for the statistics with estimated values of order
$10^{-6}\,\mu {\rm K}^4$.  Similar errors are obtained from the OEHD maps. 

A possible explanation for the lack of correlation in the temperature
maps is due to the low observed value of the quadrupole. We therefore
repeat the analysis after removing the best-fit quadrupole\footnote{We
  fit the quadrupole to the masked maps using a modified version of
  the \healpix\ routine {\texttt{remove\_dipole}}. } from the
temperature maps. The corresponding probabilities, estimated from
similarly-corrected FFP10 simulations, are recorded in the second row
of Table~\ref{tab:prob_stat_2pt} and indicate that the low power in
the quadrupole alone does contribute to the absence of
large-angle correlations. \citet{copi2009} argue that all modes below
$\ell \leq 5$ contribute to this, by cancellation with each other and
with higher order modes.

\begin{figure*}[htp!]
\begin{center}
\includegraphics[width=\textwidth]{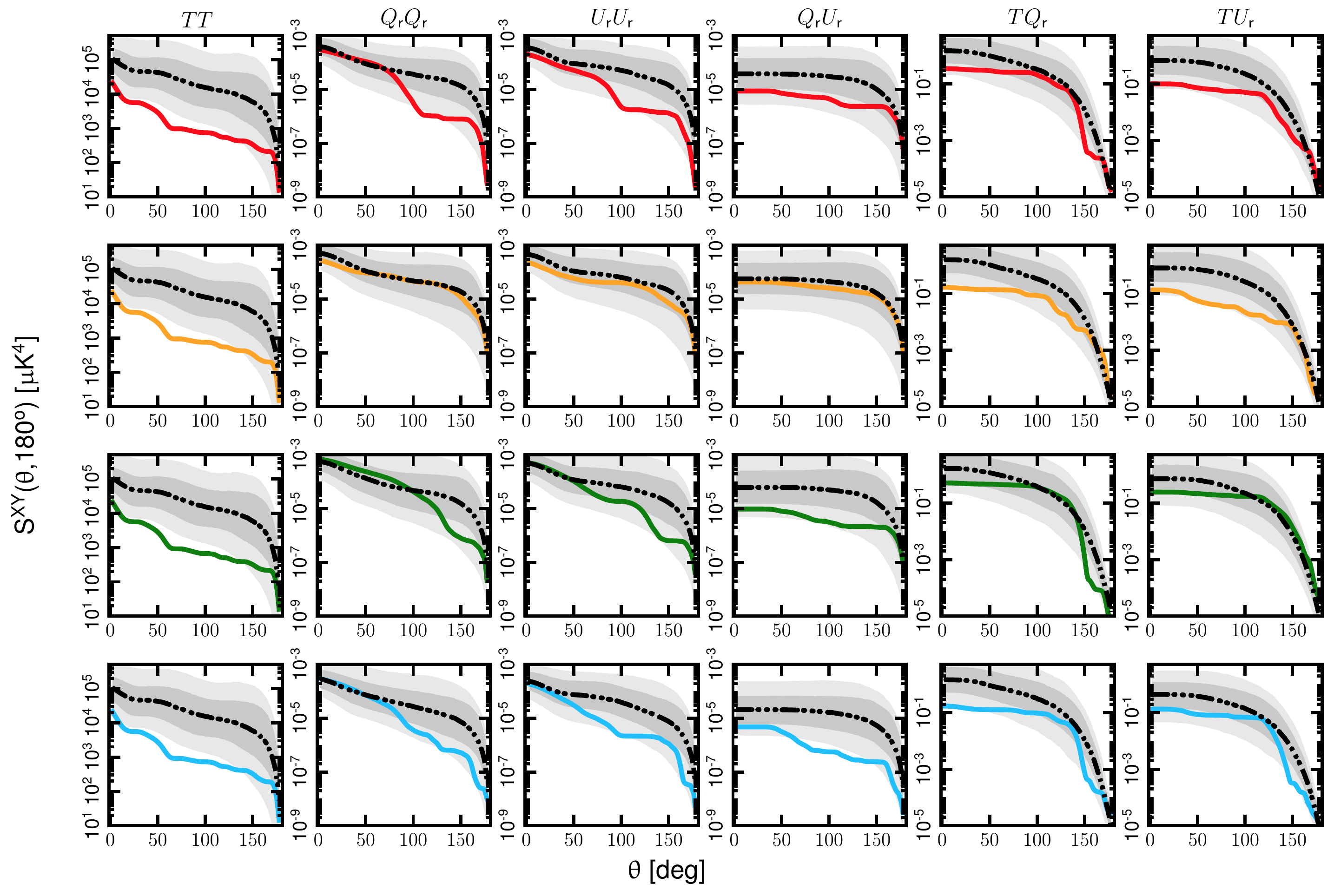}
\caption{
$S^{XY}(\theta,180^\circ)$ statistic for the \nside\ = 64
  \Planck\ CMB 2018 temperature and polarization maps. Results are
  shown for the \commander, \nilc, \sevem, and \smica\
  maps (first, second, third, and fourth rows, respectively). The solid
  lines correspond to the data, while the black three dots-dashed lines indicate the median determined from
  the corresponding FFP10 simulations, and the shaded dark and light grey areas indicate the
  corresponding 68\,\% and 95\,\% confidence areas, respectively.
}
\label{fig:stat_data}
\end{center}
\end{figure*}

\begin{table}[tp] 
\begingroup
\newdimen\tblskip \tblskip=5pt
\caption{Values for the $S^{XY}_{1/2} \equiv S^{XY}(60^\circ,
  180^\circ)$ statistic (in units of $\mu {\rm K}^4$) for the \Planck\
  2018 data with resolution parameter $N_{\rm side}=64$.} 
\label{tab:stat_2pt}
\nointerlineskip
\vskip -3mm
\footnotesize
\setbox\tablebox=\vbox{
   \newdimen\digitwidth 
   \setbox0=\hbox{\rm 0} 
   \digitwidth=\wd0 
   \catcode`*=\active 
   \def*{\kern\digitwidth}
   \newdimen\signwidth 
   \setbox0=\hbox{+} 
   \signwidth=\wd0 
   \catcode`!=\active 
   \def!{\kern\signwidth}
\halign{\hbox to 0.8in{#\leaderfil}\tabskip 4pt&
\hfil#\hfil&
\hfil#\hfil&
\hfil#\hfil&
\hfil#\hfil\tabskip 0pt\cr
\noalign{\doubleline\vskip -1pt}
\omit&\multispan4\hfil $S^{XY}_{1/2}$ [$\mu {\rm K}^4$] \hfil\cr
\noalign{\vskip -4pt}
\omit&\multispan4\hrulefill\cr
\omit\hfil Statistic\hfil&{\tt Comm.}&{\tt NILC}&{\tt SEVEM}&{\tt
  SMICA}\cr 
\noalign{\vskip 3pt\hrule\vskip 3pt}
 $TT$& 1209.2& 1156.6& 1146.2& 1142.4\cr
 $Q_\mathrm{r} Q_\mathrm{r}$& \num{8.3e-05}& \num{8.6e-05}& \num{0.00019}& \num{4.9e-05}\cr
 $U_\mathrm{r} U_\mathrm{r}$& \num{3.9e-05}& \num{4.8e-05}& \num{5.9e-05}& \num{1.6e-05}\cr
 $T Q_\mathrm{r}$& \num{0.26}& \num{0.13}& \num{0.45}& \num{0.13}\cr
 $T U_\mathrm{r}$& \num{0.065}& \num{0.044}& \num{0.2}& \num{0.081}\cr
 $Q_\mathrm{r} U_\mathrm{r}$& \num{6.4e-06}& \num{3.6e-05}& \num{7.1e-06}& \num{2.1e-06}\cr
 \noalign{\vskip 3pt\hrule\vskip 3pt}}}
\endPlancktable                    
\endgroup
\end{table}

\begin{table}[tp] 
\begingroup
\newdimen\tblskip \tblskip=5pt
\caption{Probabilities of obtaining values for the $S^{XY}_{1/2}$ statistic
  for the \Planck\ fiducial $\Lambda$CDM model at least as 
  large as the observed values of the statistic for the \Planck\ 2018
  CMB maps with resolution parameter \nside\ = 64, estimated 
  using the \commander, \nilc, \sevem, 
  and \smica\ maps.  The second row shows results for each temperature
  map after removing the corresponding best-fit quadrupole.
  In this table high probabilities would correspond to anomalously low
  $S^{XY}_{1/2}$ values.}
\label{tab:prob_stat_2pt}
\nointerlineskip
\vskip -3mm
\footnotesize
\setbox\tablebox=\vbox{
   \newdimen\digitwidth 
   \setbox0=\hbox{\rm 0} 
   \digitwidth=\wd0 
   \catcode`*=\active 
   \def*{\kern\digitwidth}
   \newdimen\signwidth 
   \setbox0=\hbox{+} 
   \signwidth=\wd0 
   \catcode`!=\active 
   \def!{\kern\signwidth}
\halign{\hbox to 1.05in{#\leaderfil}\tabskip 1em&
\hfil#\hfil&
\hfil#\hfil&
\hfil#\hfil&
\hfil#\hfil\tabskip 0pt\cr
\noalign{\doubleline\vskip -1pt}
\omit&\multispan4\hfil Probability [\%]\hfil\cr
\noalign{\vskip -4pt}
\omit&\multispan4\hrulefill\cr
\omit\hfil Statistic \hfil& {\tt Comm.}& {\tt NILC}& {\tt SEVEM}& {\tt SMICA}\cr
\noalign{\vskip 3pt\hrule\vskip 3pt}
$TT$& >99.9& >99.9& >99.9& >99.9\cr
$TT$ (no quadr.)& !96.0& !96.1& !96.1& !96.2\cr
$Q_\mathrm{r} Q_\mathrm{r}$& !42.8& !50.5& !30.3& !57.0\cr
$U_\mathrm{r} U_\mathrm{r}$& !74.5& !71.7& !68.9& !89.0\cr
$T Q_\mathrm{r}$& !83.5& !94.0& !74.6& !94.8\cr
$T U_\mathrm{r}$& !94.5& !97.8& !79.1& !88.6\cr
$Q_\mathrm{r} U_\mathrm{r}$& !88.8& !59.7& !94.4& !97.7\cr
 \noalign{\vskip 3pt\hrule\vskip 3pt}}}
\endPlancktable                    
\endgroup
\end{table}

A potential criticism of the $S^{XY}_{1/2}$ statistic relates to the
a~posteriori choice of the range of separation angles to delineate the
interesting region of behaviour of the correlation function. As in
\citetalias{planck2014-a18}, we consider the generalized statistic
$S^{XY}(\theta, 180^\circ)$ and compute it for all values of
$\theta$ both for the data and simulations.
Figure~\ref{fig:stat_data} indicates that the only excursion
outside of the 95\,\% confidence regions determined from simulations
is observed for the temperature data when the lower bound of the integral is
$\theta\,{\approx}\,60^\circ$, for all component-separation methods. No
equivalent excursions are seen for any statistics involving
polarization data, which also show notable variations between
component-separation methods. We provide a more quantitative
assessment as follows.
For each value of $\theta$, we determine the number of simulations
with a higher value of $S^{XY}(\theta,180^\circ)$, and hence infer the
most significant value of the statistic and the separation angle that
it corresponds to. However, since such an analysis is sensitive to the
look-elsewhere effect, we define a global statistic to evaluate the
true significance of the result. Specifically, we repeat the procedure
for each simulation, and search for the largest probability
irrespective of the value of $\theta$ at which it occurs. The fraction
of these probabilities lower than the maximum probability found for
the data defines a global {\it p}-value. As seen in
Table~\ref{tab:prob_stat_global_2pt}, this corresponds to values of
order 99\,\% for all of the component-separated temperature maps, and
much smaller global {\it p}-values for quantities involving the
polarization data. 
Note that the results for the temperature data are slightly more
significant than those for the 2015 data set
(i.e., 99\,\% compared to
98\,\%). More importantly, the {\it p}-values for the
temperature maps after removing the best-fit quadrupole drop to
around 86\,\%, indicating again that the anomalous value of the
$S^{TT}$ statistic is connected to the observed quadrupole.

\begin{table}[tp] 
\begingroup
\newdimen\tblskip \tblskip=5pt
\caption{Global {\it p}-value for the $S^{XY}(\theta, 180^\circ)$ statistic
  for the \Planck\ fiducial $\Lambda$CDM model at most as 
  large as the observed values of the statistic for the \Planck\ 2018
  CMB maps with resolution parameter $N_{\rm side}=64$.
  The second row shows results for each temperature
  map after removing the corresponding best-fit quadrupole.
  In this table high probabilities would correspond to anomalously low
  $S^{XY}(\theta, 180^\circ)$ values.}
\label{tab:prob_stat_global_2pt}
\nointerlineskip
\vskip -3mm
\footnotesize
\setbox\tablebox=\vbox{
   \newdimen\digitwidth 
   \setbox0=\hbox{\rm 0} 
   \digitwidth=\wd0 
   \catcode`*=\active 
   \def*{\kern\digitwidth}
   \newdimen\signwidth 
   \setbox0=\hbox{+} 
   \signwidth=\wd0 
   \catcode`!=\active 
   \def!{\kern\signwidth}
\halign{\hbox to 1.05in{#\leaderfil}\tabskip 1em&
\hfil#\hfil&
\hfil#\hfil&
\hfil#\hfil&
\hfil#\hfil\tabskip 0pt\cr
\noalign{\doubleline\vskip -1pt}
\omit&\multispan4\hfil Probability [\%]\hfil\cr
\noalign{\vskip -4pt}
\omit&\multispan4\hrulefill\cr
\omit\hfil Statistic \hfil& {\tt Comm.}& {\tt NILC}& {\tt SEVEM}& {\tt SMICA}\cr
\noalign{\vskip 3pt\hrule\vskip 3pt}
 $TT$& !98.8& !98.8& !98.8& !99.0\cr
 $TT$ (no quadr.)& !85.5& !86.2& !85.7& !87.2\cr
 $Q_\mathrm{r} Q_\mathrm{r}$& !94.2& !57.1& !74.5& !85.7\cr
 $U_\mathrm{r} U_\mathrm{r}$& !89.6& !60.3& !82.1& !85.2\cr
 $T Q_\mathrm{r}$& !80.0& !84.8& !95.8& !87.0\cr
 $T U_\mathrm{r}$& !71.0& !82.9& !34.6& !88.0\cr
 $Q_\mathrm{r} U_\mathrm{r}$& !72.5& !48.7& !87.3& !97.1\cr
 \noalign{\vskip 3pt\hrule\vskip 3pt}}}
\endPlancktable                    
\endgroup
\end{table} 

As pointed out by \citet{copi2013b}, the cross-correlation between
temperature and polarization can be used to determine whether the
$S^{TT}$ value is low due to a statistical fluke in the context of a
\LCDM\ cosmology. Specifically, if this hypothesis were true, the
$S^{TQ}$ statistic would also likely be small on large angular scales,
given that the polarization signal is partially correlated with
temperature. \citet{copi2013b} have determined that the $S^{TQ}$
statistic would be smaller than 1.403\,$\mu {\rm K}^4$ over the angular
separation range $[48^\circ, 120^\circ]$ in 99\,\% of their
constrained realizations based on the properties of the \WMAP\
seven-year data and assuming the statistical fluke hypothesis to be
true. A value of the measured statistic exceeding this limit would then
allow the hypothesis to be ruled out.  Table~\ref{tab:stat_2pt_tq}
indicates that the values of the $S^{TQ}(48^\circ,120^\circ)$
statistic measured with the \Planck\ data are significantly smaller, so
that we cannot rule out the statistical fluke hypothesis.

\begin{table}[tp] 
\begingroup
\newdimen\tblskip \tblskip=3pt
\caption{Values for the $S^{TQ}(48^\circ,
  120^\circ)$ statistic (in $\mu {\rm K}^4$) and the probability of
  obtaining values for the \Planck\ fiducial $\Lambda$CDM model at
  least as large as the observed values for the \Planck\ 
  2018 CMB maps with resolution parameter $N_{\rm side}=64$.} 
\label{tab:stat_2pt_tq}
\nointerlineskip
\vskip -3mm
\footnotesize
\setbox\tablebox=\vbox{
   \newdimen\digitwidth 
   \setbox0=\hbox{\rm 0} 
   \digitwidth=\wd0 
   \catcode`*=\active 
   \def*{\kern\digitwidth}
   \newdimen\signwidth 
   \setbox0=\hbox{+} 
   \signwidth=\wd0 
   \catcode`!=\active 
   \def!{\kern\signwidth}
\halign{\hbox to 1.0in{#\leaderfil}\tabskip 4pt&
\hfil#\hfil&
\hfil#\hfil\tabskip 0pt\cr
\noalign{\doubleline\vskip -1pt}
\omit\hfil Method \hfil& \hfil $S^{TQ}(48^\circ,120^\circ)$ [$\mu {\rm K}^4$] \hfil& Probability [\%]\cr
\noalign{\vskip 3pt\hrule\vskip 3pt}
\commander& 0.20& 83.8\cr
 \nilc& 0.11& 93.3\cr
 \sevem& 0.26& 82.9\cr
 \smica& 0.07& 97.0\cr
 \noalign{\vskip 3pt\hrule\vskip 3pt}}}
\endPlancktable                    
\endgroup
\end{table}

\citet{muir:2018} have noted that the low value of the
large-angle correlation statistic $S^{TT}_{1/2}$ can be connected
to the temperature 2-point correlation function for antipodean points,
$C^{TT}_2(180^\circ)$.  This is motivated by fact that, as seen in
Fig.~\ref{fig:2pt_data}, the otherwise largely flat
$C^{TT}_2(\theta)$ drops to negative values close to
$\theta = 180^\circ$.  Here, we extend the analysis of the
temperature correlation function to
the functions including polarization information,
$C^{XY}_2(180^\circ)$.

Table~\ref{tab:prob_cpi} presents the probabilities of finding larger
values of the 2-point correlation functions at $180^\circ$ from the data than
those estimated using the FFP10 simulations. The results determined
from the temperature maps are consistent with those reported by
\citet{muir:2018}, i.e., around 89\,\% (although note that the sense of
the {\it p}-values differs between the papers). When the
best-fit quadrupole is removed, the significance falls to around
60\,\%, indicating that the low value of the quadrupole can impact
the statistic. We do not observe any anomalous behaviour 
for the functions evaluated at this angular separation that include
polarization fields.

\begin{table}[tp] 
\begingroup
\newdimen\tblskip \tblskip=5pt
\caption{Probabilities of getting the $C^{XY}_2(180^\circ)$ statistic
  for the \Planck\ fiducial $\Lambda$CDM model at least as 
  large as the observed values of the statistic for the \Planck\ 2018
  CMB maps with resolution parameter $N_{\rm side}=64$. 
  The second row shows results for the temperature maps with the 
  best-fit quadrupole removed.
  In this table high probabilities would correspond to anomalously low
  $C^{XY}_2(180^\circ)$ values..}
\label{tab:prob_cpi}
\nointerlineskip
\vskip -3mm
\footnotesize
\setbox\tablebox=\vbox{
   \newdimen\digitwidth 
   \setbox0=\hbox{\rm 0} 
   \digitwidth=\wd0 
   \catcode`*=\active 
   \def*{\kern\digitwidth}
   \newdimen\signwidth 
   \setbox0=\hbox{+} 
   \signwidth=\wd0 
   \catcode`!=\active 
   \def!{\kern\signwidth}
\halign{\hbox to 1.05in{#\leaderfil}\tabskip 4pt&
\hfil#\hfil&
\hfil#\hfil&
\hfil#\hfil&
\hfil#\hfil\tabskip 0pt\cr
\noalign{\doubleline\vskip -1pt}
\omit&\multispan4\hfil Probability [\%]\hfil\cr
\noalign{\vskip -4pt}
\omit&\multispan4\hrulefill\cr
\omit\hfil Statistic \hfil& {\tt Comm.}& {\tt NILC}& {\tt SEVEM}& {\tt SMICA}\cr
\noalign{\vskip 3pt\hrule\vskip 3pt}
 $TT$&                        !89.6& !89.2& !89.5& !89.2\cr
 $TT$ (no quadr.)&            !60.8& !60.7& !60.5& !60.5\cr
 $Q_\mathrm{r} Q_\mathrm{r}$& !34.0& !*6.8& !41.6& !37.3\cr
 $U_\mathrm{r} U_\mathrm{r}$& !62.2& !85.8& !57.2& !48.9\cr
 $T Q_\mathrm{r}$&            !15.8& !14.9& !28.8& !10.7\cr
 $T U_\mathrm{r}$&            !83.6& !79.4& !91.9& !76.2\cr
 $Q_\mathrm{r} U_\mathrm{r}$& !80.4& !88.8& !86.0& !65.8\cr
 \noalign{\vskip 3pt\hrule\vskip 3pt}}}
\endPlancktable                    
\endgroup
\end{table}

\subsection{Hemispherical asymmetry} 
\label{sec:npoint_asymmetry}

In this section, we reassess the asymmetry between the real-space
$N$-point correlation functions computed on hemispheres, reported
previously for the \WMAP\ \citep{eriksen2005} and \Planck\ 
temperature maps \citepalias{planck2013-p09,planck2014-a18}. We
pay special attention to testing this asymmetry using the \Planck\
polarization data. Several different 2-point and 3-point functions
drawn from permutations of the $T$, $Q$, and $U$ maps are tested.

As in Sect.~\ref{sec:npoint_correlation}, we analyse the CMB estimates
at a resolution of \mbox{\nside\ = 64} and use the same configurations
of the $N$-point functions. However, here the functions are not
averaged over the full sky and depend on a choice of specific
direction, so they constitute tools for studying statistical isotropy
rather than non-Gaussianity \citep{Ferreira1997}.

\begin{figure*}[htbp!]
\begin{center}
\includegraphics[width=0.5\textwidth]{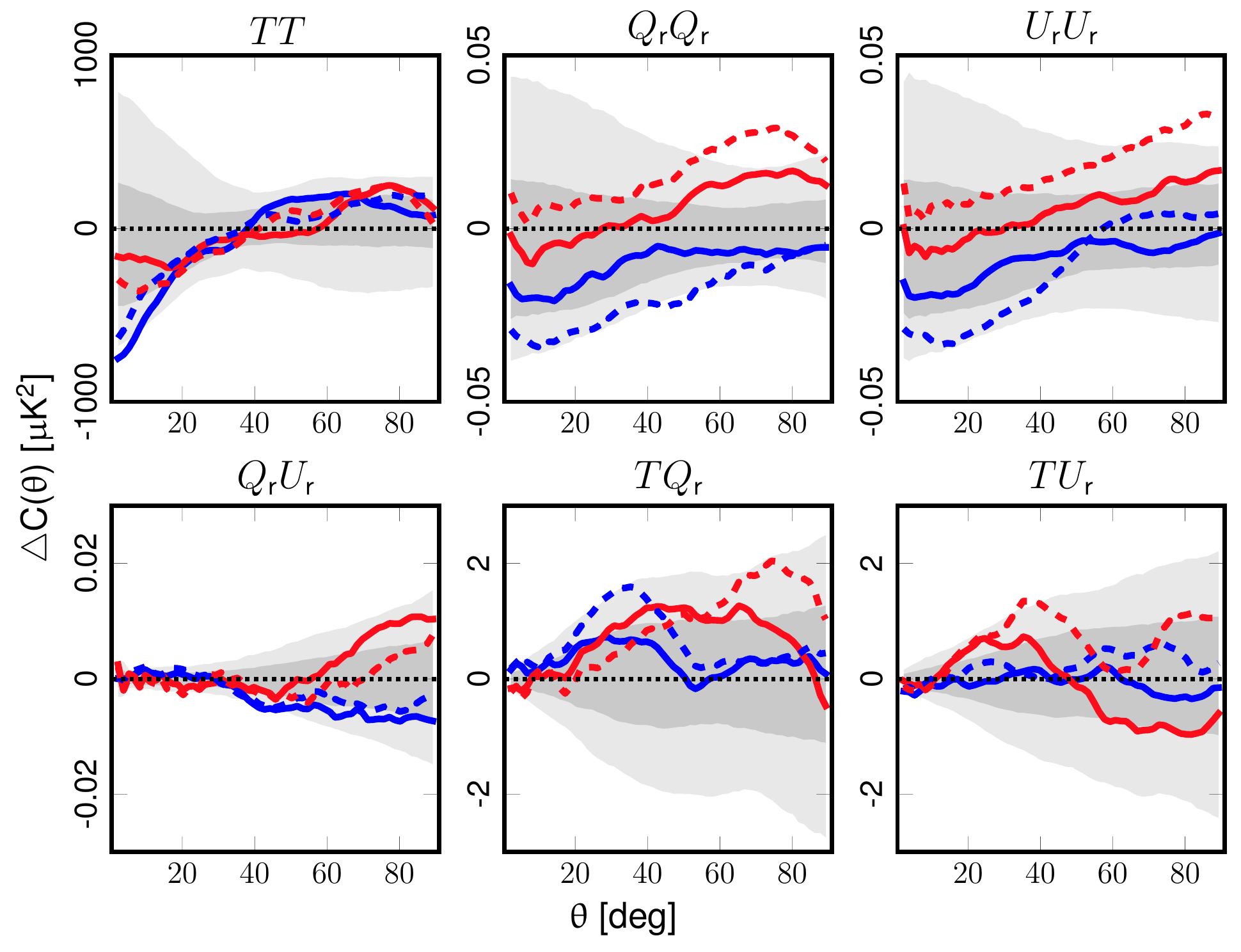}\\
\includegraphics[width=0.8\textwidth]{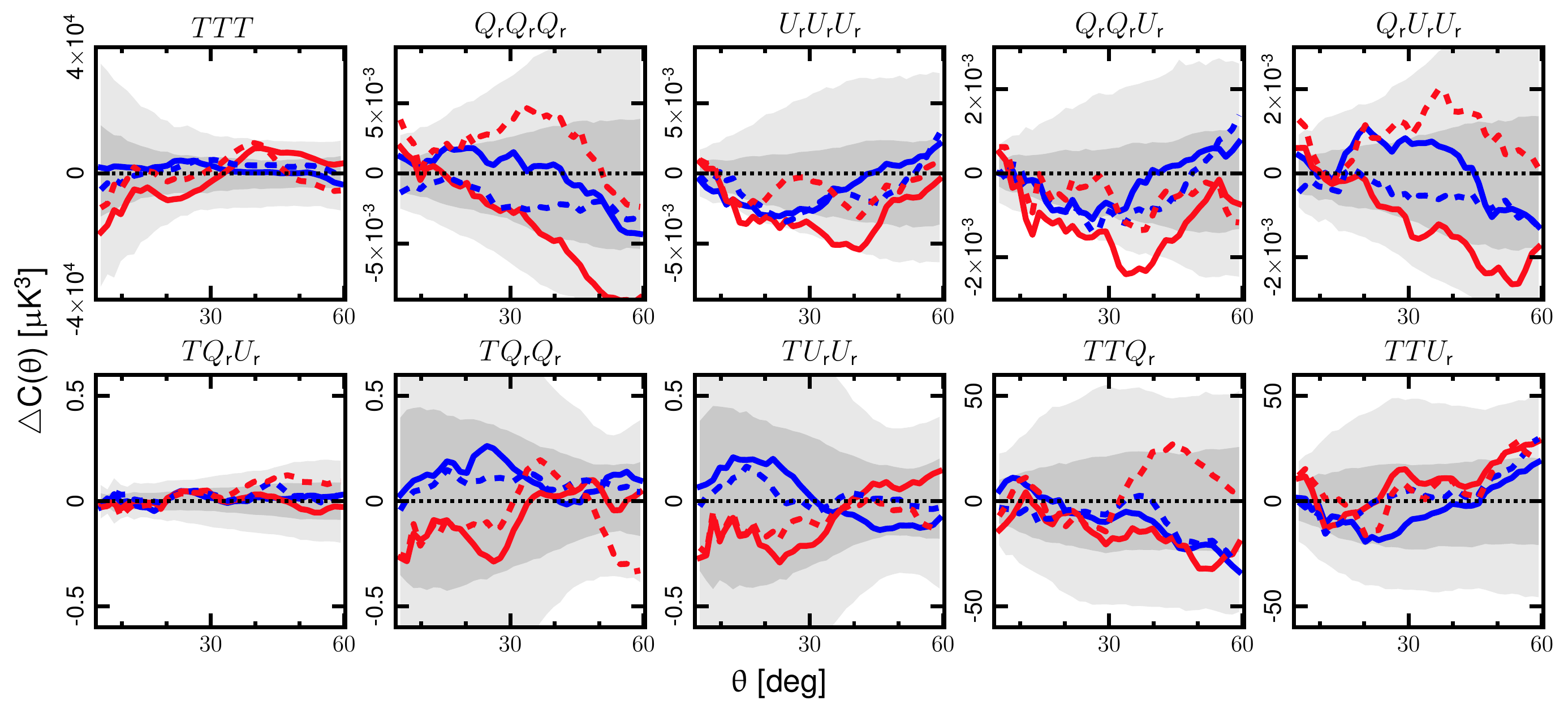}\\
\includegraphics[width=0.8\textwidth]{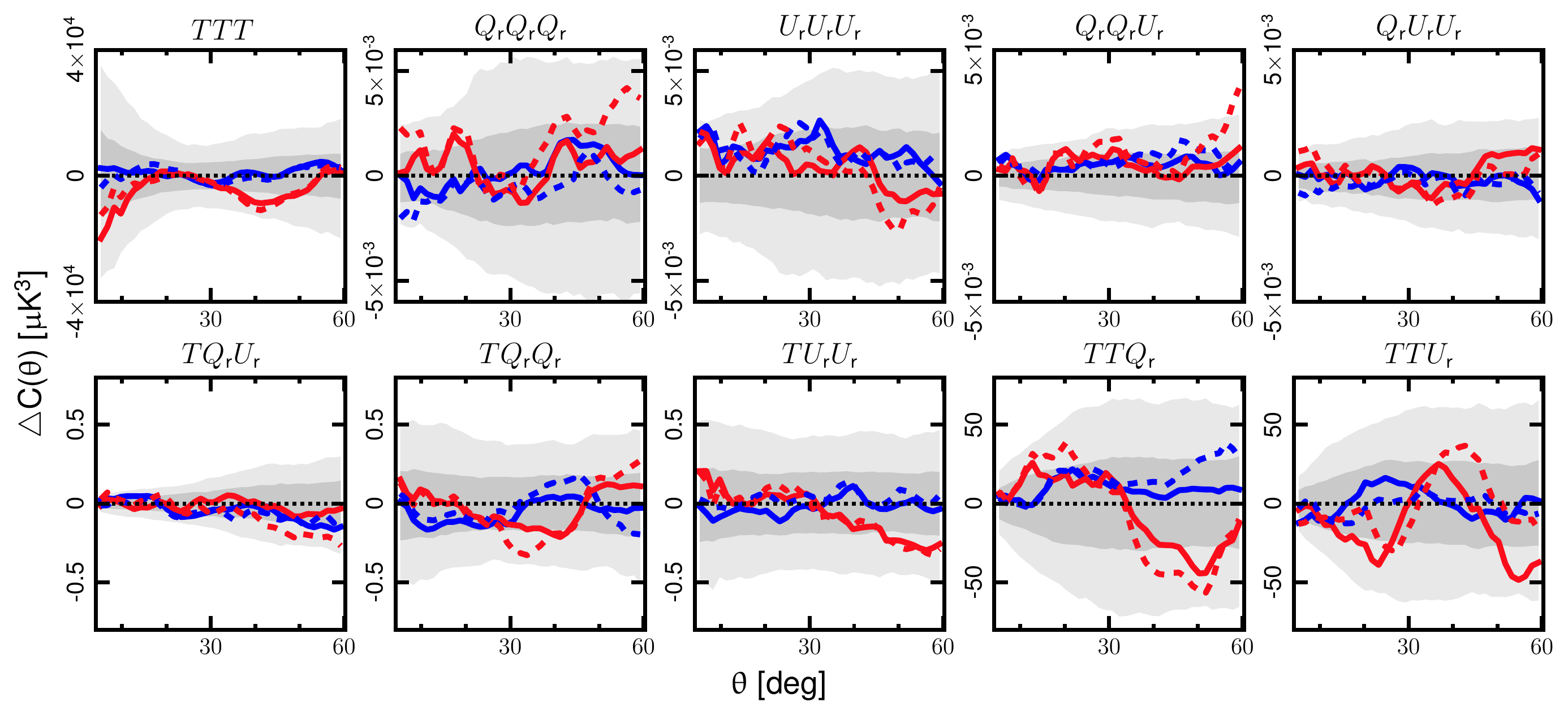}
\caption{Difference of the $N$-point correlation functions determined
  from the $N_{\rm side} = 64$
  \Planck\ CMB \commander\ 2018 temperature and polarization maps
  and the corresponding means estimated from FFP10 MC simulations.
  Results are shown for the 2-point (upper panels), pseudo-collapsed
  3-point (middle panels), and equilateral 3-point (lower
  panels) functions.
  The blue and red dashed lines correspond to the functions computed on
  the negative and positive hemispheres, respectively, determined in the
  dipole-modulation coordinate frame. The blue and red solid lines
  correspond to the functions computed on 
  the northern and southern ecliptic hemispheres, respectively. The
  shaded dark and light grey 
  regions indicate the 68\,\% and 95\,\% confidence regions,
  respectively, determined for the negative DM hemisphere. See
  Sect.~\ref{sec:npoint_correlation} for the definition of the
  separation angle $\theta$.}
\label{fig:npt_data_commander_asymm}
\end{center}
\end{figure*}

We test the asymmetry for two separate cases: (i)
the hemispheres determined in the ecliptic
coordinate frame (i.e., those hemispheres separated by the ecliptic
equator); and (ii) those determined by the dipole-modulation (DM)
analysis in Sect.~\ref{sec:dipmod_qml}.  In the latter case, the positive and
negative hemispheres are defined as those for which the DM amplitude
is positive or negative, respectively. The DM direction adopted
here, $(l,b) = (221^\circ, -20^\circ)$, then corresponds to the pole
of the positive hemisphere, taken to be the average over the results
determined from the four component-separated maps in
Table~\ref{tab:lowellmod}, for the data combination
$TT,\, TE,\, EE$. We do not include an additional analysis of the
hemispheric split defined by the
Doppler-boost direction as presented in \citetalias{planck2014-a18}, since
the observed asymmetry described there was less significant than observed for
the ecliptic and DM frames.  The results for cases (i) and (ii) are
presented in Fig.~\ref{fig:npt_data_commander_asymm}, where, as in
\citetalias{planck2014-a18}, we compute differences between the
$N$-point functions for the data and the mean values estimated from
the FFP10 MC simulations.

As in Sect.~\ref{sec:npoint_correlation}, 
we quantify the agreement of the correlation functions of the
CMB estimates with the fiducial cosmological model using a $\chi^2$
statistic defined by Eq.~(\ref{eqn:chisqr_stat}), where the
correlation function, mean, and covariance matrix are computed for a
corresponding hemisphere.
If we consider that the $\chi^2$ statistic itself can act as a measure
of fluctuation level, then asymmetry between the two measured
hemispheres can be quantified by the ratio of the corresponding
$\chi^2$ values. The probabilities of obtaining values of the $\chi^2$
statistic for the \Planck\ fiducial $\Lambda$CDM model at least as
large as the observed values are given in
Tables~\ref{tab:prob_npt_asymm_ecl} and \ref{tab:prob_npt_asymm_dm}
for the ecliptic and DM reference frames, respectively.  The
corresponding probabilities for the ratio of the $\chi^2$ values are
given in Tables~\ref{tab:prob_npt_asymm_ratio_ecl} and
\ref{tab:prob_npt_asymm_ratio_dm}.  Since we do not have any
predictions concerning the behaviour of a given hemisphere, in the
case of the $\chi^2$ ratios we provide the complementary probabilities
of the 2-tailed statistic.

\begin{table}[htbp!] 
\begingroup
\newdimen\tblskip \tblskip=5pt
\caption{Probabilities of obtaining values for the $\chi^2$ statistic
  of the $N$-point functions for
  the \Planck\ fiducial $\Lambda$CDM model at least as
  large as the observed values for the {\tt Commander}, {\tt NILC}, {\tt SEVEM},
  and {\tt SMICA}  temperature and polarization $Q$ and $U$ maps 
  estimated on northern and southern ecliptic hemispheres.
  In this table high probabilities represent significant differences
  between the two hemispheres.}
\label{tab:prob_npt_asymm_ecl}
\nointerlineskip
\vskip -3mm
\footnotesize
\setbox\tablebox=\vbox{
   \newdimen\digitwidth 
   \setbox0=\hbox{\rm 0} 
   \digitwidth=\wd0 
   \catcode`*=\active 
   \def*{\kern\digitwidth}
   \newdimen\signwidth 
   \setbox0=\hbox{+} 
   \signwidth=\wd0 
   \catcode`!=\active 
   \def!{\kern\signwidth}
\halign{\hbox to 1.0in{#\leaderfil}\tabskip 4pt&
\hfil#\hfil&
\hfil#\hfil&
\hfil#\hfil&
\hfil#\hfil&
\hfil#\hfil\tabskip 0pt\cr
\noalign{\doubleline\vskip -1pt}
\omit&\omit&\multispan4 \hfil Probability [\%]\hfil\cr
\noalign{\vskip -4pt}
\omit&\omit& \multispan4 \hrulefill\cr
\omit\hfil Function\hfil&\hfil Hemisphere \hfil&{\tt Comm.}&{\tt NILC}&{\tt SEVEM}&{\tt SMICA}\cr
\noalign{\vskip 3pt\hrule\vskip 6pt}
\multispan6\hfil 2-point functions\hfil\cr
\noalign{\vskip 3pt}
$TT$& Northern& 30.5& 30.7& 29.6& 30.9\cr
\omit& Southern& 80.2& 81.5& 80.5& 80.7\cr
$Q_\mathrm{r} Q_\mathrm{r}$& Northern& 92.4& 36.9& 87.9& 82.5\cr
\omit& Southern& 24.3& 52.6& 61.7& 70.0\cr
$U_\mathrm{r} U_\mathrm{r}$& Northern& 44.4& 38.3& 45.8& 24.1\cr
\omit& Southern& 45.2& 12.7& 87.6& 69.5\cr
$T Q_\mathrm{r}$& Northern& 36.4& 41.1& 39.2& 40.3\cr
\omit& Southern& 72.6& 82.1& 63.5& 57.5\cr
$T U_\mathrm{r}$& Northern& 53.8& 27.0& 97.1& 53.8\cr
\omit& Southern& 84.5& 67.1& 43.2& 53.5\cr
$Q_\mathrm{r} U_\mathrm{r}$& Northern& 67.0& 28.6& 25.9& 18.4\cr
\omit& Southern& 45.5& 62.5& 92.1& 59.5\cr
\noalign{\vskip 6pt\hrule\vskip 6pt}
\multispan6\hfil Pseudo-collapsed 3-point functions\hfil\cr
\noalign{\vskip 3pt}
$TTT$& Northern& 81.6& 84.5& 83.0& 83.8\cr
\omit& Southern& 19.7& 18.5& 18.8& 17.7\cr
$Q_\mathrm{r} Q_\mathrm{r} Q_\mathrm{r}$& Northern& 81.7& 88.6& 49.8& 84.1\cr
\omit& Southern& 26.6& 51.4& 22.3& 21.0\cr
$U_\mathrm{r} U_\mathrm{r} U_\mathrm{r}$& Northern& 72.3& 56.1& 51.6& 45.3\cr
\omit& Southern& *5.6& *5.6& *8.8& *2.1\cr
$TTQ_\mathrm{r}$& Northern& 74.7& 53.2& 51.4& 68.5\cr
\omit& Southern& 27.8& 24.7& 35.8& 19.6\cr
$TTU_\mathrm{r}$& Northern& 98.0& 84.7& 98.8& 96.9\cr
\omit& Southern& 23.7& 28.0& 26.4& 44.1\cr
$TQ_\mathrm{r} Q_\mathrm{r}$& Northern& 69.2& 81.2& 58.7& 53.1\cr
\omit& Southern& *2.4& *5.4& 12.9& *1.4\cr
$TU_\mathrm{r} U_\mathrm{r}$& Northern& 94.0& 99.6& 92.5& 86.0\cr
\omit& Southern& 20.4& 50.3& 27.2& 20.1\cr
$TQ_\mathrm{r} U_\mathrm{r}$& Northern& 85.0& 74.4& 87.4& 91.5\cr
\omit& Southern& 60.8& 24.6& 23.6& 23.9\cr
$Q_\mathrm{r} Q_\mathrm{r} U_\mathrm{r}$& Northern& 63.3& 60.1& 86.7& 55.8\cr
\omit& Southern& *2.8& *5.5& *5.6& *2.7\cr
$Q_\mathrm{r} U_\mathrm{r} U_\mathrm{r}$& Northern& 55.0& 53.0& 37.0& 70.9\cr
\omit& Southern& 30.2& 52.4& 16.3& 31.8\cr
\noalign{\vskip 6pt\hrule\vskip 6pt}
\multispan6\hfil Equilateral 3-point functions\hfil\cr
\noalign{\vskip 3pt}
$TTT$& Northern& 94.3& 94.3& 94.4& !94.3\cr
 \omit& Southern& 81.5& 81.4& 81.3& !80.4\cr
$Q_\mathrm{r} Q_\mathrm{r} Q_\mathrm{r}$& Northern& 82.0& 41.2& 89.2& !68.7\cr
 \omit& Southern& 17.7& 15.3& *4.3& !*3.6\cr
$U_\mathrm{r} U_\mathrm{r} U_\mathrm{r}$& Northern& 57.5& 91.4& 32.9& !82.0\cr
 \omit& Southern& 40.8& 15.2& 12.4& !*9.4\cr
$TTQ_\mathrm{r}$& Northern& 99.5& 99.8& 98.1& >99.9\cr
 \omit& Southern& 32.2& 12.8& 34.7& !39.4\cr
$TTU_\mathrm{r}$& Northern& 95.2& 99.5& 86.4& !94.6\cr
 \omit& Southern& 23.4& 42.2& 27.7& !40.1\cr
$TQ_\mathrm{r} Q_\mathrm{r}$& Northern& 91.9& 72.9& 94.5& !99.1\cr
 \omit& Southern& 57.2& 59.9& 54.5& !12.4\cr
$TU_\mathrm{r} U_\mathrm{r}$& Northern& 82.4& 96.2& 78.2& !83.4\cr
 \omit& Southern& 72.1& 91.5& 80.2& !54.1\cr
$TQ_\mathrm{r} U_\mathrm{r}$& Northern& 91.3& 79.3& 93.8& !94.6\cr
 \omit& Southern& 68.9& 55.4& 73.3& !50.2\cr
$Q_\mathrm{r} Q_\mathrm{r} U_\mathrm{r}$& Northern& 93.6& 76.9& 43.7& !87.5\cr
 \omit& Southern& 32.5& 45.5& 10.6& !11.7\cr
$Q_\mathrm{r} U_\mathrm{r} U_\mathrm{r}$& Northern& 93.4& 74.2& 52.7& !86.4\cr
 \omit& Southern& 36.0& 12.4& 60.2& !13.1\cr
 \noalign{\vskip 6pt\hrule\vskip 3pt}}}
\endPlancktable                    
\endgroup
\end{table}

\begin{table}[htbp!]
\begingroup
\newdimen\tblskip \tblskip=5pt
\caption{Probabilities of obtaining values for the $\chi^2$ statistic
  of the $N$-point functions for
  the \Planck\ fiducial $\Lambda$CDM model at least as large as the observed
  values of the statistics for the {\tt Commander}, {\tt NILC}, {\tt SEVEM},
  and {\tt SMICA}  temperature and polarization $Q$ and $U$ maps 
  estimated on negative and positive hemispheres defined by the DM
  reference frame.
  In this table high probabilities represent significant differences
  between the two hemispheres.}
\label{tab:prob_npt_asymm_dm}
\nointerlineskip
\vskip -3mm
\footnotesize
\setbox\tablebox=\vbox{
   \newdimen\digitwidth 
   \setbox0=\hbox{\rm 0} 
   \digitwidth=\wd0 
   \catcode`*=\active 
   \def*{\kern\digitwidth}
   \newdimen\signwidth 
   \setbox0=\hbox{+} 
   \signwidth=\wd0 
   \catcode`!=\active 
   \def!{\kern\signwidth}
\halign{\hbox to 1.2in{#\leaderfil}\tabskip 4pt&
\hfil#\hfil&
\hfil#\hfil&
\hfil#\hfil&
\hfil#\hfil&
\hfil#\hfil\tabskip 0pt\cr
\noalign{\doubleline\vskip -1pt}
\omit&\omit&\multispan4 \hfil Probability [\%]\hfil\cr
\noalign{\vskip -4pt}
\omit&\omit& \multispan4 \hrulefill\cr
\omit\hfil Function\hfil&\hfil Hemisphere \hfil&{\tt Comm.}&{\tt NILC}&{\tt SEVEM}&{\tt SMICA}\cr
\noalign{\vskip 3pt\hrule\vskip 6pt}
\multispan6\hfil 2-point functions\hfil\cr
\noalign{\vskip 3pt}
$TT$& Negative&                       46.4& 45.1& 45.2& 46.6\cr
\omit& Positive&                      56.8& 57.1& 57.6& 58.0\cr
$Q_\mathrm{r}Q_\mathrm{r}$& Negative& 25.8& 24.6& 77.5& 14.1\cr
\omit& Positive&                      *6.3& 98.4& 14.5& 11.9\cr
$U_\mathrm{r}U_\mathrm{r}$& Negative& 43.4& 11.1& 78.7& 23.9\cr
\omit& Positive&                      17.7& 15.0& 10.2& *3.2\cr
$TQ_\mathrm{r}$& Negative&            29.6& 52.9& 20.8& 49.7\cr
\omit& Positive&                      *1.9& *3.3& *0.5& *2.5\cr
$TU_\mathrm{r}$& Negative&            80.2& 89.0& 81.4& 79.2\cr
\omit& Positive&                      *9.7& *4.2& 17.6& *2.7\cr
$Q_\mathrm{r}U_\mathrm{r}$& Negative& 33.8& *4.0& 13.0& 17.2\cr
\omit& Positive&                      10.9& *4.3& 15.8& *1.2\cr
\noalign{\vskip 6pt\hrule\vskip 6pt}
\multispan6\hfil Pseudo-collapsed 3-point functions\hfil\cr
\noalign{\vskip 3pt}
$TTT$& Negative&                                  96.1& 96.6& 96.4& 96.3\cr
\omit& Positive&                                  *5.2& *4.9& *5.0& *5.1\cr
$Q_\mathrm{r}Q_\mathrm{r}Q_\mathrm{r}$& Negative& 91.9& 92.4& 59.0& 99.1\cr
\omit& Positive&                                  *4.5& 18.8& *2.0& *1.1\cr
$U_\mathrm{r}U_\mathrm{r}U_\mathrm{r}$& Negative& 22.3& 42.5& 13.5& 78.6\cr
\omit& Positive&                                  *6.3& *7.9& *4.2& *5.6\cr
$TTQ_\mathrm{r}$& Negative&                       77.0& 43.4& 78.1& 85.4\cr
\omit& Positive&                                  50.4& 60.9& 44.8& 39.4\cr
$TTU_\mathrm{r}$& Negative&                       95.9& 88.5& 87.3& 97.2\cr
\omit& Positive&                                  *6.8& *8.7& *4.6& *5.5\cr
$TQ_\mathrm{r}Q_\mathrm{r}$& Negative&            95.7& 76.9& 99.7& 94.2\cr
\omit& Positive&                                  *3.2& *8.3& 12.1& *5.5\cr
$TU_\mathrm{r}U_\mathrm{r}$& Negative&            79.6& 63.7& 56.3& 66.0\cr
\omit& Positive&                                  81.6& 92.0& 82.0& 42.6\cr
$TQ_\mathrm{r}U_\mathrm{r}$& Negative&            76.3& 72.1& 80.9& 94.7\cr
\omit& Positive&                                  55.7& 25.5& 58.7& 14.3\cr
$Q_\mathrm{r}Q_\mathrm{r}U_\mathrm{r}$& Negative& 68.7& 30.2& 71.8& 44.3\cr
\omit& Positive&                                  *3.4& *1.9& *1.1& *0.9\cr
$Q_\mathrm{r}U_\mathrm{r}U_\mathrm{r}$& Negative& 96.5& 90.3& 48.7& 94.2\cr
\omit& Positive&                                  17.1& 16.1& 15.9& *6.6\cr
\noalign{\vskip 6pt\hrule\vskip 6pt}
\multispan6\hfil Equilateral 3-point functions\hfil\cr
\noalign{\vskip 3pt}
$TTT$& Negative&                                  95.8& 95.9& 95.2& 95.8\cr
\omit& Positive&                                  76.4& 73.9& 74.5& 73.9\cr
$Q_\mathrm{r}Q_\mathrm{r}Q_\mathrm{r}$& Negative& 77.4& 22.3& 57.4& 68.9\cr
\omit& Positive&                                  *9.0& *6.9& *3.3& *3.7\cr
$U_\mathrm{r}U_\mathrm{r}U_\mathrm{r}$& Negative& 87.3& 54.9& 37.7& 79.0\cr
\omit& Positive&                                  22.2& 19.3& 26.9& 11.8\cr
$TTQ_\mathrm{r}$& Negative&                       94.7& 82.1& 97.5& 98.7\cr
\omit& Positive&                                  18.4& 15.0& 12.1& 18.9\cr
$TTU_\mathrm{r}$& Negative&                       95.3& 97.5& 87.6& 96.0\cr
\omit& Positive&                                  38.7& 37.8& 13.4& 49.1\cr
$TQ_\mathrm{r}Q_\mathrm{r}$& Negative&            75.7& 68.9& 64.7& 84.9\cr
\omit& Positive&                                  22.0& 15.5& 40.0& *4.7\cr
$TU_\mathrm{r}U_\mathrm{r}$& Negative&            68.0& 86.3& 67.4& 68.3\cr
\omit& Positive&                                  54.3& 70.5& 33.9& 26.8\cr
$TQ_\mathrm{r}U_\mathrm{r}$& Negative&            59.7& 30.0& 64.1& 63.3\cr
\omit& Positive&                                  26.1& 57.5& 24.6& 13.2\cr
$Q_\mathrm{r}Q_\mathrm{r}U_\mathrm{r}$& Negative& 92.6& 30.8& 51.4& 74.3\cr
\omit& Positive&                                  *3.2& 22.8& *1.9& *0.9\cr
$Q_\mathrm{r}U_\mathrm{r}U_\mathrm{r}$& Negative& 80.1& 66.0& 82.4& 67.4\cr
\omit& Positive&                                  15.1& 15.5& 42.9& *6.1\cr
 \noalign{\vskip 6pt\hrule\vskip 3pt}}}
\endPlancktable                    
\endgroup
\end{table}

\begin{table}[htbp!]
\begingroup
\newdimen\tblskip \tblskip=5pt
\caption{Probabilities of obtaining values for the ratio of $\chi^2$ of
  the $N$-point functions for
  the \Planck\ fiducial $\Lambda$CDM model at least as large as the observed
  values of the statistics for the {\tt Commander}, {\tt NILC}, {\tt SEVEM},
  and {\tt SMICA}  temperature and polarization $Q$ and $U$ maps 
  estimated on northern and southern ecliptic hemispheres.
  In this table high probabilities represent significant differences
  between the two hemispheres.}
\label{tab:prob_npt_asymm_ratio_ecl}
\nointerlineskip
\vskip -3mm
\footnotesize
\setbox\tablebox=\vbox{
   \newdimen\digitwidth 
   \setbox0=\hbox{\rm 0} 
   \digitwidth=\wd0 
   \catcode`*=\active 
   \def*{\kern\digitwidth}
   \newdimen\signwidth 
   \setbox0=\hbox{+} 
   \signwidth=\wd0 
   \catcode`!=\active 
   \def!{\kern\signwidth}
\halign{\hbox to 1.2in{#\leaderfil}\tabskip 4pt&
\hfil#\hfil&
\hfil#\hfil&
\hfil#\hfil&
\hfil#\hfil\tabskip 0pt\cr
\noalign{\doubleline\vskip -1pt}
\omit&\multispan4 \hfil Probability [\%]\hfil\cr
\noalign{\vskip -4pt}
\omit& \multispan4 \hrulefill\cr
\omit\hfil Function\hfil&{\tt Comm.}&{\tt NILC}&{\tt SEVEM}&{\tt SMICA}\cr
\noalign{\vskip 3pt\hrule\vskip 6pt}
\multispan5\hfil 2-point functions\hfil\cr
\noalign{\vskip 3pt}
$TT$& 73.2& 74.8& 73.5& 73.5\cr
$Q_\mathrm{r} Q_\mathrm{r}$& 89.3& 22.4& 51.9& 27.4\cr
$U_\mathrm{r} U_\mathrm{r}$& 14.5& 48.1& 63.2& 62.7\cr
$T Q_\mathrm{r}$& 50.5& 61.7& 41.3& 23.8\cr
$T U_\mathrm{r}$& 62.1& 63.0& 89.8& *1.0\cr
$Q_\mathrm{r} U_\mathrm{r}$& 33.7& 43.1& 91.6& 62.2\cr
\noalign{\vskip 6pt\hrule\vskip 6pt}
\multispan5\hfil Pseudo-collapsed 3-point functions\hfil\cr
\noalign{\vskip 3pt}
$TTT$& 77.0& 81.0& 79.9& 80.8\cr
$Q_\mathrm{r} Q_\mathrm{r} Q_\mathrm{r}$& 73.3& 68.2& 44.6& 85.4\cr
$U_\mathrm{r} U_\mathrm{r} U_\mathrm{r}$& 93.8& 89.0& 76.9& 90.0\cr
$TTQ_\mathrm{r}$& 63.7& 40.2& 20.4& 65.6\cr
$TTU_\mathrm{r}$& 96.9& 76.8& 97.5& 87.9\cr
$TQ_\mathrm{r} Q_\mathrm{r}$& 90.2& 91.8& 62.2& 87.3\cr
$TU_\mathrm{r} U_\mathrm{r}$& 90.7& 92.4& 90.5& 85.1\cr
$TQ_\mathrm{r} U_\mathrm{r}$& 52.3& 73.1& 89.2& 87.5\cr
$Q_\mathrm{r} Q_\mathrm{r} U_\mathrm{r}$& 96.0& 88.7& 98.7& 91.4\cr
$Q_\mathrm{r} U_\mathrm{r} U_\mathrm{r}$& 40.8& *1.1& 45.2& 49.4\cr
\noalign{\vskip 6pt\hrule\vskip 6pt}
\multispan5\hfil Equilateral 3-point functions\hfil\cr
\noalign{\vskip 3pt}
$TTT$& 40.0& 39.9& 40.8& 42.5\cr
$Q_\mathrm{r} Q_\mathrm{r} Q_\mathrm{r}$& 84.1& 42.7& 97.1& 92.8\cr
$U_\mathrm{r} U_\mathrm{r} U_\mathrm{r}$& 27.4& 94.7& 36.6& 91.4\cr
$TTQ_\mathrm{r}$& 97.6& 99.3& 88.7& 99.5\cr
$TTU_\mathrm{r}$& 92.5& 98.2& 80.9& 85.1\cr
$TQ_\mathrm{r} Q_\mathrm{r}$& 47.2& 12.2& 56.3& 98.2\cr
$TU_\mathrm{r} U_\mathrm{r}$& 21.8& 15.5& 33.2& 45.6\cr
$TQ_\mathrm{r} U_\mathrm{r}$& 43.0& 39.5& 32.4& 69.7\cr
$Q_\mathrm{r} Q_\mathrm{r} U_\mathrm{r}$& 89.5& 46.9& 64.1& 93.4\cr
$Q_\mathrm{r} U_\mathrm{r} U_\mathrm{r}$& 85.0& 84.2& 13.5& 91.1\cr
 \noalign{\vskip 6pt\hrule\vskip 3pt}}}
\endPlancktable                    
\endgroup
\end{table}

\begin{table}[htbp!]
\begingroup
\newdimen\tblskip \tblskip=5pt
\caption{Probabilities of obtaining values for the ratio of $\chi^2$ of
  the $N$-point functions for the \Planck\ fiducial $\Lambda$CDM model at
  least as large as the observed values of the statistics for the
  {\tt Commander}, {\tt NILC}, {\tt SEVEM}, and {\tt SMICA} temperature
  and polarization $Q$ and $U$ maps estimated on negative and positive
  hemispheres defined by the DM reference frame.
  In this table high probabilities represent significant differences
  between the two hemispheres.}
\label{tab:prob_npt_asymm_ratio_dm}
\nointerlineskip
\vskip -3mm
\footnotesize
\setbox\tablebox=\vbox{
   \newdimen\digitwidth 
   \setbox0=\hbox{\rm 0} 
   \digitwidth=\wd0 
   \catcode`*=\active 
   \def*{\kern\digitwidth}
   \newdimen\signwidth 
   \setbox0=\hbox{+} 
   \signwidth=\wd0 
   \catcode`!=\active 
   \def!{\kern\signwidth}
\halign{\hbox to 1.3in{#\leaderfil}\tabskip 4pt&
\hfil#\hfil&
\hfil#\hfil&
\hfil#\hfil&
\hfil#\hfil\tabskip 0pt\cr
\noalign{\doubleline\vskip -1pt}
\omit&\multispan4 \hfil Probability [\%]\hfil\cr
\noalign{\vskip -4pt}
\omit& \multispan4 \hrulefill\cr
\omit\hfil Function\hfil&{\tt Comm.}&{\tt NILC}&{\tt SEVEM}&{\tt SMICA}\cr
\noalign{\vskip 3pt\hrule\vskip 6pt}
\multispan5\hfil 2-point functions\hfil\cr
\noalign{\vskip 3pt}
\hglue 1em$TT$ & 12.8 & 15.7 & 16.0 & 15.3 \cr
\hglue 1em$Q_\mathrm{r} Q_\mathrm{r}$ & 57.3 & 95.9 & 86.0 & 12.8 \cr
\hglue 1em$U_\mathrm{r} U_\mathrm{r}$ & 49.8 & *2.6 & 89.1 & 60.4 \cr
\hglue 1em$T Q_\mathrm{r}$ & 67.3 & 80.5 & 73.0 & 82.7 \cr
\hglue 1em$T U_\mathrm{r}$ & 87.5 & 96.4 & 81.2 & 94.5 \cr
\hglue 1em$Q_\mathrm{r} U_\mathrm{r}$ & 47.4 & *2.1 & 13.7 & 62.8 \cr
\noalign{\vskip 6pt\hrule\vskip 6pt}
\multispan5\hfil Pseudo-collapsed 3-point functions\hfil\cr
\noalign{\vskip 3pt}
 \hglue 1em$TTT$ & 98.7 & 98.9 & 98.7 & 98.8 \cr
 \hglue 1em$Q_\mathrm{r} Q_\mathrm{r} Q_\mathrm{r}$ & 98.9 & 92.7 & 93.2 & 99.8 \cr
 \hglue 1em$U_\mathrm{r} U_\mathrm{r} U_\mathrm{r}$ & 49.3 & 52.9 & 31.5 & 92.6 \cr
 \hglue 1em$TTQ_\mathrm{r}$ & 52.2 & 16.0 & 57.0 & 73.3 \cr
 \hglue 1em$TTU_\mathrm{r}$ & 99.0 & 95.4 & 96.0 & 99.3 \cr
 \hglue 1em$TQ_\mathrm{r} Q_\mathrm{r}$ & 99.2 & 89.4 & 99.4 & 99.0 \cr
 \hglue 1em$TU_\mathrm{r} U_\mathrm{r}$ & *0.7 & 47.4 & 28.2 & 30.8 \cr
 \hglue 1em$TQ_\mathrm{r} U_\mathrm{r}$ & 25.6 & 56.6 & 38.8 & 95.4 \cr
 \hglue 1em$Q_\mathrm{r} Q_\mathrm{r} U_\mathrm{r}$ & 92.4 & 75.1 & 97.4 & 90.5 \cr
 \hglue 1em$Q_\mathrm{r} U_\mathrm{r} U_\mathrm{r}$ & 95.8 & 81.1 & 51.7 & 97.7 \cr
\noalign{\vskip 6pt\hrule\vskip 6pt}
\multispan5\hfil Equilateral 3-point functions\hfil\cr
\noalign{\vskip 3pt}
 \hglue 1em$TTT$ & 52.5 & 57.3 & 54.2 & 56.7 \cr
 \hglue 1em$Q_\mathrm{r} Q_\mathrm{r} Q_\mathrm{r}$ & 88.8 & 41.1 & 88.3 & 91.9 \cr
 \hglue 1em$U_\mathrm{r} U_\mathrm{r} U_\mathrm{r}$ & 85.1 & 49.6 & 10.4 & 83.0 \cr
 \hglue 1em$TTQ_\mathrm{r}$ & 93.8 & 84.3 & 98.7 & 98.2 \cr
 \hglue 1em$TTU_\mathrm{r}$ & 85.9 & 92.3 & 89.6 & 82.3 \cr
 \hglue 1em$TQ_\mathrm{r} Q_\mathrm{r}$ & 77.8 & 75.8 & 26.7 & 94.5 \cr
 \hglue 1em$TU_\mathrm{r} U_\mathrm{r}$ & *9.5 & 22.7 & 39.3 & 57.2 \cr
 \hglue 1em$TQ_\mathrm{r} U_\mathrm{r}$ & 45.5 & 42.7 & 51.0 & 71.3 \cr
 \hglue 1em$Q_\mathrm{r} Q_\mathrm{r} U_\mathrm{r}$ & 99.1 & *3.8 & 86.3 & 95.9 \cr
 \hglue 1em$Q_\mathrm{r} U_\mathrm{r} U_\mathrm{r}$ & 86.5 & 68.1 & 61.6 & 84.3 \cr
 \noalign{\vskip 6pt\hrule\vskip 3pt}}}
\endPlancktable                    
\endgroup
\end{table} 

Looking at the results in
Tables~\ref{tab:prob_npt_asymm_ecl}--\ref{tab:prob_npt_asymm_ratio_dm},
we no longer observe the high significance level for the
pseudo-collapsed 3-point function of the temperature map 
in the northern ecliptic hemisphere, as 
reported for the 2013 and 2015 \Planck\ data sets
\citepalias{planck2013-p09,planck2014-a18}. This may be a
consequence of the use of different masks in the various analyses, 
or of the improved treatment of poorly
determined modes in the estimated correlation matrix used
for the computation of the $\chi^2$ statistic, as described in
Sect.~\ref{sec:npoint_correlation}.
The largest asymmetry, at a significance of
around 97--99\,\%, is observed for the
$TTQ_\mathrm{r}$ equilateral 3-point function (although not for the
\sevem\ map). 
This comes from the very high probabilities found for the northern
ecliptic hemisphere. It is also worth noting that the {\it p}-values for 
the \smica\ $TU_\mathrm{r}$ 2-point function and the 
\nilc\ $Q_\mathrm{r} U_\mathrm{r} U_\mathrm{r}$ 
pseudo-collapsed 3-point function are very similar for the two 
hemispheres; as a consequence, the probabilities for the corresponding
$\chi^2$ ratios are very small, typically of order 1\,\%. We
cannot offer any explanation as to the similarity of 
these correlation functions in the two
hemispheres. However, one should recognize that a~posteriori 
choices for the smoothing scale and reference frames defining the
hemispheres will lead to an overestimation of the significance of the
results. In addition, the large number of correlation-function
configurations considered will increase the likelihood of finding some
apparently anomalous behaviour, even for statistically isotropic data. 

In the DM reference frame, the 3-point temperature correlation
functions in the negative hemisphere are somewhat significant,
reaching a level of 98--99\,\% for the pseudo-collapsed case.  Similar
behaviour is also observed for some of the component-separated maps
for the $Q_\mathrm{r}Q_\mathrm{r}Q_\mathrm{r}$, $TTU_\mathrm{r}$ and
$TQ_\mathrm{r}Q_\mathrm{r}$ pseudo-collapsed 3-point functions, and
for a few configurations of the equilateral 3-point function.  Of
course, the inconsistency in {\it p}-values for different
component-separated maps argues against results of cosmological
significance, but rather indicates the presence of different levels of
foreground or systematic-effect residuals, depending on the
component-separation method.

\subsection{Point-parity asymmetry}
\label{sec:pointparity}

Now we turn to another way in which statistical isotropy can be broken at
large scales.
The CMB temperature anisotropy field on the sky can be written
as the sum of parity-symmetric $T^+(\vec{\hat{n}})$ and 
parity-antisymmetric $T^-(\vec{\hat{n}})$ functions defined as
\begin{equation}
T^{\pm}(\vec{\hat{n}}) = \frac{1}{2} [T(\vec{\hat{n}}) \pm T(-\vec{\hat{n}})] \, ,
\end{equation}
where for a given direction $\vec{\hat{n}}$, the antipodal point is $- \vec{\hat{n}}$.
Under the parity transformation
$T^{(P)}(\vec{\hat{n}})=P[T(\vec{\hat{n}})] = T(- \vec{\hat{n}})$, it can be
shown that the corresponding $a_{\ell m}$ coefficients transform such
that $a^{(P)}_{\ell m} = (-1)^\ell a_{\ell m}$. As a consequence, 
$T^+(\vec{\hat{n}})$ and $T^-(\vec{\hat{n}})$ are comprised of spherical
harmonics with only even or odd $\ell$-modes, respectively. 

For the polarization field, a similar approach can be followed by
expanding the $Q$ and $U$ stokes parameter maps as
\begin{equation}
\begin{split}
Q^{\pm}(\vec{\hat{n}}) &= \frac{1}{2} [Q(\vec{\hat{n}}) \pm Q(-\vec{\hat{n}})] \, ,\\
U^{\pm}(\vec{\hat{n}}) &= \frac{1}{2} [U(\vec{\hat{n}}) \mp U(-\vec{\hat{n}})] \, .
\end{split}
\end{equation}
It can then be shown that these functions are described by $E$- and
$B$-mode angular power spectra with only even or odd $\ell$-modes,
since under the parity transformation
$a^{(P)}_{E,\ell m} = (-1)^\ell a_{E,\ell m}$ and
$a^{(P)}_{B,\ell m} = (-1)^{\ell+1} a_{B,\ell m}$.

On the largest angular scales, $2<\ell<30$, corresponding to the
Sachs-Wolfe plateau of the temperature power spectrum, we expect the
angular power spectra of $T^+(\vec{\hat{n}})$ and $T^-(\vec{\hat{n}})$ to
be of comparable amplitude.
However, an odd point-parity preference has been observed both in
various \wmap\ data releases \citep{LandOdd, kim2010, Gruppuso2010},
and confirmed in the \Planck\ 2013 and 2015 studies.
Furthermore, \citet{Land2005} have shown that this cannot be
  easily ascribed to the presence of residual Galactic foreground emission because of
  its even point-parity nature.
Here we investigate the parity asymmetry of the \Planck\ 2018
temperature data at \nside = 32 using the same estimator adopted in
\citetalias{planck2014-a18}, which is defined as follows:

\begin{linenomath*}
\begin{equation}\label{eq:pointparity}
R^\mathrm{TT}(\ell_\mathrm{max}) = \frac{C^\mathrm{TT}_+(\ell_\mathrm{max})}{C^\mathrm{TT}_-(\ell_\mathrm{max})} \, ,
\end{equation}
\end{linenomath*}
where $C^\mathrm{TT}_+(\ell_\mathrm{max})$ and $C^\mathrm{TT}_-(\ell_\mathrm{max})$ are given by

\begin{linenomath*}
\begin{equation}\label{eq:TTpointparity}
C^\mathrm{TT}_{+,-} = \frac{1}{\ell_\mathrm{tot}^{+,-}}\sum_{\ell=2,\ell_\mathrm{max}}^{+,-} \frac{\ell(\ell+1)}{2 \pi}C^\mathrm{TT}_{\ell} \, ,
\end{equation}
\end{linenomath*}
with $\ell_\mathrm{tot}^{+,-}$ being the total number of even ($+$) or odd
($-$) multipoles included in the sum up to $\ell_\mathrm{max}$, and $C^\mathrm{TT}_{\ell}$,
is the temperature angular power spectrum computed using a
quadratic maximum-likelihood (QML)
estimator \citep{Gruppuso2010, Molinari2014}.

\begin{figure}
\centering
\includegraphics[scale=0.45]{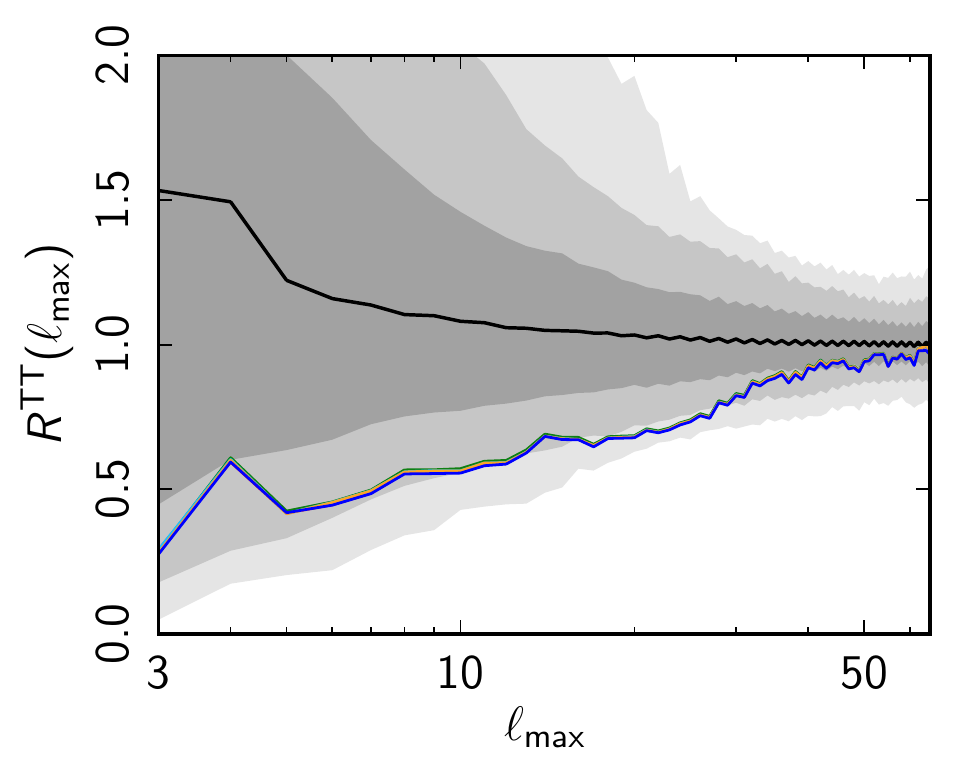}
\includegraphics[scale=0.45]{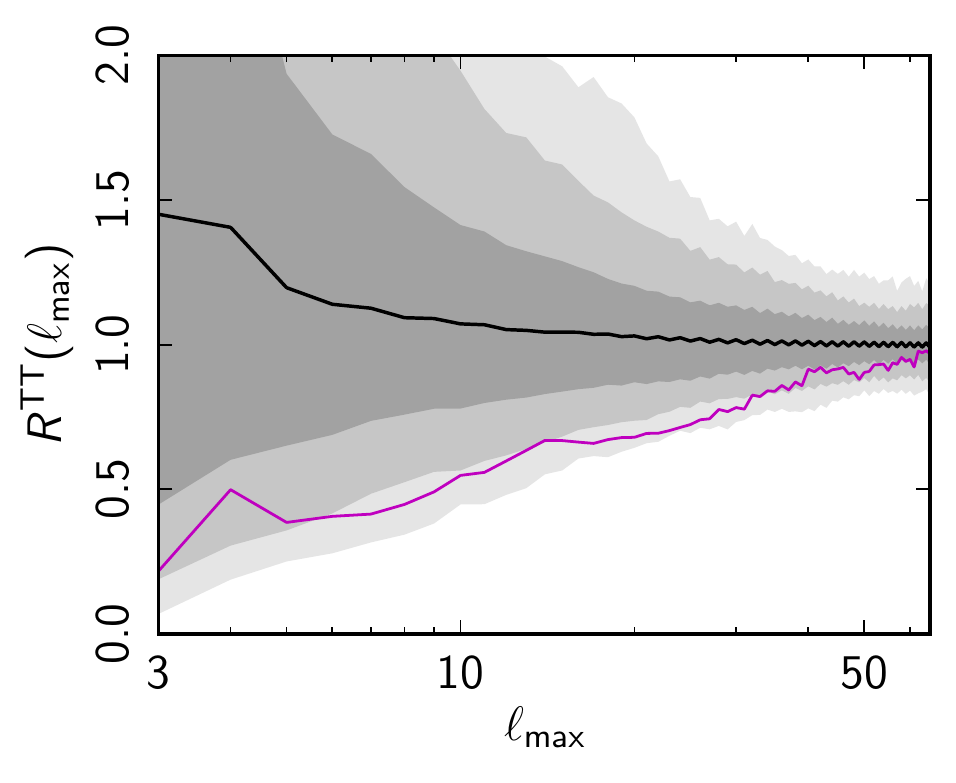} \\
\includegraphics[scale=0.45]{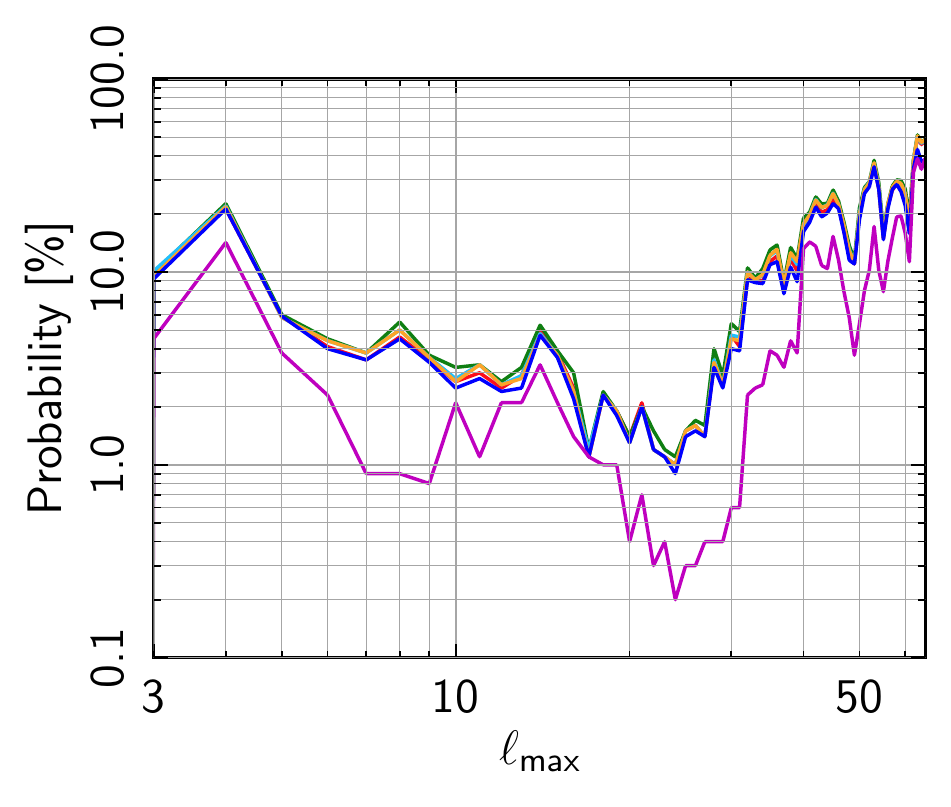}
\includegraphics[scale=0.45]{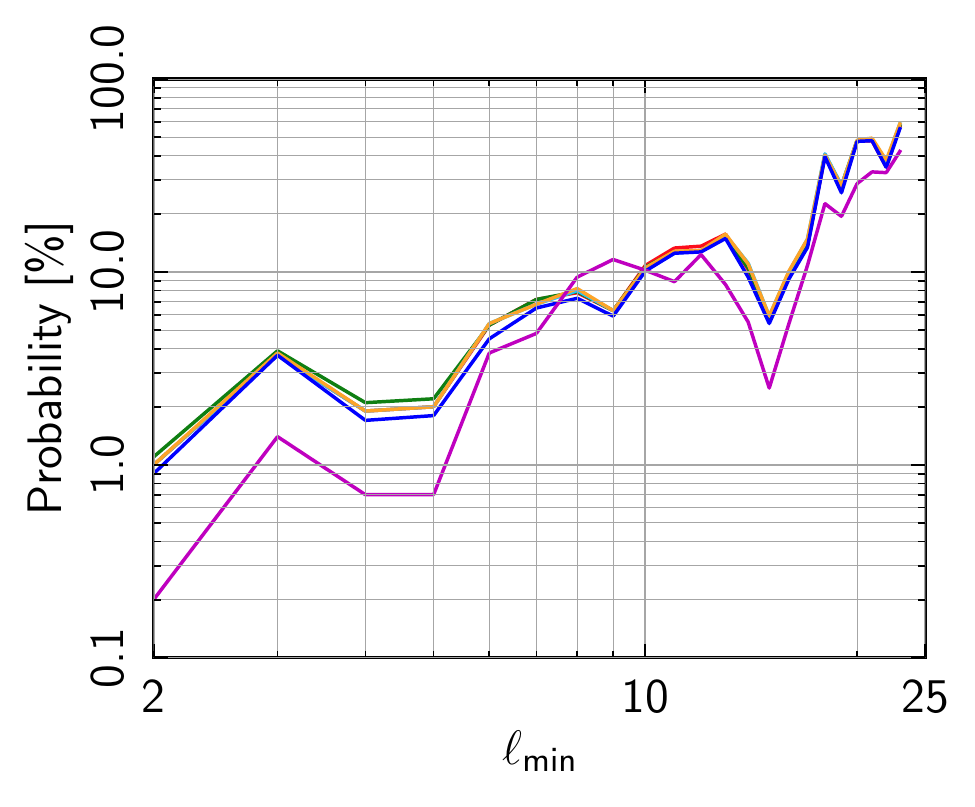}
\caption{\emph{Upper panels}: Ratio $R^\mathrm{TT}(\ell_\mathrm{max})$ of
  the \Planck\ 2018 data determined at \nside\ = 32. The shaded grey
  regions indicate the distribution of the statistic derived from the
  \smica\ MC simulations, with the dark,
  lighter, and light grey bands corresponding to the 1, 2, and 3$\,\sigma$
  confidence levels, respectively. The left panel presents the ratio computed
  from the component-separated maps, \commander\
  (red), \nilc\ (orange), \sevem\ (green), and \smica\ (cyan),
  after application of the common mask. Note that these results
  substantially overlap each other. The right panel presents the
  ratio computed when the \lowlcomm\ map (magenta)
  is analysed with the corresponding mask. \emph{Lower panels:} 
  The lower-tail probability estimator as a function of
  $\ell_\mathrm{max}$ (left), and $\ell_\mathrm{min}$ with
  a fixed $\ell_\mathrm{max} = 24$ (right).  For these lower panels
  small probabilities would correspond to anomalously odd parity.}
\label{fig:TTpointparity}
\end{figure}

The top-left panel of Fig.~\ref{fig:TTpointparity} presents the ratio,
$R^\mathrm{TT}(\ell_\mathrm{max})$, determined from the 2018
component-separated maps after applying the common mask, together with
the distribution of the \smica\ MC simulations that are representative
of the expected behaviour of the statistic in a Universe with no
preferred large-scale parity. We additionally consider an additional estimate
of the CMB sky determined using the \commander\ component-separation
methodology that has been optimized for large angular scale
analyses. This map is also used in the \Planck\ low-$\ell$ likelihood
analysis \citep{planck2016-l05}, thus we refer to it as
\lowlcomm. This corresponding result is shown in the top right panel
of the figure. The results from the component-separation products are
in good agreement, and indicate an odd-parity preference for the
multiple range considered in this test.  The lower-left panel of
Fig.~\ref{fig:TTpointparity} shows the lower-tail probability as a
function of $\ell_\mathrm{max}$ for the
$R^\mathrm{TT}(\ell_\mathrm{max})$ ratios shown in the upper panels.
The profiles of the cleaned CMB maps are very consistent, and indicate
a lower-tail probability of about 1\,\% over the range of multipoles
$\ell_\mathrm{max}= 20$--30. This is higher than the typical value
found in \citetalias{planck2014-a18}, a difference that we ascribe to
changes in the common mask. Indeed, if the 2015 maps are analysed
after applying the 2018 common mask, then a lower-tail probability of
about 1\,\% is also found. Similarly, if the 2015 common mask is applied
to the 2018 data, then a probability of 0.3\,\% is found at
$\ell_\mathrm{max} = 27$, in good agreement with the earlier results.

The \lowlcomm\ map shows good agreement with the standard
component-separated maps when the 2018 common mask is
applied. However, if the confidence mask specifically associated with the
low-$\ell$ likelihood is used, then, as seen in the top right and
lower panels of Fig.~\ref{fig:TTpointparity}, lower probabilities are
seen over the $\ell_\mathrm{max}$ range of 20--30, with a minimum
lower-tail probability of 0.2\,\% determined for
$\ell_\mathrm{max} =24$.  This behaviour is in agreement with previous
studies \citep[see, for example,][]{Gruppuso2018}, which indicate that
considering a reduced sky coverage, where the analysed regions are at
higher Galactic latitudes, diminishes the odd point parity preference.
In addition, we also observe probabilities of about 0.9\,\% in the
multipole region 7--9.  At the spectral level, the different parity
preference with respect to the 2018 common mask and the low-$\ell$
mask seems to be associated with shifts in the amplitudes of
multipoles at $\ell = 3$ and 7.

In order to quantify the impact of any a posteriori effects on the
significance levels of the \lowlcomm\ analysis, we count the number of
MC simulations with a lower-tail probability equal to, or lower than,
0.2\,\%, for at least one $\ell_\mathrm{max}$ value over a specific multipole
range. For $\ell_\mathrm{max}$ in the range 3--64, 16 out of the 999 MC maps
have this property, implying that, even considering the look-elsewhere
effect, an odd-parity
preference is observed with a lower-tail probability of about 1.6\,\%.

In the lower-right panel of Fig.~\ref{fig:TTpointparity}, the lower-tail
probability as a function of $\ell_\mathrm{min}$ is presented when
$\ell_\mathrm{max}$ is constrained to be 24, corresponding to the lowest
probability found for the \lowlcomm\ analysis. It is apparent that the
probability increases as the lowest multipoles are omitted,
demonstrating that the anomaly is mostly driven by the largest scales.

We then extend the point-parity analysis to include polarization data,
specifically considering the $TE$ and $EE$ power spectra. As in the
temperature analysis, the data are compared with MC simulations 
that describe the expected parity preference for the $\Lambda$CDM model.
However, the ratio estimator cannot be used in this case, since the
signal-to-noise is lower than for the temperature maps; this means that
estimates of the power spectra may become negative, so that
numerical problems can arise if the denominator of
Eq.~\eqref{eq:pointparity} approaches zero when summing values up to 
$\ell_\mathrm{max}$. To avoid this, a less sensitive but more robust
estimator is considered: 

\begin{linenomath*}
\begin{equation}\label{eq:Diffpointparity}
D^X(\ell_\mathrm{max}) = C^X_+(\ell_\mathrm{max})-C^X_-(\ell_\mathrm{max}),
\end{equation}
\end{linenomath*}
where $X$ corresponds to either the $TE$ or $EE$ spectra, 
and $C^X_{+,-}(\ell_\mathrm{max})$ is defined analogously 
to the temperature case in Eq.~\eqref{eq:TTpointparity}.

In order to minimize the effect of incomplete knowledge of the
instrumental noise, and given that some noise mismatch has been
observed between the data and simulations \citep{planck2016-l04}, we
determine the power spectrum using the cross-quasi-QML estimator that forms the
basis of the HFI low-$\ell$ likelihood in polarization
\citep{planck2016-l05}. Here, the signal covariance matrix is
constructed according to the FFP10 fiducial model, while the noise is
specified using the FFP8\footnote{FFP8 is the name given to the
  previous version of full focal-plane simulations utilized for the
  analysis of the 2015 \Planck\ maps \citep{planck2014-a14}. }
143-GHz noise-covariance matrix.
This is clearly a suboptimal description of the noise present
in the component-separated maps, but only affects the variance of the
estimated power spectra. However, since our statistical tests do not require
this quantity, but are based on the dispersion of the MC simulations,
the results are unaffected by the choice of the noise-covariance
matrix. To verify this, we also analyse the \sevem\ 143-GHz cleaned
map, the noise properties of which should be more closely matched by
this covariance matrix. The consistency of results between the various
component-separated maps indicate that no bias arises from this
choice. Finally, to test consistency, we apply the cross-estimator
to both the HM and OE data sets.

Examination of the $EE$ and $TE$ power spectra determined from the component-separated maps indicates that the largest angular scales, in
particular $\ell = 2$, are likely to be affected by the presence of
residual systematics, probably arising mainly from the ADC nonlinearity
\citep{planck2016-l03} and their correlation with foregrounds.  These
effects are not fully captured by the current set of MC
simulations. As a consequence, the inclusion of the $\ell = 2$
multipole in the point-parity analysis results in probabilities that
reach values of 98.5\,\% for certain component-separated products. In
what follows, we limit the multipole range of interest to
$\ell = 3$--30.

\begin{figure}
\centering
\includegraphics[scale=0.45]{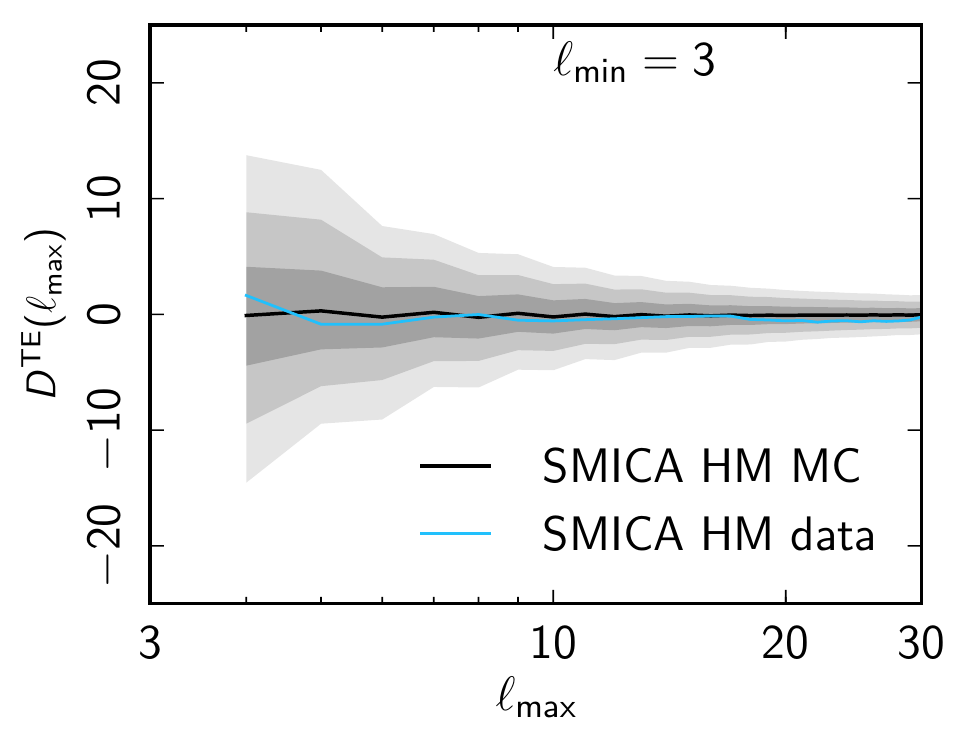}
\includegraphics[scale=0.45]{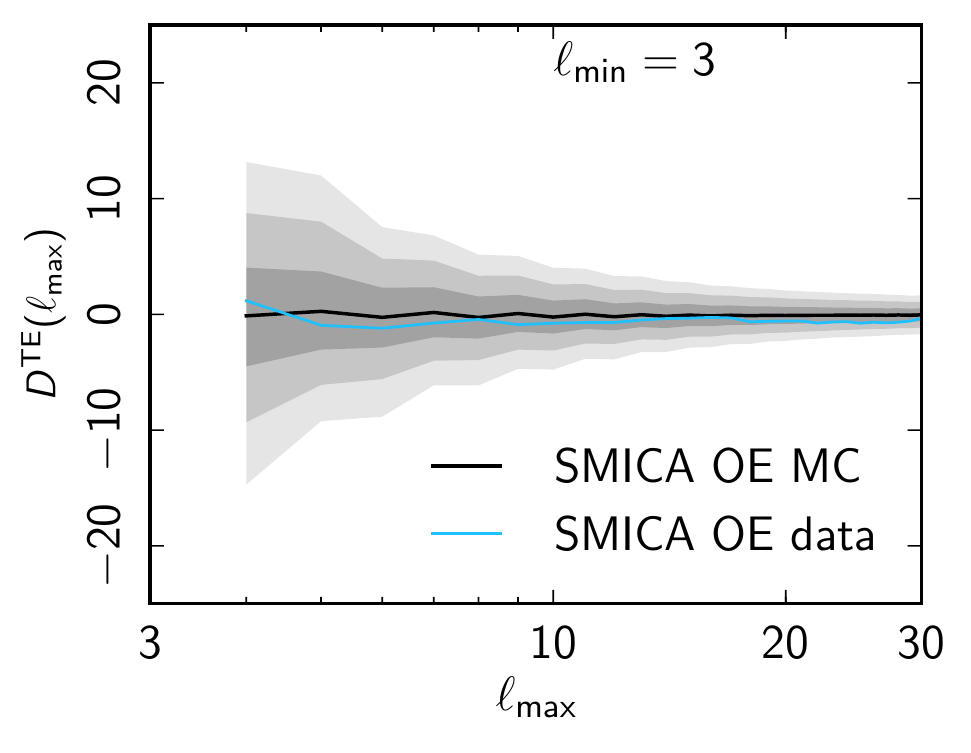} \\
\includegraphics[scale=0.44]{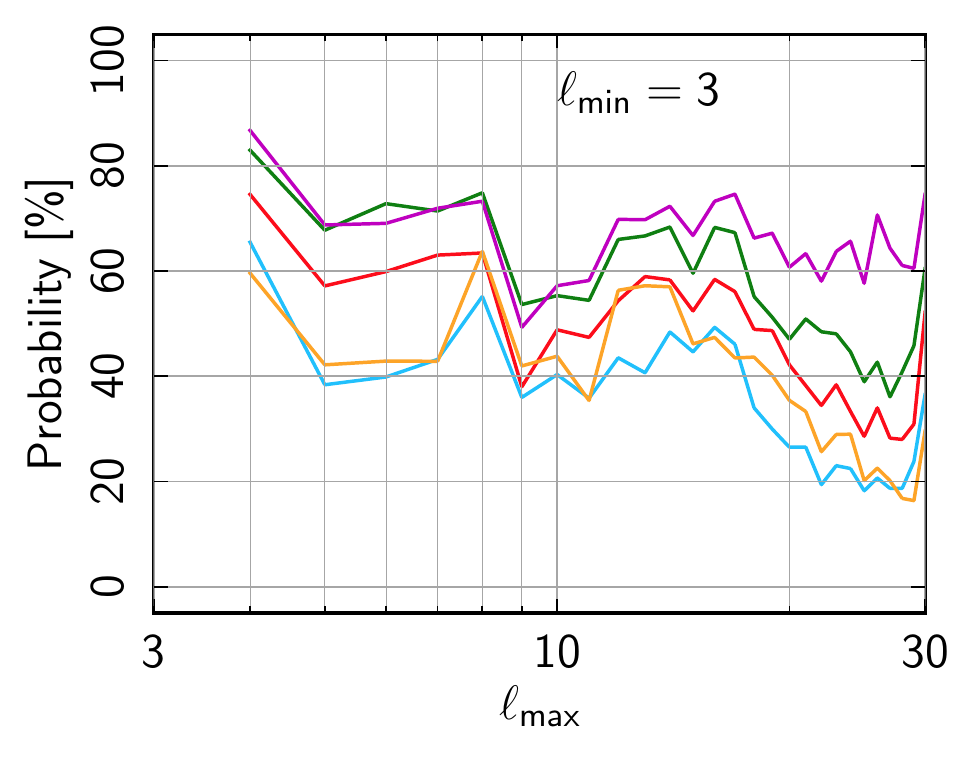} 
\includegraphics[scale=0.44]{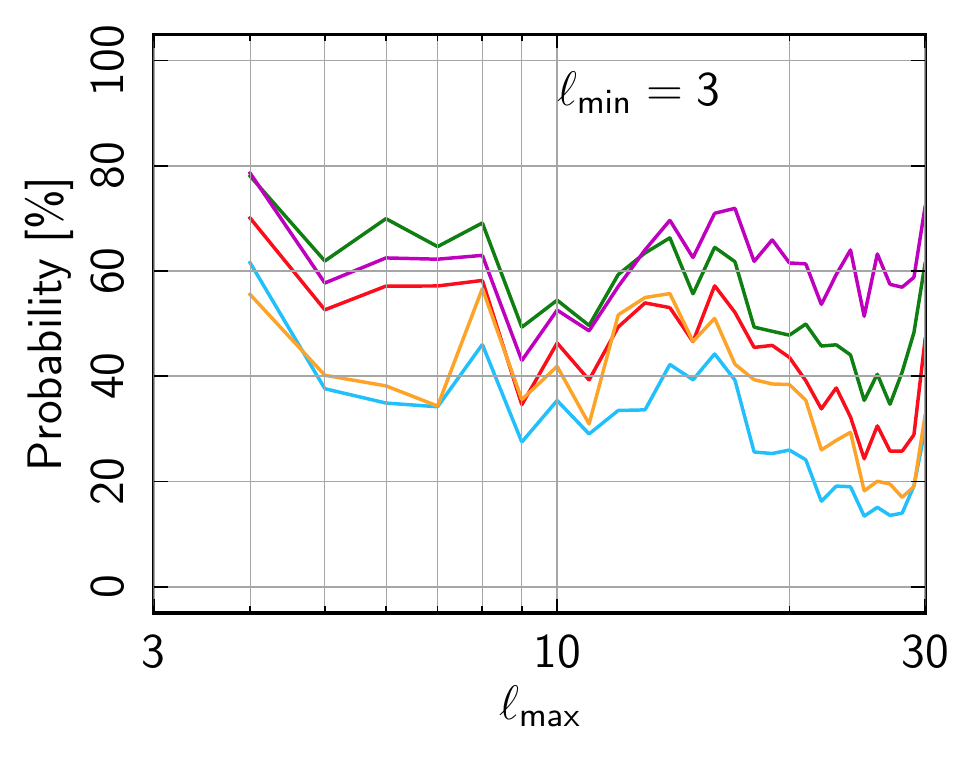} \\
\caption{\emph{Upper panels:}
    $D^\mathrm{TE}(\ell_\mathrm{max})$ for the \smica\ HM (left panel)
    and OE (right panel) data.
    The shaded grey regions indicate the distribution of the statistic
    derived from the corresponding MC simulations as in
    Fig.~\ref{fig:TTpointparity}. \emph{Lower panels:} The lower-tail
    probability of the polarization estimators as a function of
    $\ell_\mathrm{max}$ for the HM data (left panel) and OE data
    (right panel) for all component-separated methods.  Results for
    the \sevem\ 143-GHz cleaned maps are also shown (magenta line).
    For these lower panels 
    small probabilities would correspond to anomalously odd parity.}
\label{fig:PPpointparity_TE}
\end{figure}

\begin{figure}
\centering
\includegraphics[scale=0.44]{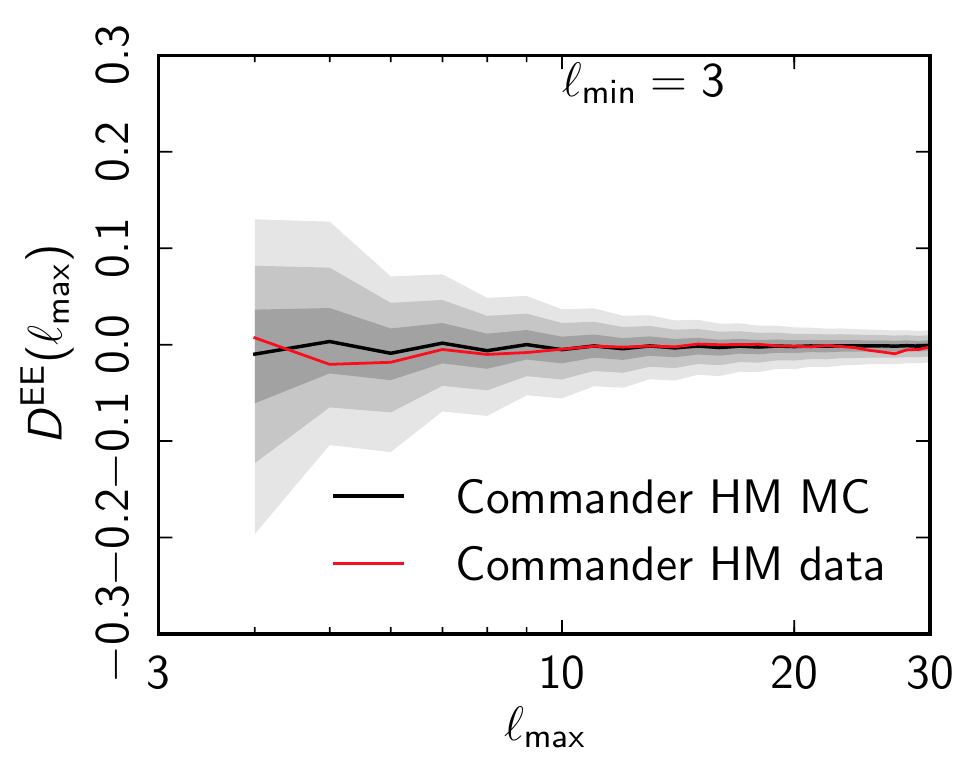}
\includegraphics[scale=0.44]{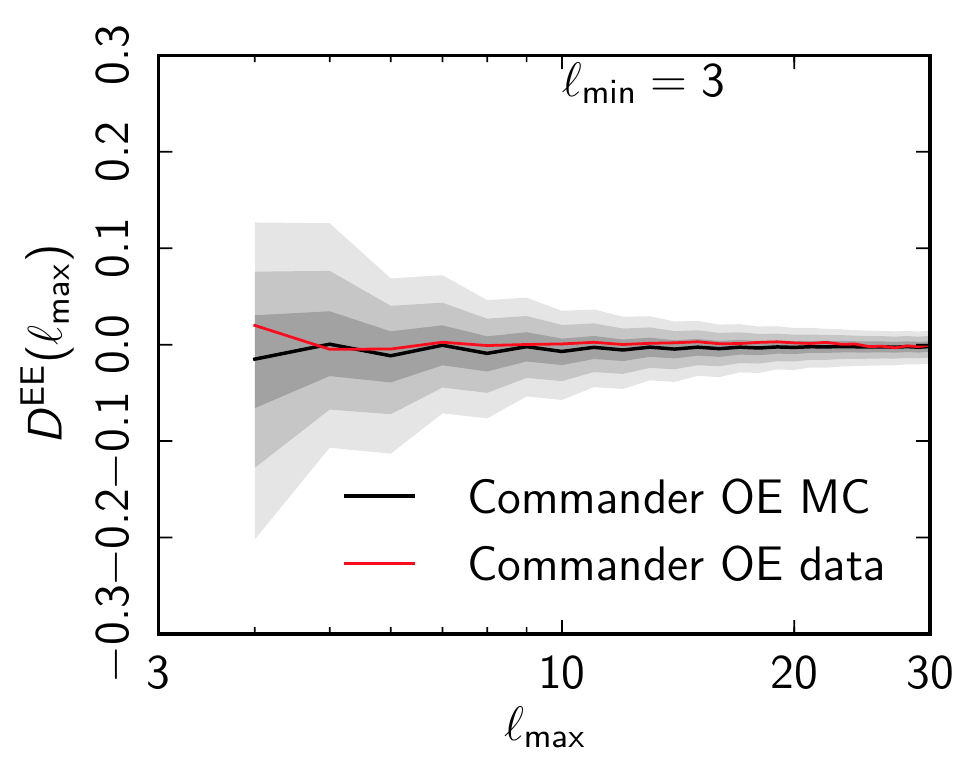} \\
\includegraphics[scale=0.44]{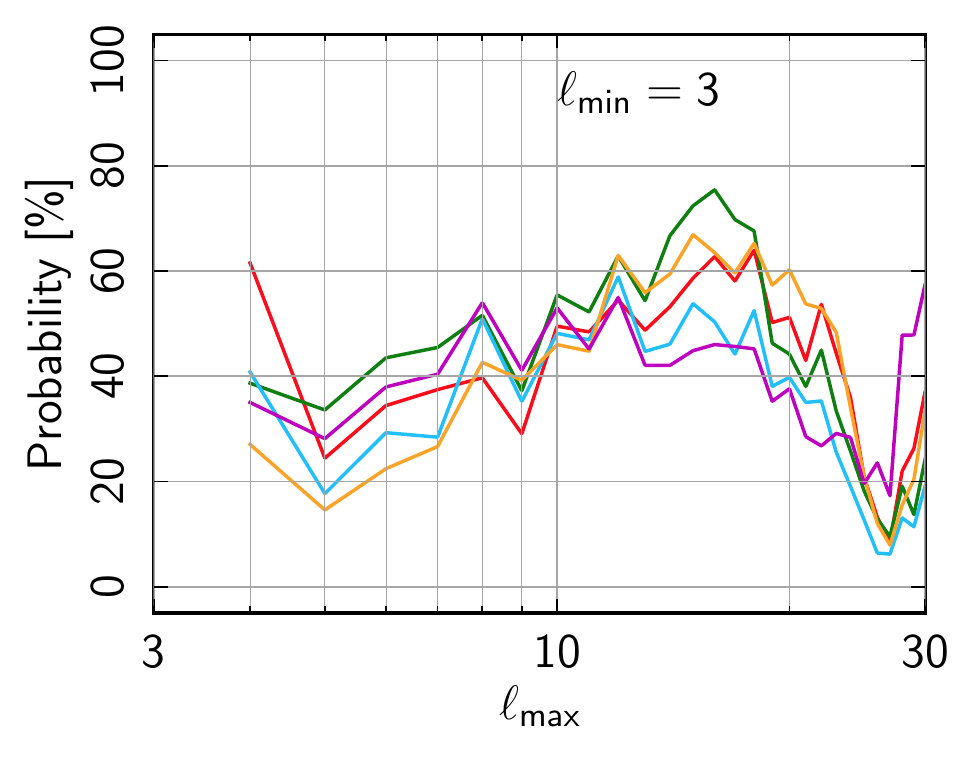} 
\includegraphics[scale=0.44]{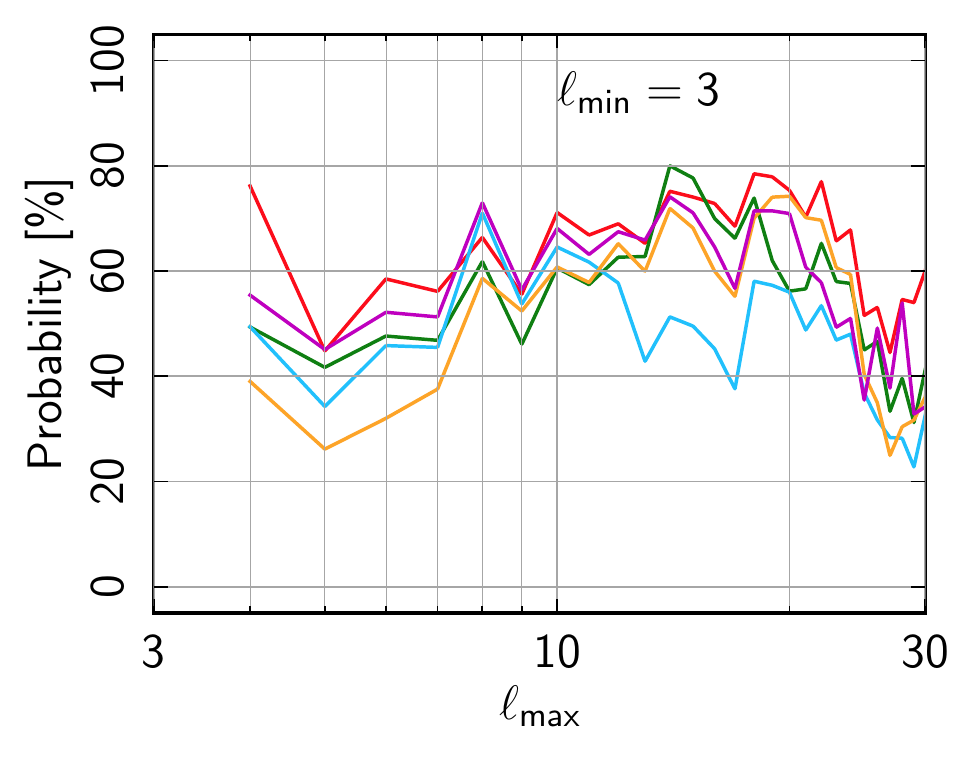} \\
\caption{\emph{Upper panels:} $D^\mathrm{EE}(\ell_\mathrm{max})$ for
  the \commander\ HM (left panel) and OE (right panel) data. The
  shaded grey regions indicate the distribution of the statistic
  derived from the corresponding MC simulations as in
  Fig.~\ref{fig:TTpointparity}.  \emph{Lower panels:} The lower-tail
  probability of the polarization estimators as a function of
  $\ell_\mathrm{max}$, as described in Fig.~\ref{fig:PPpointparity_TE}.
  For these lower panels 
  small probabilities would correspond to anomalously odd parity.}
\label{fig:PPpointparity_EE}
\end{figure}

Figure~\ref{fig:PPpointparity_TE} presents the results for the
$D^\mathrm{TE}(\ell_\mathrm{max})$ estimator. The upper panels show,
as examples, the distributions determined from the \smica\ data and
corresponding MC simulations. Results from the other component
separated maps behave similarly. In the lower panels, we show the
lower-tail probability as a function of $\ell_\mathrm{max}$ for the HM and OE
data. Results from the component-separated maps show similar trends,
but with a range of probabilities. Nevertheless, they are compatible with
the MC simulations and show a mild tendency towards decreasing
probability with higher $\ell_\mathrm{max}$, although the lowest value
observed is of order 20\,\%. Results from the HM and OE data splits are
in good agreement.

Figure~\ref{fig:PPpointparity_EE} shows equivalent results for the
$D^\mathrm{EE}(\ell_\mathrm{max})$ estimator, though with \commander\
taken as a representative example of the behaviour of the component-separated maps.
As in the $TE$ analysis, the data and MC simulations are consistent,
although the {\it p}-values decrease over the $\ell_\mathrm{max}$ range of
20 to 30, reaching minima of about 6\,\% at $\ell_\mathrm{max}=27$ for the HM data; 
however, the OE results do not indicate such a trend.

It is interesting to note that the $C^{XY}_2(180^\circ)$ statistic
introduced in Sect.~\ref{sec:n_point_anomalies} is related to the
measurement of parity asymmetry \citep[see also][]{muir:2018}.
In particular, when the functions are expressed in terms of the angular
power spectra, it can be seen that the following relationships hold:
\begin{linenomath*}
\begin{align}
C^{TT}_2(180^\circ) &= \sum^{\infty}_{\ell=2} (-1)^\ell\, { 2 \ell +1
  \over 4 \pi} C^{TT}_\ell \ ; \\
C^{Q_\mathrm{r} Q_\mathrm{r}}_2(180^\circ) &= \sum^{\infty}_{\ell=2} (-1)^\ell\, { 2 \ell +1
  \over 8 \pi} \left( C^{EE}_\ell - C^{BB}_\ell \right)\ ; \\
C^{U_\mathrm{r} U_\mathrm{r}}_2(180^\circ) &= - \sum^{\infty}_{\ell=2} (-1)^\ell\, { 2 \ell +1
  \over 8 \pi} \left( C^{EE}_\ell - C^{BB}_\ell \right)\ ; \\
C^{Q_\mathrm{r} U_\mathrm{r}}_2(180^\circ) &= - \sum^{\infty}_{\ell=2} (-1)^\ell\, { 2 \ell +1
  \over 4 \pi} C^{EB}_\ell\ .
\label{eqn:cpi}
\end{align}
\end{linenomath*}
The expected values for the remaining correlation functions are zero,
i.e., $C^{TQ_\mathrm{r}}_2(180^\circ) = C^{TU_\mathrm{r}}_2(180^\circ) = 0$. The factor
$(-1)^\ell$ splits the sum into contributions
from even (parity-symmetric) and odd (parity-antisymmetric) multipoles
with opposite signs. The functions are therefore differences between
parity-even and parity-odd multipole band-powers, and can be compared
to Eq.~\eqref{eq:Diffpointparity}. 
However, the expressions are not identical, since, 
in the latter case, the power spectrum is weighted by the
factor $\ell(\ell+1)$, as compared to the $(2\ell+1)$ weighting used in
the former case. 
Nevertheless, the results presented in Table~\ref{tab:prob_cpi} seem to be
consistent with those in this section. 

In summary, no anomalous lower-tail probability is found. However, the
low signal-to-noise of the \Planck\ polarization data over this range
of scales is a limiting factor for this analysis. Future
higher sensitivity observations are certainly required in order to
investigate the point parity asymmetry in polarization.

\subsection{Peak distribution asymmetry}
\begin{figure*}
  \centering
  \begin{tabular}{cc}
  \includegraphics[width=88mm]{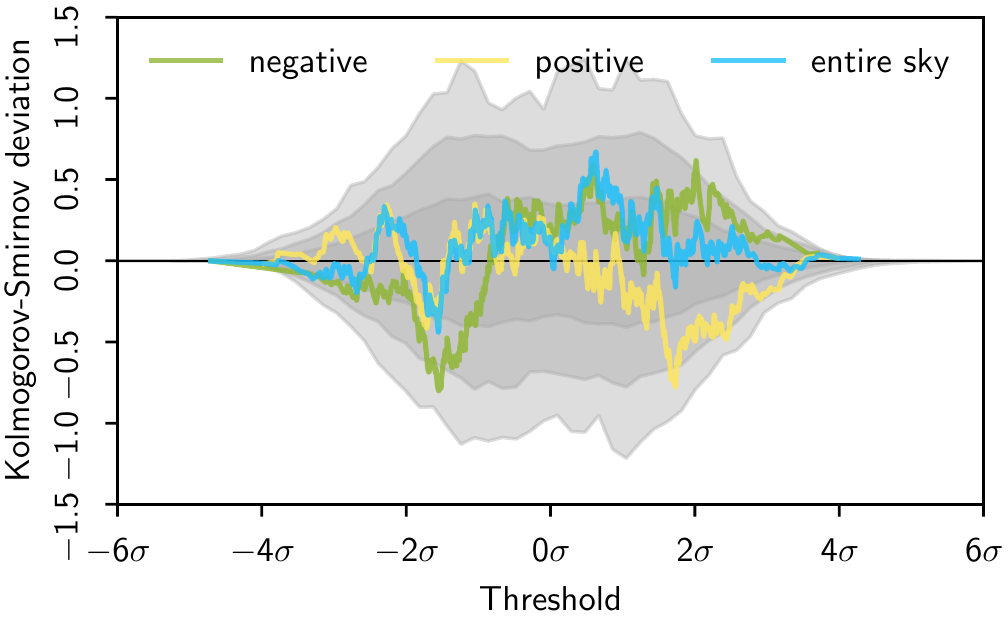} &
  \includegraphics[width=88mm]{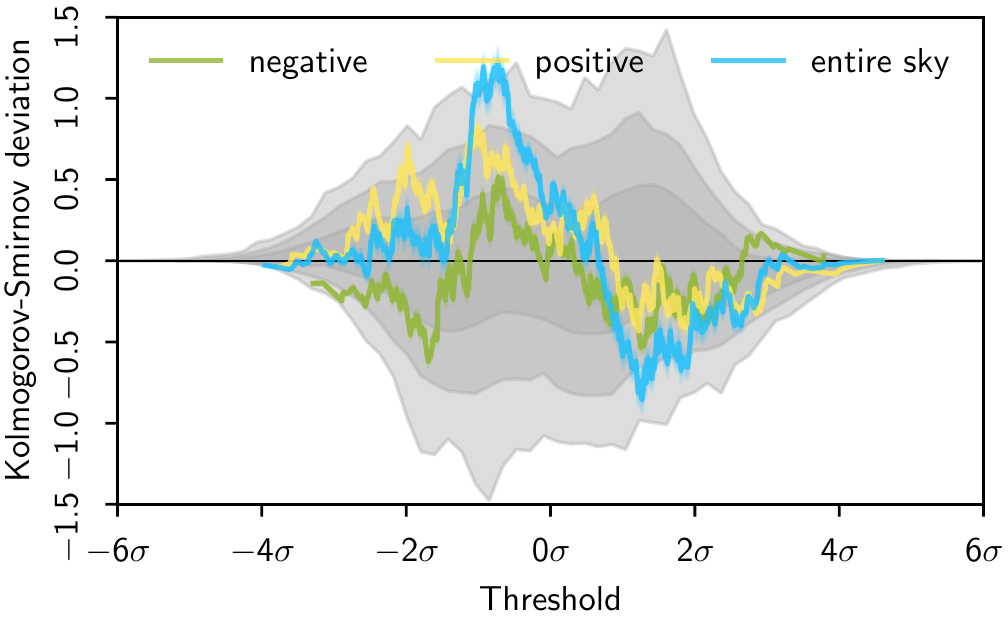} \\
  \includegraphics[width=88mm]{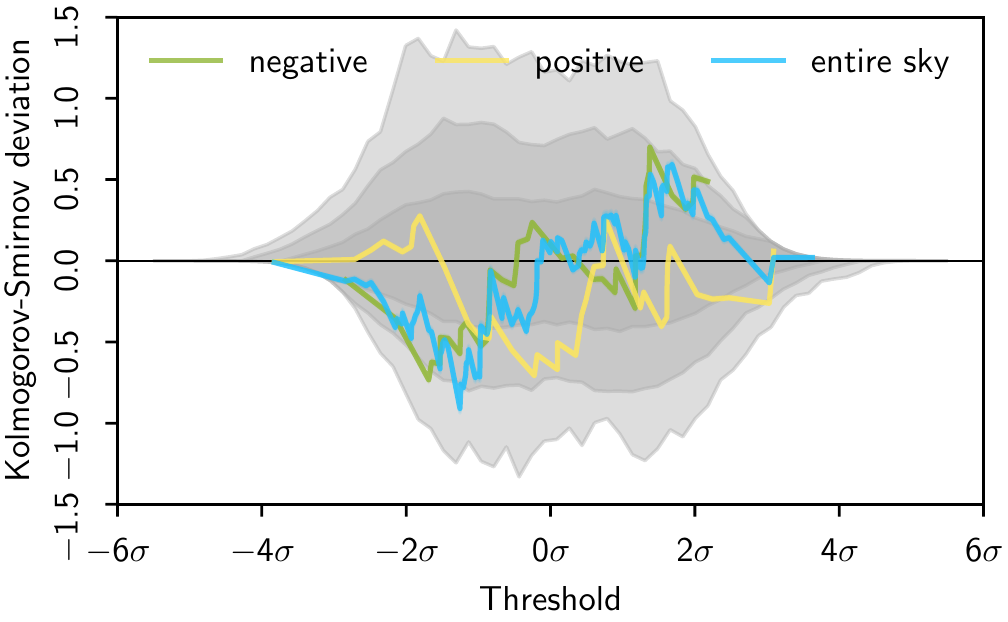} &
  \includegraphics[width=88mm]{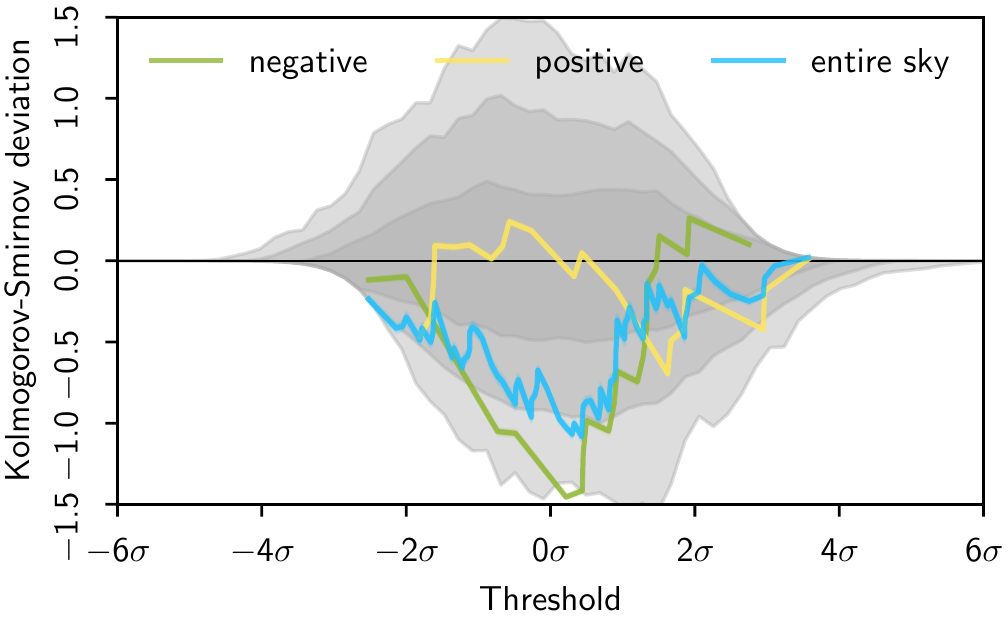}
  \end{tabular}
  \caption{ Kolmogorov-Smirnov deviation of the peak distribution for
    the \smica\ temperature
    $T$ (left) and the reconstructed $E$-mode polarization (right) maps in $70\deg$ radius
    discs centred on the positive and negative asymmetry directions
    as determined by the \smica\ $TT, TE, EE$ QML estimator in
    Sect.~\ref{sec:dipmod_qml}.  The top panels correspond
    to maps filtered with GAUSS kernels of $120\arcm$ and the bottom
    panels to filtering with $600\arcm$ FWHM.
  }
  \label{fig:peaks:asym}
\end{figure*}

Localized anomalies on the CMB sky can be searched for by testing how
the statistical properties of local extrema (or peaks) vary in patches
as a function of location.  Since we expect the asymmetry measured by
the QML estimator in Sect.~\ref{sec:dipmod_qml} to be mirrored by a
slight difference in the temperature and polarization peak
distributions in the corresponding positive and negative hemispheres,
we present and examine these quantities here. Note that the
test is not powerful enough in itself to determine a modulation
direction, but relies on the dipole orientation results presented in
Table~\ref{tab:lowellmod}.

We extract peaks from the filtered maps discussed in
Sect.~\ref{sec:peakstat} within the $70\deg$ radius discs centred on
the positive and negative asymmetry directions determined by the
\smica\ $TT, TE, EE$ QML estimator (see
Table~\ref{tab:lowellmod}). Separate peak CDFs are then constructed
for these directions, then compared to the full-sky peak distribution,
and the median CDF of the simulations. The results are presented in
Fig.~\ref{fig:peaks:asym}, which indicates the Kolmogorov-Smirnov
deviation from the median simulation CDF of the full sky. No
significant difference between the $E$-mode peak distributions for the
positive and negative directions is observed when the data are filtered
on a $120\arcm$ scale. For a $600\arcm$ filtering scale, a marginal
difference between the data and simulations is observed in the
negative direction, as previously observed for the temperature results
in \citetalias{planck2014-a18}.

\subsection{The Cold Spot and other large-scale peaks in temperature and polarization}
\label{sec:large_scale_peaks}

\begin{figure}[htbp!]
\begin{center}
\includegraphics[width=0.5\textwidth]{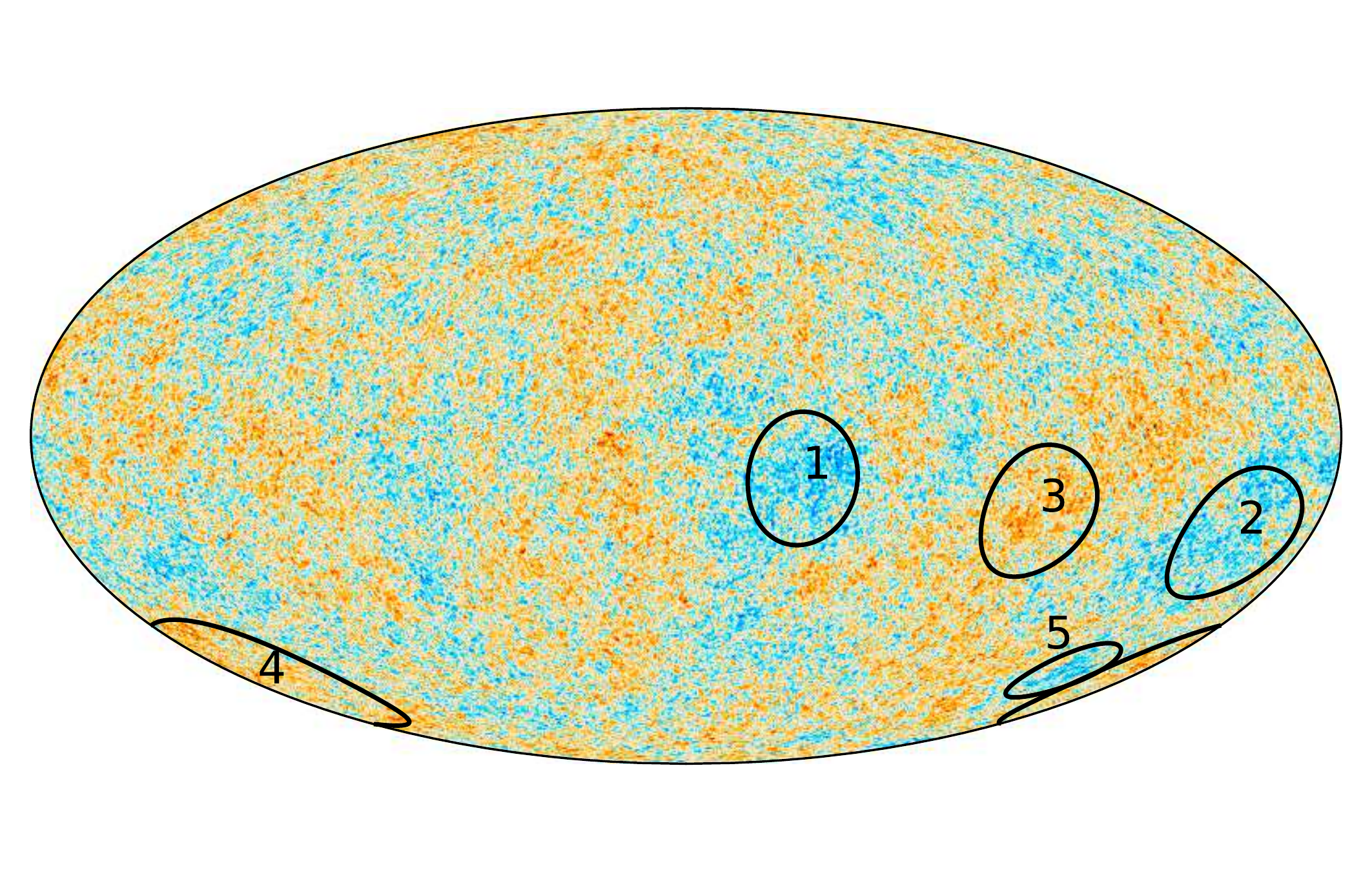}
\end{center}
\vspace{-0.75truecm}
\caption{Locations of the large-scale maxima or minima, numbered 1 to 5,
considered in the analysis and indicated on an inpainted version of
the \commander\ map at $N_{\rm side} = 2048$.
In particular, Peak~5 corresponds to the \cs.
}
\label{fig:peaks_loc}
\end{figure}

The Cold Spot was first detected in the \wmap\ temperature data as an
anomalously cold region in terms of the spherical Mexican-hat wavelet
(SMHW) coefficients \citep{vielva2004,cruz2005} and later confirmed
with \Planck\ data \citepalias[and references
therein]{planck2013-p09, planck2014-a18}.  The shape and the local
properties of the \cs\ have been studied using a variety of
statistical approaches 
(e.g., \citealt{cayon2005}; \citetalias{planck2014-a18})
Recently, \citet{marcos-caballero2017b} analysed the local properties
of the \cs\ (and other large-scale peaks) via multipolar
profiles that characterize the local shape in terms of the discrete
Fourier transform of the azimuthal angle. In that paper, the
anomalous nature of the Cold Spot is identified by being one of the
most prominent peaks in curvature, as quantified by comparison
to the predictions of \LCDM.\footnote{Note that the curvature or
Laplacian of a field smoothed with a Gaussian of a given scale is
equivalent to the SMHW coefficient at that scale.} Its internal
structure was then subsequently studied by \citet{chiang2018} in the
context of an analysis of the variation of the acoustic peak positions
in different parts of the sky. It was found that the \cs\ region shows
a large synchronous shift of the peaks towards smaller multipoles.

In this section, we analyse the polarization pattern of the \cs,
considered as a minimum in the temperature field when smoothed with a
Gaussian of standard deviation $R=5^\circ$, to search for evidence
as to whether its origin is primordial or otherwise.
\citet{vielva2011} and \citet{fernandez-cobos2013} provide
forecasts on whether given experimental configurations 
can distinguish between these hypotheses,
and at what level of significance.
In addition to the \cs, four additional large-scale peaks are
considered, selected
as the most anomalous structures on large angular scales,
here taken to be extrema with a filter of $R=10^\circ$
\citep[see][for more details]{marcos-caballero2017a,marcos-caballero2017b}.
Since these peaks dominate the temperature field on large angular
scales, the particular value of the smoothing scale used in their
identification is not critical for locating the peaks. The scale
$R=10^\circ$ is chosen in order to highlight these structures while
preserving their geometrical properties, such as curvature and
eccentricity. These features correspond to two maxima and two minima,
shown as Peaks~1 to 4 in Fig.~\ref{fig:peaks_loc}, all located in the
southern Galactic hemisphere. Given their locations, they are expected
to make a significant contribution to the hemispheric-power
asymmetry.
In addition, the peaks should induce a particular pattern in the
polarization field, due to the correlation between the {temperature and
$E$-mode polarization}.

Our analysis of the peaks is based on the calculation of the $T$ and
$Q_{\rm r}$ angular profiles. Since these are computed by averaging the signal
over azimuthal angle, the analysis only characterizes the circularly
symmetric part of each peak. However, the $T$ and $Q_{\rm r}$ profiles, in
the absence of any $TB$ correlation and neglecting the eccentricity of
the peaks, contain all of the temperature and polarization information
that we are interested in. 

In order to have a complete characterization of the peaks, we have to
calculate derivatives of the fields up to second order. Since we are
interested in the azimuthally symmetric signal, only the peak height
and the Laplacian are relevant to the $Q_{\rm r}$ profiles. Once these two
quantities are calculated from the temperature smoothed at a given
scale $R$, their values are used to construct the expected theoretical
profile in the polarization field \citep[see][for technical
details]{marcos-caballero2016}.

In order to analyse the large-scale peaks, the profiles are calculated
by averaging the temperature and polarization fields over
azimuthal angle in different bins of the radial distance $\theta$. We
use intervals in $\theta$ from 0 to 30\deg, with a width of
1\deg. In the case of polarization, the $Q$ and $U$ components are
rotated to obtain the $Q_\mathrm{r}$ and $U_\mathrm{r}$ Stokes
parameters. The mean profile and the covariance matrix are discretized
in the same way as the observed peak profiles.
Assuming that the CMB field is Gaussian, the statistical distribution
of the profiles, which is obtained by conditioning the values of the
peak height and the Laplacian, is also Gaussian. In the absence of
noise, the covariance matrix corresponds to the cosmic variance,
modified to account for the fact that the derivatives at the centre of
the peak are fixed to the measured values. In the case of the
temperature profiles, this contribution introduces large correlations
between different angular bins, which leads to a problem when a mask
is applied to the data. Incomplete sky coverage modifies the
correlation between the bins of the data, producing a disagreement
between the data and the theoretical predictions.
In order to properly characterize the masked data, we adopt the
methodology developed in \citet{marcos-caballero2016} and generate
$2000$ CMB simulations that are constrained to have a peak located at
the same position as every large-scale peak under consideration. Since
the covariance matrix does not depend on the value of the peak height
and curvature, we use the same simulations for each of the large-scale
peaks with $R=10^\circ$. However, the covariance matrix depends on the
scale of the peak, and thus separate simulations are employed for the
\cs.
Since the mean profiles are not affected by the mask, we use the
analytical expressions in \citet{marcos-caballero2016} to calculate
the theoretical models instead of the average value of the
simulations. This procedure allows us to reduce the sample variance
due to the finite set of realizations.

\begin{figure*}[htbp!]
\begin{center}
\includegraphics[width=0.35\textwidth]{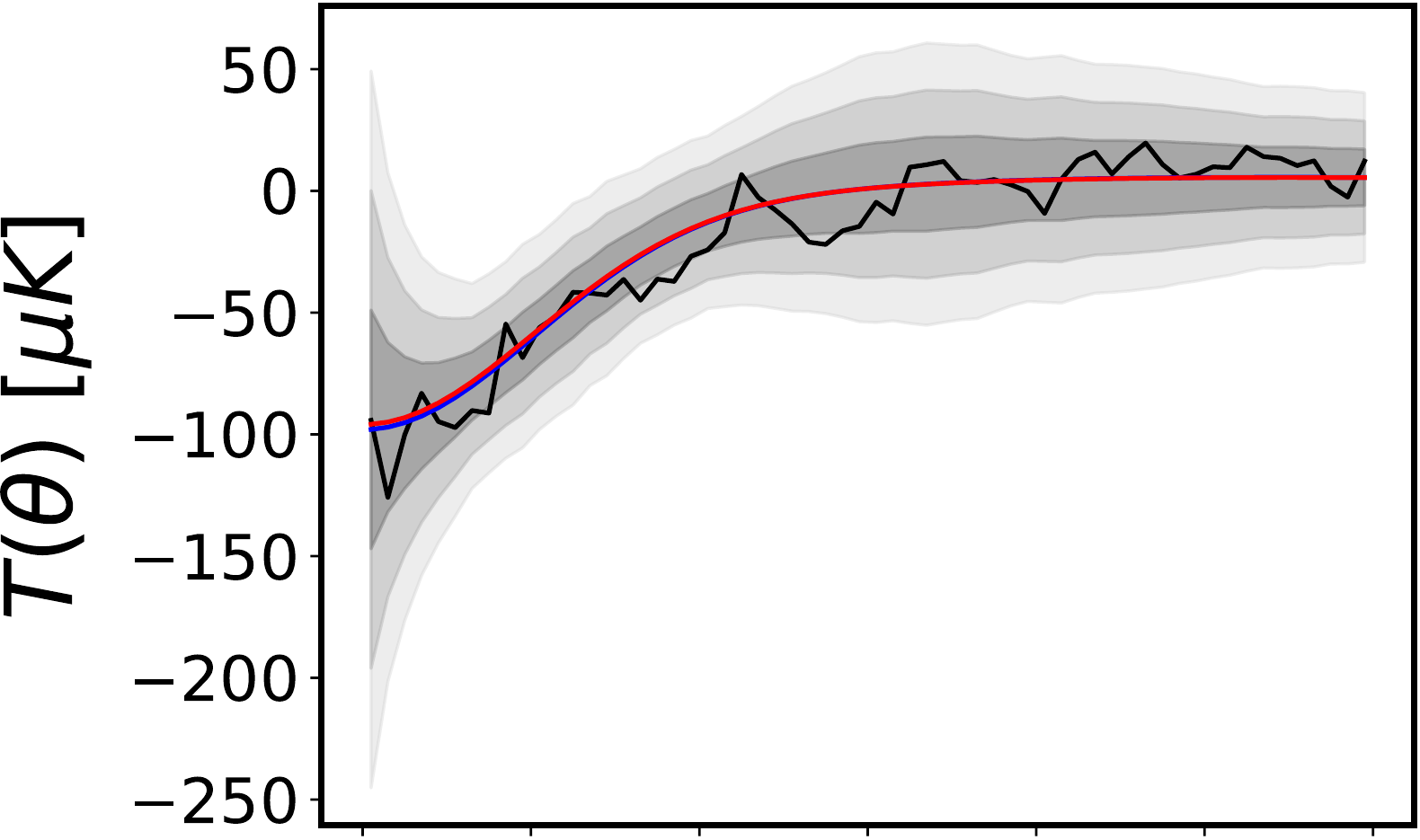}
\hspace*{0.1\textwidth}
\includegraphics[width=0.33\textwidth]{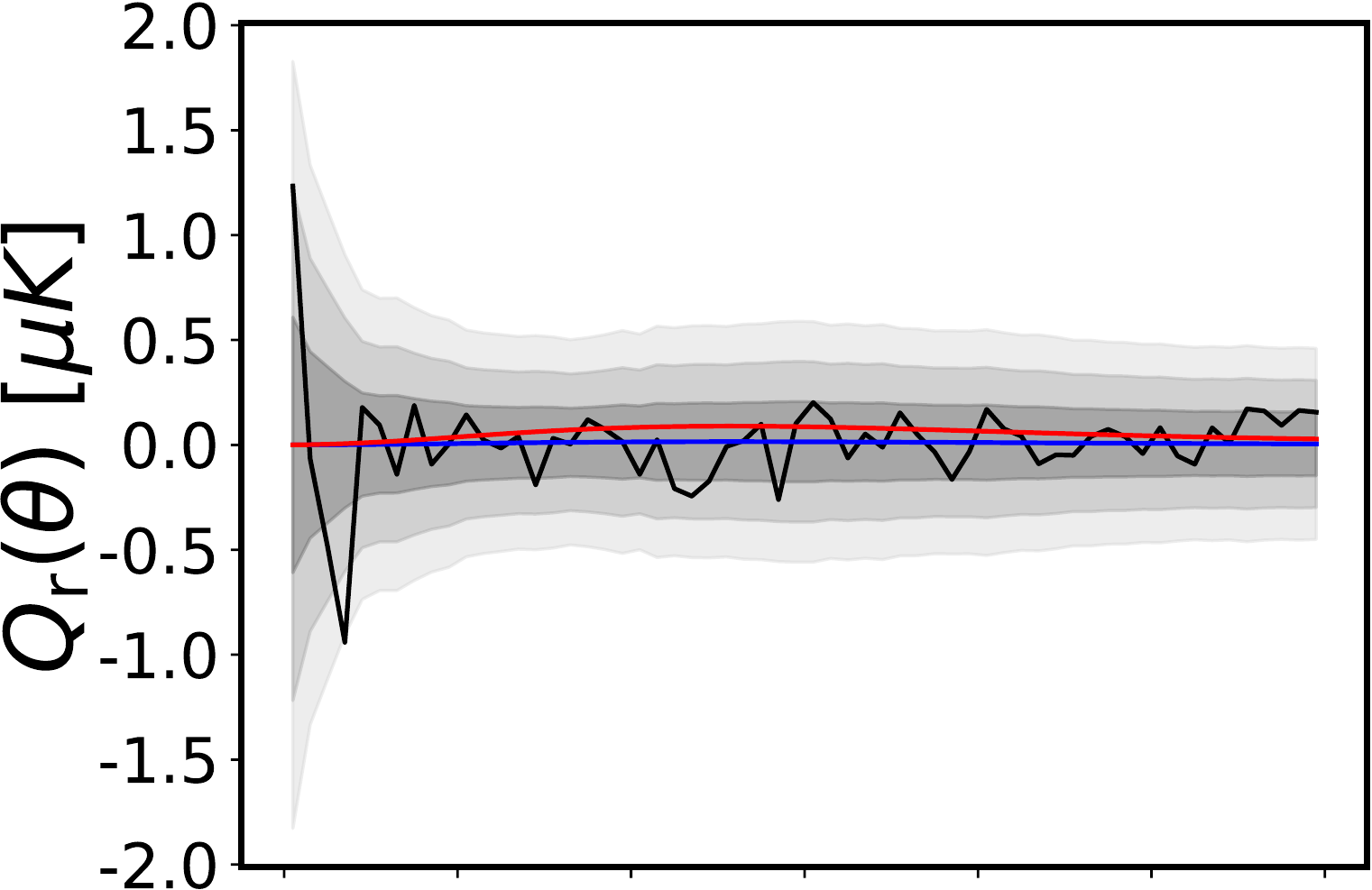}
\\
\includegraphics[width=0.35\textwidth]{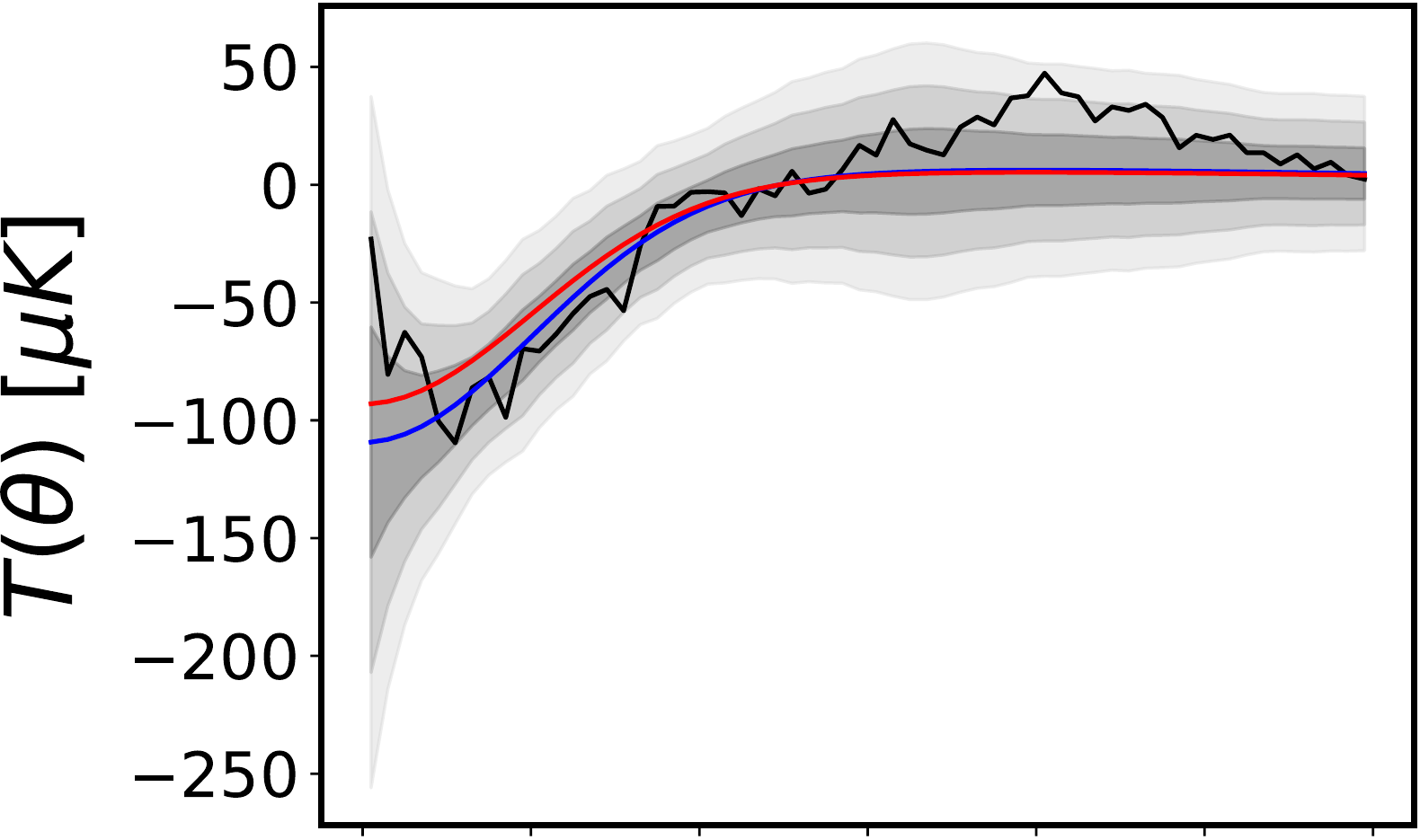}
\hspace*{0.1\textwidth}
\includegraphics[width=0.33\textwidth]{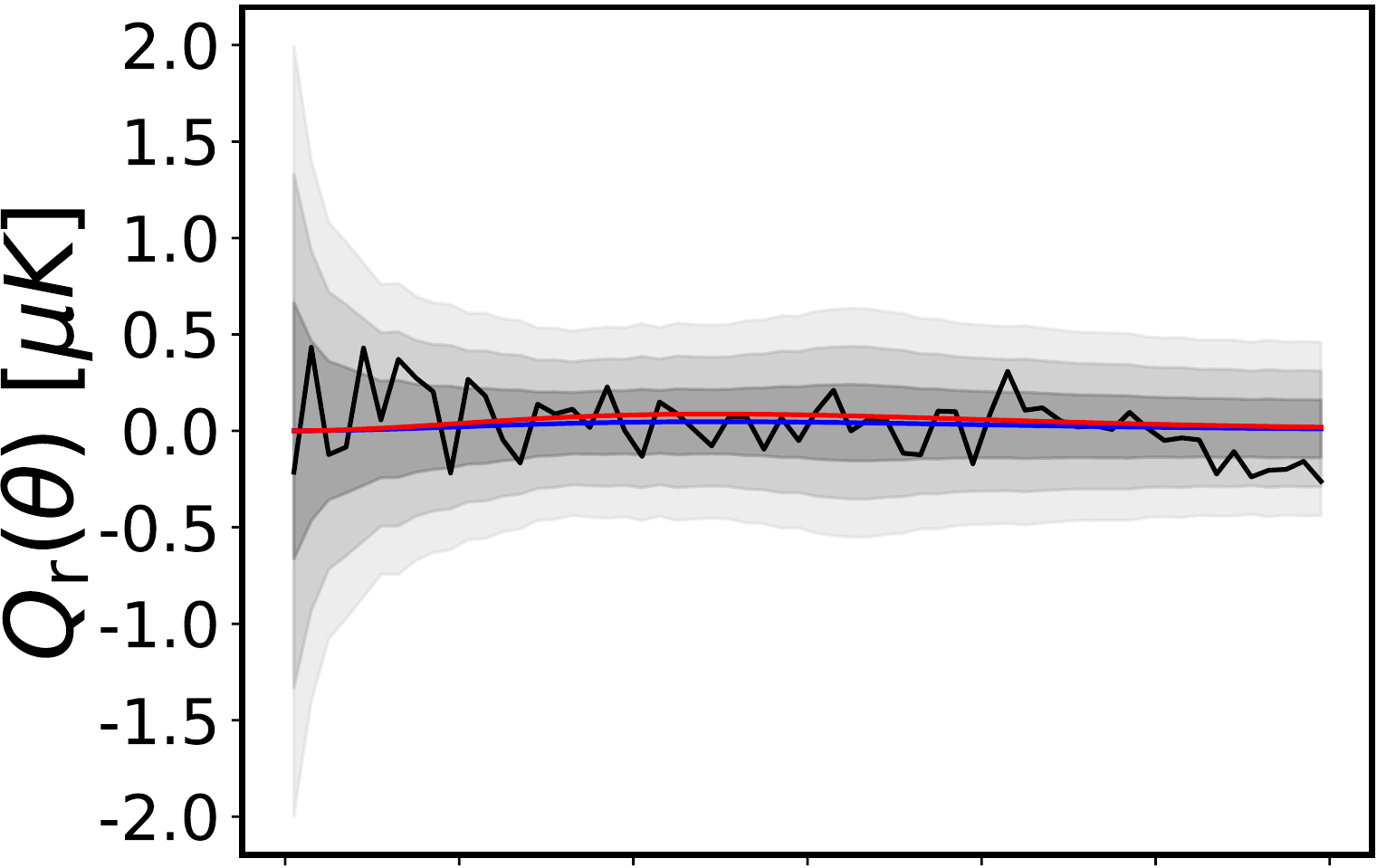}
\\
\hspace*{0.01\textwidth}\includegraphics[width=0.34\textwidth]{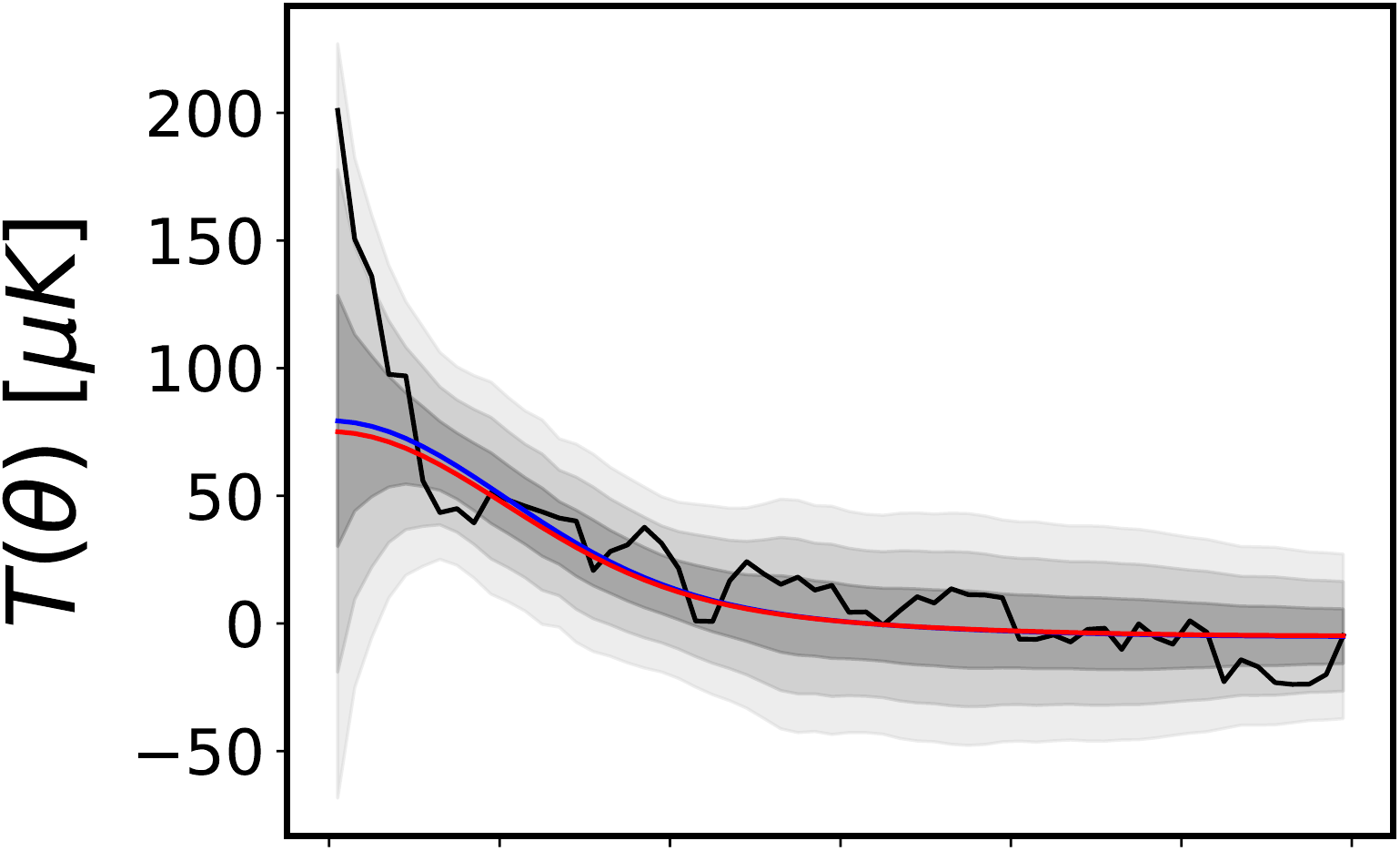}
\hspace*{0.1\textwidth}
\includegraphics[width=0.33\textwidth]{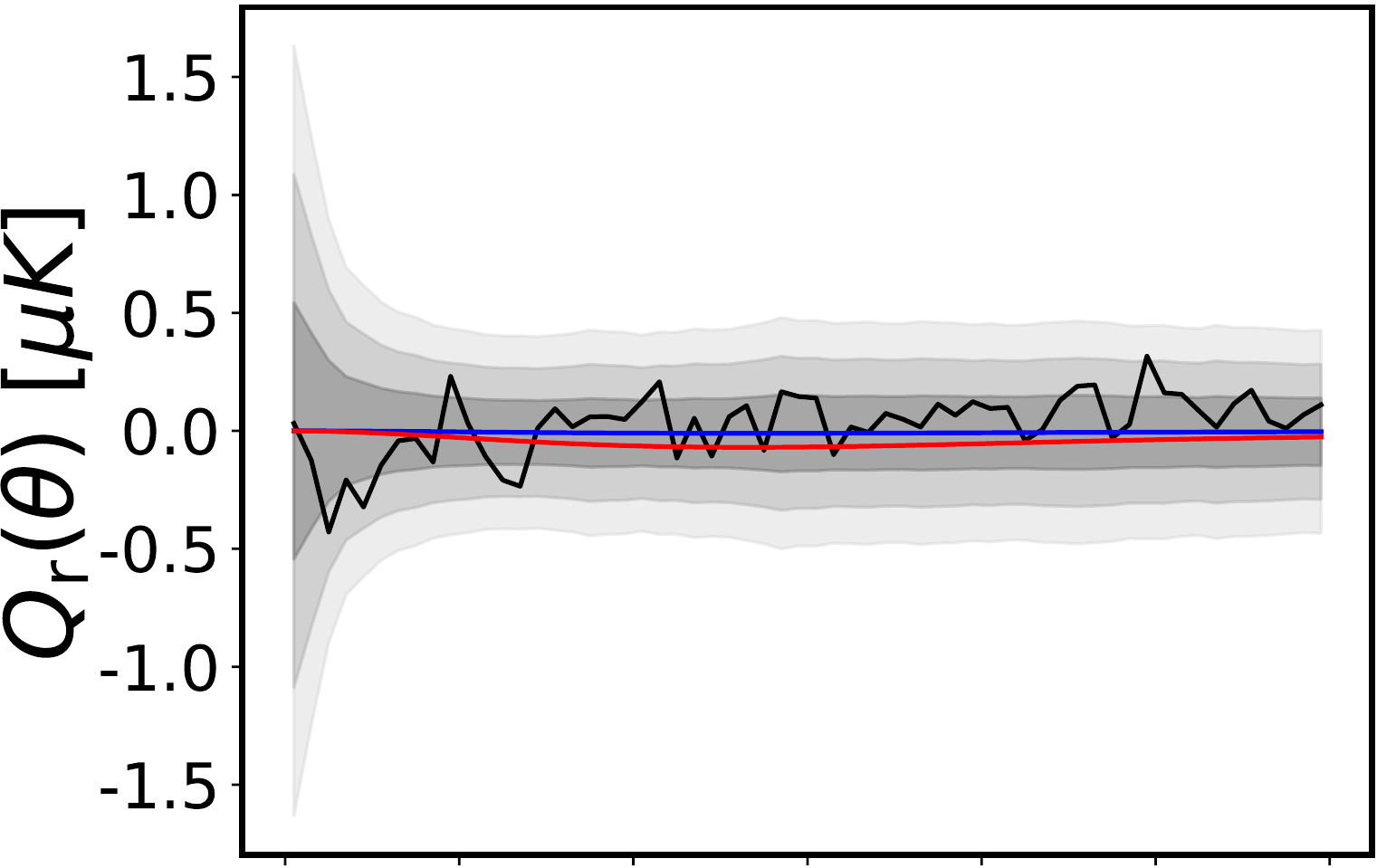}
\\
\hspace*{0.01\textwidth}\includegraphics[width=0.34\textwidth]{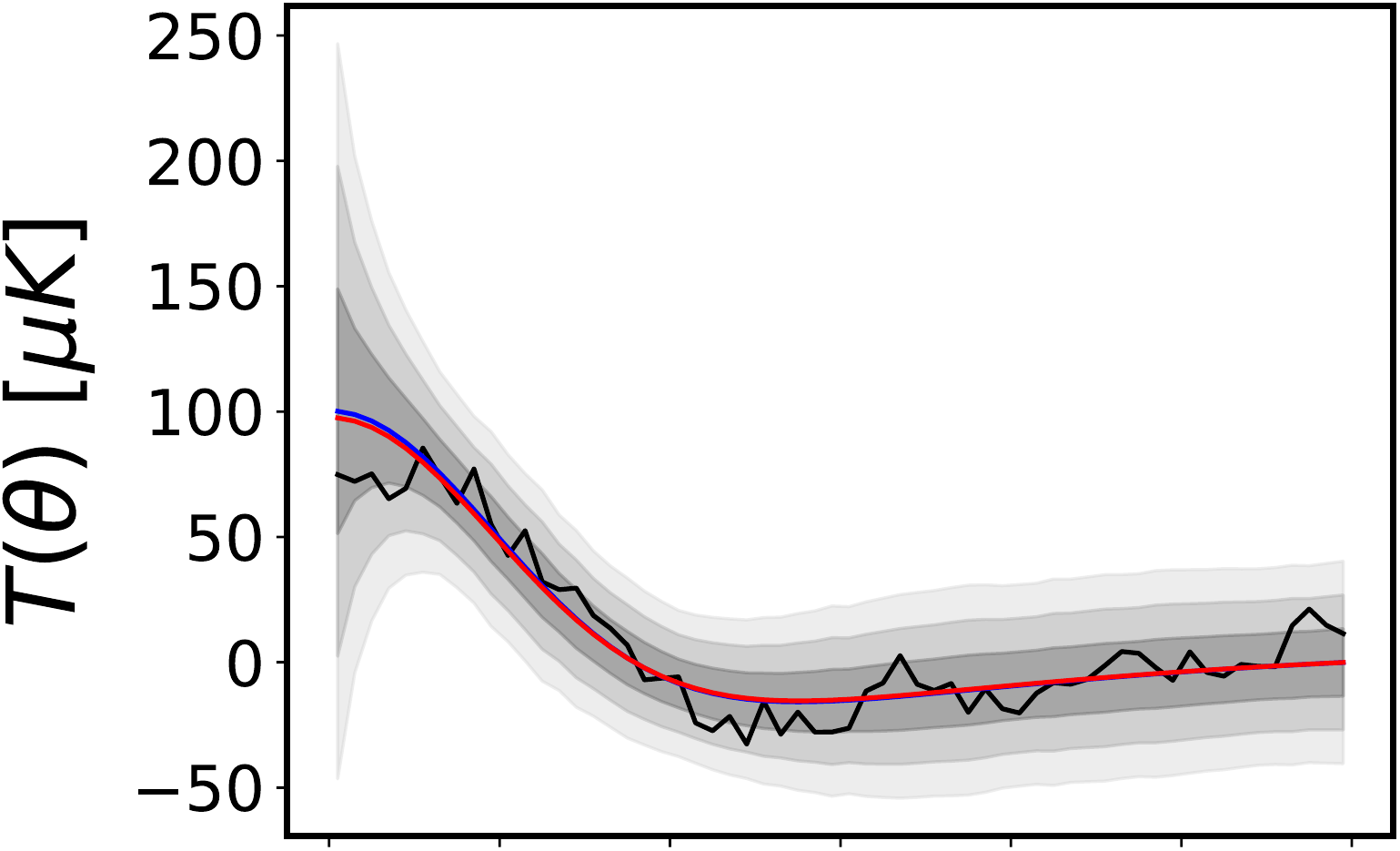}
\hspace*{0.1\textwidth}
\includegraphics[width=0.33\textwidth]{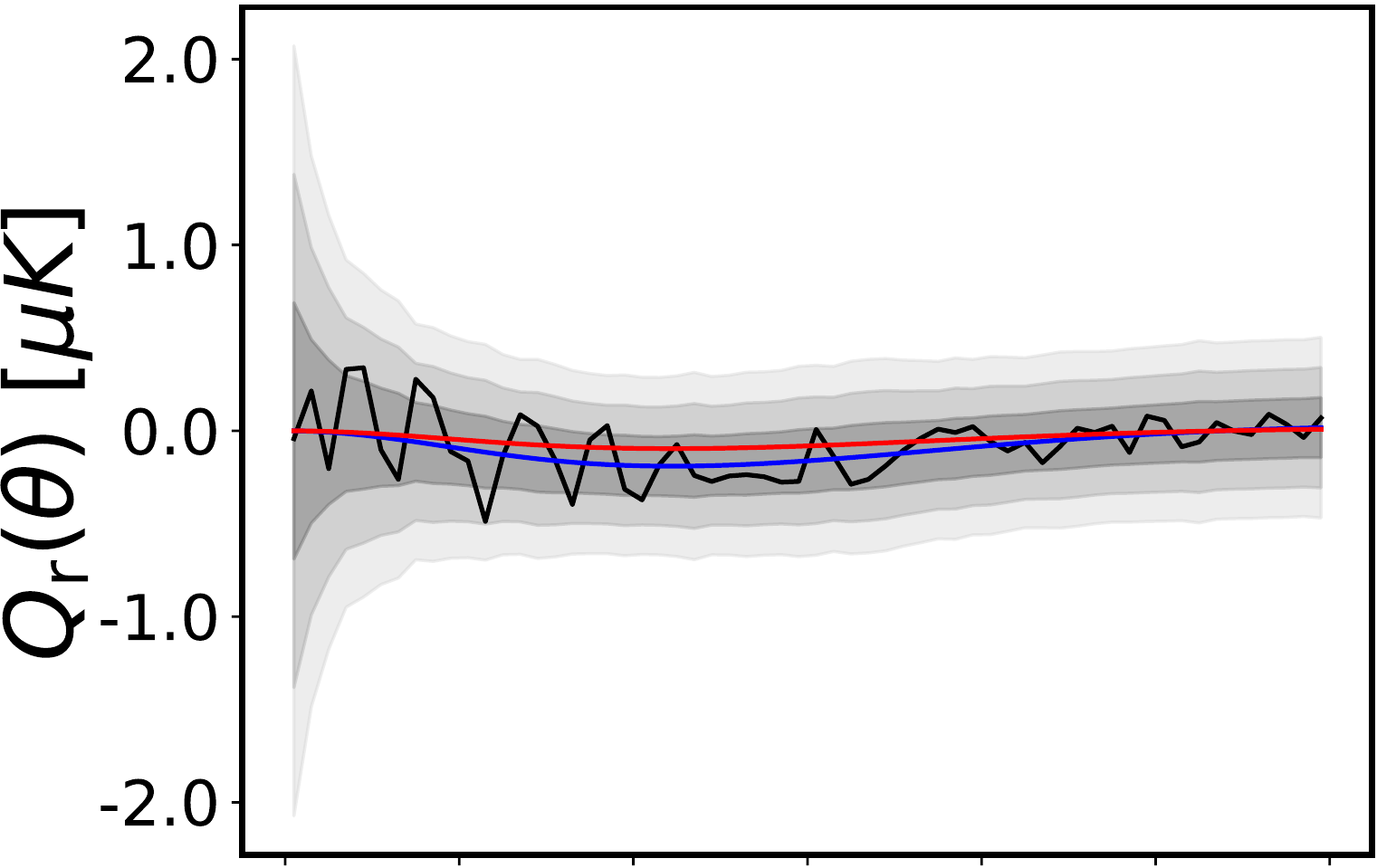}
\\
\includegraphics[width=0.35\textwidth]{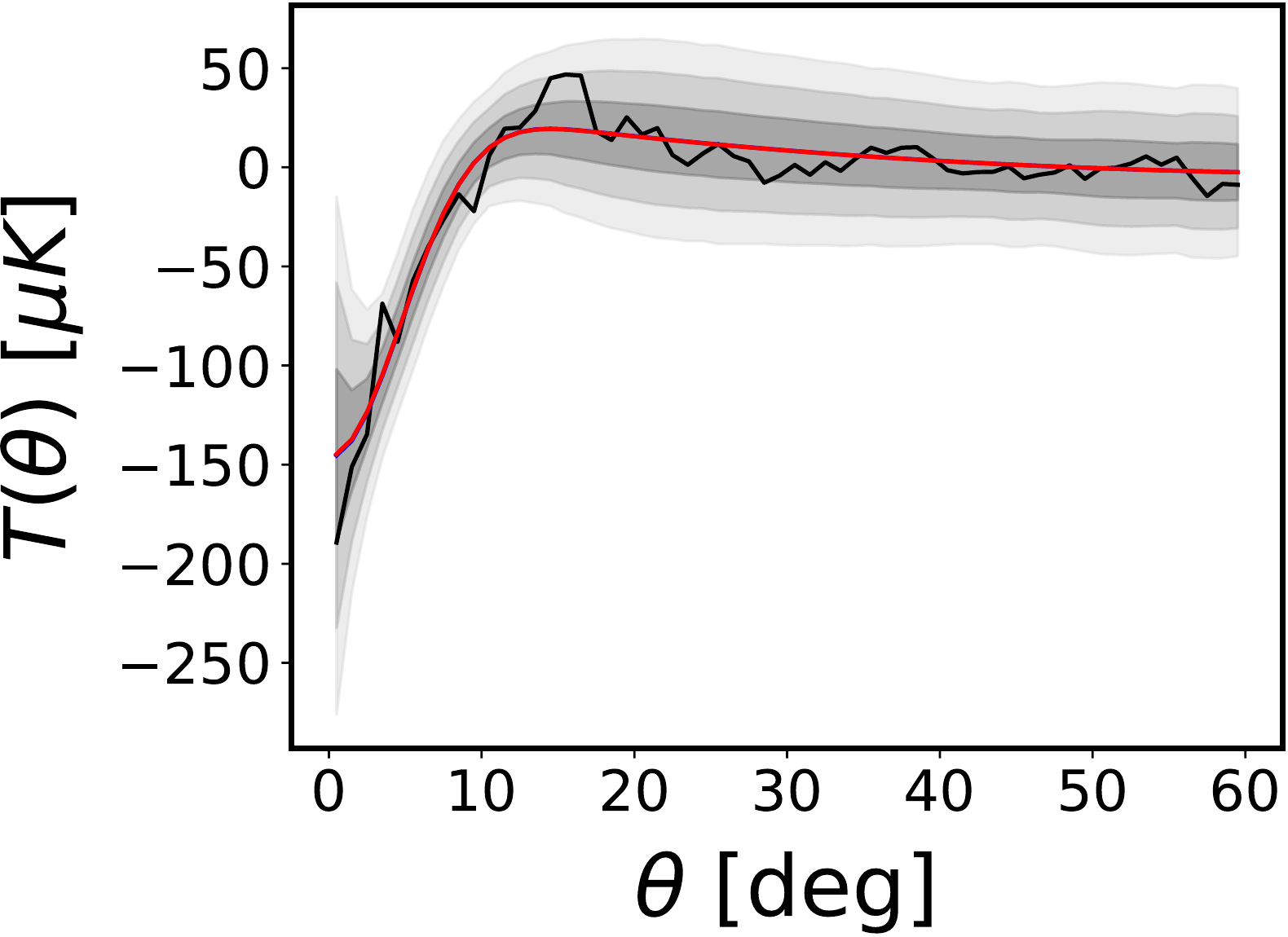}
\hspace*{0.1\textwidth}
\includegraphics[width=0.33\textwidth]{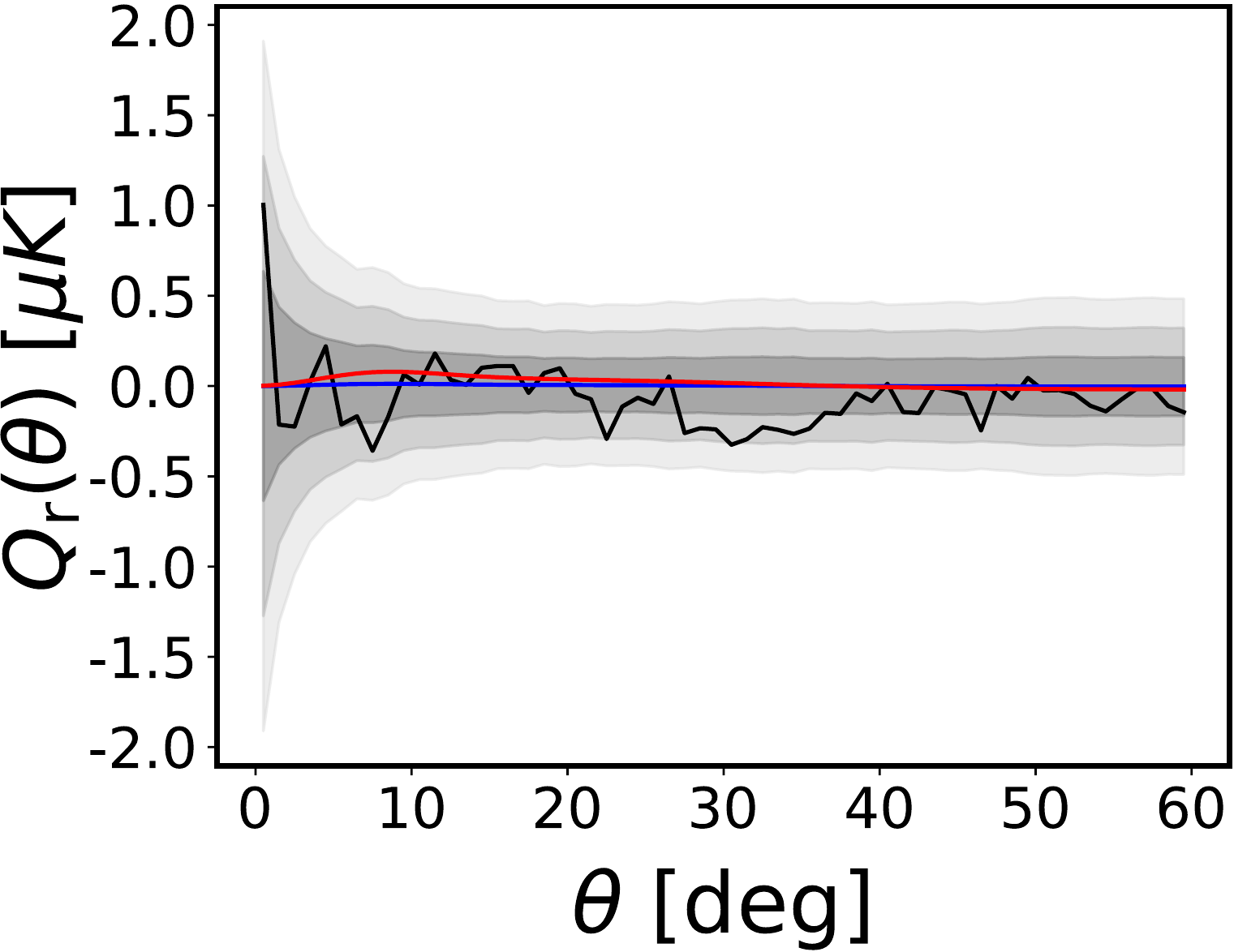}
\\
\end{center}
\caption{
Temperature (left column) and $Q_{\rm r}$ (right column) profiles of the large-scale Peaks~1 (top panel)
  through 5 (bottom panel), as shown in Fig.~\ref{fig:peaks_loc}.  The
  black solid lines represent the profiles obtained from the \sevem\
  CMB map.  The red curves correspond to the expected theoretical
  profiles, while the blue lines represent the theoretical profiles
  rescaled by the estimated amplitude $A$ for each profile.
  The shaded regions represent
  the $\pm1$, $\pm2$ and $\pm3\,\sigma$ confidence levels.}
\label{fig:tqr_profiles_lsp}
\end{figure*}

In polarization, the contribution of the peak to the covariance is
small compared with the cosmic variance \citep{marcos-caballero2016}
and is ignored in subsequent calculations.  The polarization analysis
is then carried out at \nside\ = 512 with the common mask applied.
The cosmic-variance part of the covariance matrix is calculated
theoretically from the angular power spectrum
\citep{marcos-caballero2016,marcos-caballero2017b}, whereas the noise
part is estimated from the FFP10 noise simulations.  In order to take
into account the possible contribution of the anisotropic noise, the
covariances are calculated at the specific locations of each
peak. The consistency of the noise between data and simulations
is verified by computing the profiles in the OEHD and HMHD maps.
The noise values in the profiles at each peak location are
compatible with those obtained from the simulations. Since we are
analysing only large scales, the noise can be ignored in the case of
the temperature profiles.

The temperature and polarization signals induced by the peaks can be
estimated by rescaling the theoretical profiles by a factor $A$. 
The value of $A$ is obtained using the maximum likelihood estimator
assuming that the profiles are Gaussian. In this case, it is
possible to show that the distribution of $A$ is also Gaussian
with the following mean and standard deviation:
\begin{equation}
A = \frac{\sum_{i,j} \hat{X}(\theta_i) C^{-1}_{ij}
  X(\theta_j)}{\sum_{ij} X(\theta_i) C^{-1}_{ij} X(\theta_j)} \ ;
\label{eqn:A_scale_factor}
\end{equation}
\begin{equation}
\sigma_A = \frac{1}{\sqrt{\sum_{i,j} X(\theta_i) C^{-1}_{ij}
    X(\theta_j)}} \ .
\end{equation}
Here $\hat{X}(\theta_i)$ and $X(\theta_i)$ are the measured profile
and the corresponding theoretical expectation for the angular bin
centred at $\theta_i$, respectively. In these equations, $X$
represents the $T$ or the $Q_{\rm r}$ profile, depending on the case we are
analysing. Since we are interested in large-scale peaks, angles up to
$60^\circ$ are considered in the analysis, with a bin width of
$1^\circ$. The matrix $C_{ij}$ is the covariance of the angular bins,
which includes both cosmic variance and noise.

Forecasts regarding the ability of polarization measurements to
distinguish between the standard model and other alternatives make
use of the Fisher discriminant
\citep{vielva2011,fernandez-cobos2013}. This method is based on the
overlapping of the probability distributions of the amplitude $A$
for different hypotheses. In the case of the Cold Spot, the standard
model and alternatives are compared, assuming zero correlation between
temperature and polarization. The predicted significance levels for
detection of the polarization signal of the Cold Spot in these
papers is higher than presented here, since a different
characterization of the theoretical profile is used; while the model
we assume is based on the temperature and the Laplacian observed in
the data, the mean value of the model in
\citet{fernandez-cobos2013} was calculated using only the Laplacian
(i.e., SMHW coefficient). Moreover, the Cold Spot amplitude is
characterized with respect to the observed standard deviation, not
using the theoretical expectation, as in our analysis. This
difference in the theoretical models results in different
significance levels for the forecasts. However, the normalization of
the theoretical profiles is not relevant when the significance of
the hypothesis ($A/\sigma_A$) is calculated from the observed
amplitude $A$. 

The $T$ and $Q_{\rm r}$ profiles determined from the four component
separation methods are shown in 
Fig.~\ref{fig:tqr_profiles_lsp}.  The corresponding profile
amplitudes are given in Table~\ref{tab:amplitudes_lsp}.
The temperature results are compatible
with the analysis of the same peaks performed in
\citet{marcos-caballero2017b}.
Regarding the polarization analysis, the covariance of
$Q_\mathrm{r}$ is not affected by the presence of a peak in the
field, which results in larger uncertainties for polarization than
for temperature.
Indeed, the signal-to-noise ratio for the $Q_{\rm r}$
amplitude factors are only about 1, and the polarization signal of the
peaks cannot be detected; more sensitive polarization data would be
required to further test the model. The values of the amplitude factors measured
from the peak profiles are consistent with the standard \LCDM\ model.

\begin{table*}[htbp!]
\begingroup
\newdimen\tblskip \tblskip=5pt
\caption{Amplitudes $A$ estimated from Eq.~\eqref{eqn:A_scale_factor}
  of the $T$ and $Q_{\rm r}$ profiles for the large-scale peaks
  considered here. The \cs\ corresponds to Peak~5.}
\label{tab:amplitudes_lsp}
\nointerlineskip
\vskip -3mm
\footnotesize
\setbox\tablebox=\vbox{
   \newdimen\digitwidth 
   \setbox0=\hbox{\rm 0} 
   \digitwidth=\wd0 
   \catcode`*=\active 
   \def*{\kern\digitwidth}
   \newdimen\signwidth 
   \setbox0=\hbox{+} 
   \signwidth=\wd0 
   \catcode`!=\active 
   \def!{\kern\signwidth}
\halign{\hbox to 1.6in{#\leaderfil}\tabskip 4pt&
\hfil#\hfil&
\hfil#\hfil&
\hfil#\hfil&
\hfil#\hfil\tabskip 0pt\cr
\noalign{\doubleline\vskip -1pt}
\omit&\multispan4 \hfil Amplitudes\hfil\cr
\noalign{\vskip -4pt}
\omit&\multispan4\hrulefill\cr
\omit\hfil Peak\hfil&{\tt Comm.}&{\tt NILC}&{\tt SEVEM}&{\tt SMICA}\cr
\noalign{\vskip 3pt\hrule\vskip 3pt}
\omit&\multispan4 \hfil$T$ profiles\hfil\cr
\noalign{\vskip 3pt}
Peak 1& $!1.099\pm0.088$& $!1.014\pm0.084$& $!1.022\pm0.081$& $!1.096\pm0.083$\cr
Peak 2& $!1.183\pm0.086$& $!1.205\pm0.084$& $!1.175\pm0.084$& $!1.187\pm0.079$\cr
Peak 3& $!1.090\pm0.078$& $!1.097\pm0.075$& $!1.057\pm0.077$& $!1.057\pm0.073$\cr
Peak 4& $!1.022\pm0.016$& $!1.023\pm0.016$& $!1.027\pm0.016$& $!1.029\pm0.016$\cr
Peak 5& $!1.005\pm0.003$& $!1.005\pm0.003$& $!1.005\pm0.003$& $!1.005\pm0.003$\cr
\noalign{\vskip 3pt\hrule\vskip 3pt}
\omit&\multispan4 \hfil$Q_{\rm r}$ profiles\hfil\cr
\noalign{\vskip 3pt}
Peak 1& $-1.330\pm1.102$& $-0.200\pm1.146$& $!0.164\pm1.117$& $-0.034\pm1.134$\cr
Peak 2& $!0.978\pm1.088$& $!1.523\pm1.088$& $!0.446\pm1.124$& $!0.791\pm1.077$\cr
Peak 3& $!0.129\pm1.261$& $!0.090\pm1.298$& $!0.196\pm1.270$& $-0.331\pm1.311$\cr
Peak 4& $!0.691\pm0.957$& $!0.150\pm1.011$& $!1.981\pm0.994$& $!0.516\pm0.927$\cr
Peak 5& $-0.232\pm0.951$& $-0.170\pm1.035$& $!0.152\pm1.019$& $-0.598\pm1.023$\cr
 \noalign{\vskip 3pt\hrule\vskip 3pt}}}

\endPlancktable                    
\endgroup
\end{table*} 

\section{Dipole modulation and directionality}
\label{sec:dipmod}

In this section, we examine dipolar violations of isotropy.  
In \citetalias{planck2014-a18}, we performed a non-exhaustive 
series of tests to try to elucidate the nature of the 
observed asymmetry in temperature, and we follow this approach again here, 
both to reconfirm the previous temperature results, and to search for
evidence of equivalent signatures in the polarization data.

Despite warnings about the use of the 2015 \Planck\ polarization maps
for the study of isotropy and statistics, several papers have
attempted to fit dipolar-modulation models to the
data. \citet{Aluri2017} applied a local-variance estimator to
low-resolution $E$-mode maps derived from the 2015 \commander\
solution, and found a power asymmetry at the level of
around $3\,\%$ over the range $\ell=20$--240, with a preferred direction 
broadly aligned with the CMB dipole.\footnote{Referred
  to as the ``Solar dipole'' in accompanying \Planck\ papers \citep[see
  also section~2.1 of][]{planck2016-l01}.}
The $\ell$ range
was selected on the basis that an apparent noise mismatch between the
data and the FFP8 simulations could be minimized by the
application of a simple scaling factor. However, we note that this
mismatch is almost certainly due to the absence of modelled systematic
effects in the FFP8 simulations, rather than an actual miscalibration
of the instrumental noise.  Conversely, \citet{Ghosh2018}
applied a pixel-based method to maps of the squared polarization
amplitude, but found no evidence for the presence of a
dipolar-modulation signal. They suggested that this may be due to residual
systematics masking a real effect, since strong clustering of
the dipole directions inferred from the FFP8 simulations was also observed.

These results indicate the necessity to use improved simulations that
characterize the data more completely, and to adapt the estimators to
eliminate bias resulting from the anisotropic noise and systematic effect
residuals present in both data and simulations. As usual, we test for
inconsistencies between the data and simulations using null tests
based on the HMHD and OEHD data splits.

All the tests in this section involve the fitting of a dipole.  In 
Sects.~\ref{sec:varasym} and \ref{sec:powerasymmetry}, we fit a dipole 
explicitly to a map of power on the sky.  On the other hand, in 
Sect.~\ref{sec:dipmod_qml} we measure the coupling of $\ell$ to 
$\ell \pm 1$ modes in the CMB multipole covariance matrix.  There are 
differences in how we combine the fitted dipoles, which determine 
the particular form of dipolar asymmetry that the test will be sensitive to.

The tests can also be distinguished by whether they are based on amplitude 
or direction.  The approaches of Sects.~\ref{sec:varasym} and 
\ref{sec:dipmod_qml} are both sensitive to the dipole modulation 
{\em amplitude}.  In particular, in Sect.~\ref{sec:varasym} we examine 
the data for dipolar modulation of the pixel-to-pixel variance, while in 
Sect.~\ref{sec:dipmod_qml} we search for modulation of the angular 
power spectra.  These approaches differ mainly in terms of their $\ell$ 
weighting.  {\em Directionality} in the data is the subject of 
Sect.~\ref{sec:powerasymmetry}, with the directions determined by 
band-power dipole fits.

Given these differences in the approaches in this section, it is important 
to keep in mind that the results cannot usually be directly compared, even 
though all probe some aspect of dipolar asymmetry.

\subsection{Variance asymmetry}
\label{sec:varasym}

We first study the \Planck\ 2018 temperature and polarization maps
using the local-variance method introduced by \citet{Akrami2014}.
Despite its relative simplicity, the local-variance estimator serves
as a powerful method for detecting violations of statistical isotropy
if they are of a type corresponding to a large-scale power asymmetry, as
observed in the \Planck\ 2015 temperature data \citepalias{planck2014-a18}.

Here, we closely follow the previous temperature analysis, but extend
its application to the scalar $E$-mode polarization data. The method
can briefly be described as follows. We define a set of discs of
various sizes uniformly distributed on the sky. The centres of the
discs are defined to be the pixel centroids of a \healpix\ map at some
specific low resolution, here taken to be $N_{\mathrm{side}}=16$ for
both temperature and polarization analyses. Since it is important to
cover the entire sky, we do not work with discs of radii smaller than
$4\deg$ (given our choice of $N_{\mathrm{side}}$ for the centroids).
These combinations of \nside\ and disc radii have been shown
to be adequate for detecting large-scale power asymmetry in CMB
temperature data, and allow us to focus on the question of whether a
corresponding large-scale anomalous power asymmetry exists in the
polarization data.
For each sky map (data and simulations), we first remove the monopole
and dipole components from the masked map and then compute the
variance of the fluctuations for each disc of a given size using only
the unmasked pixels. This results in local-variance maps at the
\healpix\ resolution of $N_{\mathrm{side}}=16$. We also estimate the
expected average and variance of the variances on each disc from the
simulations, and then subtract the average variance map from both the
observed and simulated local-variance maps. Finally, we fit a dipole
to each of the local-variance maps using the \healpix\ {\ttfamily
  remove\_dipole} routine, where each pixel is weighted by the inverse
of the variance of the variances that we have computed from the
simulations at that pixel. Note that, at all stages of the analysis,
we work only with those discs for which more than $10\,\%$ of the area
is unmasked. In this way, we define an amplitude and direction for the
variance asymmetry of each map. We then compare the local-variance
amplitudes for the observed data to those of the simulations,
containing statistically isotropic CMB realizations, and assess the
level of anisotropy in the data. Since \citet{Akrami2014} have shown
that the amplitudes of higher multipoles in such fits to temperature
data are consistent with statistically isotropic simulations, we work
only with the dipole amplitudes of the local-variance maps here.
However, although we do not consider higher-multipole asymmetries,
such features may exist in local-variance maps estimated from the
polarization data.

Table~\ref{tab:varasym1} presents the results for a local-variance
analysis of the full-resolution, $N_{\mathrm{side}}=2048$, \Planck\
2018 temperature data
over a range of disc radii, $4\deg\leq r_\text{disc}\leq20\deg$. The
{\it p}-values are measured as the fractional number of simulations
with local-variance dipole amplitudes larger than those inferred from
the data. The results are consistent for all the component-separated
maps, and also with those of the \Planck\ 2015 analysis, although a
higher level of significance is seen here. In particular, none of the
simulations yield local-variance dipole amplitudes larger than those
determined from the data for the $4\deg$, $6\deg$, and $8\deg$ discs.
This illustrates that the \Planck\ temperature sky is asymmetric if
one chooses to focus on variance over this range of angular scale.
We also see the expected increase in {\it p}-values as the disc radius
is increased to values larger than $8\deg$.
  
\begin{figure}[htbp!]
\begin{center}
\mbox{\epsfig{figure=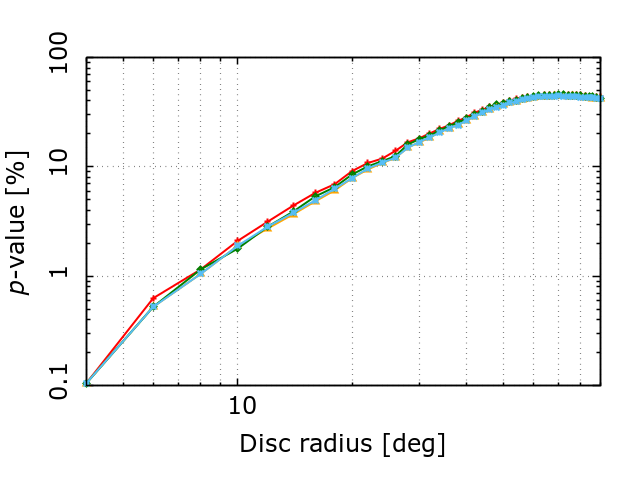,width=0.95\linewidth}}\\
\mbox{\epsfig{figure=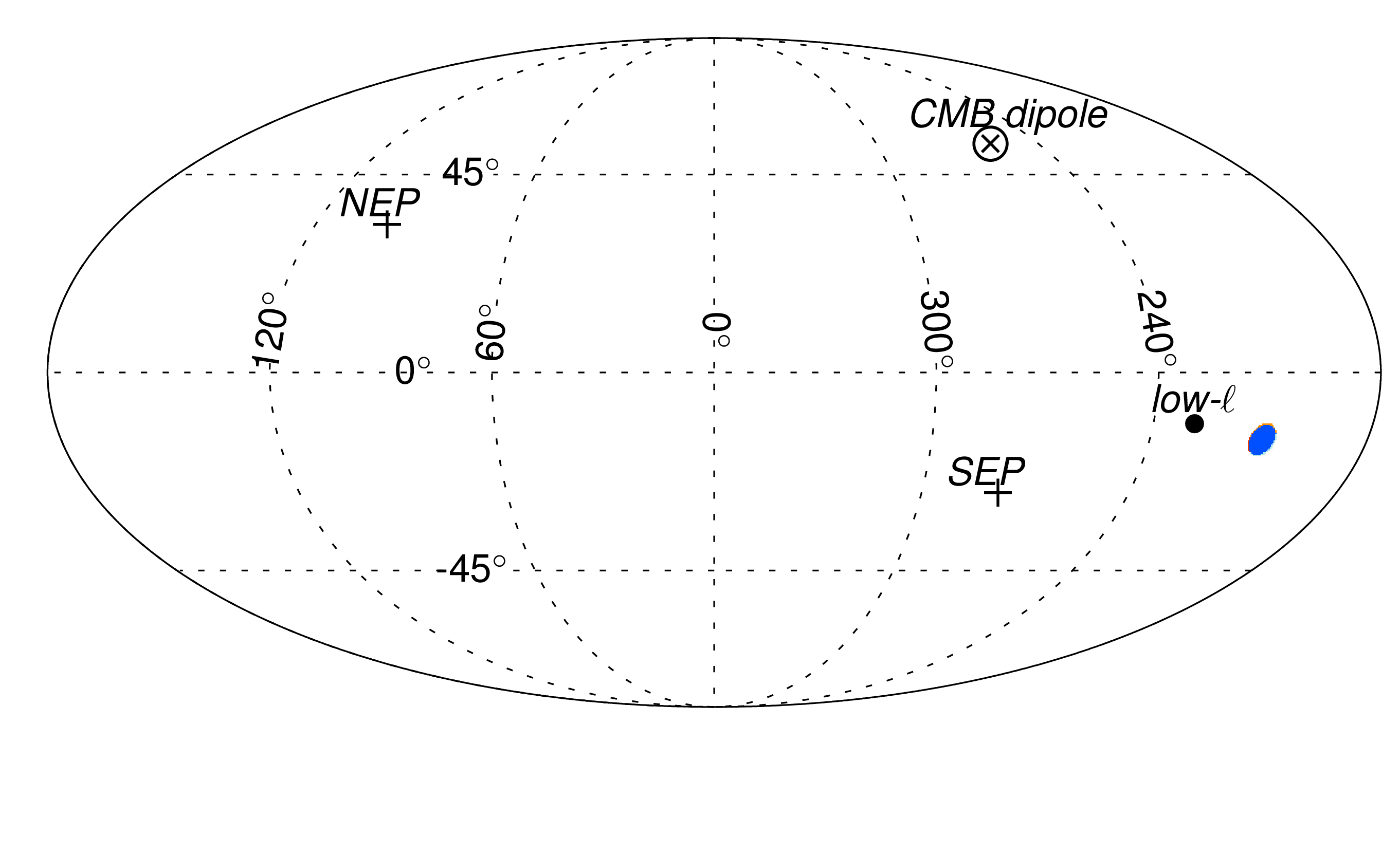,width=1\linewidth}}
\end{center}
\vspace{-1.0truecm}
\caption{{\it Upper panel:} {\it p}-values for variance asymmetry
  measured as the fraction of simulations with local-variance dipole
  amplitudes larger than those inferred from the
  data, for low-resolution ($N_{\mathrm{side}}=64$) temperature maps.
  The {\it p}-values are given as a function of disc radius and for
  the four component-separated
  temperature maps, \commander\ (red), \nilc\ (orange), \sevem\ (green),
  and \smica\ (blue). {\it Lower panel:} Corresponding local-variance
  dipole directions for the four component-separation temperature
  maps, and for 4\deg\ discs with the lowest {\it p}-values. Note that
  the four directions match almost perfectly, so that the symbols
  essentially overlap. 
  For reference, we also show the CMB dipole direction, the north
  ecliptic pole (NEP), the south ecliptic pole (SEP), and the
  preferred dipolar modulation axis (labelled as ``low-$\ell$'')
  derived from the temperature data in Sect.~\ref{sec:dipmod_qml}. 
}
\label{fig:varasym1}
\end{figure}

Since our focus is on large angular scales, we repeat the
local-variance analysis for low-resolution, $N_{\mathrm{side}}=64$,
temperature maps. Figure~\ref{fig:varasym1} (upper panel) shows the
{\it p}-values for the component-separated maps as a function of disc
radius, over the range $4\deg\leq r_\text{disc}\leq90\deg$. 
The preferred low-$\ell$ modulation direction determined from the
temperature data in Sect.~\ref{sec:dipmod_qml} is also indicated.
Excellent agreement is found between the component-separation methods. The
overall level of significance is lower compared to the full-resolution
results, but high significance (low {\it
  p}-value) results are found for small disc radii. The lowest {\it
  p}-values correspond to the smallest disc
radius, i.e., $4\deg$ discs, and the {\it p}-values increase with
disc size. Figure~\ref{fig:varasym1} 
(lower panel) also shows the preferred directions for the four maps 
when 4\deg\ discs are used. The observed directions are
in excellent agreement with each other, with the results of the
full-resolution analysis, and with the findings of the \Planck\ 2015
analysis.
The values of the $4\deg$-disc local-variance dipole directions for
the four component-separation methods are provided in
Table~\ref{tab:varasym2}
for both the high-resolution and low-resolution cases.
  
\begin{table}[htbp!]  
\begingroup	
\newdimen\tblskip \tblskip=5pt 
\caption{{\it p}-values for variance asymmetry measured using different
  discs for the full-resolution ($N_{\mathrm{side}}=2048$) {\Planck}
  2018 \commander, \nilc, \sevem, and \smica\ temperature
  solutions. The values represent the fraction of simulations with
  local-variance dipole amplitudes larger than those inferred from the
  data, and hence small $p$-values correspond to anomalously large
  variance asymmetry.}
\label{tab:varasym1}
\nointerlineskip 
\vskip -3mm 
\footnotesize  
\setbox\tablebox=\vbox{ 
   \newdimen\digitwidth 
   \setbox0=\hbox{\rm 0} 
   \digitwidth=\wd0 
   \catcode`!=\active 
   \def!{\kern\digitwidth}
   \newdimen\signwidth 
   \setbox0=\hbox{>} 
   \signwidth=\wd0 
   \catcode`*=\active 
   \def*{\kern\signwidth}
\halign{\hbox to 1.0in{#\leaderfil}\tabskip 4pt&
\hfil#\hfil\tabskip 8pt&
\hfil#\hfil\/\tabskip 8pt&
\hfil#\hfil\/\tabskip 8pt&
\hfil#\hfil\/\tabskip 0pt\cr
\noalign{\doubleline}
\noalign{\vskip -1pt}
\omit&\multispan4 \hfil {\it p}-value [\%]\hfil\cr
\noalign{\vskip -4pt}
\omit&\multispan4\hrulefill\cr
\omit \hfil Disc radius [deg] \hfil&\omit\hfil {\tt Comm.}
\hfil&\omit\hfil \nilc \hfil&\omit\hfil \sevem
\hfil&\omit\hfil \smica \hfil\cr   
\noalign{\vskip 3pt\hrule\vskip 3pt}
4 & $<$0.1&  $<$0.1&  $<$0.1&  $<$0.1\cr
6&  $<$0.1&  $<$0.1&  $<$0.1&  $<$0.1\cr
8&  $<$0.1&  $<$0.1&  $<$0.1&  $<$0.1\cr
10& *0.1& *0.1& *0.1& *0.1\cr
12& *0.3& *0.2&  *0.3&  *0.1\cr
14& *0.7&  *0.6&  *0.7&  *0.6\cr
16& *0.8&  *0.8&  *0.8&  *0.8\cr
18& *1.5&  *1.3&  *1.6&  *1.3\cr
20& *2.0&  *1.7&  *1.9&  *1.7\cr
\noalign{\vskip 3pt\hrule\vskip 3pt}}} %
\endPlancktable	
\endgroup %
\end{table}

\begin{table}[htbp!]
\begingroup	
\newdimen\tblskip \tblskip=5pt 
\caption{Local-variance dipole directions for the variance asymmetry
  of the four component-separated temperature maps. All directions
  quoted here are for 4\deg\ discs, and for full-resolution
  ($N_{\mathrm{side}}=2048$), as well as downgraded, low-resolution
  ($N_{\mathrm{side}}=64$) maps.}
 \label{tab:varasym2}
\nointerlineskip 
\vskip -3mm 
\footnotesize  
\setbox\tablebox=\vbox{
\halign{\hbox to 0.9in{#\leaderfil}\tabskip 6pt&
\hfil#\hfil\tabskip=5pt&\hfil#\hfil\/\tabskip=0pt&\hfil#\hfil\cr
\noalign{\doubleline}
\omit& \multispan2 \hfil $(l,b)$ [deg] \hfil\cr
\noalign{\vskip -4pt}
\omit&\multispan2\hrulefill\cr
\omit\hfil Data\hfil& $N_{\mathrm{side}}=2048$& $N_{\mathrm{side}}=64$\cr
\noalign{\vskip 3pt\hrule\vskip 3pt}
{\tt Commander}& \hfil $(205,-20)$\hfil& \hfil $(209,-15)$\hfil\cr
\nilc& \hfil $(205,-20)$\hfil& \hfil $(209,-15)$\hfil\cr
\sevem& \hfil $(205,-19)$\hfil& \hfil $(209,-14)$\hfil\cr
\smica& \hfil $(205,-19)$\hfil& \hfil $(209,-15)$\hfil\cr
\noalign{\vskip 3pt\hrule\vskip 3pt}}} %
\endPlancktable	
\endgroup %
\end{table}

The analysis of the polarization data is significantly more subtle
than for temperature because of the inherently low signal-to-noise. We
first validate the technique by applying it to polarization
simulations. The analysis shows that the direct application of the
method to isotropic simulations of $E$-mode polarization signal
returns local-variance dipole directions that are not uniformly
distributed.
This arises from the strongly anisotropic, correlated, and
non-Gaussian structure of the \Planck\ polarization noise, which needs
to be corrected before any statistical method of anisotropy detection
is applied to the data. This is not an issue for the temperature maps,
since the signal-to-noise ratio is large.

The following procedure is adopted to minimize the
effects of the anisotropic noise. We first preprocess the $E$-mode
polarization maps (both the data and simulations) pixel by pixel
and apply a bias correction and weighting. We consider the following
types of transformation to the maps:
\begin{linenomath*}
\begin{equation}
X_i \rightarrow (X_i-M_i) / \sigma_i^p.
\label{eq:varasym_weighting}
\end{equation}
\end{linenomath*}
Here, $X_i$ is the value of the map at pixel $i$, $M_i$ is the mean at
pixel $i$ computed from the simulations, $\sigma_i$ is the standard
deviation at pixel $i$ (again computed from the simulations), and
$p\ge1$ is an integer.

\begin{figure}[htbp!]
\begin{center}
\mbox{\epsfig{figure=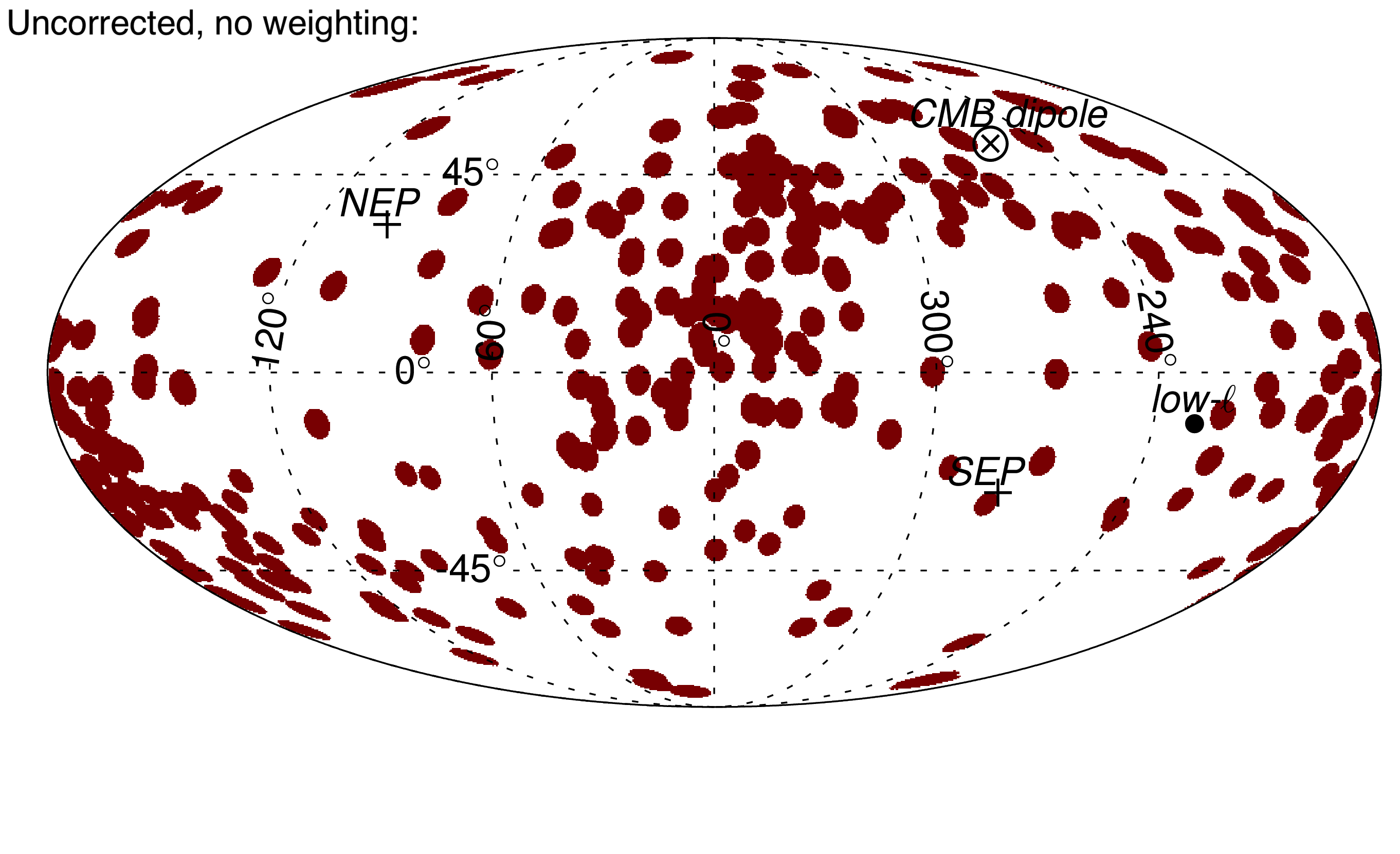,width=1\linewidth}}\\[-0.75truecm]
\mbox{\epsfig{figure=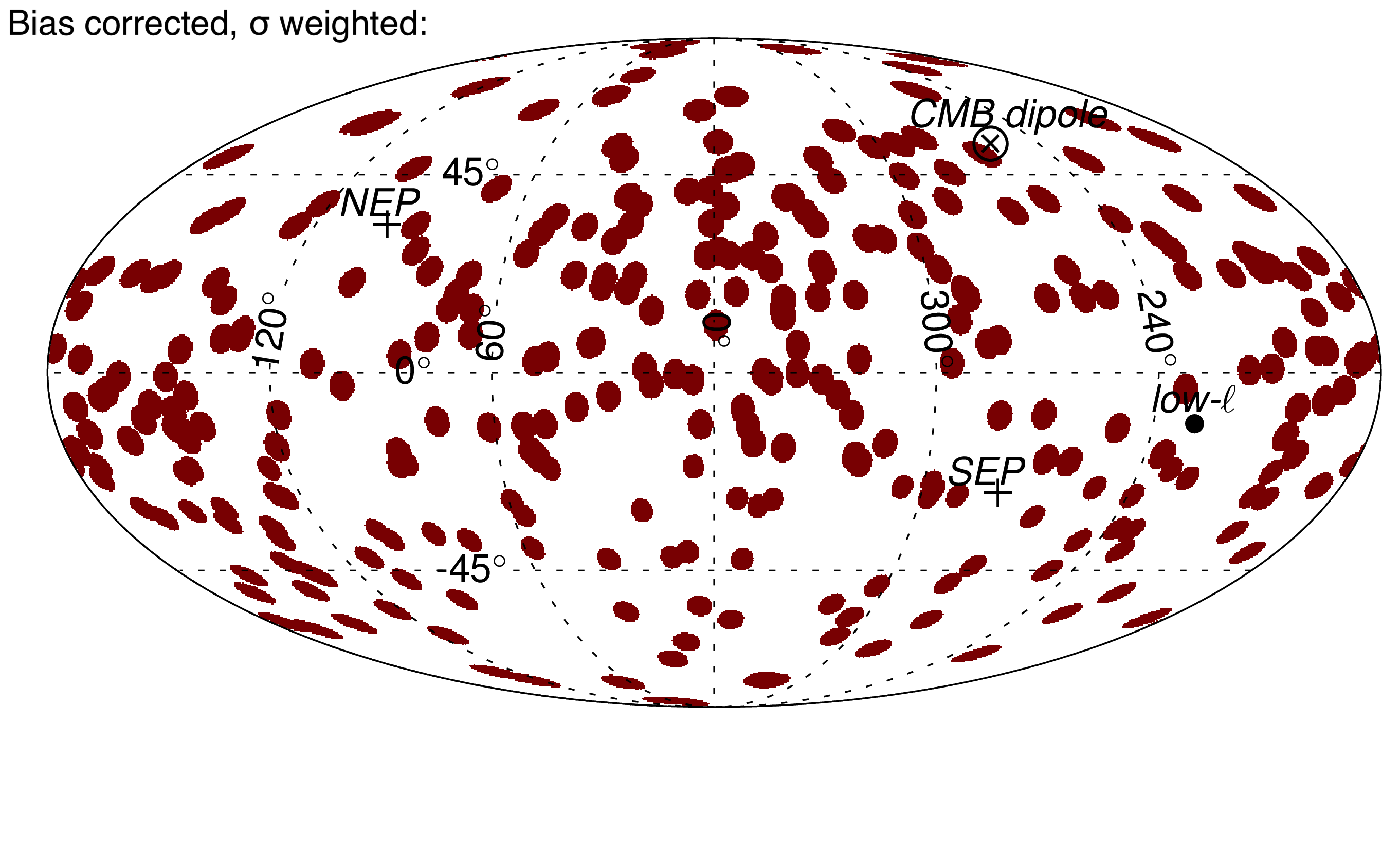,width=1\linewidth}}\\[-0.75truecm]
\mbox{\epsfig{figure=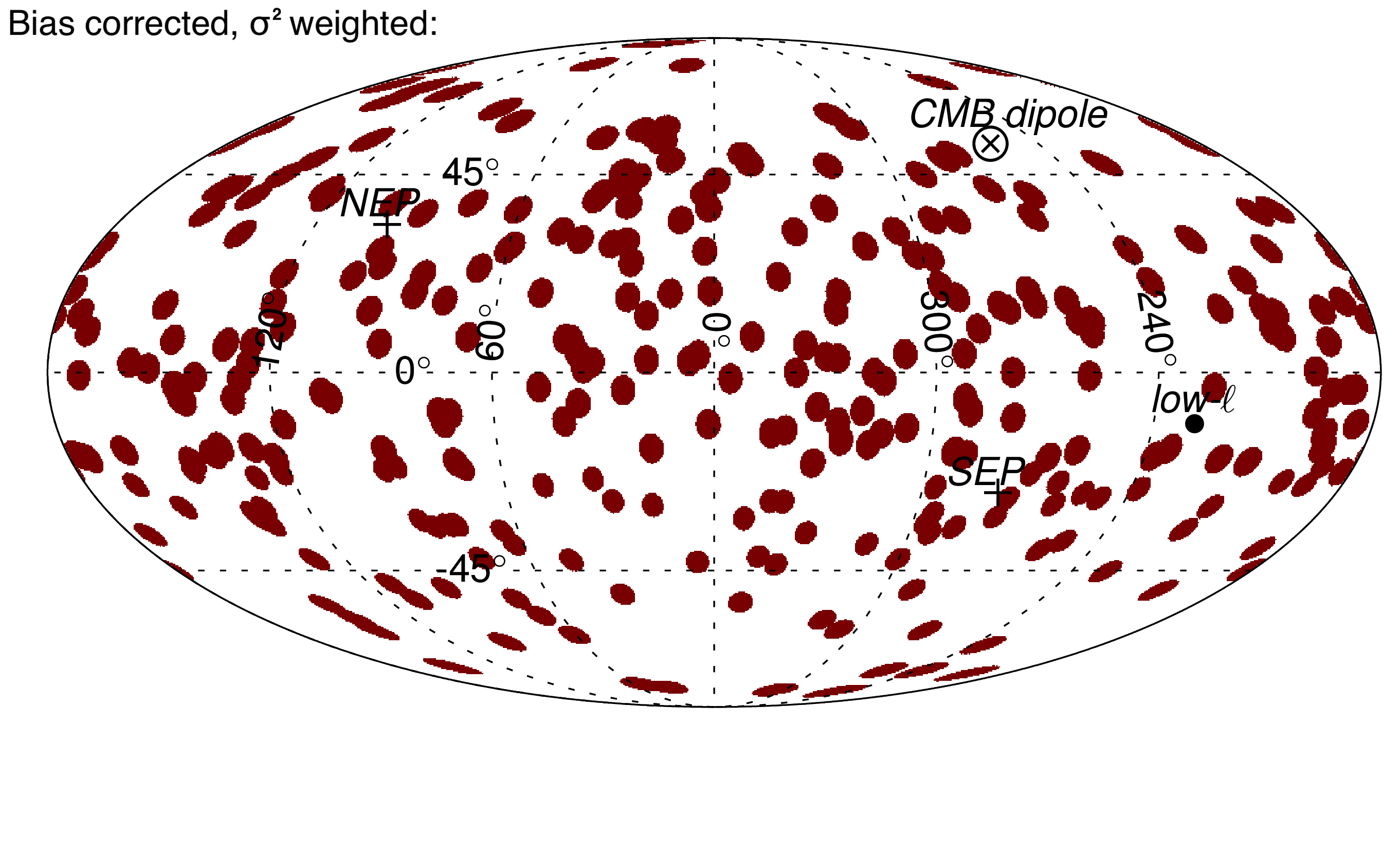,width=1\linewidth}}
\end{center}
\vspace{-1.0truecm}
\caption{Impact of bias correction and weighting on the uniformity of
  dipole directions for $E$-mode polarization HMHD simulation maps.
  {\it Upper panel:} local-variance dipole directions for the
  \commander\ simulations of $E$-mode polarization maps, obtained for
  4\deg\ discs, when no bias correction and weighting have been
  applied to the maps. {\it Middle panel:}
 The same as in the upper panel, but when the transformation $X_i
\rightarrow (X_i-M_i) / \sigma_i$ has been applied to the polarization
maps before applying the local-variance estimator. {\it Lower panel:}
The same as in the upper and middle panels, but when the
transformation $X_i \rightarrow (X_i-M_i) / \sigma_i^2$ has been
applied to the maps (i.e., with a square in the denominator). 
The results for the other component-separation
methods are very similar to the ones presented here.
  For reference, we also show the CMB dipole direction, the north
  ecliptic pole (NEP), the south ecliptic pole (SEP), and the
  preferred dipolar modulation axis (labelled as ``low-$\ell$'')
  derived from the temperature data in Sect.~\ref{sec:dipmod_qml}. 
}  \label{fig:varasym2}
\end{figure}

Figure~\ref{fig:varasym2} shows the distribution of the local-variance
dipole directions for an analysis of the HMHD $E$-mode simulation maps
when unmodified, and after 
transformation according to Eq.~\eqref{eq:varasym_weighting} with
$p=1$ or 2. Since we are interested in large-scale anomalies, 
we restrict the study to a resolution of $N_{\mathrm{side}}=64$, and
provide results for the \commander\ data and 4\deg\ discs only; the results
for the other component-separation methods are very similar. The untransformed
case clearly shows that the noise does not yield uniformly
distributed dipole directions, while by applying the
transformation given by Eq.~\eqref{eq:varasym_weighting} a significant
improvement in the uniformity is observed. Since the non-uniformity for 
simulated maps including signal arises from the noise, this transformation
also improves the uniformity of the recovered dipole directions.
Additionally, we see slightly more improvement for $p=2$,
and adopt this weighting for the analysis of the
polarization data.

\begin{figure}[htbp!]
\begin{center}
\mbox{\epsfig{figure=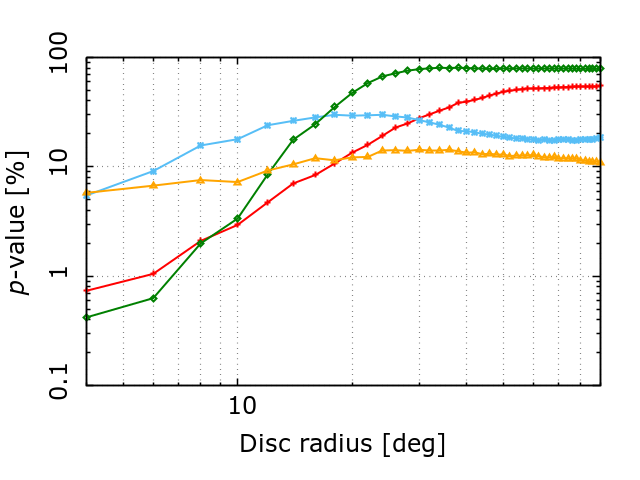,width=0.95\linewidth}}\\
\mbox{\epsfig{figure=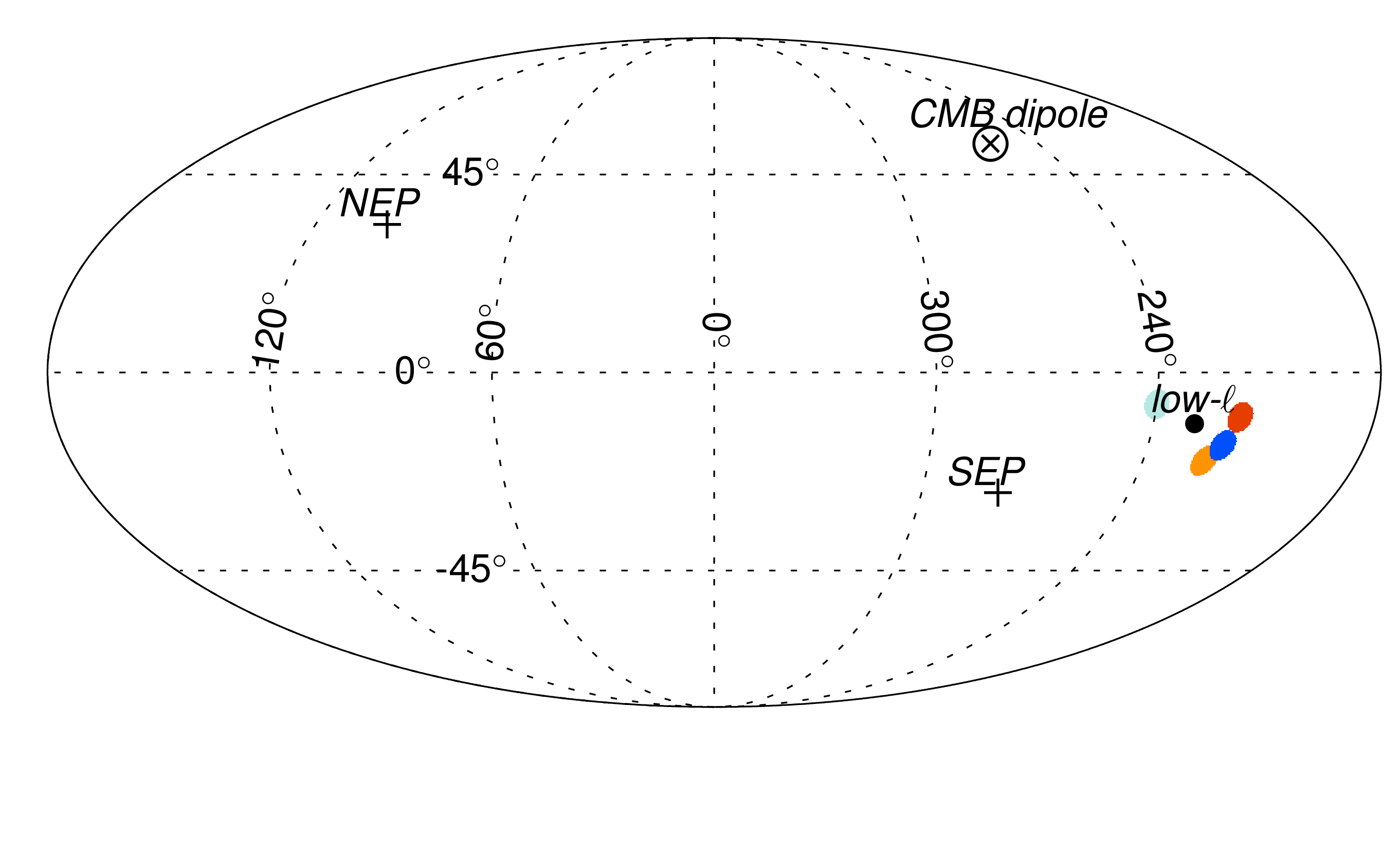,width=1\linewidth}}
\end{center}
\vspace{-1.0truecm}
\caption{{\it Upper panel:} {\it p}-values for variance asymmetry
  measured as the number of simulations with local-variance dipole
  amplitudes larger than those inferred from the data, as a
  function of disc radius for the four component-separated
  $E$-mode polarization maps, \commander\ (red), \nilc\ (orange), \sevem\ (green),
  and \smica\ (blue), at the \healpix\ resolution of $N_{\mathrm{side}}=64$. {\it Lower panel:}
  corresponding local-variance dipole directions for the four
  component-separation $E$-mode polarization maps, and for 4\deg\
  discs.
  For reference, we also show the CMB dipole direction, the north
  ecliptic pole (NEP), the south ecliptic pole (SEP), and the
  preferred dipolar modulation axis (labelled as ``low-$\ell$'')
  derived from the temperature data in Sect.~\ref{sec:dipmod_qml}. 
}  \label{fig:varasym3}
\end{figure}

Figure~\ref{fig:varasym3} (upper panel) presents the {\it p}-values
obtained by applying the local-variance estimator to the four
component-separated $E$-mode polarization maps.
Even though we typically see an increase in {\it p}-values between
smaller and larger disc radii, as observed for the temperature data,
the detailed behaviour differs. Moreover, the curves seem to be
divided into two groups, one including \commander\ and \sevem,
and the other consisting of \smica\ and \nilc.
This might reflect differences in the component-separation approaches,
particularly given that the former methods operate in the pixel
domain, and the latter in the harmonic domain. There is also a likely
connection with a variation in residual systematic effects present in
the component-separated maps, depending on the component-separation
methodology applied. Furthermore, these differences, and particularly
the relatively high values for the \smica\ and \nilc\ maps, argue
against a detection of cosmological power asymmetry in the polarization data.

Nevertheless, the preferred directions we have found for the $E$-mode
polarization data, as shown in the lower panel of
Fig.~\ref{fig:varasym3}, are intriguingly close to those determined
for the temperature data. In order to quantify the significance of
this alignment, we have computed the separation angles between the
$E$-mode and temperature-variance dipole directions for the data as
well as for all the simulations, and computed {\it p}-values defined
as the fraction of simulations with $T$-$E$ separation angles smaller
than those inferred from the data. Again, the results cannot be
interpreted as evidence of power asymmetry in polarization, and the
\sevem\ result is a notable outlier. Table~\ref{tab:varasym3}
summarizes these results, providing the coordinates of the preferred
local-variance dipole directions and corresponding {\it p}-values for
the four component-separation methods, together with the 
separation angles between temperature and $E$-mode dipole directions
and their associated {\it p}-values.

Despite the bias correction and weighting procedure employed for
reducing the non-uniformity of the dipole directions in simulations,
it is clear that the treatment of anisotropic noise plays an important
role in our analysis of the polarization data, and its impact on the
variance asymmetry results may need further consideration. While the
close alignment of the temperature and polarization directions
could simply be a coincidence, future data sets may offer additional
insight.

\begin{table}[htbp!]
\begingroup	
\newdimen\tblskip \tblskip=5pt 
\caption{Local-variance dipole directions and {\it p}-values for the
  variance asymmetry of the four component-separated $E$-mode
  polarization maps at the \healpix\ resolution of
  $N_{\mathrm{side}}=64$. The {\it p}-values represent the fraction of
  simulations with local-variance dipole amplitudes larger than those
  inferred from the data. All directions quoted here are for 4\deg\
  discs. We also show the separation angles and degrees of
  alignment between the preferred directions inferred from
  temperature, $T$,  and $E$-mode polarization data, with the {\it p}-values measured as the
  fraction of simulations with separation angles smaller than those
  inferred from the data.
}
 \label{tab:varasym3}
\nointerlineskip 
\vskip -3mm 
\footnotesize  
\setbox\tablebox=\vbox{
\halign{\hbox to 0.9in{#\leaderfil}\tabskip 6pt&
\hfil#\hfil\tabskip=5pt&\hfil#\hfil\tabskip=5pt&\hfil#\hfil\tabskip=5pt&\hfil#\hfil\/\tabskip=0pt&\hfil#\hfil\cr
\noalign{\doubleline}
\omit& \omit& \omit& \multispan2 \hfil Alignment with $T$ \hfil \cr
\noalign{\vskip -4pt}
\omit& \omit& \omit&\multispan2\hrulefill\cr
\omit\hfil Data\hfil& {\it p}-value [\%]& $(l,b)$ [deg] & $\cos\alpha^{\rm a}$& {\it p}-value [\%]\cr
\noalign{\vskip 3pt\hrule\vskip 3pt}
{\tt Commander}& 0.7& \hfil $(217,-10)$\hfil & 0.99 & 0.9\cr
\nilc& 5.8& \hfil $(222,-19)$\hfil & 0.97 & 1.9\cr
\sevem& 0.4& $(240,-7)$ & 0.86 & 6.9\cr
\smica& 5.5& \hfil $(219,-16)$\hfil & 0.99 & 0.9\cr
\noalign{\vskip 3pt\hrule\vskip 3pt}}} %
\endPlancktable	
\endgroup %
{\footnotesize $^{\rm a}$ $\alpha$ is the separation angle between the
  preferred directions computed for the temperature and $E$-mode
  polarization data.}
\end{table}

\subsection{Dipole modulation: QML analysis}
\label{sec:dipmod_qml}

In this section, we use the QML estimator originally introduced in
\citet{Moss2011} to assess the level of dipole asymmetry in the CMB
sky, and further extend the analysis to polarization data.
We note, however, that \citet{Contreras2017a} found that \Planck\
polarization data are unlikely to be able to distinguish between
dipole-modulated skies and skies consistent with \LCDM\ (although this
statement is somewhat model dependent). Furthermore, the analysis we
carry out here is a purely phenomenological one performed in multipole
space, with no attempt to connect to any real-space modulation;
however, several physical $k$-space models are considered in a companion
paper \citep{planck2016-l10}.

We employ a version of the estimator proposed in
\citetalias{planck2014-a18}, which was further developed in \citet{Zibin2015}
and \citet{Contreras2017a}.  In particular, we use
\begin{linenomath*}
\begin{eqnarray}
 \tilde{X}^{WZ}_0 &=& \frac{6\sum_{\ell m} \delta C^{WZ}_{\ell
 \ell+1}A_{\ell m} S^{(WZ)}_{\ell m\,\ell+1\, m+M}}
 {\sum_{\ell}\left(\delta C^{WZ}_{\ell \ell +1}\right)^2(\ell +
 1)F^{(W}_{\ell}F^{Z)}_{\ell+1}},
 \label{eq:est0}\\
 \tilde{X}^{WZ}_{+1} &=& \frac{6\sum_{\ell m} \delta C^{WZ}_{\ell
 \ell+1}B_{\ell m} S^{(WZ)}_{\ell m\,\ell+1\, m+M}}
 {\sum_{\ell}\left(\delta C^{WZ}_{\ell \ell +1}\right)^2(\ell +
 1)F^{(W}_{\ell}F^{Z)}_{\ell+1}},
\label{eq:est1}
\end{eqnarray}
\end{linenomath*}
with
\begin{linenomath*}
\begin{equation}
  S^{WZ}_{\ell m\ell'm'} \equiv W^*_{\ell m} Z_{\ell'm'} - \left<W^*_{\ell
  m} Z_{\ell'm'}\right>.
  \label{eq:transformed_mfc}
\end{equation}
\end{linenomath*}
Here the $\delta C_{\ell \ell'}$ are determined by the model of
modulation; for this case we have assumed scale-invariant power modulation
out to some $\ell_\mathrm{max}$ and
thus $\delta C_{\ell \ell'} = 2(C_{\ell} + C_{\ell'})$.  The remaining terms are:
$\tilde{X}_M$, the spherical harmonic transform of the modulation
amplitude and direction; $WZ =$ $TT$, $TE$, or $EE$; $W_{\ell m}$ and $Z_{\ell
m}$, which are inverse-covariance filtered data; $F^{W}_\ell \approx
\left< W_{\ell m} W^*_{\ell m} \right>$; and the last term on the
right-hand-side of
Eq.~\eqref{eq:transformed_mfc} denotes the mean-field correction (details,
including the precise form of $F^{W}_\ell$ can be found in appendix~A.1 of
\citealt{planck2014-a17} and appendix~B of \citealt{Contreras2017a}).
The parentheses in the superscripts indicate symmetrization over the
enclosed variables. The coupling coefficients are given by
\begin{linenomath*}
\begin{eqnarray}
 A_{\ell m} &=& \sqrt{\frac{(\ell+1)^2 - m^2}{(2\ell+1)(2\ell+3)}}, \\
 B_{\ell m} &=& \sqrt{\frac{(\ell+m+1)(\ell+m+2)}{2(2\ell+1)(2\ell+3)}}.
\end{eqnarray}
\end{linenomath*}
The $\tilde{X}_M$ can be transformed into amplitudes of modulation along
each Cartesian axis, since they are simply the spherical harmonic
decomposition of a dipole.  Likewise we can write the modulation in terms
of its spherical coordinates (amplitude and direction) as
\begin{linenomath*}
\begin{align}
 \tilde{A} &= \sqrt{\Delta \tilde{X}^2_0 + 2|\Delta \tilde{X}_1|^2},
\label{eq:amplitude} \\
 \tilde{\theta} &= \cos^{-1}{\left(\frac{\Delta \tilde{X}_0}{\tilde{A}}\right)},
\label{eq:thetadir} \\
 \tilde{\phi} &= -\tan^{-1}{\left(\frac{\mathrm{Im}{[\Delta \tilde{X}_1]}}{\mathrm{Re}{[
\Delta \tilde{X}_1]}}\right)}.
\label{eq:phidir}
\end{align}
\end{linenomath*}
Note that for $E$-mode polarization, these coupling coefficients
neglect corrections on the very largest scales ($\ell \lesssim 10$)
due to the non-local definition of $E$-modes. This gives rise to a
slightly different coupling induced in $E$ as compared to
$T$ by a dipole modulation of the primordial fluctuations
\citep{Contreras2017a}. This has little effect on the comparison of the data
with simulations, however, since the simulations are treated in the same way. A
potentially more significant effect is a mismatch in
the optical depth between data and simulations; below we estimate this to have
essentially negligible influence.

The estimators of Eqs.~\eqref{eq:est0} to \eqref{eq:est1} can be
combined with inverse-variance weighting over all data combinations
($TT$, $TE$, $EE$) to obtain a combined minimum-variance estimator,
given by
\begin{linenomath*}
\begin{equation}
\Delta \tilde{X}_M =
  \frac{\sum_{WZ} \Delta\tilde{X}^{WZ}_M\left(\sigma_X^{WZ}\right)^{-2}}
       {\sum_{WZ}\left(\sigma_X^{WZ}\right)^{-2}}.
  \label{eq:fullest}
\end{equation}
\end{linenomath*}
We calculate the variance from the scatter of simulations,
although they agree closely with the Fisher errors given in
\citetalias{planck2014-a18}.

Here we test for an $\ell$-space asymmetry in temperature
and polarization. The model considered is a scale-invariant modulation
from $\ell_{\rm min}\,{=}\,2$ out to a variable maximum multipole,
$\ell_{\rm max}$.  It is important to stress that, when we combine
temperature and polarization, the phenomenological, $\ell$-space
approach we adopt here is very simplistic and does {\it not\/} capture the
behaviour of the modulated fluctuations arising from the different
$k$--$\ell$ kernels for $T$ and $E$ modes \citep{Contreras2017a}.  For
example, modulated fluctuations that exhibit a 7\,\% dipolar asymmetry
in $T$ to $\ell\,{\approx}\,65$ are not expected to produce an $E$-mode
asymmetry of the same amplitude and over the same scales.  Therefore,
strictly speaking, the ``model'' being tested when we combine $TT$,
$TE$, and $EE$ has no physical basis; a more physical, $k$-space approach
is considered in the companion paper \citet{planck2016-l10}.

\begin{figure*}[htbp!]
\begin{center}
\mbox{\includegraphics[width=0.495\hsize]{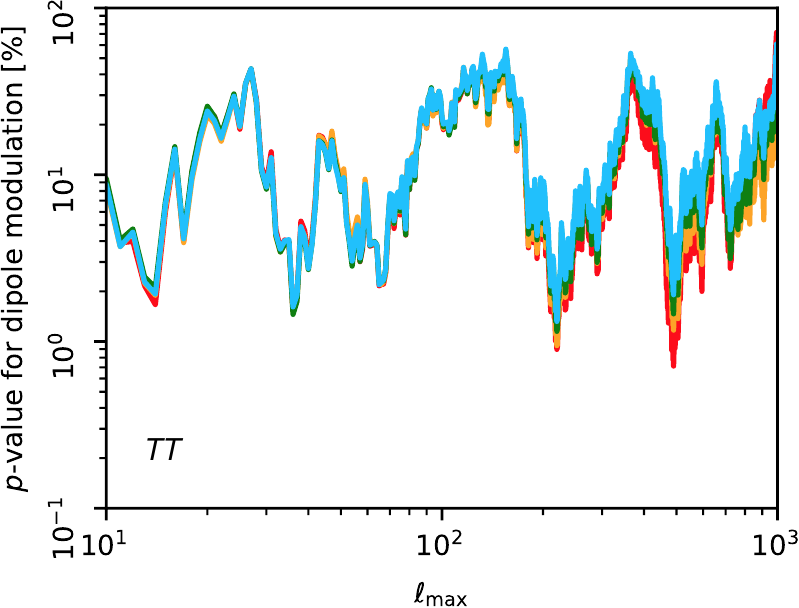}}
\mbox{\includegraphics[width=0.495\hsize]{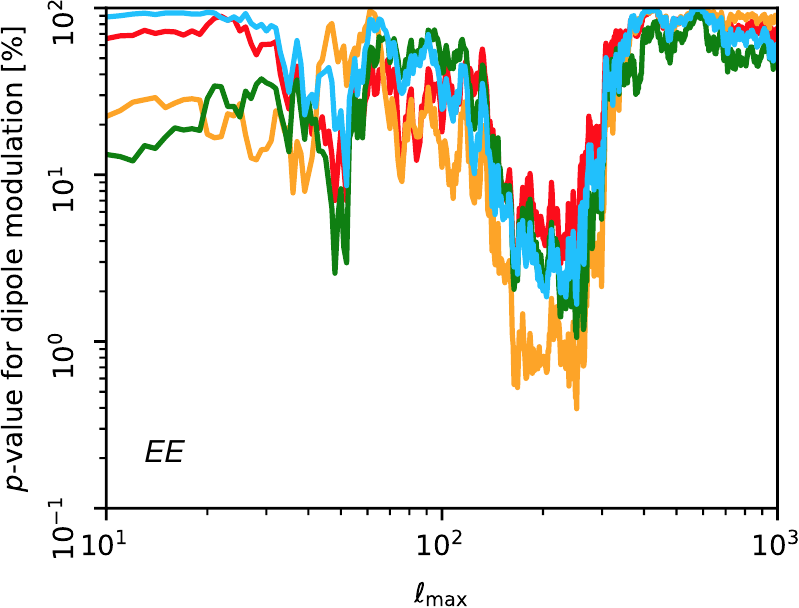}}
\mbox{\includegraphics[width=0.495\hsize]{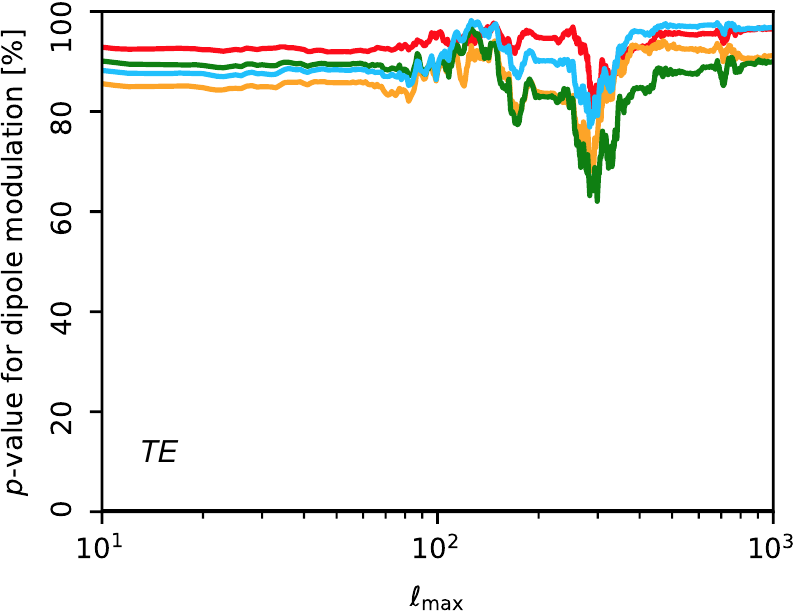}}
\mbox{\includegraphics[width=0.495\hsize]{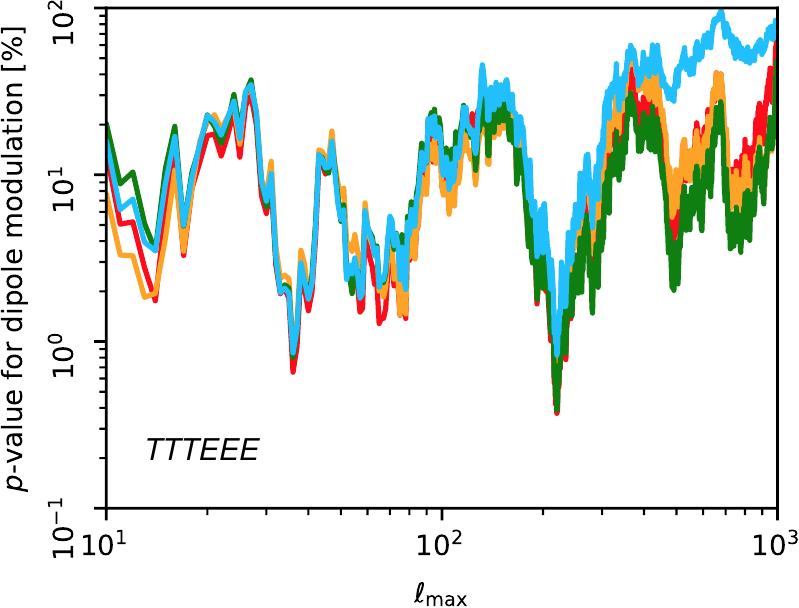}}
\end{center}
\caption{Probability determined from the QML analysis for a Monte Carlo
simulation to have a larger dipole-modulation amplitude than the \commander\
(red), \nilc\ (orange), \sevem\ (green), and \smica\ (blue) data sets, with
$\ell_{\min} = 2$. The top-left, top-right, and bottom-left panels use $TT$,
$EE$, and $TE$, while the final panel is for all data combined.
Small $p$-values here correspond to anomalously large dipole-modulation
amplitudes.
We emphasize that the statistic here is cumulative and apparent trends in the
curves can be misleading.}
\label{fig:pvaluesdipmodTTEE}
\end{figure*}

In Fig.~\ref{fig:pvaluesdipmodTTEE} we show the {\it p}-values of the
full-mission data compared to the FFP10 simulations, considering $TT$
(top left), $EE$ (top right), or $TE$ (bottom left) data
independently. While the $TT$ results are largely consistent with our
previous analysis \citepalias[differences are within the
expected scatter given the different analysis
masks,][]{planck2014-a18}, our $EE$ and $TE$ results are new. The
polarization-only results show mildly significant asymmetry to
$\ell_{\rm max}\,{\approx}\,250$, and are featureless elsewhere.
This is minimally affected by the mismatch in $\tau$ between the data
and simulations. The effect of excluding the $\ell \lesssim 10$ data (where
$\tau$ is most relevant) from the analysis modifies the amplitude estimation by
of order 10\,\% for scales $\ell \gtrsim 50$. Thus a 10\,\% mismatch
in $\tau$ would (at most) correspond to a 1\,\% error in the amplitude,
which decreases at smaller scales. This error is negligible compared
to the noise contribution on these scales.
The temperature-polarization cross-correlation results are rather
featureless in the full range of $\ell_{\rm max}$ considered.  In the
$\ell_{\rm max}\,{\approx}\,250$ region \nilc\ shows systematically lower
{\it p}-values compared to the other methods, although the modulation
amplitudes are still statistically consistent. This
difference could be attributable to a percent level {\it p}-value
excess of asymmetry in the OEHD data observed in the same
region; however, such an excess does not appear in the HMHD data. No
other combinations of OEHD and HMHD data show any significant excess
of asymmetry.

In the bottom right panel of Fig.~\ref{fig:pvaluesdipmodTTEE} we combine
temperature and polarization (including $TE$) data and show the {\it p}-values
of the data compared to simulations. Only the {\it p}-value dip at $\ell_{\rm
max}\,{\approx}\,250$ falls lower than the corresponding dip in temperature alone.
Moreover, it is important to stress that the combined significance increasing
at that scale is not in itself evidence that the asymmetry has a genuine,
physical origin.  As pointed out in \citet{Contreras2017a}, the combination of
a statistically isotropic polarization signal with temperature data will,
simply due to Gaussian statistics, spuriously increase the significance of a
$\,{\approx}\,3\,\sigma$ temperature signal with 30\,\% probability for \Planck.
Furthermore, our phenomenological model does not properly combine
temperature and polarization in 3D $k$-space, as previously
mentioned~\citep[also see][and references therein]{Zibin2015}.

\begin{table}[htbp!]
\begingroup
\newdimen\tblskip \tblskip=5pt
\caption{Amplitude and direction of the low-$\ell$ dipole-asymmetry
  signal determined from the QML analysis for the range
  $\ell\,{=}\,2$--64. The errors
  are calculated from the FFP10 simulations. Recall that the expectation
  for noise-free Gaussian skies (the ``cosmic variance of dipole modulation,''
  see equation~50 in \citetalias{planck2014-a18}) corresponds to an
  amplitude of approximately 3\,\%.  The polarization data on their own show
  no evidence of modulation, and the addition of polarization has
  very little effect on the temperature asymmetry signal.
}
\label{tab:lowellmod}
\nointerlineskip
\vskip -3mm
\footnotesize
\setbox\tablebox=\vbox{
   \newdimen\digitwidth
   \setbox0=\hbox{\rm 0}
   \digitwidth=\wd0
   \catcode`*=\active
   \def*{\kern\digitwidth}
   \newdimen\signwidth
   \setbox0=\hbox{+}
   \signwidth=\wd0
   \catcode`!=\active
   \def!{\kern\signwidth}
\halign{ \hbox to 1.0in{#\leaderfil}\tabskip 4pt&
         \hfil#\hfil\tabskip 8pt&
         \hfil#\hfil\tabskip 0pt\cr                           
\noalign{\doubleline}
\omit& \omit& Direction\cr   
\omit\hfil Data\hfil&Amplitude& $(l, b)$ [Deg]\cr   
\noalign{\vskip 4pt\hrule\vskip 6pt}
\multispan3\hfil 2015 $TT$\hfil\cr
\noalign{\vskip 4pt}
{\tt \commander}     & $0.063^{+0.025}_{-0.013}$ & $(213, -26)\pm28$\cr
\noalign{\vskip 3pt}
{\tt \nilc}          & $0.064^{+0.027}_{-0.013}$ & $(209, -25)\pm28$\cr
\noalign{\vskip 3pt}
{\tt \sevem}         & $0.063^{+0.026}_{-0.013}$ & $(211, -25)\pm28$\cr
\noalign{\vskip 3pt}
{\tt \smica}         & $0.062^{+0.026}_{-0.013}$ & $(213, -26)\pm28$\cr
\noalign{\vskip 4pt\hrule\vskip 6pt}
\multispan3\hfil 2018 $TT$\hfil\cr
\noalign{\vskip 4pt}
{\tt \commander}     & $0.070^{+0.032}_{-0.015}$ & $(221, -22)\pm31$\cr
\noalign{\vskip 3pt}
{\tt \nilc}          & $0.069^{+0.032}_{-0.015}$ & $(221, -24)\pm31$\cr
\noalign{\vskip 3pt}
{\tt \sevem}         & $0.070^{+0.032}_{-0.015}$ & $(221, -22)\pm31$\cr
\noalign{\vskip 3pt}
{\tt \smica}         & $0.070^{+0.032}_{-0.015}$ & $(221, -22)\pm31$\cr
\noalign{\vskip 4pt\hrule\vskip 6pt}
\multispan3\hfil 2018 $EE$\hfil\cr
\noalign{\vskip 4pt}
{\tt \commander}     & $0.137^{+0.863}_{-0.422}$ & $(192, !3)\pm103$\cr
\noalign{\vskip 3pt}
{\tt \nilc}          & $0.105^{+0.900}_{-0.459}$ & $(267, -5)\pm105$\cr
\noalign{\vskip 3pt}
{\tt \sevem}         & $0.110^{+0.891}_{-0.469}$ & $(245, \,37)\pm105$\cr
\noalign{\vskip 3pt}
{\tt \smica}         & $0.066^{+0.934}_{-0.501}$ & $(215, \,11)\pm108$\cr
\noalign{\vskip 4pt\hrule\vskip 6pt}
\multispan3\hfil 2018 $TT$,$EE$\hfil\cr
\noalign{\vskip 4pt}
{\tt \commander}     & $0.072^{+0.031}_{-0.015}$ & $(217, -18)\pm29$\cr
\noalign{\vskip 3pt}
{\tt \nilc}          & $0.069^{+0.032}_{-0.015}$ & $(225, -23)\pm31$\cr
\noalign{\vskip 3pt}
{\tt \sevem}         & $0.069^{+0.032}_{-0.015}$ & $(219, -17)\pm31$\cr
\noalign{\vskip 3pt}
{\tt \smica}         & $0.068^{+0.031}_{-0.015}$ & $(221, -19)\pm31$\cr
\noalign{\vskip 4pt\hrule\vskip 6pt}
\multispan3\hfil 2018 $TT$,$TE$,$EE$\hfil\cr
\noalign{\vskip 4pt}
{\tt \commander}     & $0.072^{+0.031}_{-0.015}$ & $(218, -19)\pm29$\cr
\noalign{\vskip 3pt}
{\tt \nilc}          & $0.069^{+0.032}_{-0.015}$ & $(225, -23)\pm31$\cr
\noalign{\vskip 3pt}
{\tt \sevem}         & $0.069^{+0.032}_{-0.015}$ & $(219, -17)\pm31$\cr
\noalign{\vskip 3pt}
{\tt \smica}         & $0.068^{+0.032}_{-0.015}$ & $(221, -19)\pm31$\cr
\noalign{\vskip 3pt\hrule\vskip 4pt}}}
\endPlancktable                    
\endgroup
\end{table}

\begin{table}[htbp!]
\begingroup
\newdimen\tblskip \tblskip=5pt
\caption{Amplitude and direction of the low-$\ell$ dipole
  modulation signal determined from the QML analysis for the range
  $\ell=2$--220. The errors
  are calculated from the FFP10 simulations. This $\ell$ range corresponds to
  the lowest {\it p}-value for the $TT$, $TE$, $EE$ data.}
\label{tab:midellmod}
\nointerlineskip
\vskip -3mm
\footnotesize
\setbox\tablebox=\vbox{
   \newdimen\digitwidth
   \setbox0=\hbox{\rm 0}
   \digitwidth=\wd0
   \catcode`*=\active
   \def*{\kern\digitwidth}
   \newdimen\signwidth
   \setbox0=\hbox{+}
   \signwidth=\wd0
   \catcode`!=\active
   \def!{\kern\signwidth}
\halign{ \hbox to 1.0in{#\leaderfil}\tabskip 4pt&
         \hfil#\hfil\tabskip 8pt&
         \hfil#\hfil\tabskip 0pt\cr                           
\noalign{\doubleline}
\omit& \omit& Direction\cr   
\omit\hfil Data\hfil&Amplitude& $(l, b)$ [Deg]\cr   
\noalign{\vskip 4pt\hrule\vskip 6pt}
{\tt \commander}& $0.023^{+0.008}_{-0.004}$& $(220, -5)\pm25$\cr
\noalign{\vskip 3pt}
{\tt \nilc}     & $0.022^{+0.008}_{-0.004}$& $(228, -2)\pm25$\cr
\noalign{\vskip 3pt}
{\tt \sevem}    & $0.023^{+0.008}_{-0.004}$& $(224, -5)\pm25$\cr
\noalign{\vskip 3pt}
{\tt \smica}    & $0.022^{+0.008}_{-0.004}$& $(226, -2)\pm26$\cr
\noalign{\vskip 3pt\hrule\vskip 4pt}}}
\endPlancktable                    
\endgroup
\end{table}

In Table~\ref{tab:lowellmod} we show the measured amplitude and
direction of dipolar modulation in the oft-quoted $\ell\,{=}\,2$--64
range, with and without polarization data. Polarization clearly has
very little effect on the modulation parameters in this region, and
$TE$ in particular has a nearly negligible effect at this scale.
It is worth recalling that the expectation for purely Gaussian skies
is a dipole-modulation amplitude of approximately 3\,\% (see equation~50 in
\citetalias{planck2014-a18}).
In Table~\ref{tab:midellmod} we show the amplitude and direction of
dipolar modulation for the combined $TT$, $TE$, and $EE$ data in the
$\ell\,{=}\,2$--220 range, which is the range with the lowest
{\it p}-value. The fact that the amplitude for this $\ell$-range is smaller
than that for $\ell\,{=}\,2$--64 is consistent with the
prediction for statistically isotropic skies \citepalias[as noted
in][the amplitude typical of modulation should decrease as
$1/\ell_{\rm max}$]{planck2014-a18}. The proximity of the directions observed for the
two scales, $\ell\,{\leq}\,64$ and $\ell\,{\leq}\,220$, was also noted in
\citetalias{planck2014-a18}, where tests of directionality were performed. There
we showed that the two directions are correlated at a slightly higher level than
seen in simulations, but that this can be traced to the low-$\ell$ anomaly
on the larger scales where the dipole-modulation amplitude is larger.
Removing angular scales $\ell < 100$ eliminates the significance of
the modulation entirely.

In \citetalias{planck2014-a18} we applied a look-elsewhere correction
to the corresponding results and demonstrated that if we allow the
range of $\ell_{\rm max}$ to vary (as we have done here) then small {\it
  p}-values (of order 0.1--1\,\%) occur of order 10\,\% of the
time. Repeating that analysis here would yield a similar result. We
have found that, as expected, the \Planck\ polarization data have not
been able to refute or confirm the original signal found in
temperature.  The polarization data alone also appear to be consistent
with statistical isotropy. Better polarization data are required to
further test whether there might be a physical origin for the original
temperature-modulation signature.

\subsection{Angular clustering of the power distribution}
\label{sec:powerasymmetry}

Despite the lack of evidence for any strong anomaly in the amplitude
of dipole modulation discussed in the previous sub-sections, 
in \citetalias{planck2014-a18}, we confirmed the apparent
presence of a deviation from statistical isotropy in the \Planck\ data
using an angular-clustering analysis,
as previously seen in \citetalias{planck2013-p09}.
In particular, some alignment of preferred
directions determined from maps of the temperature power distribution
on the sky was observed over a wide range of angular scales.

Specifically, by calculating the power spectrum locally in patches, for
various multipole ranges, and then fitting dipoles to maps of these
band-power estimates,
it was found that the directions were aligned at the 2--$3\,\sigma$ level up
to $\ell\,{=}\,1500$ when compared to simulations.  In the standard
cosmological model, although such maps are expected to exhibit dipolar
distributions of power due to Gaussian random fluctuations, the
associated directions should be independent random variables.
Evidence for the close correlation and alignment of directions on
different angular scales then appears to be a signature of broken statistical
isotropy.
Since we do not observe a significant amplitude of dipole
modulation over similar angular scales, then the result of finding clustering
of directions seems mysterious---it is hard to imagine a concrete (e.g.,
inflationary) model that would
cluster the directions without affecting the amplitude of the modulation
\citep[but see][]{hansen2018}.
Nevertheless, regarding this as a purely empirical
question, it is important to repeat the directional-clustering analysis
and broaden its scope to include polarization.

\begin{figure}[htbp!]
\centering
\includegraphics[scale=0.25]{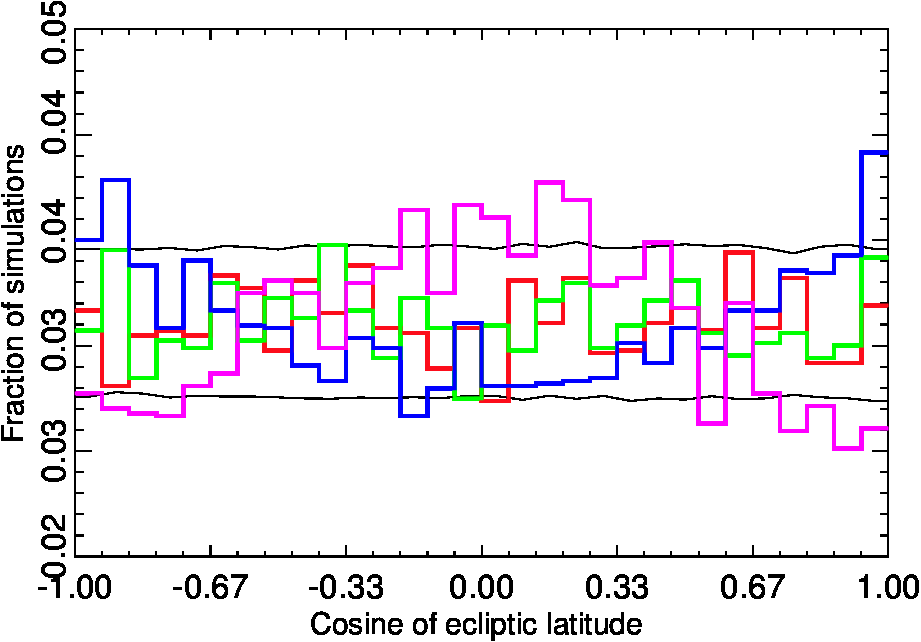}
\caption[Binned power distribution]{Distribution of directions for power
  dipoles estimated from
  independent 100-multipole bins for simulated HM \commander\ maps
  with the common mask applied.
  The fraction of dipoles pointing
  towards a given ecliptic latitude direction is shown, where the data
  are binned by the cosine of the direction, since it is this quantity
  that is uniformly distributed on the sky for a Gaussian random
  field.  The horizontal black lines show the expected $95\,\%$
  deviation taken from a uniform distribution. The coloured lines show
  the distributions for temperature at $\ell\,{<}\,800$ (red line), temperature
  at $\ell\,{>}\,800$ (green line), $EE$ polarization at $\ell\,{<}\,800$
  (blue line), and $EE$ polarization at $\ell\,{>}\,800$ (magenta line).  The
  temperature results are determined using uniform weighting, while the $EE$
  results have inverse-variance weighting applied.}
\label{fig:uniformity}
\end{figure}


Here, we repeat the analysis using the \Planck\ 2018 data and extend
it to include polarization measurements. The local power spectra,
$TT$, $TE$, and $EE$, are estimated directly from maps of the
temperature and Stokes $Q$ and $U$ parameters. Since the analysis is
sensitive to differences between the noise properties of the data and
simulations, we consider only cross-spectra computed between the two
HM or OE maps for each component-separation method.

We adopt the same approach for the estimation of the dipole alignment
as described in detail in \citetalias{planck2014-a18}, a brief summary
of which follows.
\begin{enumerate}
\item Local $TT$, $TE$ and $EE$ power spectra are estimated \citep[using the
  {\tt MASTER} approach,][]{hivon2002} directly
  from the $T$, $Q$, and $U$ maps at $N_\mathrm{side}\,{=}\,2048$ for the 12
  patches of the sky corresponding to the $N_\mathrm{side}\,{=}\,1$
  {\tt HEALPix} base pixels. Leakage between the $E$ and $B$ modes due to
  incomplete sky coverage is removed in the power-spectrum estimation during
  the inversion of the full $TEB$ kernel. The same results would be expected
  using the $E$ maps, but the extended masks required for these maps would
  increase the uncertainty of the results.  The spectra are binned over
  various (even) bin sizes between $\Delta\ell\,{=}\,8$ and
  $\Delta\ell\,{=}\,32$.\footnote{In order to maintain
  consistency with previous Planck Collaboration papers, some of the figures
  will show results using 100-multipole bins.  These 100-$\ell$ bins were
  constructed from the 16-$\ell$ bins using the prescription described in
  \citetalias{planck2013-p09} and \citet{hansen2009}.
  Although this means that there are not exactly 100 multipoles in each bin,
  they will nevertheless be referred to as ``100-$\ell$ bins.'' Note that
  these bins are used for illustrative purposes in the figures only, while the
  analysis always uses smaller bins.}

\item For each power-spectrum multipole bin, an $N_\mathrm{side}\,{=}\,1$
  {\tt HEALPix} map with the local power distribution is constructed.
\item The best-fit dipole amplitude and direction is estimated from
  this map using inverse-variance weighting, where the variance is
  determined from the local spectra computed from the simulations. As
  in previous papers 
  \citepalias{planck2013-p09,planck2014-a18},
  the fitted dipole amplitudes are found
  to be consistent with those determined from simulations.  
\item A measure of the alignment of the different multipole blocks is
  then constructed. We use the mean of the cosine of the angles
  between all pairs of dipoles. This is essentially the Rayleigh
  statistic (RS) and we will refer to it as such, although it differs
  by not including any amplitude information. Smaller values
  of the RS correspond to less clustering. 
\item The clustering as a function of $\ell_\mathrm{max}$ is then
  assessed using {\it p}-values determined as follows.  We first
  construct the RS using all multipoles up to $\ell_\mathrm{max}$. The
  {\it p}-value is then given by the fraction of simulations with a
  higher RS than for the data for this $\ell_\mathrm{max}$. A small
  {\it p}-value therefore means that there are few simulations that
  exhibit as strong clustering as the data. Note that the {\it
    p}-values are highly correlated because the RS as a function of
  $\ell_\mathrm{max}$ is cumulative.
\item After calculating results for each bin size, we calculate the
  variance-weighted mean of the power spectra over all bin sizes (the
  $C_\ell$ for a given bin size is weighted by $1/\sqrt{N_{\rm b}}$ where
  $N_{\rm b}$ is the bin size). In this way, we marginalize over bin sizes
  to obtain local power spectra and thereby RS values for each single
  multipole.
\end{enumerate}

Since the test is based on the angular correlation between power
dipoles (dipoles of the angular distribution of power as described
above) and not on the absolute direction, the results should be
unaffected by any preferred directions in the data caused by its noise
properties, as long as these are matched by the noise properties of
the simulations.  Nevertheless, we have tested the uniformity of the
estimated dipole directions in the simulated data. We find
that the $EE$ polarization directions are strongly influenced by the
high noise level, with a strong bias towards the ecliptic plane. In
order to reduce the noise bias on direction, we weight the $Q$ and $U$
maps by their inverse noise variance before estimating the local
spectra.
Figure~\ref{fig:uniformity} shows the histogram of the ecliptic
latitudes of the power dipoles determined from simulated maps
constructed in 100-multipole bins. The horizontal black lines indicate
the $95\,\%$ confidence interval obtained from uniformly distributed
directions. The coloured lines show the distribution of directions
from temperature and $EE$ polarization dipoles estimated on small and
large scales, using inverse-variance weighting in the latter case.
Deviation from a uniform distribution is seen for the polarization
directions, most notably for the small-scale $EE$ results (magenta
line).
The ecliptic longitude values fall consistently within the $95\,\%$
confidence interval determined from uniformly distributed directions
for both temperature and polarization, although some modulation is
seen.

\begin{figure*}[htbp!]
\centering
\includegraphics[scale=0.1]{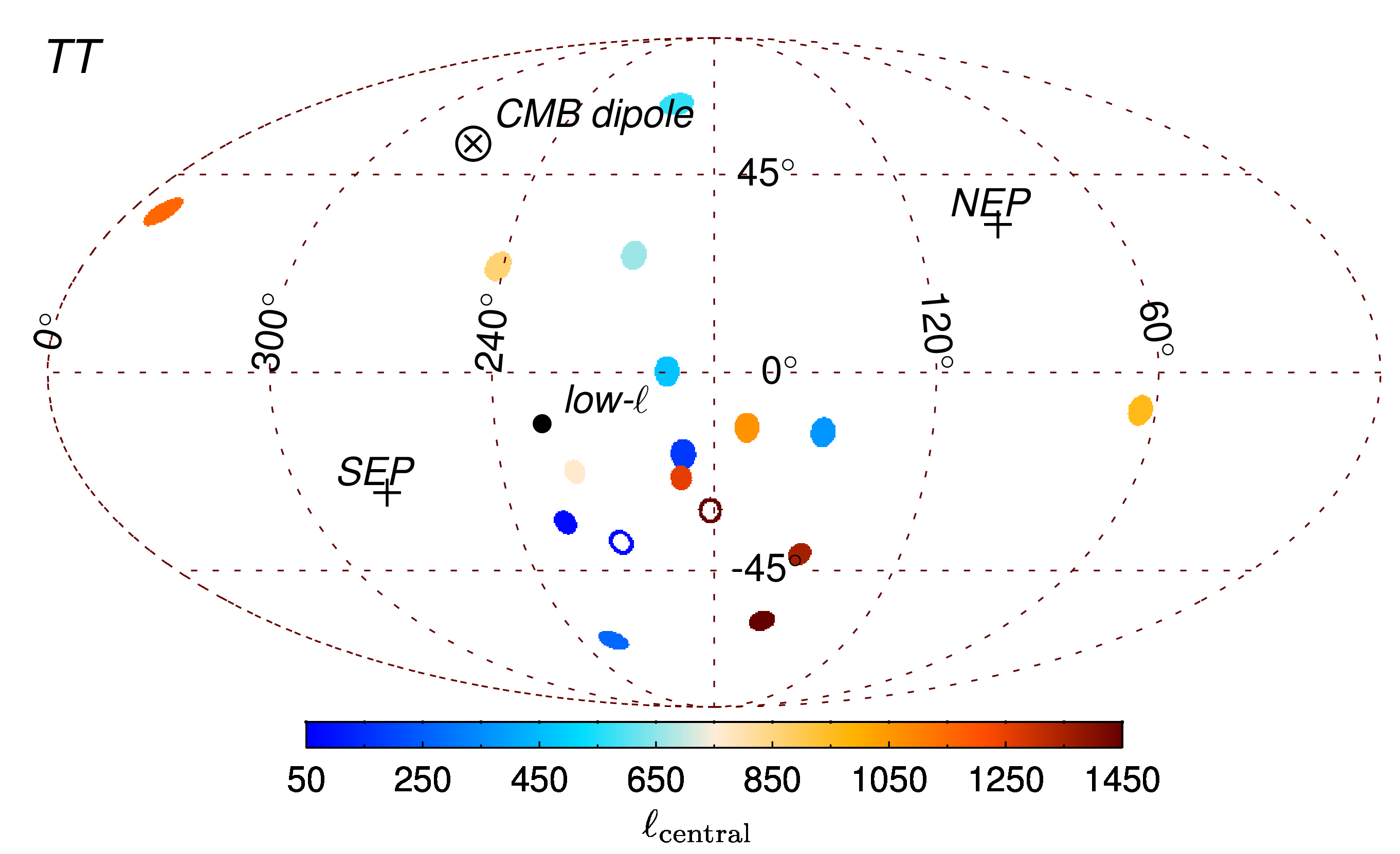}
\includegraphics[scale=0.25]{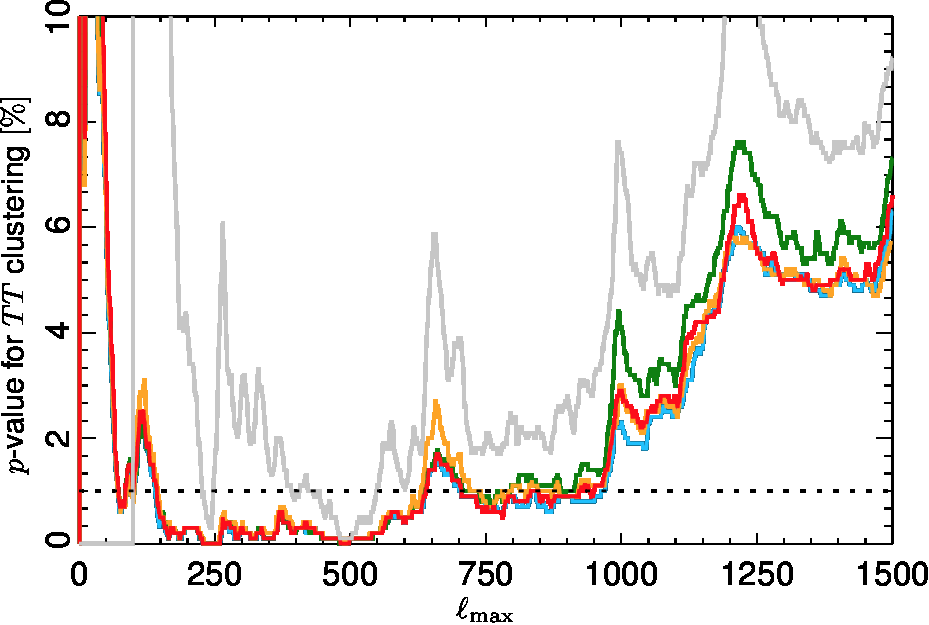} \\
\includegraphics[scale=0.1]{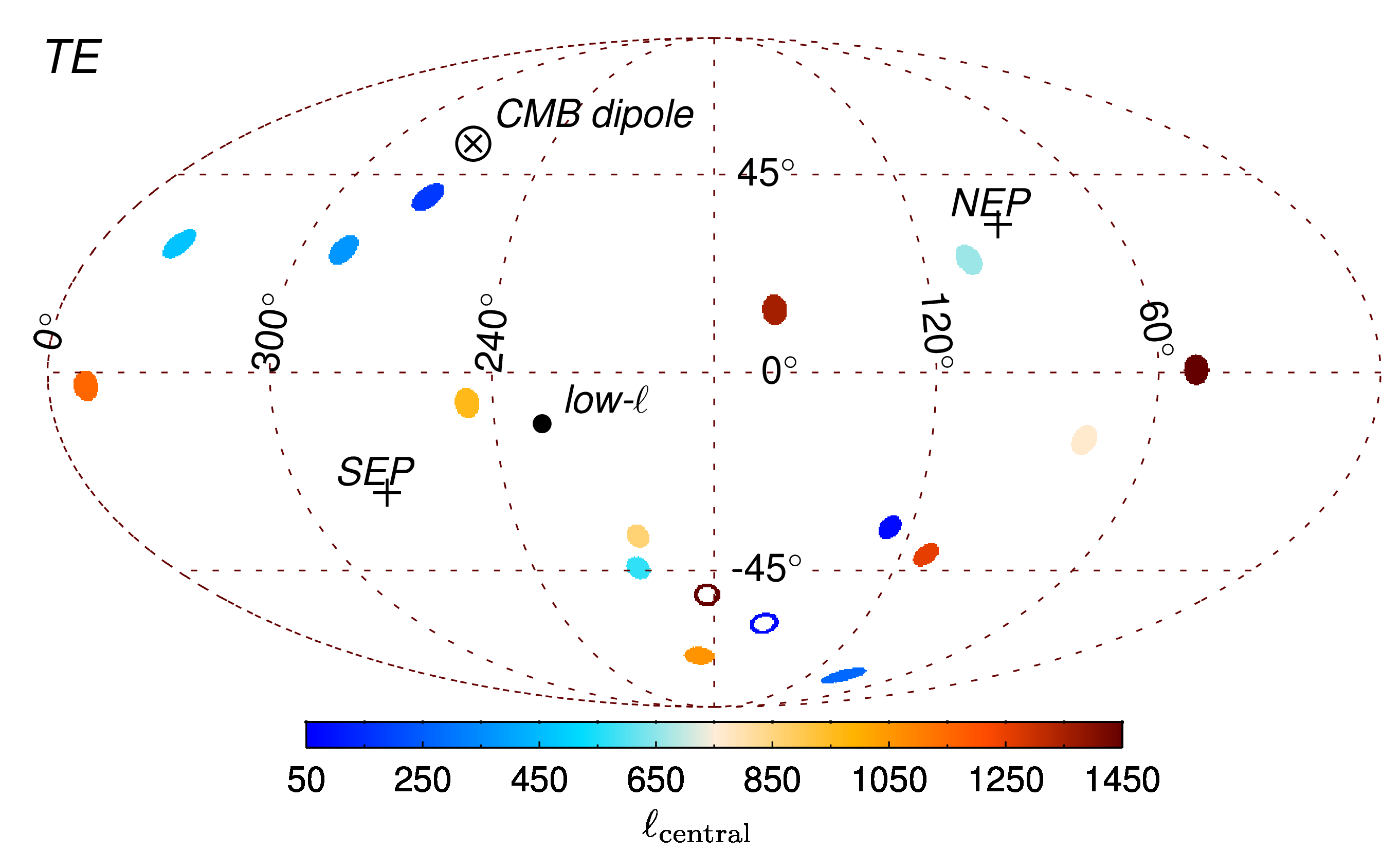}
\includegraphics[scale=0.25]{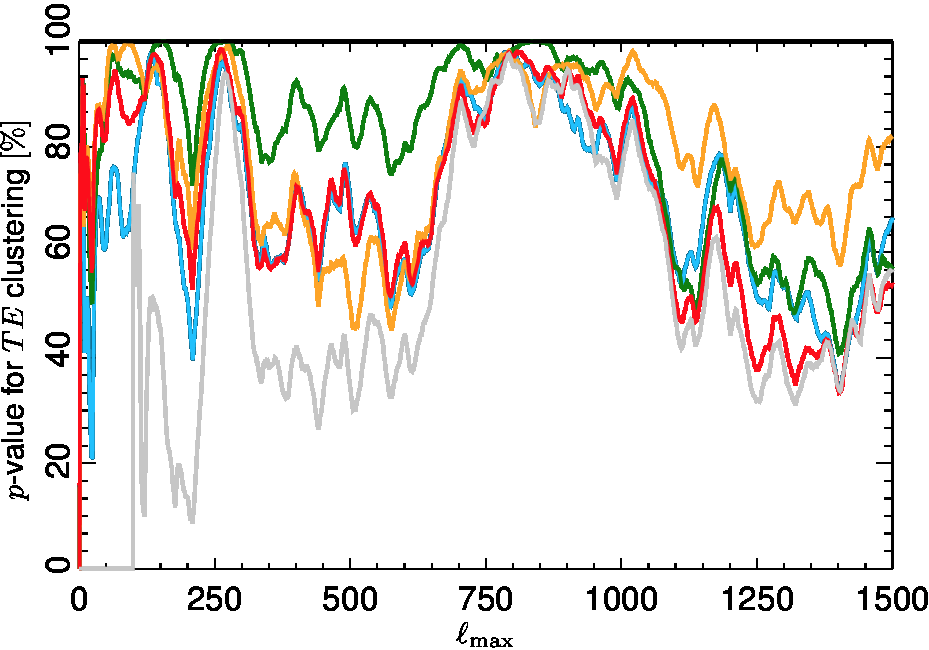} \\
\includegraphics[scale=0.1]{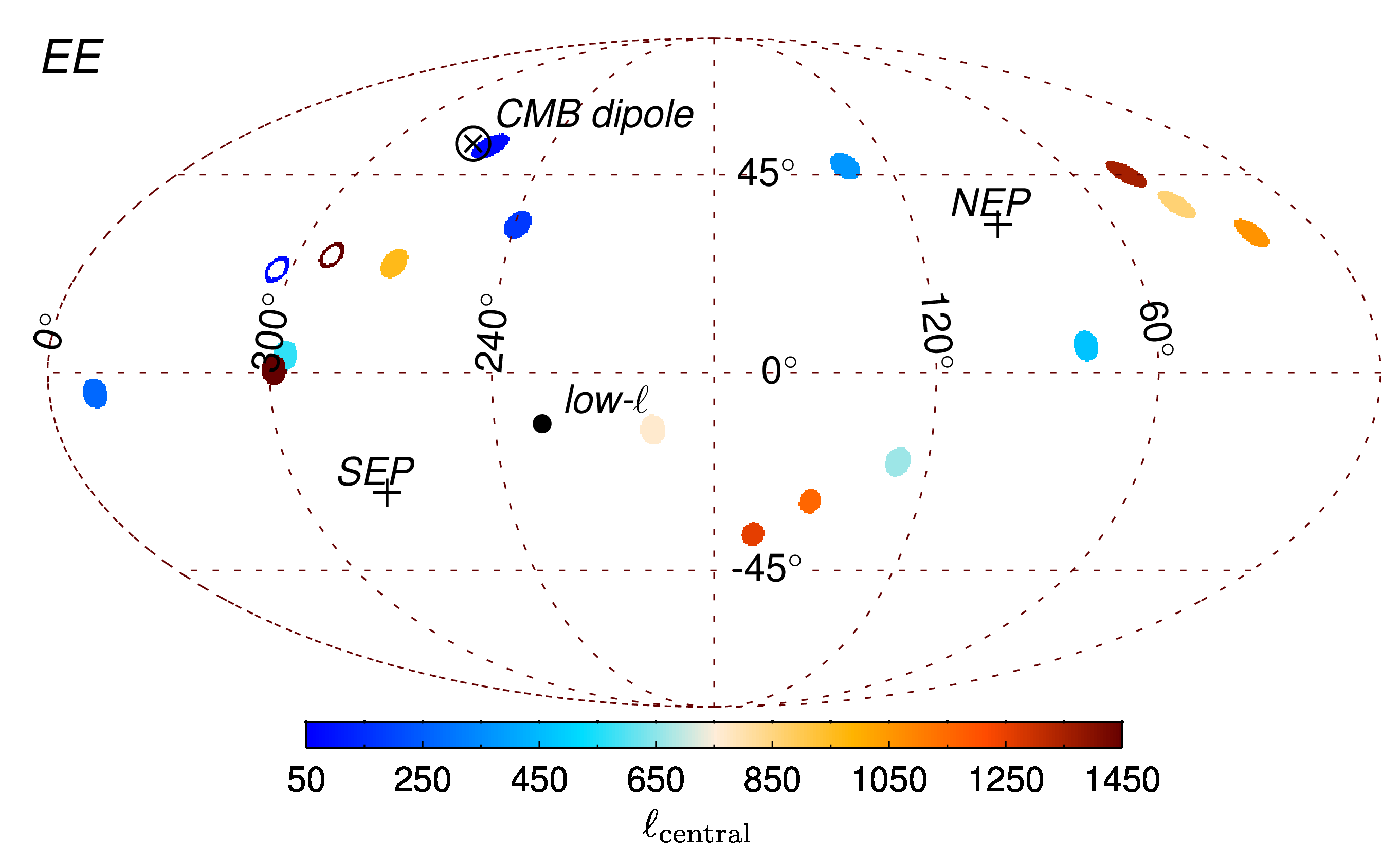}
\includegraphics[scale=0.25]{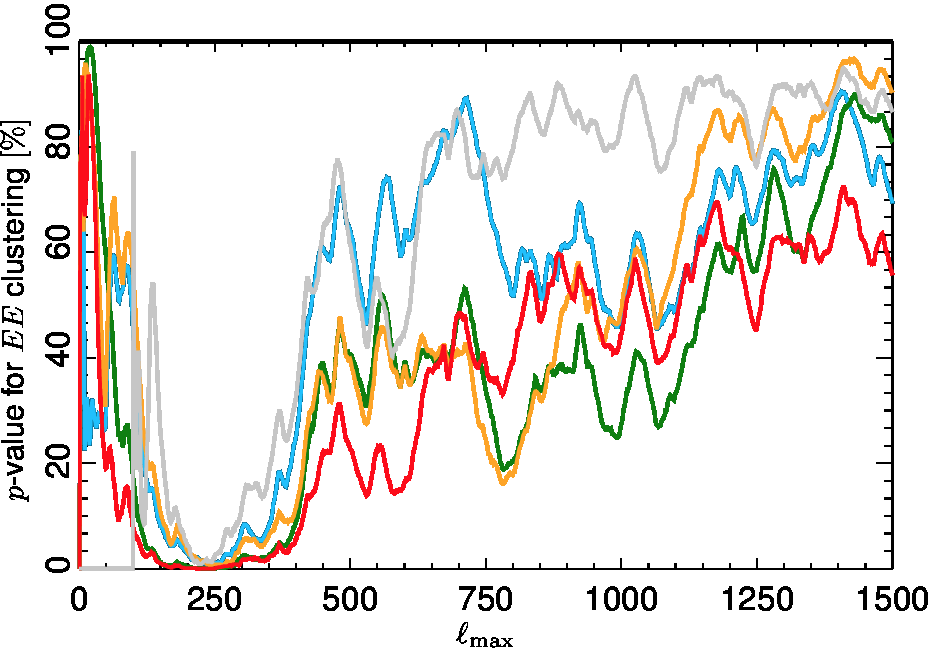} \\
\caption{
{\it Left panels}: Dipole directions for independent 100-multipole bins of the
  local power-spectrum distribution from $\ell=2$ to $1500$ in the \commander\
  OE maps, with the OE common mask applied.
  The preferred directions for maps of specific multipole bins are
  colour coded according to the central value of the bin, $\ell_{\rm central}$,
  as shown in the colour bar.
  Note that the maps have been
  rotated about the Galactic north-south axis, so that Galactic longitude
  $l=180\deg$ is in the centre of the map. The top panel shows the directions
  for the $TT$ power spectrum, the middle panel for the $TE$ power spectrum
  and the bottom panel for the $EE$ power spectrum.  
  In all panels, we also show the CMB dipole direction, the north
  ecliptic pole (NEP), the south ecliptic pole (SEP), and the
  preferred dipolar modulation axis (labelled as ``low-$\ell$'')
  derived from the temperature data in Sect.~\ref{sec:dipmod_qml}. 
  The average directions determined
  from the two multipole ranges $\ell=2$--300 and $\ell=750$--1500
  are shown as blue and dark red (open) rings, respectively.
  The mean error on the derived directions that results from masking
  the data is $44\deg$ in the range $\ell\,{<}\,800$ and $62\deg$ in the
  range $\ell\,{>}\,800$ for temperature, but $78\deg$ in the range
  $\ell\,{<}\,800$ and $91\deg$ in the range $\ell\,{>}\,800$ for $EE$
  polarization.
   {\it Right panels}: Derived {\it p}-values for the angular
  clustering of the power distribution in OE maps as a function of
  $\ell_\mathrm{max}$, determined for \commander\ (red), \nilc\
  (orange), \sevem\ (green), and \smica\ (blue), based on FFP10
  simulations using the OE common mask.  These are the same colours
  used throughout the paper (e.g., see Fig.~\ref{Variance_S2Nratio}),
  while the grey line shows the
  \commander\ results when excluding the first 100 multipoles in the
  analysis. These {\it p}-values are based on the fraction of
  simulations with a higher RS, determined over the $\ell$ range up to
  the given $\ell_\mathrm{max}$, compared to the data,
  hence small $p$-values would correspond to anomalously aligned
  dipole directions.
  The results shown here have been marginalized over bin sizes in the range
  $\Delta\ell\,{=}\,8$ to $\Delta\ell\,{=}\,32$.  
}
\label{fig:asymdir_oe}
\end{figure*}

\begin{figure*}[htbp!]
\centering
\includegraphics[scale=0.1]{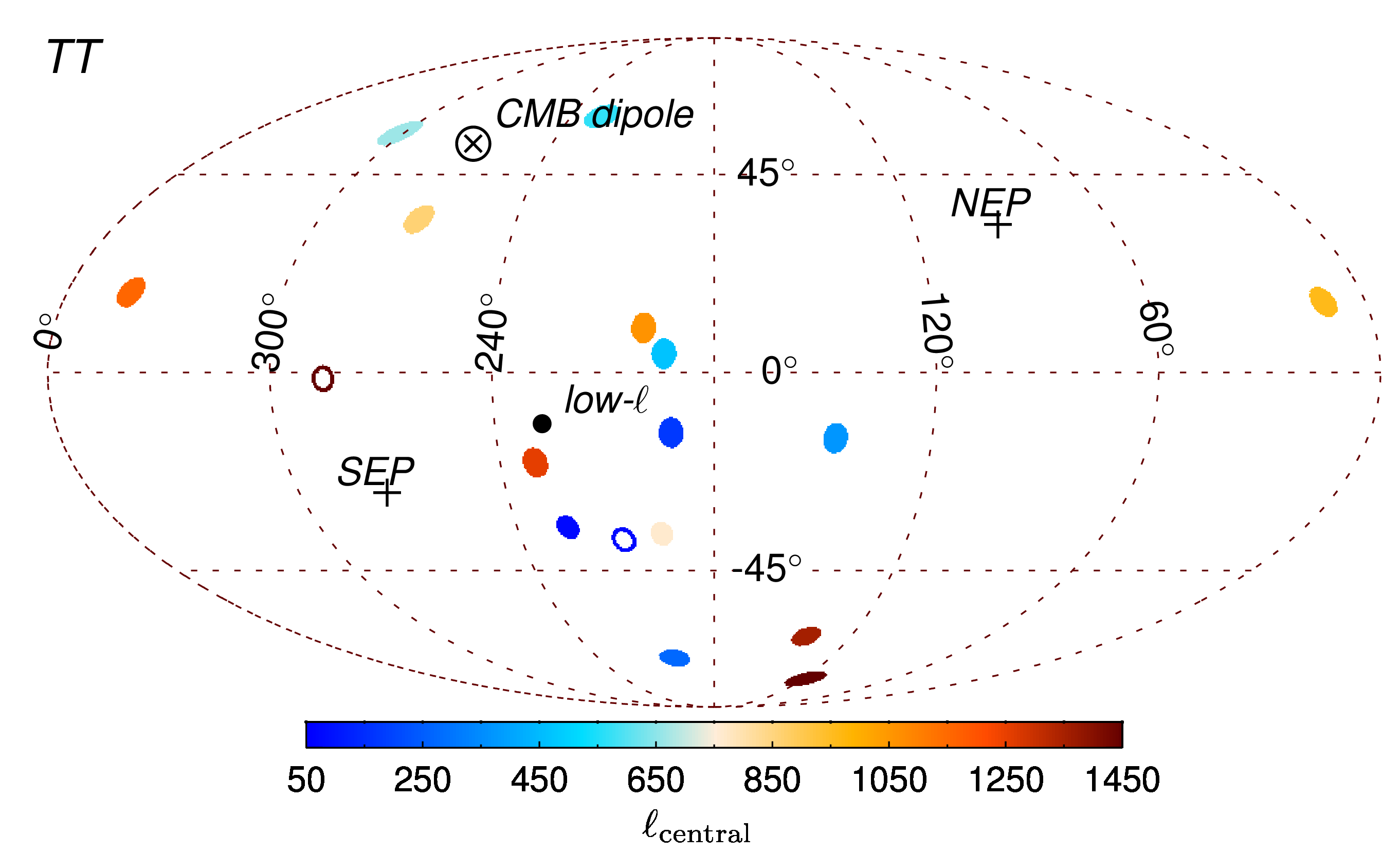}
\includegraphics[scale=0.25]{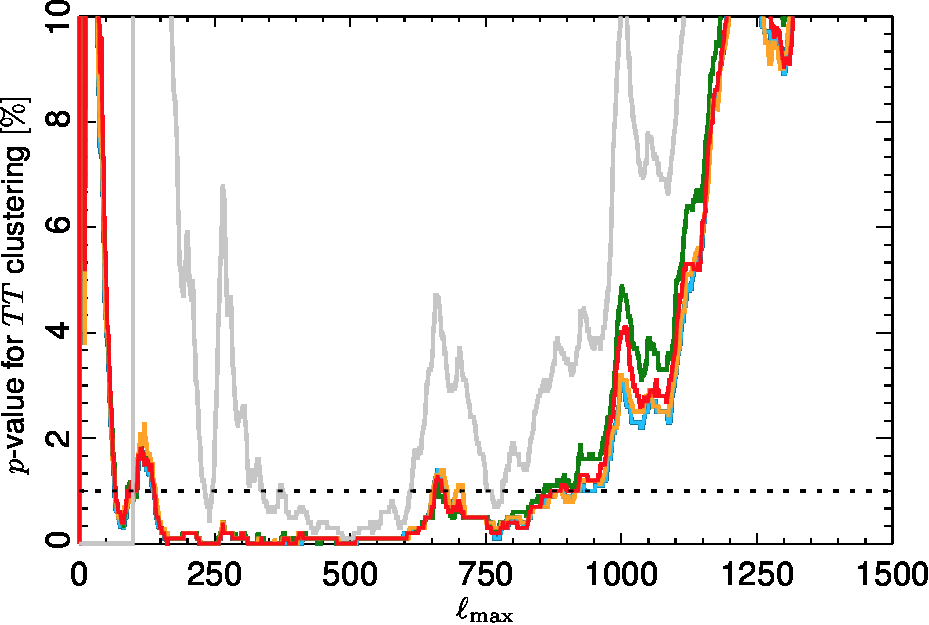} \\
\includegraphics[scale=0.1]{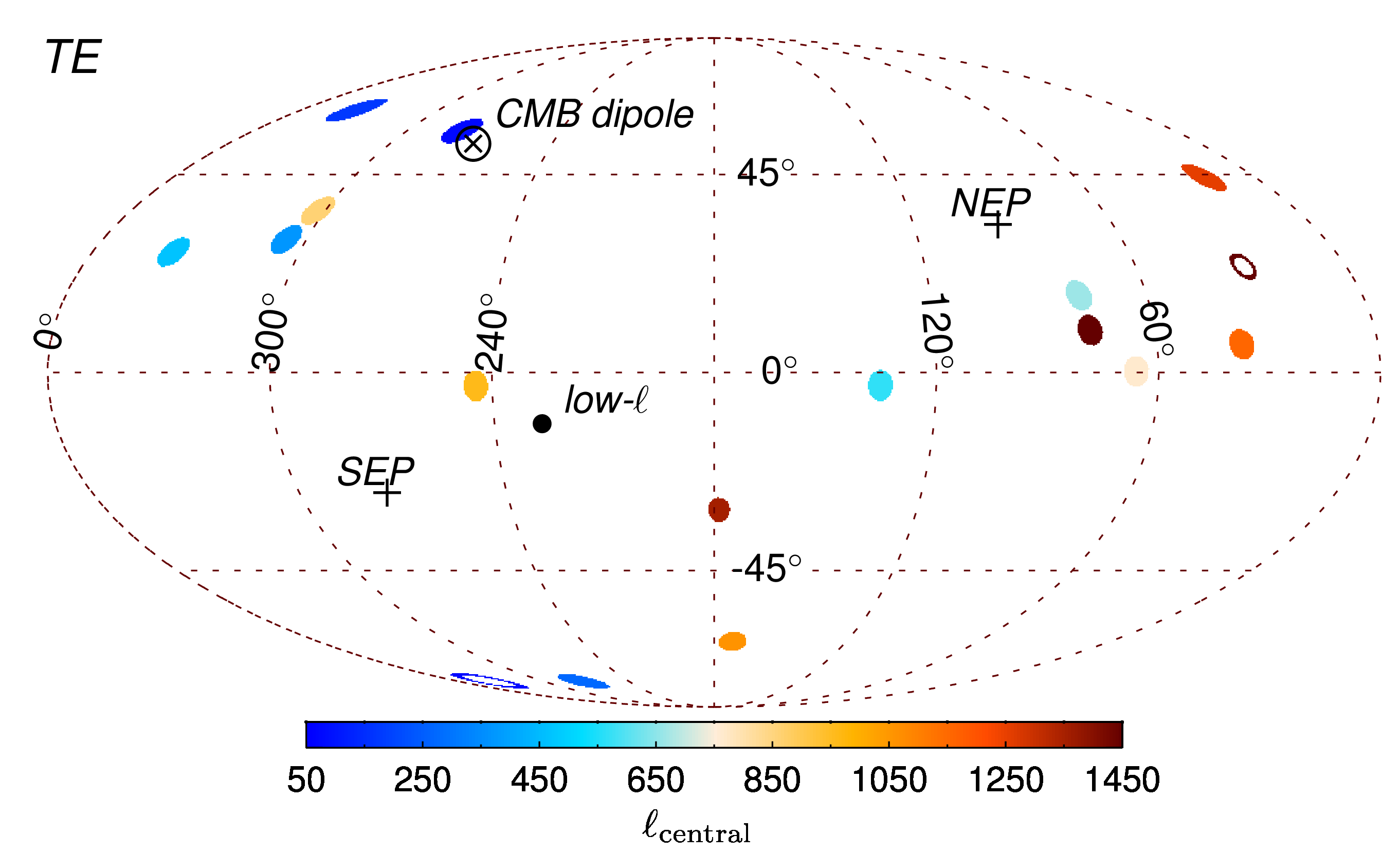} 
\includegraphics[scale=0.25]{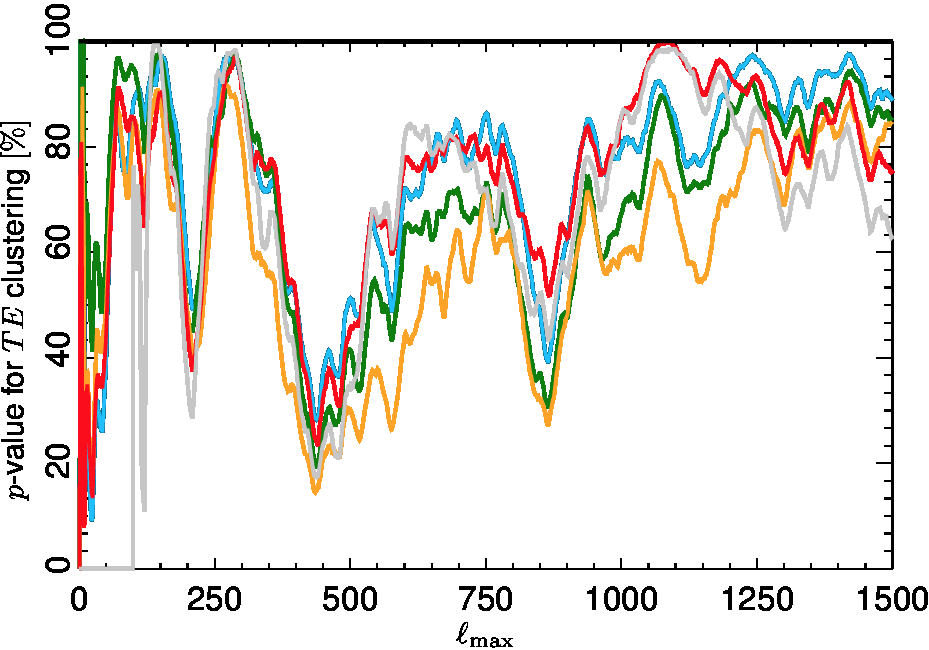} \\
\includegraphics[scale=0.1]{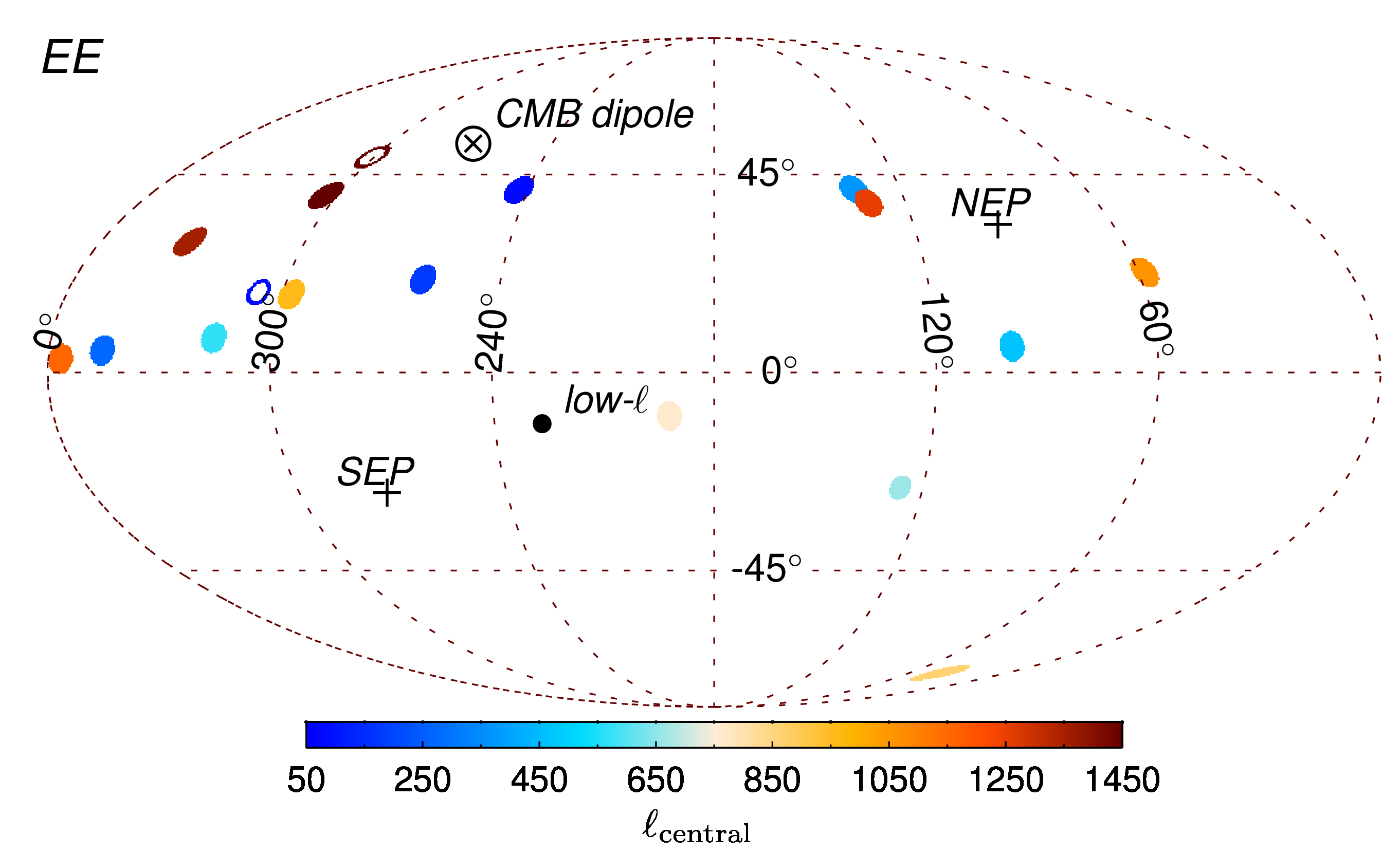}
\includegraphics[scale=0.25]{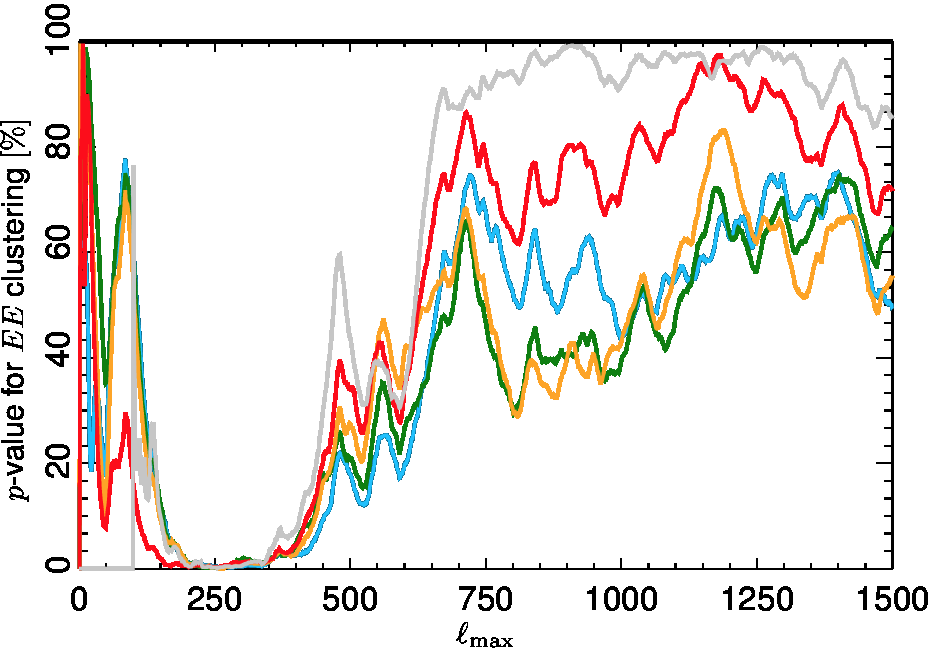} \\
\caption{As for Fig.~\ref{fig:asymdir_oe}, but for the
  HM maps, with the HM common mask applied.
}
\label{fig:asymdir_hm}
\end{figure*}

The three panels on the left of Fig.~\ref{fig:asymdir_oe}
show the $TT$, $TE$, and $EE$ dipole directions determined for
fifteen successive 100-multipole bins from the
OE split of the \commander\ data with the OE common mask applied.
This particular binning has been chosen for visualization purposes. The
temperature results (top panel) are consistent with those of
\citetalias{planck2014-a18} (although note that, in order to highlight
the clustering, the plots in the current paper are rotated 180\deg\
about the Galactic north-south axis, so that the Galactic longitude of
$l=180\deg$ is at the centre of the image).  
The preferred low-$\ell$ modulation direction determined from the
temperature data in Sect.~\ref{sec:dipmod_qml} is also indicated.
The right panels of Fig.~\ref{fig:asymdir_oe} presents the
corresponding {\it p}-values for the power asymmetry as a function of
$\ell_\mathrm{max}$.  The significance of the temperature alignment as
a function of $\ell_\mathrm{max}$ is consistent with earlier results
up to $\ell_{\rm max}\,{\approx}\,1000$.
We see that from $\ell_{\rm max}\,{\approx}\,150$ to
1000, the {\it p}-values are below $1\,\%$ for all
values of the maximum multipole
for all methods. However, from $\ell_{\rm max}\,{\approx}\,1000$, the
{\it p}-values increase rapidly.  Figure~\ref{fig:asymdir_hm} presents
the equivalent results for the HM data set.

The application of the OE- or HM-unobserved pixel masks is necessary
for the analysis of the component-separated maps in order to avoid
complications related to inpainting in these pixels.  However, the
\commander\ data set is the exception here, since it applies per-pixel
inverse-noise weighting per frequency channel, so that unobserved pixels in a
given channel are simply given zero weight in the parametric
fits. Thus a valid CMB estimate is provided for such pixels, at the
expense of higher noise.  This is particularly relevant when comparing
results to the 2015 analysis, since the number of unobserved pixels
has increased significantly between the 2015 and 2018 data sets.
Indeed, it seems apparent that the application of the unobserved pixel
masks has significantly increased the error on direction for
$\ell_{\rm max}\,{>}\,1000$, resulting in a corresponding change in
{\it p}-values for these multipoles.

\begin{figure}[htbp!]
\centering
\includegraphics[scale=0.25]{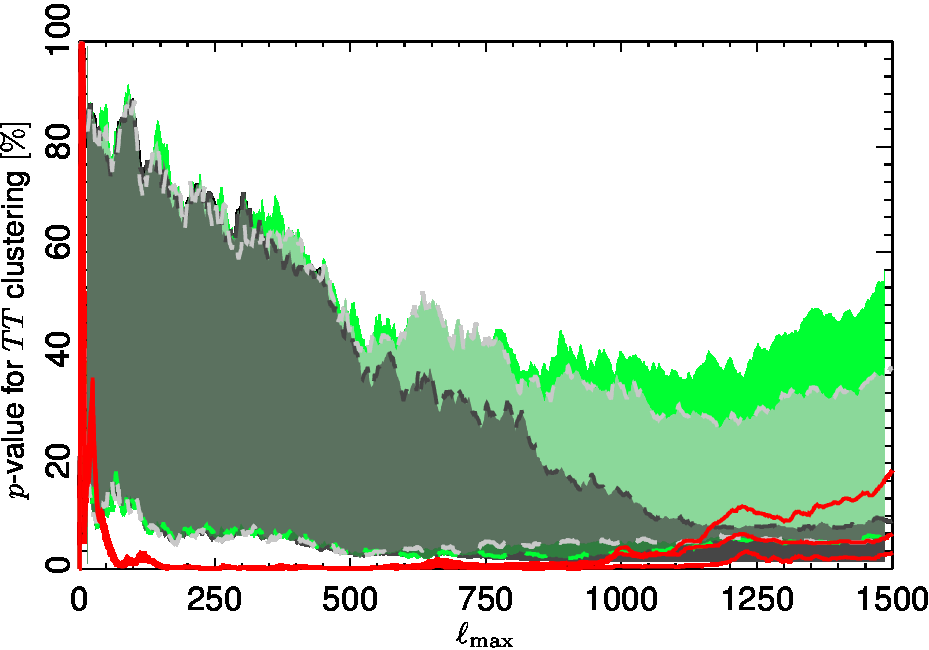}
\caption{Modulation results at high $\ell$.
  The shaded bands show the $68\,\%$ spread of {\it p}-values for
  asymmetry taken from \commander\ simulations with a mean {\it
    p}-value less than $10\,\%$ for $\ell\,{>}\,1000$. The dark green band
  (dark green boundaries) represents results computed from the OE data
  set using the full-mission common mask, the grey
  band (with grey boundaries) to the same data set analysed with the
  OE common mask applied,
  and the light green band (with light green boundaries) corresponds
  to the HM data set analysed using the HM common
  mask. The red solid lines show the same
  three cases for the data: lower line, full-mission common mask; middle line,
  OE common mask; and upper line, HM common mask.}
\label{fig:diffmask}
\end{figure}

To test the change in significance for $\ell_{\rm max}\,{>}\,1000$ due to the
inclusion of the unobserved pixel masks, we investigate all
simulations with low {\it p}-values for $\ell_{\rm max}\,{>}\,1000$
and check if a
similar trend can be seen. As a trade-off between ensuring a low {\it p}-value
and having a sufficient number of simulations to obtain
reliable statistics, we select those simulations with a mean {\it p}-value
$\langle p\rangle<10\,\%$ for $\ell\,{>}\,1000$. In
Fig.~\ref{fig:diffmask} we show the $68\,\%$ spread of these {\it p}-values
obtained from OE \commander\ simulations
with the full-mission and OE common masks applied, and the HM
\commander\ simulations with the HM
common mask applied.  The red solid lines show the corresponding
results for the data. We see a similar trend in the
simulations as observed in the data: the {\it p}-values
increase for smaller scales as a result of the higher uncertainty on
direction caused by the unobserved pixel masks.

\begin{figure*}[htbp!]
\centering
\includegraphics[scale=0.1]{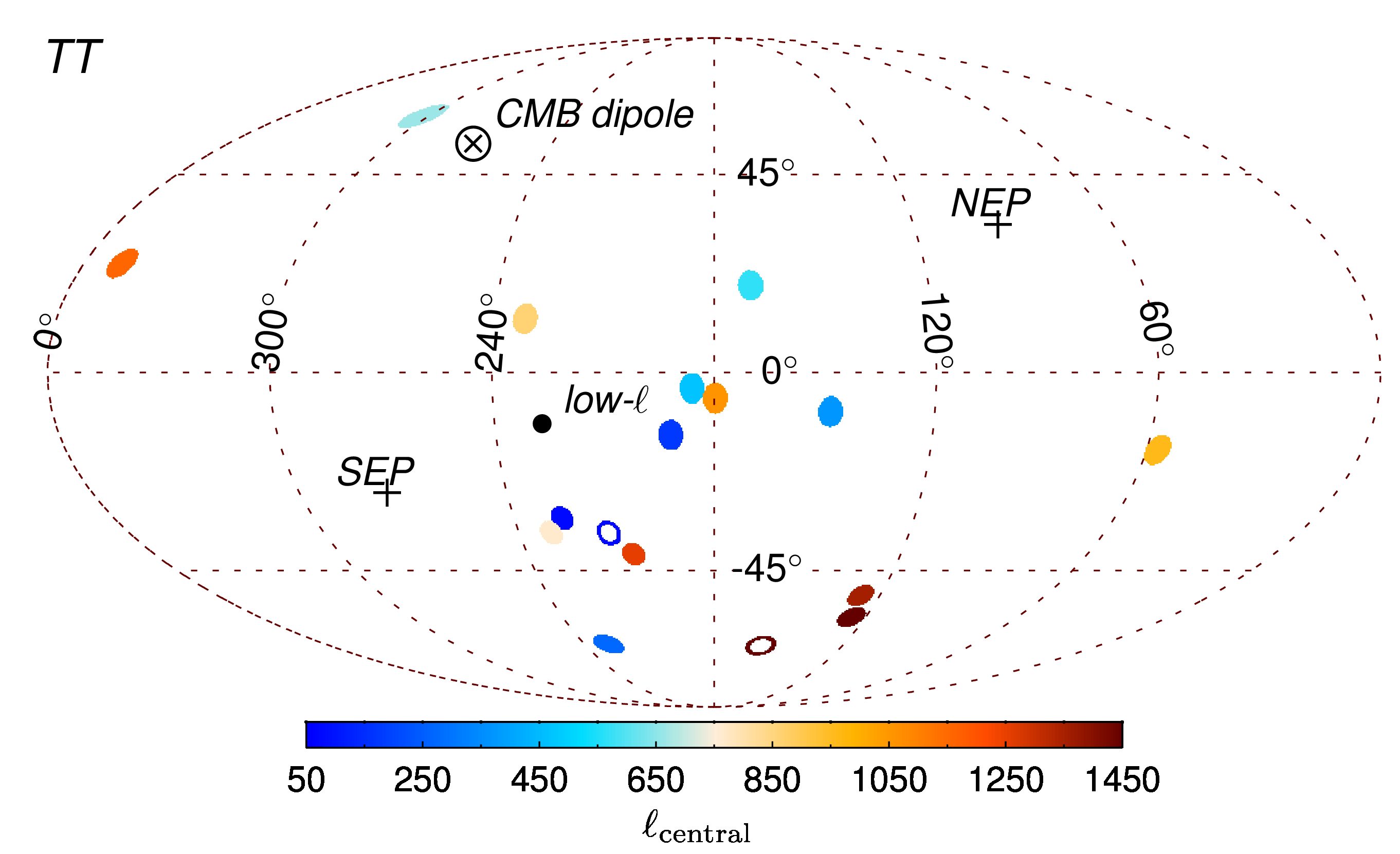}
\includegraphics[scale=0.25]{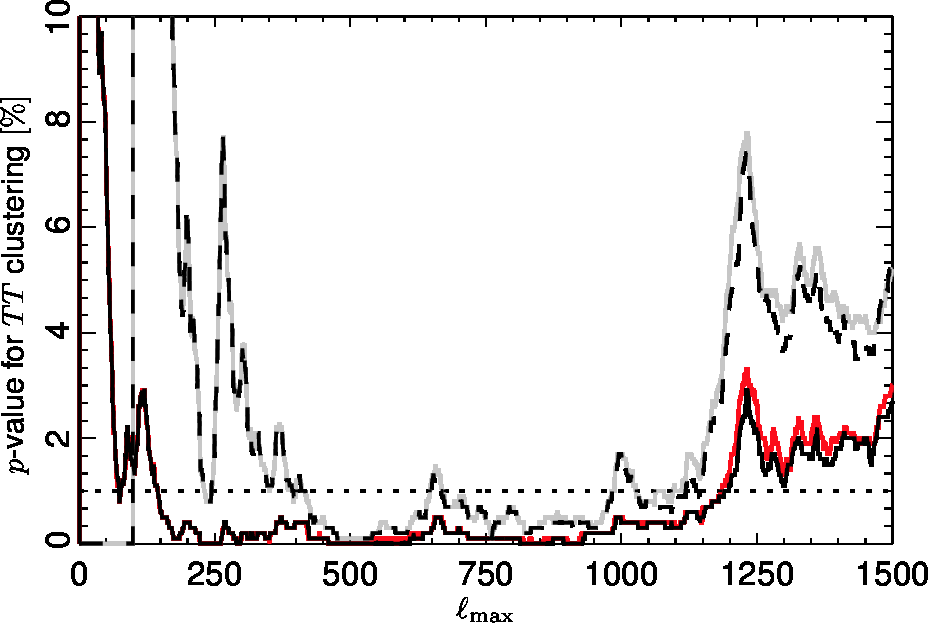} \\
\includegraphics[scale=0.1]{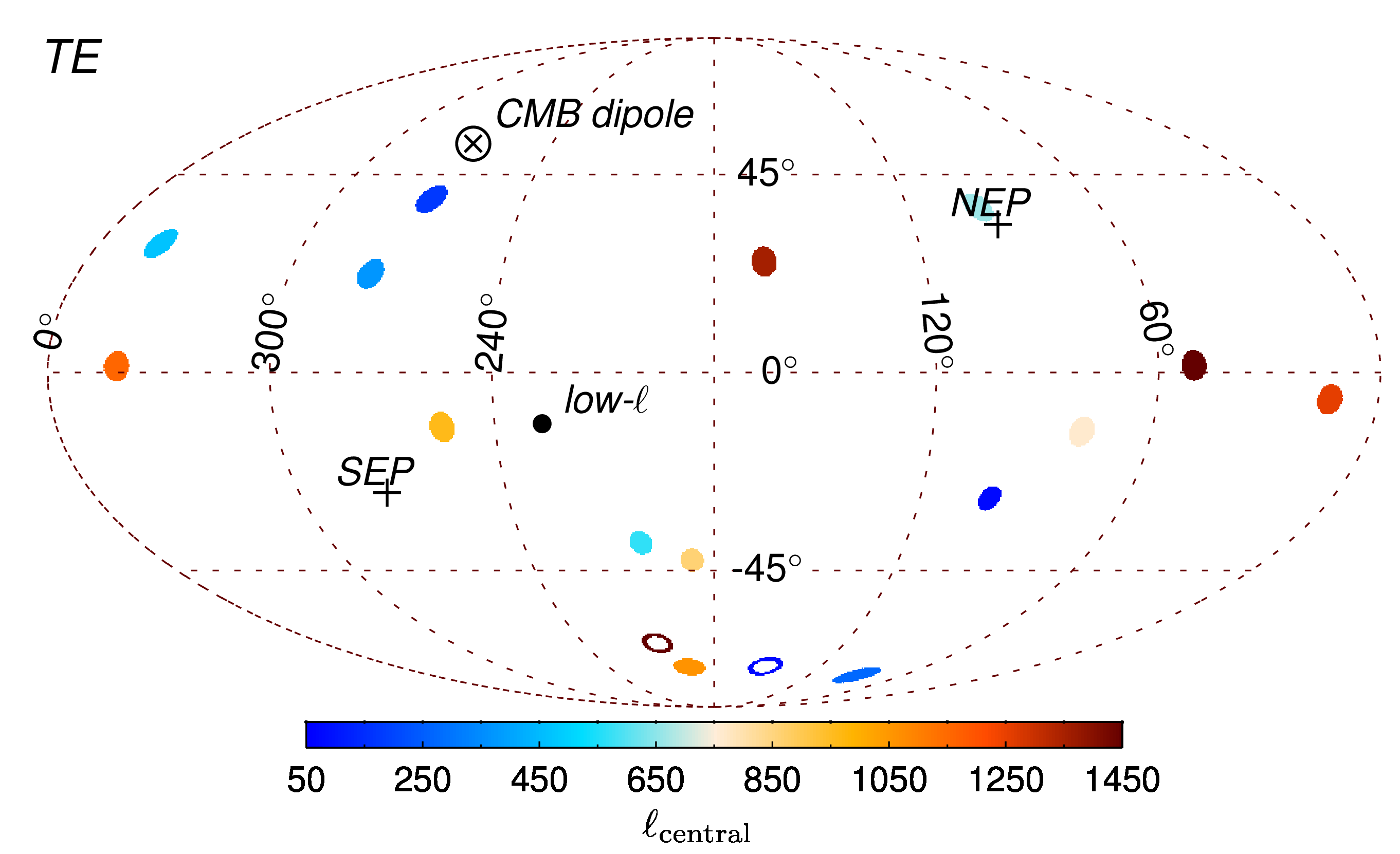}
\includegraphics[scale=0.25]{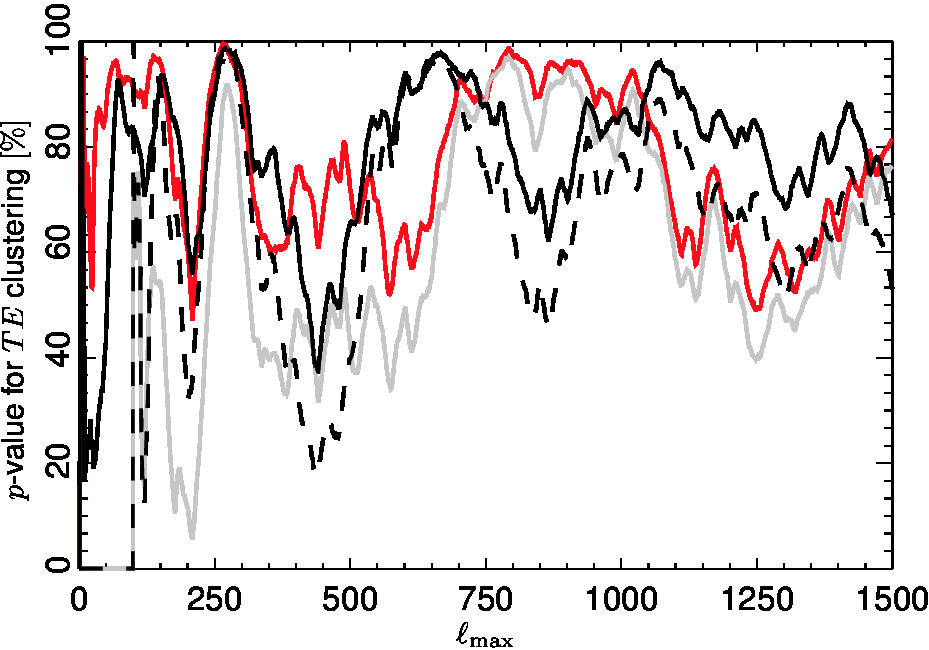} \\
\includegraphics[scale=0.1]{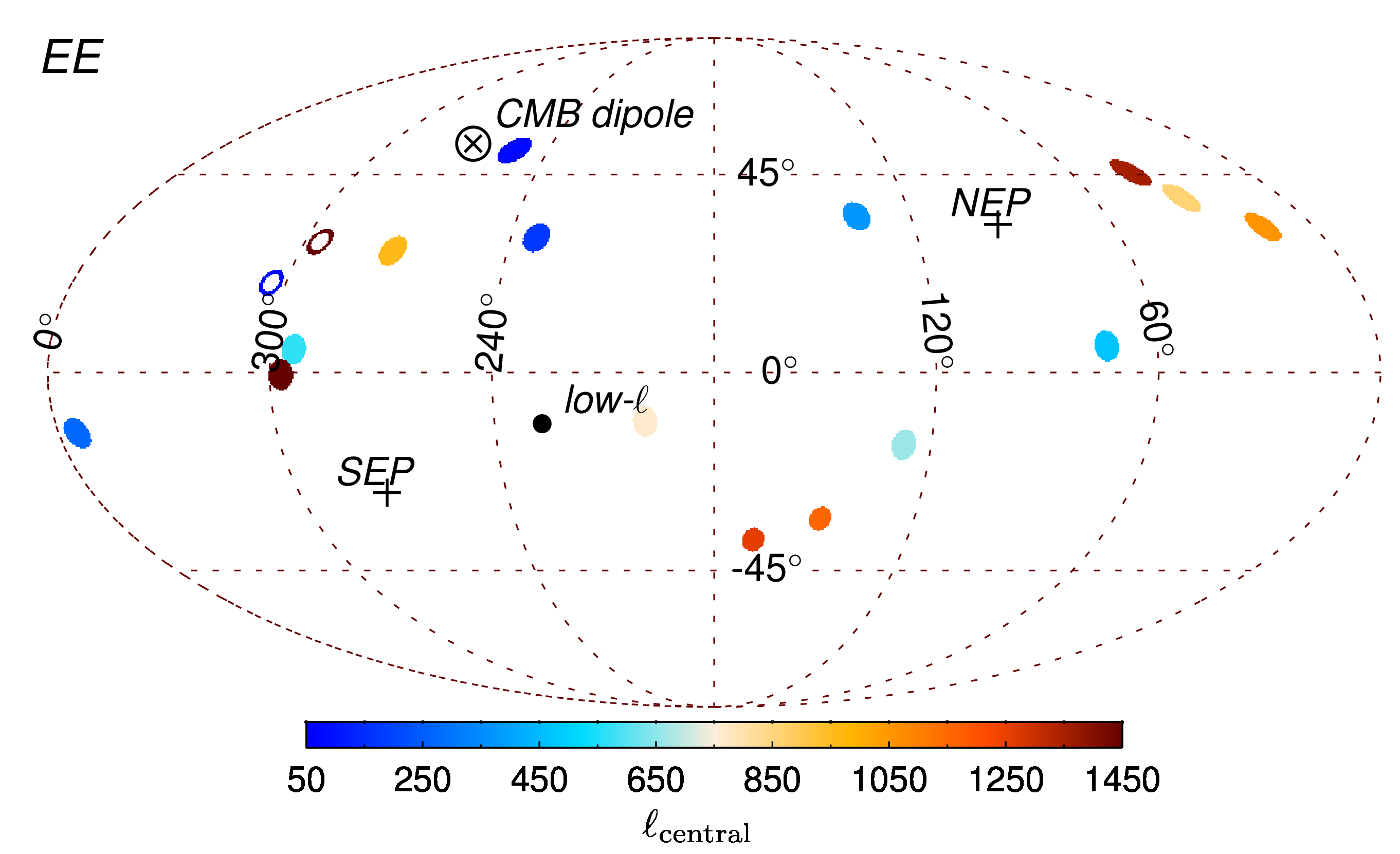}
\includegraphics[scale=0.25]{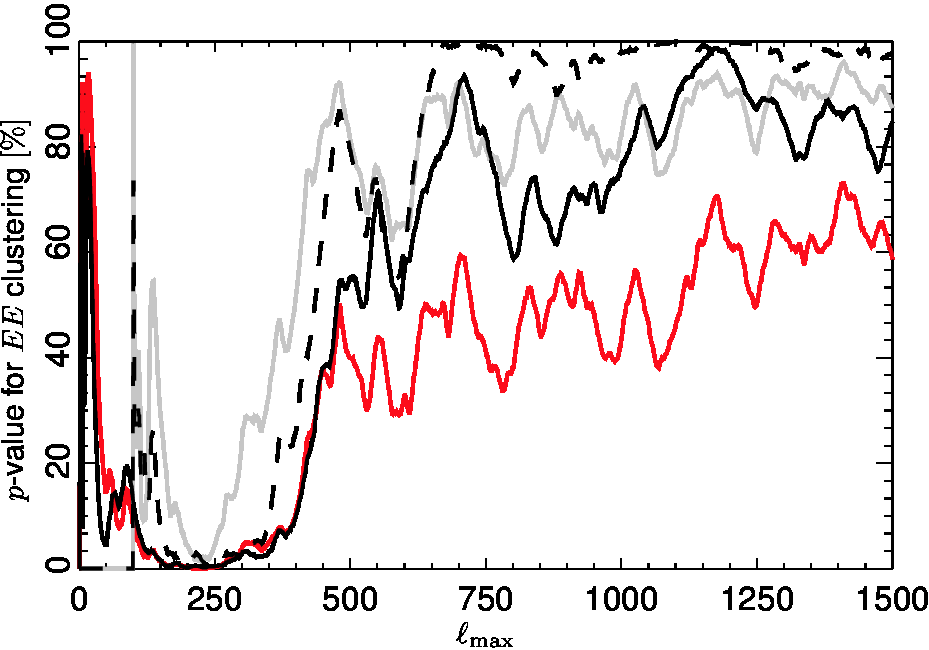} \\
\caption{{\it Left panels}: As for Fig.~\ref{fig:asymdir_oe}, but for
  the \commander\ OE maps with the full-mission common mask applied.
  The mean error on the derived direction
  that results from masking the data is $39\deg$ in the range $\ell\,{<}\,800$
  and $50\deg$ in the range $\ell\,{>}\,800$ for temperature, but $78\deg$ in
  the range $\ell\,{<}\,800$ and $91\deg$ in the range $\ell\,{>}\,800$
  for $EE$ polarization.  {\it Right panels}: Derived {\it p}-values
  for the angular clustering of the power distribution in
  OE maps as a function of $\ell_\mathrm{max}$, determined for
  \commander\ (red line) based on simulations with the full-mission
  common mask applied.  The grey line
  shows the \commander\ results when excluding the first 100
  multipoles in the analysis, the black solid line shows results for the
  \commander\ HM split, and the black dashed line corresponds to the grey
  line for the \commander\ HM split, in all cases with the common mask
  applied.
  }
\label{fig:asymdir}
\end{figure*}

In Fig.~\ref{fig:asymdir} we show the results for the \commander\ OE
analysis when the full-mission common mask has been applied. We see
that, for $\ell\,{<}\,1000$, the results are largely consistent with the
previous analysis when the unobserved pixel mask was also applied.
From $\ell\,{\approx}\,150$ to 1150, the {\it p}-values are
consistently below $1\,\%$ for all multipoles.  This is in good
agreement with the \commander\ results in \citetalias{planck2014-a18},
although we note that the latter results were computed using the
2015 common mask that did not include unobserved pixels.
For the $EE$ polarization signal, some alignment seems to be indicated,
reaching $p<1\,\%$ at $\ell\,{\approx}\,150$, which corresponds
to the multipole range where the alignment in temperature also starts to
be seen. The $EE$ alignment will be discussed in more detail below.

We note that for $\ell_\mathrm{max}\,{<}\,100$ the temperature {\it p}-values
are not consistent with the detection of a low-$\ell$
asymmetry/modulation, as seen in Sect.~\ref{sec:varasym}. However,
over this $\ell$ range, there are very few bins and the variance of
the RS might therefore be too high for the effect to be visible.  We
further test the multipole dependence of the alignment by restricting the
analysis to multipoles above $\ell_\mathrm{min} = 100$. The grey lines
in Fig.~\ref{fig:asymdir} show the \commander\ results in this case,
and indicate that clustering at the $p\,{<}\,1\,\%$ level is
still found for temperature. The clustering significance can therefore
not be solely attributed to large-angular-scale features. Note also
that the dip in {\it p}-values for the $EE$ polarization at
$\ell\,{\approx}\,200$ is still seen even with the restriction
$\ell_\mathrm{min}\,{>}\,100$.

It is also apparent from Fig.~\ref{fig:asymdir} that
some {\it p}-values are close to $100\,\%$, in particular for $TE$.
A high {\it p}-value means a low value for the RS
statistic, and hence it seems worthwhile asking if such low values are
unusual.  In fact we find that scanning over the range of $\ell_{\rm max}$,
the maximum $p$-value of the RS
statistic for the $TE$ data is exceeded in 20--40\,\% of the simulations
(for the example of \commander, for the various masks and data splits),
and hence does not appear to be anomalous.

\begin{figure}[htbp!]
\centering
\includegraphics[scale=0.25]{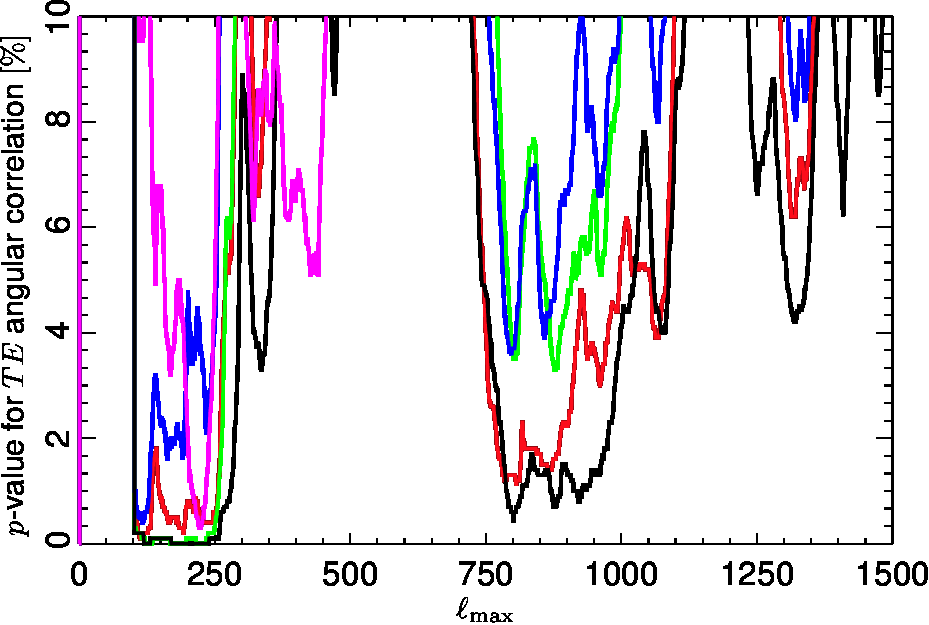}
\caption{Derived {\it p}-values for the angular correlation of $TT$
  and $EE$ dipole directions in \commander\ maps as a function of
  $\ell_\mathrm{max}$, determined using the HM split with the
  full-mission common mask
  (black), the HM split with the HM common mask (green),
  the OE map with the full-mission common mask (red),
  and the OE map with the OE common mask (blue),
  based on FFP10 simulations. The magenta line shows the corresponding
  correlation between $TE$ and $EE$ directions for the \commander\ HM
  split with the common mask. The {\it p}-values are based on the fraction
  of simulations with a higher RS, determined over the $\ell$ range up
  to the given $\ell_\mathrm{max}$, compared to the data,
  hence small $p$-values would correspond to anomalously aligned
  dipole directions.  The results
  shown here have been marginalized over bin sizes in the range
  $\Delta\ell\,{=}\,8$ to $\Delta\ell\,{=}\,32$.}
\label{fig:TE_cross_pvalues}
\end{figure}

In order to further study the alignment in $EE$ polarization for
$\ell\,{<}\,300$, we tested whether the directions of the $EE$ dipoles in
this range are correlated with the directions for the $TT$ dipoles. Here we
made a small change in the statistic as already described: in point 4 in the
above description of the method, we instead use the mean of the cosine of the
angles between all pairs of dipoles, where one dipole is always taken
from $EE$ and the other always from $TT$. In this way, the statistic
measures the cross-correlation between $TT$ and $EE$ directions. In
Fig.~\ref{fig:TE_cross_pvalues} we show the {\it p}-values for the
cross-correlation statistic as a function of multipole. Note that a
strong correlation between the position of the $TT$ and $EE$ dipoles
are detected for the same multipole range where the $EE$ polarization
dipoles (and $TT$ dipoles) are clustered.

We have seen that the $EE$ directions appear clustered for 
$\ell_{\rm max}\approx200$. This is also the case for $TT$, and so we would
like to test whether the two preferred directions are also related.
What Fig.~\ref{fig:TE_cross_pvalues} shows is that not only are the $TT$
and $EE$ dipoles clustered among themselves, but that they also appear
to be clustered towards the same direction. 
If $TT$ and $EE$ were independent, this would be highly unexpected, even if 
$TT$ and $EE$ were separately clustered. However, we know that the $TE$ 
spectrum is non-zero, giving rise to a correlation between $T$ and $E$.
In order to test whether such a $TT$--$EE$ directional correlation is expected 
in the case when both $TT$ and $EE$ are clustered individually, we perform 
the following test.  We examine all simulations having a
minimum {\it p}-value of less than $1\,\%$ for both $TT$ and $EE$ in
overlapping multipole ranges (similar to what we observed in the data). 
Only two simulations have overlapping $TT$ and $EE$ multipole ranges using 
this criterion, but neither of them has a significant correlation between 
$TT$ and $EE$ directions.
While this may seem to suggest that the effect in $EE$ is not just due
to the already known directional clustering in $TT$ (and the $TE$ correlation),
it is hard to draw firm conclusions without many more simulations.

\begin{figure}[htbp!]
\centering
\includegraphics[scale=0.1]{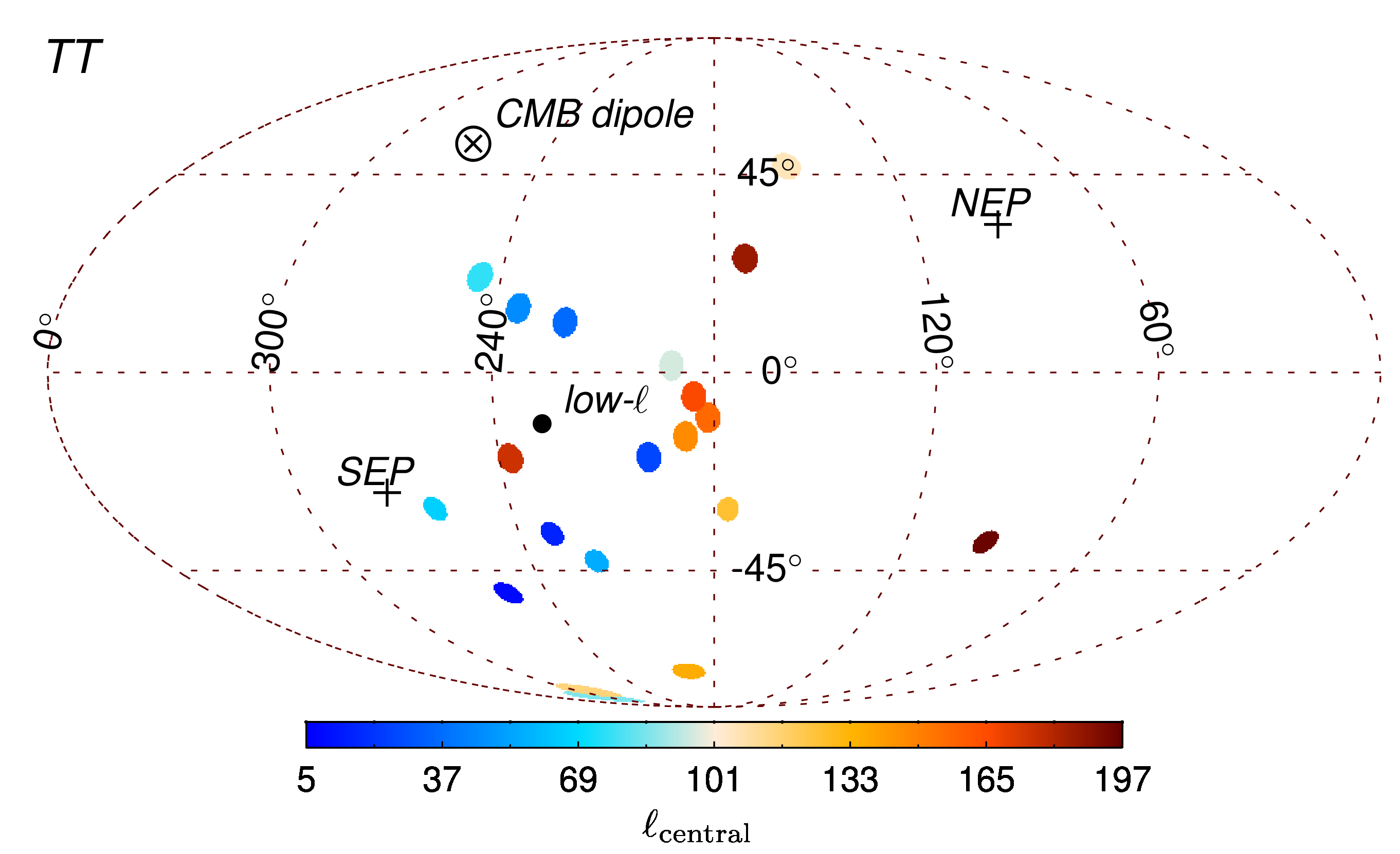}
\includegraphics[scale=0.1]{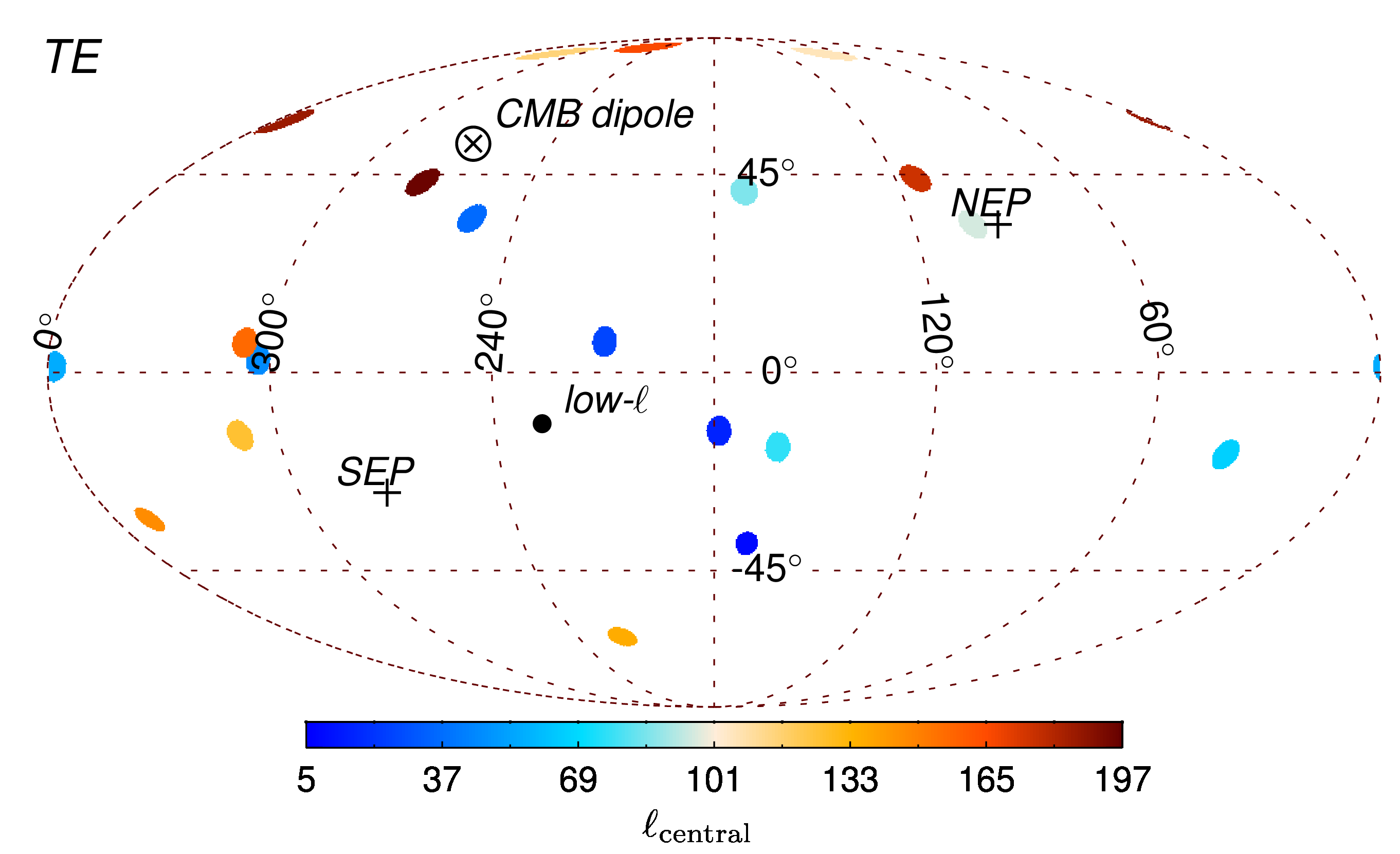}
\includegraphics[scale=0.1]{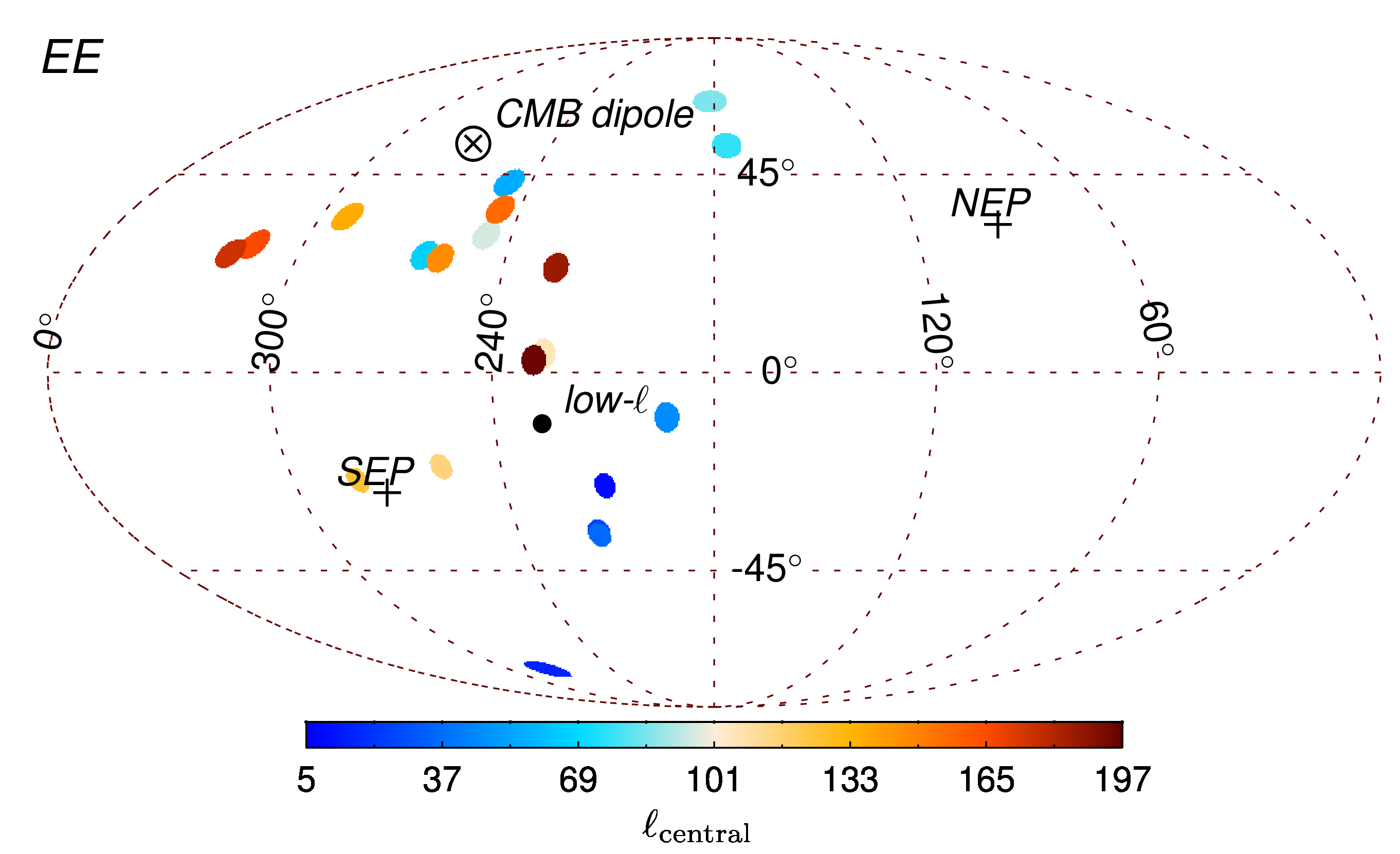}
\caption{Dipole directions for independent 10-multipole bins of the
  local power-spectrum distribution from $\ell=2$ to $200$ in the
  \commander\ HM maps with the full-mission common mask applied.
  This is a finer binning in order to investigate the directional
  clustering of different multipole ranges. Note that the
  maps have been rotated about the Galactic north-south axis, such that
  Galactic longitude $l=180\deg$ is in the centre of the map. The top,
  middle, and bottom plots show maps based on the $TT$, $TE$, and $EE$
  spectra, respectively.  
  In all panels, we also show the CMB dipole direction, the north
  ecliptic pole (NEP), the south ecliptic pole (SEP), and the
  preferred dipolar modulation axis (labelled as ``low-$\ell$'')
  derived from the temperature data in Sect.~\ref{sec:dipmod_qml}.  }
\label{fig:asymdir_10bin}
\end{figure}

We investigate this further in Fig.~\ref{fig:asymdir_10bin} where we
show the dipole directions for blocks of 10 multipoles for $\ell\,{<}\,200$
for $TT$, $TE$, and $EE$. There is some hint here of the
correlation between $TT$ and $EE$ directions
seen in Fig.~\ref{fig:TE_cross_pvalues}.
Note that in Fig.~\ref{fig:TE_cross_pvalues} this angular
correlation appears stronger for the HM split than for the OE split; given
the large errors on direction for polarization, differences in the noise
properties (including systematic effects) for the HM and OE split may
be responsible. The middle panel in Fig.~\ref{fig:asymdir_10bin} shows the
$TE$ directions. Clearly the $TE$ directions are more scattered than
$TT$ and $EE$, as expected from Fig.~\ref{fig:asymdir}. The directions for
several multipole blocks in the range $\ell\,{<}\,100$ coincide quite well with
the $TT$ and $EE$ directions, but most of the 10-multipole blocks point in
different directions. To further compare the angular clustering of $TE$
dipoles with $TT$ and $EE$ dipoles, we construct the statistic
measuring the correlation between $TE$ and $TT$ directions, as well as
between $TE$ and $EE$ directions, in the same way as described above
for the correlation between $TT$ and $EE$. We find no significant
correlations between the $TT$ and $TE$ directions, but the correlation
between $TE$ and $EE$ again shows some sign of similar
multipole ranges where $EE$ is aligned and the $TT$ and $EE$ directions
correlate. This is shown as a magenta line in
Fig.~\ref{fig:TE_cross_pvalues}. 
The correlation between $TE$ and $EE$
directions at $\ell\,{\approx}\,200$ may arise as a result of the fact
that $TT$ and $EE$ are aligned, as well as the fact that $TT$ and $EE$
directions are also correlated at exactly these multipoles. However, since
none of the simulations show a high significance for all three of these
statistics in one common multipole range for one given simulation, we are
unable to test this with simulations. Nevertheless, we might have expected a
similar correlation between $TT$ and $TE$ if this was really the case.

In \citetalias{planck2014-a18}, we made a more detailed study of the total
significance of the alignment based on the {\it p}-values in figures like
Fig.~\ref{fig:asymdir}. Specifically, we compared the mean and minimum
{\it p}-values (over a range of $\ell_{\rm max}$)
computed from the data with the value from simulations.
Here we repeat this study of ``global statistics'' for selected cases of
particular interest (see \citealt{Bennett2011} and
\citetalias{planck2014-a18} for discussion of how this is an attempt to
assess the overall significance without making choices a posteriori).
The results are shown in Table~\ref{tab:globstat}.  For temperature,
we see that none of the simulations have lower mean {\it p}-values
than the data. This is valid for most masks or foreground-subtraction
methods.  Because the data have a minimum {\it p}-value of 0 for all
masks and methods, only an upper bound on significance can be
set. Considering only $\ell_\mathrm{min}\,{>}\,100$, the {\it p}-values are
still low. 

Turning now to polarization,
for $TE$, the numbers are completely consistent with simulations,
while for $EE$, the results are unstable to the choice of masks and
component-separation methods. The mean {\it p}-value is always
consistent with simulations, which is not unexpected given the short
multipole range where the {\it p}-values are low (as discussed above).
However, for the minimum {\it p}-value, the lowest values seen
in the multipole range $\ell=100$--300 are zero for some masks and some
methods. This also occurs in some simulations and hence only an upper bound
on the significance can be set. A similar conclusion can be made for
the $TT$/$EE$ angular correlation, but the results are in this case
more stable with choice of mask and component-separation method.

\begin{table}[htbp!]
\begingroup
\newdimen\tblskip \tblskip=5pt
  \caption{Significance of the angular clustering of the power distribution.
  We indicate the actual mean/min {\it p}-value of the data, determined from
  Figs.~\ref{fig:asymdir_oe}, \ref{fig:asymdir_hm}, \ref{fig:asymdir},
  and \ref{fig:TE_cross_pvalues} and written as a fraction of the
  number of simulations used to assess the values, together with
  the percentage of simulations with a lower mean/minimum {\it
    p}-value than the data.
  \label{tab:globstat}}
\vskip -3mm
\footnotesize
\setbox\tablebox=\vbox{
   \newdimen\digitwidth
   \setbox0=\hbox{\rm 0}
   \digitwidth=\wd0
   \catcode`*=\active
   \def*{\kern\digitwidth}
   \newdimen\signwidth
   \setbox0=\hbox{+}
   \signwidth=\wd0
   \catcode`!=\active
   \def!{\kern\signwidth}
   \newdimen\opwidth
   \setbox0=\hbox{<}
   \opwidth=\wd0
   \catcode`?=\active
   \def?{\kern\opwidth}
\halign{\hbox to 1.2in{#\leaderfil}\tabskip -5pt&
\hfil#\hfil\tabskip 4pt&
\hfil#\hfil&
\hfil#\hfil&
\hfil#\hfil\tabskip 0pt\cr
\noalign{\doubleline}
\omit& *Mean& *$\%$ of sims& *Min.& *$\%$ of sims\cr
\omit\hfil Data\hfil& *{\it p}-value& *(mean)& *{\it p}-value& *(min.)\cr
\noalign{\vskip 3pt\hrule\vskip 6pt}
\omit&\multispan4 \hfil $TT$\hfil\cr
\noalign{\vskip 3pt}
{\tt Commander}$^\mathrm{a}$&   *1.06& *<0.10& ?*0.00& *<2.50\cr
{\tt Commander}$^\mathrm{ae}$&  *4.52& *?0.40& ?*0.10& *?3.60\cr
{\tt Commander}$^\mathrm{b}$&   *1.13& *<0.10& ?*0.00& *<2.50\cr
{\tt Commander}$^\mathrm{be}$&  *4.66& *?0.40& ?*0.10& *?3.60\cr
{\tt Commander}$^\mathrm{c}$&   *3.87& *?0.10& ?*0.00& *<2.60\cr
{\tt Commander}$^\mathrm{ce}$&  *9.80& *?1.30& ?*0.10& *?3.90\cr
{\tt Commander}$^\mathrm{d}$&   *2.44& *<0.10& ?*0.00& *<2.50\cr
{\tt Commander}$^\mathrm{de}$&  *6.69& *?0.90& ?*0.10& *?3.50\cr
{\tt NILC}$^\mathrm{d}$&        *2.42& *<0.10& ?*0.00& *<2.20\cr
{\tt NILC}$^\mathrm{de}$&       *7.48& *?1.10& ?*0.10& *?3.80\cr
{\tt SEVEM}$^\mathrm{d}$&       *2.74& *?0.10& ?*0.00& *<2.30\cr
{\tt SEVEM}$^\mathrm{de}$&      *7.62& *?1.00& ?*0.10& *?3.30\cr
{\tt SMICA}$^\mathrm{d}$&       *2.27& *<0.10& ?*0.00& *<2.40\cr
{\tt SMICA}$^\mathrm{de}$&      *6.51& *?0.90& ?*0.10& *?3.70\cr
\noalign{\vskip 3pt\hrule\vskip 6pt}
\omit&\multispan4 \hfil $TE$\hfil\cr
\noalign{\vskip 3pt}
{\tt Commander}$^\mathrm{a}$&   78.56& ?92.99& ?17.32& ?84.48\cr
{\tt Commander}$^\mathrm{b}$&   76.80& ?91.49& ?47.05& ?99.50\cr
{\tt Commander}$^\mathrm{c}$&   74.15& ?88.09& ?*8.81& ?64.46\cr
{\tt Commander}$^\mathrm{d}$&   69.83& ?81.68& ?33.53& ?97.00\cr
\noalign{\vskip 3pt\hrule\vskip 6pt}
\omit&\multispan4 \hfil $EE$\hfil\cr
\noalign{\vskip 3pt}
{\tt Commander}$^\mathrm{a}$&   56.85& ?60.46& ?*0.10& ?*4.20\cr
{\tt Commander}$^\mathrm{ae}$&  73.40& ?87.19& ?*0.70& ?13.11\cr
{\tt Commander}$^\mathrm{b}$&   39.21& ?31.13& ?*0.00& *<2.60\cr
{\tt Commander}$^\mathrm{be}$&  70.86& ?83.38& ?*1.40& ?19.92\cr
{\tt Commander}$^\mathrm{c}$&   53.18& ?52.65& ?*0.30& *?7.31\cr
{\tt Commander}$^\mathrm{ce}$&  65.31& ?72.07& ?*0.10& *?4.30\cr
{\tt Commander}$^\mathrm{d}$&   35.93& ?26.73& ?*0.00& *<2.80\cr
{\tt Commander}$^\mathrm{de}$&  65.73& ?73.87& ?*0.90& ?14.51\cr
{\tt NILC}$^\mathrm{d}$&        46.65& ?43.74& ?*0.70& ?11.21\cr
{\tt NILC}$^\mathrm{de}$&       57.82& ?61.16& ?*0.20& ?*4.80\cr
{\tt SEVEM}$^\mathrm{d}$&       38.41& ?30.33& ?*0.00& *<2.80\cr
{\tt SEVEM}$^\mathrm{de}$&      64.88& ?73.67& ?*0.40& *?6.91\cr
{\tt SMICA}$^\mathrm{d}$&       53.17& ?53.55& ?*0.80& ?12.31\cr
{\tt SMICA}$^\mathrm{de}$&      70.14& ?82.28& ?*0.70& ?11.81\cr
\noalign{\vskip 3pt\hrule\vskip 6pt}
\omit&\multispan4 \hfil $TT/EE$ angular corr.\hfil\cr
\noalign{\vskip 3pt}
{\tt Commander}$^\mathrm{a}$&   18.24& ?*4.50& ?*0.00& *<1.30\cr
{\tt Commander}$^\mathrm{b}$&   22.32& ?*7.41& ?*0.10& ?*2.30\cr
{\tt Commander}$^\mathrm{c}$&   35.21& ?21.42& ?*0.00& *<1.30\cr
{\tt Commander}$^\mathrm{d}$&   36.04& ?21.22& ?*0.40& ?*5.11\cr
\noalign{\vskip 3pt\hrule\vskip 3pt}}}
\endPlancktable                    
\tablenote {{\rm a}} HM split with common mask.\par
\tablenote {{\rm b}} OE split with common mask.\par
\tablenote {{\rm c}} HM split with HM-unobserved pixel mask.\par
\tablenote {{\rm d}} OE split with OE-unobserved pixel mask.\par
\tablenote {{\rm e}} Only multipoles above $\ell_\textrm{min}=100$ are considered.\par
\endgroup
\end{table}

\section{Conclusions}
\label{sec:conclusions}

In this paper, we present a study of the statistical isotropy
and Gaussianity of the CMB using the \Planck\ 2018 temperature and
polarization data.  The \Planck\ 2015 release essentially corresponded
to the limit of our ability to probe CMB anomalies with
temperature fluctuations alone.  The use of large-angular-scale
polarization measurements enables largely {\em independent} tests of
these peculiar features.  In principle, this can reduce or eliminate
the subjectivity and ambiguity in interpreting their statistical
significance.

As in previous work, we follow a model-independent approach and focus
on null-hypothesis testing by calculating and reporting {\it p}-values
for a number of statistical tests.
These tests are performed on maps of the CMB anisotropy that originate
from the four component-separation methods, \commander, \nilc, \sevem,
and \smica, described in \citet{planck2016-l04}. For polarization
studies, we consider both maps of the Stokes parameters, $Q$ and $U$,
and of the $E$-mode signal generated using a novel method described in
Appendix~\ref{sec:EB_maps}.
The consistency of the results determined from the
component-separated maps is an important indicator of the
cosmological origin of any significant {\it p}-values, or otherwise.

The temperature results are consistent with previous findings in
\citetalias{planck2013-p09} and
\citetalias{planck2014-a18}. Specifically, the observed
fluctuations are largely compatible with Gaussian statistics and
statistical isotropy, with some indications of departures from the
expectations of \LCDM\ in a few cases. Such signatures are well known;
thus, in this summary, we focus on the properties of the polarization data.

  In Sect.~\ref{sec:non_gaussianity}, we examine aspects of the
Gaussianity of the polarized CMB fluctuations.  Tests of 1D moments, 
$N$-point correlation functions, Minkowski functionals, peak
statistics, and the oriented and unoriented stacking of peaks yield
no indications of significant departures from
Gaussianity. In addition, no evidence is found for a low variance of
the polarized sky signal.

  Section~\ref{sec:anomalies} provides an updated study of several
previously known peculiarities.  We find no evidence in the
polarization data of a lack of large-scale angular correlations, a
hemispherical asymmetry in the behaviour of $N$-point functions or
peak distributions, a violation of point-parity symmetry, or a
polarization signature associated with the Cold Spot.

In Sect.~\ref{sec:dipmod} we perform a series of tests searching for
the signature in polarization of the well-known large-scale dipolar
power asymmetry.  Neither investigations using a variance estimator
nor via $\ell$ to $\ell \pm 1$ mode coupling find strong evidence of
this asymmetry. However, an interesting alignment of the preferred
directions of the temperature and $E$-mode dipolar modulation is found
using the variance asymmetry estimator at a modest significance,
depending on the component-separated map in question. The mode-coupling
estimator indicates that the polarization-only results
show some apparent asymmetry over scales up to 
$\ell_{\rm max}\,{\approx}\,250$, a range that overlaps the scales of interest
for the variance asymmetry. Similarly, an independent, but related,
test of directionality finds suggestions of some alignment of
directions in the $EE$ polarization signal beginning at
$\ell_{\rm max}\approx150$ and extending to $\ell_{\rm max}\approx250$. 

There are some caveats worth pointing out here.
Firstly, as described in \citet{Contreras2017a}, one could predict of order
30\,\% chance that any of the $p$-value dips would increase in significance
when even statistically {\it isotropic\/} polarization data were added.
Secondly, all of these dipole-modulation and hemispheric-asymmetry statistics
are just measuring slightly different weightings of the $\ell$ to
$\ell\pm1$ couplings on the same sky, and hence they cannot be considered to be
independent of each other.  Lastly, we are only testing phenomenological
models here, rather than physical modulation models where there is a prediction
for how scales in temperature and in polarization might be separately
modulated.  Hence it is unclear if the hints of $EE$--$TT$ dipole-modulation
alignment are what we would {\it expect\/} if there was a physical
mechanism responsible for some modulation.
Whether the hint of alignment between the temperature and the
polarization dipolar power asymmetry is more than a coincidence can
only be addressed once new data are available from forthcoming
large-scale and low-noise polarization experiments.

A notable feature of all of the polarization analyses is the variation
in {\it p}-values for a given test between the four
component-separated maps. This is a consequence of the fact that
different component-separation methods respond to noise and residual
systematic effects in different ways. However, it may also indicate
an incomplete understanding of the noise properties of the data, both
in terms of amplitude and correlations between angular scales.
This should not be considered surprising, given that \Planck\ was not
optimized for polariation measurements. 
Although remarkable progress has been made in reducing the systematic
effects that contaminated the 2015 polarization maps on large angular
scales, particularly for the HFI instrument
\citep{planck2014-a10,planck2016-l03}, thus allowing more robust
measurement of the optical-depth-to-reionization $\tau$, residual
systematics, and our ability to simulate them, can limit the kind of
statistical tests of non-Gaussianity and isotropy that can be applied
to the data.  Nevertheless, a detailed set of null tests applied to the
maps indicates that these issues do not dominate the analysis on
intermediate and large angular scales, particularly for the
statistical tests presented in this paper.

Future experiments that can measure the cosmological $E$-modes at the
cosmic-variance limit are required in order to unambiguously test for
the presence of anomalies in the polarized sky.  However, given the
amplitude of the effects seen in the \Planck\ temperature data, it may
still remain difficult to claim high significance ($>3\,\sigma$)
detections in polarization (although detailed forecasts related to this are
highly model dependent).  Nevertheless, this should not prevent us from
undertaking such searches, since any detection of anomalies in the
polarized sky signal will inevitably take us beyond the standard model
of cosmology.


\begin{acknowledgements}
\label{sec:acknowledgements}
The Planck Collaboration acknowledges the support of: ESA; CNES, and
CNRS/INSU-IN2P3-INP (France); ASI, CNR, and INAF (Italy); NASA and DoE
(USA); STFC and UKSA (UK); CSIC, MINECO, JA, and RES (Spain); Tekes,
AoF, and CSC (Finland); DLR and MPG (Germany); CSA (Canada); DTU Space
(Denmark); SER/SSO (Switzerland); RCN (Norway); SFI (Ireland);
FCT/MCTES (Portugal); ERC and PRACE (EU). A description of the \Planck\
Collaboration and a list of its members, indicating which technical or
scientific activities they have been involved in, can be found at
\href{url}{http://www.cosmos.esa.int/web/planck/planck-collaboration}.
This work has received funding from the European Union's Horizon 2020
research and innovation programme under grant agreement numbers
687312, 776282, and 772253. 
We thank Rui Shi for pointing out an error in Fig.~\ref{fig:stat_data}.
\end{acknowledgements}


\bibliographystyle{aat}
\bibliography{Planck_bib,IandS_2016,EB_maps}

\begin{appendix}

\section{\textit{E}- and \textit{B}-mode map reconstruction}
\label{sec:EB_maps}

\begin{figure*}[htbp!]
  \centering
  \includegraphics[width=\textwidth]{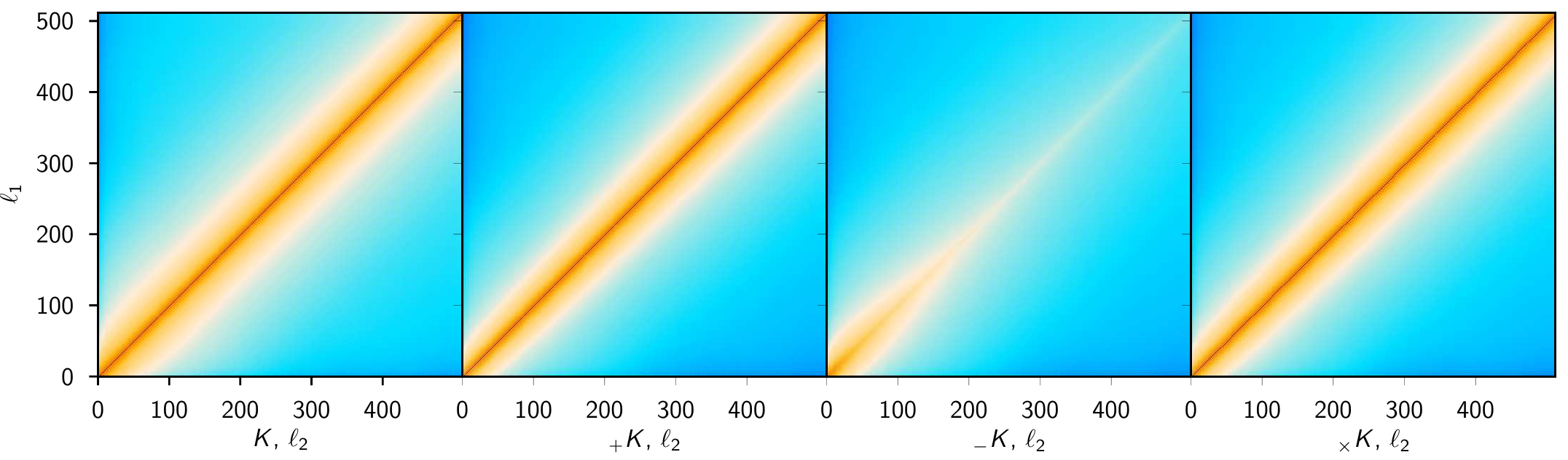}
  \includegraphics[width=\textwidth]{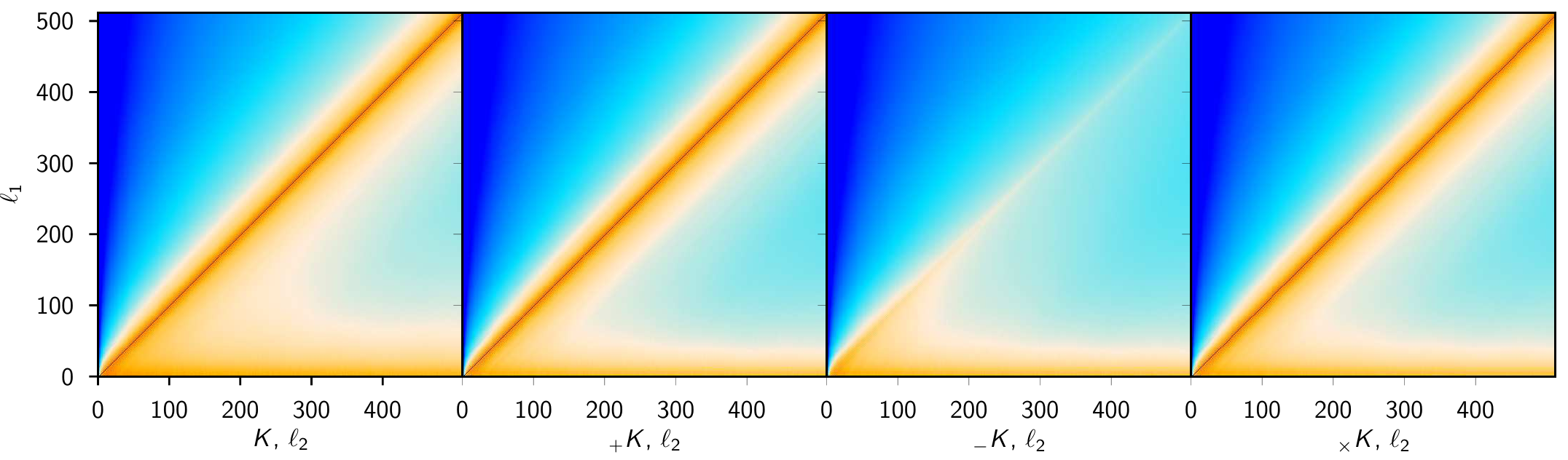}
  \includegraphics[width=\textwidth]{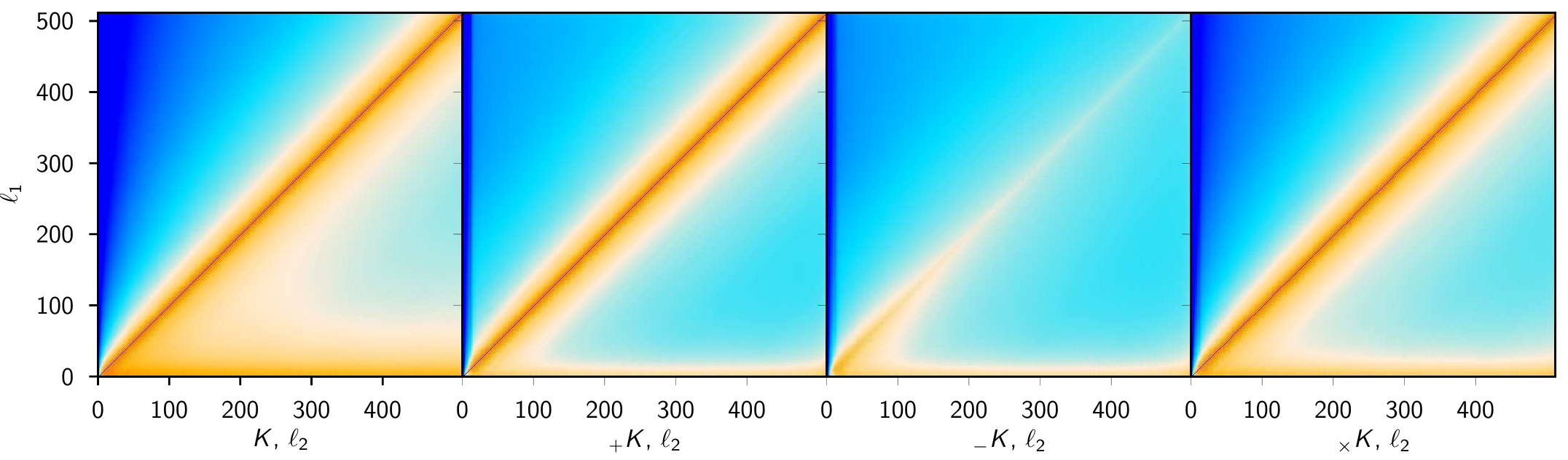}
  \caption{Mode-coupling matrices up to $\ell_{\text{max}}\,{=}\,512$ for the common
    polarization mask with point sources at $N_{\rm side}\,{=}\,1024$ for masking (top),
    inpainting (middle), and purified-inpainting (bottom) methods. The coloured
    shading represents normalized matrix elements
    $K_{\ell_1\ell_2}/\sqrt{K_{\ell_1\ell_1}K_{\ell_2\ell_2}}$
    on a logarithmic scale,
    with values spanning from $10^{-8}$ (deep blue) to $1$ (dark brown).}
  \label{fig:purified:coupling}
\end{figure*}

\begin{figure*}[htbp!]
  \centering
\usetikzlibrary{shapes.misc}
\tikzset{cross/.style={cross out, draw=black, fill=none, minimum size=2*(#1-\pgflinewidth), inner sep=0pt, outer sep=0pt}, cross/.default={2pt}}
\begin{tabular}{ccc}
Nearest neighbours of pixel 0 &
Nearest neighbours of pixel 64 &
Nearest neighbours of pixel 112 \vspace{-8pt}\\
\begin{tikzpicture}[baseline=0, scale=1.25]
\draw (-2.2,-2) rectangle (2.2,2);
\draw[step=1.0,black,dotted] (-2.2,-2) grid (2.2,2);
\draw[rotate around={    89.99999999999993    :(   -.6497964112848062E-01,   -.2410923371574706E-16)},dotted,blue] (   -.6497964112848062E-01,   -.2410923371574706E-16) ellipse (    1.827499469797501     and     1.700464916650725    );
\draw [-> ] (    .6416354467160541E-16,     .1204498186978118E-31) -- (    .4999999999999999    ,    -.8450090191339879E-17);
\draw [->>] (    .6416354467160541E-16,     .1204498186978118E-31) -- (    .5571345448026552E-16,     .5000000000000000    );
\draw [red,fill,ultra thick] (    .6416354467160541E-16,     .1204498186978118E-31) circle [radius=0.25];
\draw [white,ultra thick] (   -.2499999999999999    ,     .5110978055612544E-19) -- (    .2500000000000001    ,    -.5110978055610136E-19);
\draw [-> ] (    1.627804219348088    ,    -.7405419162521586E-16) -- (    2.117305313555149    ,    -.8164137434110681E-16);
\draw [->>] (    1.627804219348088    ,    -.7405419162521586E-16) -- (    1.627804219348088    ,     .4999999999999999    );
\draw [red,fill,ultra thick] (    1.627804219348088    ,    -.7405419162521586E-16) circle [radius=0.25];
\draw [white,ultra thick] (    1.377804219348088    ,    -.7361590927107482E-16) -- (    1.877804219348088    ,    -.7449247397935690E-16);
\draw [-> ] (    .6833716473172202    ,    -.6116515184733697    ) -- (    1.143680625664938    ,    -.7990069489021115    );
\draw [->>] (    .6833716473172202    ,    -.6116515184733697    ) -- (    .8737167920613143    ,    -.1497117522177264    );
\draw [red,fill,ultra thick] (    .6833716473172202    ,    -.6116515184733697    ) circle [radius=0.25];
\draw [white,ultra thick] (    .4520215838980064    ,    -.5169037050108642    ) -- (    .9147217107364340    ,    -.7063993319358752    );
\draw [-> ] (   -.1791494783961806    ,    -1.493501467254341    ) -- (    .1757891233150596E-01,    -1.945817488379659    );
\draw [->>] (   -.1791494783961806    ,    -1.493501467254341    ) -- (    .2803843515768814    ,    -1.302159751071797    );
\draw [red,fill,ultra thick] (   -.1791494783961806    ,    -1.493501467254341    ) circle [radius=0.25];
\draw [white,ultra thick] (   -.2770487127563502    ,    -1.263467252599885    ) -- (   -.8125024403601094E-01,    -1.723535681908798    );
\draw [-> ] (   -.8010169345578867    ,    -.8052107404979804    ) -- (   -.7958221645926090    ,    -1.302606573831314    );
\draw [->>] (   -.8010169345578867    ,    -.8052107404979804    ) -- (   -.3036211012245534    ,    -.8052107404979805    );
\draw [red,fill,ultra thick] (   -.8010169345578867    ,    -.8052107404979804    ) circle [radius=0.25];
\draw [white,ultra thick] (   -.8023224086870024    ,    -.5552141490466204    ) -- (   -.7997114604287709    ,    -1.055207331949340    );
\draw [-> ] (   -1.619031458230720    ,     .2220044572714304E-15) -- (   -2.108641918300164    ,     .2697068589298647E-15);
\draw [->>] (   -1.619031458230720    ,     .2220044572714304E-15) -- (   -1.619031458230720    ,    -.4999999999999998    );
\draw [red,fill,ultra thick] (   -1.619031458230720    ,     .2220044572714304E-15) circle [radius=0.25];
\draw [white,ultra thick] (   -1.369031458230720    ,     .1913882872927466E-15) -- (   -1.869031458230720    ,     .2526206272501142E-15);
\draw [-> ] (   -.8010169345578866    ,     .8052107404979802    ) -- (   -.7958221645926088    ,     1.302606573831314    );
\draw [->>] (   -.8010169345578866    ,     .8052107404979802    ) -- (   -1.298412767891220    ,     .8052107404979802    );
\draw [red,fill,ultra thick] (   -.8010169345578866    ,     .8052107404979802    ) circle [radius=0.25];
\draw [white,ultra thick] (   -.8023224086870023    ,     .5552141490466203    ) -- (   -.7997114604287708    ,     1.055207331949340    );
\draw [-> ] (   -.1791494783961798    ,     1.493501467254341    ) -- (    .1757891233150702E-01,     1.945817488379658    );
\draw [->>] (   -.1791494783961798    ,     1.493501467254341    ) -- (   -.6386833083692416    ,     1.684843183436885    );
\draw [red,fill,ultra thick] (   -.1791494783961798    ,     1.493501467254341    ) circle [radius=0.25];
\draw [white,ultra thick] (   -.2770487127563496    ,     1.263467252599884    ) -- (   -.8125024403601003E-01,     1.723535681908797    );
\draw [-> ] (    .6833716473172202    ,     .6116515184733700    ) -- (    1.143680625664938    ,     .7990069489021119    );
\draw [->>] (    .6833716473172202    ,     .6116515184733700    ) -- (    .4930265025731261    ,     1.073591284729013    );
\draw [red,fill,ultra thick] (    .6833716473172202    ,     .6116515184733700    ) circle [radius=0.25];
\draw [white,ultra thick] (    .4520215838980064    ,     .5169037050108645    ) -- (    .9147217107364340    ,     .7063993319358755    );
\draw (   -.6497964112848062E-01,   -.2410923371574706E-16) node[cross] {};
\end{tikzpicture}
&
\begin{tikzpicture}[baseline=0, scale=1.25]
\draw (-2.2,-2) rectangle (2.2,2);
\draw[step=1.0,black,dotted] (-2.2,-2) grid (2.2,2);
\draw[rotate around={   -59.62923283900606    :(   -.4548153919587753E-01,    .6525651720748378E-02)},dotted,blue] (   -.4548153919587753E-01,    .6525651720748378E-02) ellipse (    2.183416633625807     and     1.254520268414916    );
\draw [-> ] (    .2130110120924014E-15,     .1316771069906447E-15) -- (    .5000000000000002    ,     .1381853093436934E-15);
\draw [->>] (    .2130110120924014E-15,     .1316771069906447E-15) -- (    .2195192144454501E-15,     .5000000000000001    );
\draw [red,fill,ultra thick] (    .2130110120924014E-15,     .1316771069906447E-15) circle [radius=0.25];
\draw [white,ultra thick] (   -.2499999999999998    ,     .1294105815426318E-15) -- (    .2500000000000002    ,     .1339436324386576E-15);
\draw [-> ] (    1.717837974498484    ,    -.5863315557429447    ) -- (    2.204623979791101    ,    -.6190039358527982    );
\draw [->>] (    1.717837974498484    ,    -.5863315557429447    ) -- (    1.757657437757367    ,    -.8873919240684630E-01);
\draw [red,fill,ultra thick] (    1.717837974498484    ,    -.5863315557429447    ) circle [radius=0.25];
\draw [white,ultra thick] (    1.468513127353767    ,    -.5679707116375149    ) -- (    1.967162821643200    ,    -.6046923998483745    );
\draw [-> ] (    .7933741432198518    ,    -.8838548951958675    ) -- (    1.286174874951616    ,    -.9462335148119542    );
\draw [->>] (    .7933741432198518    ,    -.8838548951958675    ) -- (    .8614235464374009    ,    -.3909193859367495    );
\draw [red,fill,ultra thick] (    .7933741432198518    ,    -.8838548951958675    ) circle [radius=0.25];
\draw [white,ultra thick] (    .5455342396194924    ,    -.8510618748506078    ) -- (    1.041214046820211    ,    -.9166479155411271    );
\draw [-> ] (   -.1276504804706333    ,    -1.193355891724496    ) -- (    .3611023801599381    ,    -1.298501128797395    );
\draw [->>] (   -.1276504804706333    ,    -1.193355891724496    ) -- (   -.2250524339773428E-01,    -.7103929785799619    );
\draw [red,fill,ultra thick] (   -.1276504804706333    ,    -1.193355891724496    ) circle [radius=0.25];
\draw [white,ultra thick] (   -.3719941609931803    ,    -1.140477107806033    ) -- (    .1166932000519137    ,    -1.246234675642959    );
\draw [-> ] (   -.8450412803078986    ,    -.3046373700826286    ) -- (   -.3489555773445706    ,    -.3387589039017695    );
\draw [->>] (   -.8450412803078986    ,    -.3046373700826286    ) -- (   -.8131672726684616    ,     .1938212967839353    );
\draw [red,fill,ultra thick] (   -.8450412803078986    ,    -.3046373700826286    ) circle [radius=0.25];
\draw [white,ultra thick] (   -1.094492675624659    ,    -.2880843832064944    ) -- (   -.5955898849911385    ,    -.3211903569587629    );
\draw [-> ] (   -1.734156837676231    ,     .6262514017860016    ) -- (   -1.252425500201356    ,     .7156677993767271    );
\draw [->>] (   -1.734156837676231    ,     .6262514017860016    ) -- (   -1.813412280995283    ,     1.116644041987617    );
\draw [red,fill,ultra thick] (   -1.734156837676231    ,     .6262514017860016    ) circle [radius=0.25];
\draw [white,ultra thick] (   -1.980476575652037    ,     .5835128163680214    ) -- (   -1.487837099700424    ,     .6689899872039817    );
\draw [-> ] (   -.9290852922384335    ,     .9128992496940035    ) -- (   -.4416734880952983    ,     1.014423665745921    );
\draw [->>] (   -.9290852922384335    ,     .9128992496940035    ) -- (   -1.023922471304895    ,     1.399084209892842    );
\draw [red,fill,ultra thick] (   -.9290852922384335    ,     .9128992496940035    ) circle [radius=0.25];
\draw [white,ultra thick] (   -1.174150644510826    ,     .8634728082318871    ) -- (   -.6840199399660407    ,     .9623256911561199    );
\draw [-> ] (   -.1276504804706331    ,     1.193355891724496    ) -- (    .3611023801599383    ,     1.298501128797395    );
\draw [->>] (   -.1276504804706331    ,     1.193355891724496    ) -- (   -.2327957175435321    ,     1.676318804869030    );
\draw [red,fill,ultra thick] (   -.1276504804706331    ,     1.193355891724496    ) circle [radius=0.25];
\draw [white,ultra thick] (   -.3719941609931801    ,     1.140477107806033    ) -- (    .1166932000519139    ,     1.246234675642959    );
\draw [-> ] (    .8430384006825964    ,     .2944040350281708    ) -- (    1.339638160033619    ,     .3152844359524374    );
\draw [->>] (    .8430384006825964    ,     .2944040350281708    ) -- (    .8202597814924873    ,     .7936174425370791    );
\draw [red,fill,ultra thick] (    .8430384006825964    ,     .2944040350281708    ) circle [radius=0.25];
\draw [white,ultra thick] (    .5932783263502897    ,     .2834539085667721    ) -- (    1.092798475014903    ,     .3053541614895695    );
\draw (   -.4548153919587753E-01,    .6525651720748378E-02) node[cross] {};
\end{tikzpicture}
&
\begin{tikzpicture}[baseline=0, scale=1.25]
\draw (-2.2,-2) rectangle (2.2,2);
\draw[step=1.0,black,dotted] (-2.2,-2) grid (2.2,2);
\draw[rotate around={    56.16184145981165    :(    .1628307597040435    ,   -.1293348756602872    )},dotted,blue] (    .1628307597040435    ,   -.1293348756602872    ) ellipse (    1.769690646426870     and     1.369133749541781    );
\draw [-> ] (     .000000000000000    ,      .000000000000000    ) -- (    .4999999999999999    ,    -.3910070252960345E-17);
\draw [->>] (     .000000000000000    ,      .000000000000000    ) -- (   -.3910070252960345E-17,     .5000000000000000    );
\draw [red,fill,ultra thick] (     .000000000000000    ,      .000000000000000    ) circle [radius=0.25];
\draw [white,ultra thick] (   -.2500000000000000    ,    -.1490650653797202E-17) -- (    .2500000000000000    ,     .1490650653797202E-17);
\draw [-> ] (    1.634659006315523    ,    -.1879468429590463E-16) -- (    2.124074285166141    ,    -.2074143831155687E-16);
\draw [->>] (    1.634659006315523    ,    -.1879468429590463E-16) -- (    1.634659006315523    ,     .5000000000000000    );
\draw [red,fill,ultra thick] (    1.634659006315523    ,    -.1879468429590463E-16) circle [radius=0.25];
\draw [white,ultra thick] (    1.384659006315523    ,    -.2010964539949098E-16) -- (    1.884659006315523    ,    -.1747972319231828E-16);
\draw [-> ] (    .8208179001957903    ,    -.6277976881389429    ) -- (    1.317027357016820    ,    -.6563860207350647    );
\draw [->>] (    .8208179001957903    ,    -.6277976881389429    ) -- (    .8534902803056438    ,    -.1302053248028444    );
\draw [red,fill,ultra thick] (    .8208179001957903    ,    -.6277976881389429    ) circle [radius=0.25];
\draw [white,ultra thick] (    .5712915299666677    ,    -.6124161875528862    ) -- (    1.070344270424913    ,    -.6431791887249996    );
\draw [-> ] (   -.7544720669735561E-01,    -1.149042517301100    ) -- (    .4202828556144734    ,    -1.214072624639809    );
\draw [->>] (   -.7544720669735561E-01,    -1.149042517301100    ) -- (   -.1041709935864622E-01,    -.6586498770994844    );
\draw [red,fill,ultra thick] (   -.7544720669735561E-01,    -1.149042517301100    ) circle [radius=0.25];
\draw [white,ultra thick] (   -.3233007814758349    ,    -1.116352985803557    ) -- (    .1724063680811236    ,    -1.181732048798642    );
\draw [-> ] (   -.9550251968111414    ,    -1.108293015180177    ) -- (   -.4635479042620615    ,    -1.186059009167469    );
\draw [->>] (   -.9550251968111414    ,    -1.108293015180177    ) -- (   -.8854164749204181    ,    -.6193166270665328    );
\draw [red,fill,ultra thick] (   -.9550251968111414    ,    -1.108293015180177    ) circle [radius=0.25];
\draw [white,ultra thick] (   -1.202247929840211    ,    -1.071132279535716    ) -- (   -.7078024637820720    ,    -1.145453750824637    );
\draw [-> ] (   -.8677291188689022    ,     .7361400989787896E-01) -- (   -.3708027131234385    ,     .7883667802696601E-01);
\draw [->>] (   -.8677291188689022    ,     .7361400989787896E-01) -- (   -.8724039546767564    ,     .5735648357776665    );
\draw [red,fill,ultra thick] (   -.8677291188689022    ,     .7361400989787896E-01) circle [radius=0.25];
\draw [white,ultra thick] (   -1.117716797870000    ,     .7113200513479868E-01) -- (   -.6177414398678042    ,     .7609601466095924E-01);
\draw [-> ] (   -.7544720669735643E-01,     1.149042517301100    ) -- (    .4202828556144725    ,     1.214072624639809    );
\draw [->>] (   -.7544720669735643E-01,     1.149042517301100    ) -- (   -.1404773140360658    ,     1.639435157502715    );
\draw [red,fill,ultra thick] (   -.7544720669735643E-01,     1.149042517301100    ) circle [radius=0.25];
\draw [white,ultra thick] (   -.3233007814758356    ,     1.116352985803558    ) -- (    .1724063680811228    ,     1.181732048798642    );
\draw [-> ] (    .8208179001957904    ,     .6277976881389428    ) -- (    1.317027357016820    ,     .6563860207350646    );
\draw [->>] (    .8208179001957904    ,     .6277976881389428    ) -- (    .7881455200859369    ,     1.125390051475041    );
\draw [red,fill,ultra thick] (    .8208179001957904    ,     .6277976881389428    ) circle [radius=0.25];
\draw [white,ultra thick] (    .5712915299666679    ,     .6124161875528861    ) -- (    1.070344270424913    ,     .6431791887249995    );
\draw (    .1628307597040435    ,   -.1293348756602872    ) node[cross] {};
\end{tikzpicture}
\medskip\\
Nearest neighbours of pixel 0 &
Nearest neighbours of pixel 1048576 &
Nearest neighbours of pixel 2095104 \\
\begin{tikzpicture}[baseline=0, scale=1.25]
\draw (-2.2,-2) rectangle (2.2,2);
\draw[step=1.0,black,dotted] (-2.2,-2) grid (2.2,2);
\draw[rotate around={    89.99999999999993    :(   -.6859552884997133E-01,   -.7232770114724117E-16)},dotted,blue] (   -.6859552884997133E-01,   -.7232770114724117E-16) ellipse (    1.802347634847133     and     1.682089455846366    );
\draw [-> ] (    .6484720107270608E-16,     .1204498186978118E-31) -- (    .5000000000000001    ,    -.3551860784032141E-17);
\draw [->>] (    .6484720107270608E-16,     .1204498186978118E-31) -- (    .6129534028867393E-16,     .5000000000000000    );
\draw [red,fill,ultra thick] (    .6484720107270608E-16,     .1204498186978118E-31) circle [radius=0.25];
\draw [white,ultra thick] (   -.2499999999999999    ,     .3119493454215119E-23) -- (    .2500000000000001    ,    -.3119493430125155E-23);
\draw [-> ] (    1.595771023913903    ,    -.1952664383111239E-15) -- (    2.095770388130722    ,    -.1845526039508046E-15);
\draw [->>] (    1.595771023913903    ,    -.1952664383111239E-15) -- (    1.595771023913903    ,     .4999999999999998    );
\draw [red,fill,ultra thick] (    1.595771023913903    ,    -.1952664383111239E-15) circle [radius=0.25];
\draw [white,ultra thick] (    1.345771023913903    ,    -.1952664118413274E-15) -- (    1.845771023913903    ,    -.1952664647809203E-15);
\draw [-> ] (    .6764142903993017    ,    -.6106744638268438    ) -- (    1.138353958204272    ,    -.8020159367058147    );
\draw [->>] (    .6764142903993017    ,    -.6106744638268438    ) -- (    .8677559457559531    ,    -.1487346975712004    );
\draw [red,fill,ultra thick] (    .6764142903993017    ,    -.6106744638268438    ) circle [radius=0.25];
\draw [white,ultra thick] (    .4454443839944227    ,    -.5150036619313780    ) -- (    .9073841968041807    ,    -.7063452657223095    );
\draw [-> ] (   -.1872096728322434    ,    -1.474299587329184    ) -- (    .4132375003536248E-02,    -1.936238766198039    );
\draw [->>] (   -.1872096728322434    ,    -1.474299587329184    ) -- (    .2747299465767032    ,    -1.282957871146639    );
\draw [red,fill,ultra thick] (   -.1872096728322434    ,    -1.474299587329184    ) circle [radius=0.25];
\draw [white,ultra thick] (   -.2828806665921943    ,    -1.243329760397215    ) -- (   -.9153867907229242E-01,    -1.705269414261152    );
\draw [-> ] (   -.7978847510334268    ,    -.7978850046742388    ) -- (   -.7978844331420386    ,    -1.297884845728519    );
\draw [->>] (   -.7978847510334268    ,    -.7978850046742388    ) -- (   -.2978849099791462    ,    -.7978850046742390    );
\draw [red,fill,ultra thick] (   -.7978847510334268    ,    -.7978850046742388    ) circle [radius=0.25];
\draw [white,ultra thick] (   -.7978848305062991    ,    -.5478850046742515    ) -- (   -.7978846715605545    ,    -1.047885004674226    );
\draw [-> ] (   -1.595770516630908    ,     .4267184198694263E-16) -- (   -2.095769880848132    ,     .8547601423806917E-16);
\draw [->>] (   -1.595770516630908    ,     .4267184198694263E-16) -- (   -1.595770516630908    ,    -.4999999999999999    );
\draw [red,fill,ultra thick] (   -1.595770516630908    ,     .4267184198694263E-16) circle [radius=0.25];
\draw [white,ultra thick] (   -1.345770516630908    ,     .1205567200825880E-16) -- (   -1.845770516630908    ,     .7328801196562647E-16);
\draw [-> ] (   -.7978847510334267    ,     .7978850046742386    ) -- (   -.7978844331420384    ,     1.297884845728519    );
\draw [->>] (   -.7978847510334267    ,     .7978850046742386    ) -- (   -1.297884592087707    ,     .7978850046742386    );
\draw [red,fill,ultra thick] (   -.7978847510334267    ,     .7978850046742386    ) circle [radius=0.25];
\draw [white,ultra thick] (   -.7978848305062991    ,     .5478850046742513    ) -- (   -.7978846715605543    ,     1.047885004674226    );
\draw [-> ] (   -.1872096728322427    ,     1.474299587329183    ) -- (    .4132375003537192E-02,     1.936238766198038    );
\draw [->>] (   -.1872096728322427    ,     1.474299587329183    ) -- (   -.6491492922411891    ,     1.665641303511728    );
\draw [red,fill,ultra thick] (   -.1872096728322427    ,     1.474299587329183    ) circle [radius=0.25];
\draw [white,ultra thick] (   -.2828806665921937    ,     1.243329760397214    ) -- (   -.9153867907229163E-01,     1.705269414261151    );
\draw [-> ] (    .6764142903993017    ,     .6106744638268438    ) -- (    1.138353958204272    ,     .8020159367058148    );
\draw [->>] (    .6764142903993017    ,     .6106744638268438    ) -- (    .4850726350426502    ,     1.072614230082487    );
\draw [red,fill,ultra thick] (    .6764142903993017    ,     .6106744638268438    ) circle [radius=0.25];
\draw [white,ultra thick] (    .4454443839944228    ,     .5150036619313780    ) -- (    .9073841968041806    ,     .7063452657223095    );
\draw (   -.6859552884997133E-01,   -.7232770114724117E-16) node[cross] {};
\end{tikzpicture}
&
\begin{tikzpicture}[baseline=0, scale=1.25]
\draw (-2.2,-2) rectangle (2.2,2);
\draw[step=1.0,black,dotted] (-2.2,-2) grid (2.2,2);
\draw[rotate around={    63.32308002257355    :(   -.3440347691103194E-03,   -.3814512480857008E-04)},dotted,blue] (   -.3440347691103194E-03,   -.3814512480857008E-04) ellipse (    2.052517867352314     and     1.299219106541133    );
\draw [-> ] (     .000000000000000    ,      .000000000000000    ) -- (    .5000000000000000    ,    -.3937895625168156E-17);
\draw [->>] (     .000000000000000    ,      .000000000000000    ) -- (   -.3937895625168156E-17,     .5000000000000000    );
\draw [red,fill,ultra thick] (     .000000000000000    ,      .000000000000000    ) circle [radius=0.25];
\draw [white,ultra thick] (   -.2500000000000000    ,     .2620253001098478E-17) -- (    .2500000000000000    ,    -.2620253001098478E-17);
\draw [-> ] (    1.666765454793028    ,     .4556751909787970    ) -- (    2.166764643915447    ,     .4560172522922979    );
\draw [->>] (    1.666765454793028    ,     .4556751909787970    ) -- (    1.666423014667142    ,     .9556750221312513    );
\draw [red,fill,ultra thick] (    1.666765454793028    ,     .4556751909787970    ) circle [radius=0.25];
\draw [white,ultra thick] (    1.416765513360900    ,     .4555040654913842    ) -- (    1.916765396225157    ,     .4558463164662098    );
\draw [-> ] (    .8333035910578707    ,    -.3720430860729298    ) -- (    1.333303339320329    ,    -.3723228726129855    );
\draw [->>] (    .8333035910578707    ,    -.3720430860729298    ) -- (    .8335835323287505    ,     .1279568010880539    );
\draw [red,fill,ultra thick] (    .8333035910578707    ,    -.3720430860729298    ) circle [radius=0.25];
\draw [white,ultra thick] (    .5833036302197928    ,    -.3719031540911000    ) -- (    1.083303551895948    ,    -.3721830180547596    );
\draw [-> ] (   -.1084822391616451E-02,    -1.199968502672228    ) -- (    .4989143603167047    ,    -1.200872543997078    );
\draw [->>] (   -.1084822391616451E-02,    -1.199968502672228    ) -- (   -.1807810667662568E-03,    -.6999696794716863    );
\draw [red,fill,ultra thick] (   -.1084822391616451E-02,    -1.199968502672228    ) circle [radius=0.25];
\draw [white,ultra thick] (   -.2510844137456301    ,    -1.199516481847297    ) -- (    .2489147689623972    ,    -1.200420523497158    );
\draw [-> ] (   -.8338204924230398    ,    -.8279829871605943    ) -- (   -.3338210561046114    ,    -.8286079084279160    );
\draw [->>] (   -.8338204924230398    ,    -.8279829871605943    ) -- (   -.8331959158989199    ,    -.3279835488509144    );
\draw [red,fill,ultra thick] (   -.8338204924230398    ,    -.8279829871605943    ) circle [radius=0.25];
\draw [white,ultra thick] (   -1.083820297267238    ,    -.8276706126050430    ) -- (   -.5838206875788419    ,    -.8282953617161455    );
\draw [-> ] (   -1.666660017116983    ,    -.4559040621761911    ) -- (   -1.166660829209751    ,    -.4562487793371373    );
\draw [->>] (   -1.666660017116983    ,    -.4559040621761911    ) -- (   -1.666315679816316    ,     .4409576710019392E-01);
\draw [red,fill,ultra thick] (   -1.666660017116983    ,    -.4559040621761911    ) circle [radius=0.25];
\draw [white,ultra thick] (   -1.916659957767381    ,    -.4557317984323792    ) -- (   -1.416660076466585    ,    -.4560763259200030    );
\draw [-> ] (   -.8334075900287448    ,     .3721365071477345    ) -- (   -.3334078422437296    ,     .3724173777512109    );
\draw [->>] (   -.8334075900287448    ,     .3721365071477345    ) -- (   -.8336883056875325    ,     .8721363936835725    );
\draw [red,fill,ultra thick] (   -.8334075900287448    ,     .3721365071477345    ) circle [radius=0.25];
\draw [white,ultra thick] (   -1.083407550606334    ,     .3719961105529674    ) -- (   -.5834076294511555    ,     .3722769037425016    );
\draw [-> ] (   -.1084822391635283E-02,     1.199968502672774    ) -- (    .4989143603166859    ,     1.200872543997624    );
\draw [->>] (   -.1084822391635283E-02,     1.199968502672774    ) -- (   -.1988863716485890E-02,     1.699967325873315    );
\draw [red,fill,ultra thick] (   -.1084822391635283E-02,     1.199968502672774    ) circle [radius=0.25];
\draw [white,ultra thick] (   -.2510844137456489    ,     1.199516481847843    ) -- (    .2489147689623784    ,     1.200420523497704    );
\draw [-> ] (    .8328923855791276    ,     .8277751311593604    ) -- (    1.332891824433054    ,     .8283976404476779    );
\draw [->>] (    .8328923855791276    ,     .8277751311593604    ) -- (    .8322695320234440    ,     1.327774572563746    );
\draw [red,fill,ultra thick] (    .8328923855791276    ,     .8277751311593604    ) circle [radius=0.25];
\draw [white,ultra thick] (    .5828925794454130    ,     .8274637903411727    ) -- (    1.082892191712842    ,     .8280864719775481    );
\draw (   -.3440347691103194E-03,   -.3814512480857008E-04) node[cross] {};
\end{tikzpicture}
&
\begin{tikzpicture}[baseline=0, scale=1.25]
\draw (-2.2,-2) rectangle (2.2,2);
\draw[step=1.0,black,dotted] (-2.2,-2) grid (2.2,2);
\draw[rotate around={    58.81097486047001    :(    .2180196175041219    ,   -.1429025734695809    )},dotted,blue] (    .2180196175041219    ,   -.1429025734695809    ) ellipse (    1.800232568627389     and     1.353192987957232    );
\draw [-> ] (     .000000000000000    ,      .000000000000000    ) -- (    .5000000000000000    ,    -.3261531376368050E-19);
\draw [->>] (     .000000000000000    ,      .000000000000000    ) -- (   -.3261531376368050E-19,     .5000000000000000    );
\draw [red,fill,ultra thick] (     .000000000000000    ,      .000000000000000    ) circle [radius=0.25];
\draw [white,ultra thick] (   -.2500000000000000    ,    -.1161359763899660E-19) -- (    .2500000000000000    ,     .1161359763899660E-19);
\draw [-> ] (    1.746716593220663    ,    -.4532377930234244E-16) -- (    2.246715831470400    ,    -.4534445417357750E-16);
\draw [->>] (    1.746716593220663    ,    -.4532377930234244E-16) -- (    1.746716593220663    ,     .4999999999999999    );
\draw [red,fill,ultra thick] (    1.746716593220663    ,    -.4532377930234244E-16) circle [radius=0.25];
\draw [white,ultra thick] (    1.496716593220663    ,    -.4533023793348512E-16) -- (    1.996716593220663    ,    -.4531732067119976E-16);
\draw [-> ] (    .8735516970976026    ,    -.5725038033098022    ) -- (    1.373551441211597    ,    -.5727592170780554    );
\draw [->>] (    .8735516970976026    ,    -.5725038033098022    ) -- (    .8738073605371836    ,    -.7250395037836110E-01);
\draw [red,fill,ultra thick] (    .8735516970976026    ,    -.5725038033098022    ) circle [radius=0.25];
\draw [white,ultra thick] (    .6235517297476116    ,    -.5723760339730452    ) -- (    1.123551664447594    ,    -.5726315726465593    );
\draw [-> ] (   -.5850165870479454E-03,    -1.144114340214323    ) -- (    .4994147219577746    ,    -1.144625666943084    );
\draw [->>] (   -.5850165870479454E-03,    -1.144114340214323    ) -- (   -.7368985828639643E-04,    -.6441149284884716    );
\draw [red,fill,ultra thick] (   -.5850165870479454E-03,    -1.144114340214323    ) circle [radius=0.25];
\draw [white,ultra thick] (   -.2505848858594165    ,    -1.143858676766386    ) -- (    .2494148526853207    ,    -1.144370003662259    );
\draw [-> ] (   -.8745393582871775    ,    -1.143779346451262    ) -- (   -.3745398109510659    ,    -1.144291422437768    );
\draw [->>] (   -.8745393582871775    ,    -1.143779346451262    ) -- (   -.8740277816432998    ,    -.6437799353005997    );
\draw [red,fill,ultra thick] (   -.8745393582871775    ,    -1.143779346451262    ) circle [radius=0.25];
\draw [white,ultra thick] (   -1.124539227303919    ,    -1.143523433161211    ) -- (   -.6245394892704359    ,    -1.144035259741314    );
\draw [-> ] (   -.8739536559215033    ,     .5587586945266210E-03) -- (   -.3739538466191853    ,     .5590088538756826E-03);
\draw [->>] (   -.8739536559215033    ,     .5587586945266210E-03) -- (   -.8739539058369136    ,     .5005587586943862    );
\draw [red,fill,ultra thick] (   -.8739536559215033    ,     .5587586945266210E-03) circle [radius=0.25];
\draw [white,ultra thick] (   -1.123953655921472    ,     .5586336758129617E-03) -- (   -.6239536559215346    ,     .5588837132402803E-03);
\draw [-> ] (   -.5850165871533887E-03,     1.144114340214411    ) -- (    .4994147219576692    ,     1.144625666943173    );
\draw [->>] (   -.5850165871533887E-03,     1.144114340214411    ) -- (   -.1096343315914977E-02,     1.644113751940262    );
\draw [red,fill,ultra thick] (   -.5850165871533887E-03,     1.144114340214411    ) circle [radius=0.25];
\draw [white,ultra thick] (   -.2505848858595220    ,     1.143858676766475    ) -- (    .2494148526852152    ,     1.144370003662348    );
\draw [-> ] (    .8735516970975922    ,     .5725038033098021    ) -- (    1.373551441211587    ,     .5727592170780553    );
\draw [->>] (    .8735516970975922    ,     .5725038033098021    ) -- (    .8732960336580112    ,     1.072503656241243    );
\draw [red,fill,ultra thick] (    .8735516970975922    ,     .5725038033098021    ) circle [radius=0.25];
\draw [white,ultra thick] (    .6235517297476012    ,     .5723760339730450    ) -- (    1.123551664447583    ,     .5726315726465592    );
\draw (    .2180196175041219    ,   -.1429025734695809    ) node[cross] {};
\end{tikzpicture}
\\
\end{tabular}

  \caption{
    Examples of the finite-difference stencils for \healpix\ pixels
    with eight (left and centre) and seven (right) neighbours for $N_{\rm side}=8$ (top)
    and $N_{\rm side}=1024$ (bottom) in ring ordering \citep[see also][]{Bowyer2011}. Red circles represent positions
    of \healpix\ pixel centres in a gnomonic projection onto a plane tangent to the central
    pixel (i.e., looking straight down at the tangent plane), with the dotted grid aligned with local $\hat{\vec{x}}$ and $\hat{\vec{y}}$
    directions, illustrating the average pixel pitch $h = (\pi/3)^{1/2}\!/$\nside. The white bars represent the directions of polarization, specifically
the direction of the polarization ellipse for the $+Q$ polarization mode.
Single- and
    double-arrow vectors show projections of $\hat{\vec\theta}$ and $\hat{\vec\phi}$
    directions, respectively, for neighbouring pixels onto the tangent plane.
    The black cross corresponds to the average pixel position, while the blue dotted
    ellipse represents the pixel position covariance.  Although for some
positions on the sky, the polarization directions are aligned, this is not at
all true near the poles (pixel~0); hence just adding $Q$ and $U$ does not
make sense.  Additionally we can see that because the grid is distorted,
second-order finite-difference schemes need more than just nearest neighbours to work.}
  \label{fig:purified:stencils}
\end{figure*}

\subsection{Introduction and notation}
\label{sec:EB_maps:into}

\renewcommand{\a}{\vec{a}}
\renewcommand{\b}{\vec{b}}
\newcommand{\e}{\vec{e}}
\newcommand{\p}{\vec{p}}

\renewcommand{\u}{\vec{u}}
\renewcommand{\v}{\vec{v}}
\renewcommand{\w}{\vec{w}}

\renewcommand{\B}{\tens{B}}
\renewcommand{\E}{\tens{E}}
\renewcommand{\F}{\tens{F}}
\newcommand{\bbK}{\tens{K}}
\renewcommand{\M}{\tens{M}}

\newcommand{\norm}[2]{\langle\vec{#1}\cdot\vec{#2}\rangle}

Polarization of the CMB is usually measured on the sky in terms of the
Stokes parameters $Q$ and $U$ \citep[see e.g.,][]{HuWhite1997}.
However, these are not scalar
quantities and therefore not rotationally invariant. Nevertheless,
scalar $E$ and pseudo-scalar $B$ maps can be determined from the measured
quantities, as described in Sect.~\ref{sec:polarization_preamble}.
Such maps offer definite advantages for any map-based analyses, such
as those presented in the main body of this paper. Since 
they are generated via the application of non-local spherical harmonic
transformations, the $E$ and $B$ estimators are non-trivial to
construct in the typical case of partial sky coverage, resulting from
the need to mask out strong foreground contributions in the data.

Alternative sets of related scalars are occasionally used in the literature
\citep{Zaldarriaga1997, Bunn:2002df}, defined by
\begin{align}
  a^{(\mathcal{E},\mathcal{B})}_{\lm} &\equiv \sqrt{\frac{(\ell+2)!}{(\ell-2)!}}\ a^{(E,B)}_{\lm}, \nonumber\\
  {\rm and}\quad a^{(\psi_E,\psi_B)}_{\lm} &\equiv -\sqrt{\frac{(\ell-2)!}{(\ell+2)!}}\ a^{(E,B)}_{\lm}.
\end{align}
They are related to the standard $E$ and $B$ estimators in harmonic
space by a factor of
\begin{equation} \label{purified:bilaplacian}
  (\ell+2)(\ell+1)\ell(\ell-1) = \ell(\ell+1)\big[ \ell(\ell+1) - 2 \big],
\end{equation}
which corresponds to an application of the bi-Laplacian operator such that
$\mathcal{E},\mathcal{B} = -\nabla^2(\nabla^2+2)\, \psi_{E,B}$ in real
space. Unlike $E$ and $B$, $\mathcal{E}$ and $\mathcal{B}$ maps can be
derived from $Q$ and $U$ by a (local) second-derivative operator,
although the noise power at high $\ell$ is then significantly enhanced.

The central problem for the reconstruction of $E$- and $B$-mode maps
from partial sky coverage is that neither spin-0 nor spin-2 spherical
harmonics are orthogonal under the masked inner product,
\begin{equation}
  \norm{u}{v} \equiv \u^{\sf T}\cdot\M\cdot\v = \sum\limits_{i} M_{ii} u_i v_i,
\end{equation}
and thus mode mixing generally occurs. Here, we introduce a vector
notation for the polarization map, $\p \equiv (Q_i,U_i)$, and the
masking operator, $\M \equiv \text{diag}(M_i)$, which multiplies an
individual map pixel $i$ by a mask value $M_i$. To keep the notation
concise, we will denote all other linear operators similarly, e.g.,\
$\E$ and $\B$ correspond to the projection of the polarization map
onto its $E$- and $B$-mode components, although numerically this would
be implemented using spherical harmonic transforms rather than matrix
multiplications.

Mode mixing is a well-known problem for the estimation of power
spectra on a masked sky \citep[see, for example,
][]{2011MNRAS.414..823R}. In this case, the effect of masking results
in an estimated power spectrum that is a linear combination of the
full-sky quantities. For an isotropic Gaussian random field with
an ensemble average spectrum $\langle C_\ell \rangle$, the
masked sky mode averages $\langle \tilde{C}_\ell \rangle$ are
given by
\begin{align}
\langle \tilde{C}^{TT}_{\ell_1} \rangle &=
  \frac{1}{2\ell_1 + 1} \sum\limits_{\ell_2} K_{\ell_1 \ell_2} \langle C^{TT}_{\ell_2} \rangle,\nonumber\\
\langle \tilde{C}^{EE}_{\ell_1} \rangle &=
  \frac{1}{2\ell_1 + 1} \sum\limits_{\ell_2} {_{+}}K_{\ell_1 \ell_2} \langle C^{EE}_{\ell_2}\rangle
  + {_{-}}K_{\ell_1 \ell_2} \langle C^{BB}_{\ell_2}\rangle,\nonumber\\
\langle \tilde{C}^{BB}_{\ell_1} \rangle &=
  \frac{1}{2\ell_1 + 1} \sum\limits_{\ell_2} {_{-} }K_{\ell_1 \ell_2}\langle C^{EE}_{\ell_2}\rangle
  + {_{+}}K_{\ell_1 \ell_2} \langle C^{BB}_{\ell_2}\rangle,\nonumber\\
\langle \tilde{C}^{TE}_{\ell_1} \rangle &=
  \frac{1}{2\ell_1 + 1} \sum\limits_{\ell_2} {_{\times}}K_{\ell_1 \ell_2} \langle C^{TE}_{\ell_2} \rangle,
\end{align}
where the coupling matrices $K_{\ell_1 \ell_2}$ depend on the method used,
as illustrated in Fig.~\ref{fig:purified:coupling}.

For reconstruction of low-$\ell$ multipoles, maximum-likelihood estimators are widely used
\citep{Efstathiou:2003tv, Bielewicz:2004en, deOliveiraCosta:2006zj, Feeney:2011bi}. Filling
in the missing data in the CMB maps can be done using various statistical priors
\citep{Abrial:2008mz, Bucher:2011nf, Starck:2012gd, Nishizawa:2013uwa}, in particularly
using constrained Gaussian realizations. Methods targeting decomposition of polarization
into pure $E$ and $B$ modes plus ambiguous components have already been developed
\citep{Bunn:2002df, Bunn:2010kf, Bunn:2016lxi}. Some of the approaches discussed in the literature require solving
large linear algebra problems (typically via iterative solvers), and could be expensive on
high-resolution maps. With the large number of simulations that need to be processed for \Planck\ 
data analysis, numerical performance becomes a very important issue.

We consider three direct approaches to the computation of $E$- and $B$-mode maps
in the case of incomplete sky coverage in Sects.~\ref{sec:EB_maps:masking},
\ref{sec:EB_maps:inpainting}, and \ref{sec:EB_maps:purified}. The suitability of
these approaches for our purposes depends upon the uniformity of the reconstruction,
since the method-specific residuals are generally quite inhomogeneous and dependent
on the mask.

\subsection{Masking}
\label{sec:EB_maps:masking}

The simplest method to implement is the direct computation of  $E$-
and $B$-modes from the $Q$ and $U$ data after the application of a mask
that zeros the problematic pixels.
As a consequence, $E$- and $B$-mode mixing does result, with mode
coupling matrices expressible analytically in terms of Wigner-3j symbols
and the power spectrum of the mask ${\cal W}_\ell = \sum_m|W_{\ell m}|^2/(2\ell+1)$ defined by a window function $W(\hat{\vec{n}})$:
\begin{align}
K_{\ell_1 \ell_2} =&\, \frac{1}{4\pi} (2\ell_1 + 1)(2\ell_2 + 1) \sum\limits_{\ell_3} (2\ell_3 + 1)\nonumber\\
  &\ \times\left(\begin{array}{ccc} \ell_1 & \ell_2 & \ell_3 \\ 0 & 0 & 0 \end{array}\right)^2 {\cal W}_{\ell_3};\nonumber\\
_{\pm}K_{\ell_1 \ell_2} =&\, \frac{1}{8\pi} (2\ell_1 + 1)(2\ell_2 + 1) \sum\limits_{\ell_3} (2\ell_3 + 1)\nonumber\\
  &\ \times\left(\begin{array}{ccc} \ell_1 & \ell_2 & \ell_3 \\ 2 & -2 & 0 \end{array}\right)^2 \,\frac{1 \pm (-1)^{(\ell_1 + \ell_2 + \ell_3)}}{2}\ {\cal W}_{\ell_3};\nonumber\\
_{\times}K_{\ell_1 \ell_2} =&\, \frac{1}{4\pi} (2\ell_1 + 1)(2\ell_2 + 1) \sum\limits_{\ell_3} (2\ell_3 + 1)\nonumber\\
  &\ \times\left(\begin{array}{ccc} \ell_1 & \ell_2 & \ell_3 \\ 0 & 0 & 0 \end{array}\right)
  \left(\begin{array}{ccc} \ell_1 & \ell_2 & \ell_3 \\ 2 & -2 & 0 \end{array}\right) \nonumber\\
  &\ \times\,\frac{1 + (-1)^{(\ell_1 + \ell_2 + \ell_3)}}{2}\ {\cal W}_{\ell_3}.
\end{align}
The resultant mode-coupling matrices are symmetric and are shown in
Fig.~\ref{fig:purified:coupling} for the common polarization mask at
$N_{\rm side}=1024$ resolution. In practice, it is often faster to evaluate
these matrices using Monte Carlo methods rather than explicit summation
involving Wigner symbols.

As an alternative, if one masks ${\cal E}$ and ${\cal B}$ maps
instead, there is no mode mixing between $E$ and $B$. Unfortunately,
second-order derivative operators enhance the noise power, which
results in large artefacts in the reconstruction due to high-to-low-$\ell$
mode-coupling in the masking operator, unless extremely strong
apodization is applied to the mask
\citep[as described in][]{2006PhRvD..74h3002S}.

\subsection{Simple inpainting}
\label{sec:EB_maps:inpainting}

\subsubsection{Overview}

The application of a diffusive inpainting procedure to the masked
pixels of input sky maps has proven to be a satisfactory approach to
handle incomplete sky coverage when searching for evidence of
primordial non-Gaussianity in the \Planck\ temperature and polarization
data \citep{planck2013-p09a,planck2014-a19}.  To be more explicit, the
procedure works by replacing each masked pixel by the average of all
nearest-neighbour pixels, then the process is repeated over a large number of
iterations. Such an approach is straightforward to implement, but the
convergence rate of the inpainted solution is slow for the largest
scales.

To address this,
we adopt slightly improved finite-difference ``Laplacian stencils,''
as detailed in Sect.~\ref{sec:EB_maps:stencils}, and we further develop the
details of the finite-differencing multigrid approach in
Sect.~\ref{sec:EB_maps:multigrid}. 

One particular aspect of our improved method is that
when computing the average over the nearest neighbours of a given
masked pixel, their contributions are Gaussian-weighted by the distance
to their positions in the tangent plane via gnomonic projection.
In addition, basis orientation differences between the $Q$ and $U$
polarization components are properly projected (via parallel transport
in the tangent plane) to form a tensor Laplacian stencil.
We improve on the speed of the inpainting algorithm by noting that an
infinite number of relaxation iterations converges to the solution of
an elliptical Laplace equation $\nabla^2 T = 0$ with the Dirichlet
boundary conditions given by the unmasked pixels; this can be solved
in $O(n \log n)$ operations using the standard multi-grid methods of
\citet{Brandt:1977kr} adapted to the spherical \healpix\
pixelization, as described in Sect.~\ref{sec:EB_maps:multigrid}.

Mode-coupling matrices for inpainted temperature maps have been
presented in \citet{2017PhRvD..95d3532G}. Here, we extend the results
to polarization, as shown in Fig.~\ref{fig:purified:coupling}, where
the matrices have been computed for the common polarization mask at
$N_{\rm side}=1024$ resolution.
Unlike for the case of masking, the inpainting mode-coupling matrices
are not symmetric and result in excellent suppression of the low-to-high-$\ell$
mode mixing, at the expense of increased high-to-low-$\ell$ mode mixing.
Similarly to the case of pure ${\cal E}$- and ${\cal B}$- mode
masking, high-to-low-$\ell$ mixing renders the inpainting of $Q$ and $U$ maps
susceptible to the transfer of noise artefacts into
low-$\ell$ patterns, albeit to a lesser extent.

\begin{figure*}[htbp!]
  \centering
\begin{tikzpicture}[scale=0.3,place/.style={circle,draw=blue!50,fill=blue!20,thick}]
  \draw[dotted] (-3,12) -- (33,12) node[right] {$N_{\text{side}} = 128$};
  \draw[dotted] (-3,9) -- (33,9) node[right] {$N_{\text{side}} = 64$};
  \draw[dotted] (-3,6) -- (33,6) node[right] {$N_{\text{side}} = 32$};
  \draw[dotted] (-3,3) -- (33,3) node[right] {$N_{\text{side}} = 16$};
  \draw[dotted] (-3,0) -- (33,0) node[right] {$N_{\text{side}} = 8$};
  \draw[->] (-3,-2) -- (33,-2) node[midway,below] {iterations};
  \draw
  (0,12) node[place] (A1) {} --
  (1,9) node[place] (B1) {} --
  (2,6) node[place] (C1) {} --
  (3,3) node[place] (D1) {} --
  (4,0) node[place] {} --
  (5,3) node[place] (D2) {} --
  (6,0) node[place] {} --
  (7,3) node[place] (D3) {} --
  (8,6) node[place] (C2) {} --
  (9,3) node[place] (D4) {} --
  (10,0) node[place] {} --
  (11,3) node[place] (D5) {} --
  (12,0) node[place] {} --
  (13,3) node[place] (D6) {} --
  (14,6) node[place] (C3) {} --
  (15,9) node[place] (B2) {} --
  (16,6) node[place] (C4) {} --
  (17,3) node[place] (D7) {} --
  (18,0) node[place] {} --
  (19,3) node[place] (D8) {} --
  (20,0) node[place] {} --
  (21,3) node[place] (D9) {} --
  (22,6) node[place] (C5) {} --
  (23,3) node[place] (DA) {} --
  (24,0) node[place] {} --
  (25,3) node[place] (DB) {} --
  (26,0) node[place] {} --
  (27,3) node[place] (DC) {} --
  (28,6) node[place] (C6) {} --
  (29,9) node[place] (B3) {} --
  (30,12) node[place] (A2) {};
  \draw[->,very thin] (A1) to [out=5,in=175] (A2);
  \draw[->,very thin] (B1) to [out=10,in=170] (B2);
  \draw[->,very thin] (B2) to [out=10,in=170] (B3);
  \draw[->,very thin] (C1) to [out=20,in=160] (C2);
  \draw[->,very thin] (C2) to [out=20,in=160] (C3);
  \draw[->,very thin] (C4) to [out=20,in=160] (C5);
  \draw[->,very thin] (C5) to [out=20,in=160] (C6);
  \draw[->,very thin] (D1) to [out=40,in=140] (D2);
  \draw[->,very thin] (D2) to [out=40,in=140] (D3);
  \draw[->,very thin] (D4) to [out=40,in=140] (D5);
  \draw[->,very thin] (D5) to [out=40,in=140] (D6);
  \draw[->,very thin] (D7) to [out=40,in=140] (D8);
  \draw[->,very thin] (D8) to [out=40,in=140] (D9);
  \draw[->,very thin] (DA) to [out=40,in=140] (DB);
  \draw[->,very thin] (DB) to [out=40,in=140] (DC);
\end{tikzpicture}
  \caption{
    Multigrid inpainting schedule for an $N_{\text{side}} = 128$ map. Filled nodes
    represent successive iterations of diffusion steps (Eq.~\ref{multigrid:diffuse}),
    downward strokes represent calculation of the residual (Eq.~\ref{multigrid:residual})
    and its downgrade to the coarser grid, while upward strokes represent the upgrade of
    the correction (Eq.~\ref{multigrid:correct}) to the finer grid, and solid arrows
    represent the merge of the correction into the solution (Eq.~\ref{multigrid:update}).
    Eight diffusion steps are used at all grid levels except the lowest, where
    64 steps are used, which is enough to find the static solution there.}
  \label{fig:purified:multigrid}
\end{figure*}
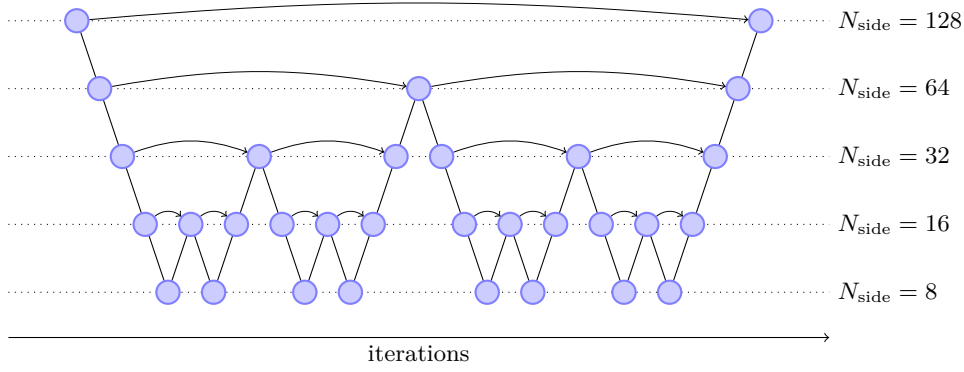

\subsubsection{Finite-difference stencils}
\label{sec:EB_maps:stencils}

Finite-difference approximations of differential operators are non-trivial to evaluate on the \healpix\ grid,
especially for polarization. In this section we discuss first- and second-order-accurate stencils
for the Laplace operator using finite differences. In general, they can be represented as a weighted
sum of the intensity and polarization for the pixel and its nearest neighbours:
\begin{equation}
  {\cal L}[I] = \frac{1}{h^2} \sum\limits_i c_i I_i,\ 
  {\cal L}[Q+iU] = \frac{1}{h^2} \sum\limits_i \tilde{c}_i (Q_i+iU_i),
\end{equation}
with real $c_i$ for scalar and complex $\tilde{c}_i$ for tensor values, and
$h^2 = \pi/3N_{\text{side}}^2$ being the average area of a pixel. We will derive the stencil weights in a
flat-sky approximation by projecting onto a tangent plane through the pixel where the derivative
operator is being evaluated.

A point on the unit sphere with spherical coordinates $\theta$ and $\phi$ has a local set of three
orthonormal vectors,
\begin{align}
  \hat{\vec{n}} &\equiv \Big(\sin\theta\cos\phi, \sin\theta\sin\phi, \cos\theta\Big),\nonumber\\
  \hat{\vec{\theta}} &\equiv \frac{\partial\hat{\vec{n}}}{\partial\theta} =
    \Big(\cos\theta\cos\phi, \cos\theta\sin\phi, -\sin\theta\Big),\nonumber\\
  \hat{\vec{\phi}} &\equiv \frac{1}{\sin\theta}\,\frac{\partial\hat{\vec{n}}}{\partial\phi} =
    \Big(-\sin\phi, \cos\phi, 0\Big),
\end{align}
associated with it, which form a basis on a tangent plane $(\hat{\vec{x}}, \hat{\vec{y}}) \equiv (\hat{\vec{\theta}}, \hat{\vec{\phi}})$ and a normal direction $\hat{\vec{z}} \equiv \hat{\vec{n}}$. Nearby points can
be projected onto a selected tangent plane via gnomonic projection, which maps the radius vector $\vec{r}$ into
\begin{equation}
  \vec{\rho} = \frac{\vec{r}}{\vec{r}\cdot\hat{\vec{n}}} - \hat{\vec{n}}, \hspace{1em}
  \vec{\rho}\cdot\hat{\vec{n}} \equiv 0,
\end{equation}
and thus introduces local coordinates on a tangent plane,
\begin{equation}
  (x,y) = (\hat{\vec{x}}\cdot\vec{\rho}, \hat{\vec{y}}\cdot\vec{\rho}).
\end{equation}
Second-order-accurate discretizations of the Laplacian operator on rectangular grids are well known
and easy to derive \citep{Patra:2006gw}; however, \healpix\ pixels are placed differently,
as is illustrated in Fig.~\ref{fig:purified:stencils}. Deformation of the grid is never small
and does not scale down with increasing \nside. In addition, differences in orientation of the
local bases are never small around the poles (as is obvious from the left column of
Fig.~\ref{fig:purified:stencils}), and care must be taken in the parallel transport of the polarization
tensor represented by the Stokes parameters $Q$ and $U$.

Under rotation of the basis $(\hat{\vec{e}}_1, \hat{\vec{e}}_2)$ used to define the
Stokes parameters $Q$ and $U$, i.e.,
\begin{align}
  \hat{\vec{e}}_1' &= \phantom{-}\cos\psi\ \hat{\vec{e}}_1 + \sin\psi\ \hat{\vec{e}}_2,\nonumber\\
  \hat{\vec{e}}_2' &= -\sin\psi\ \hat{\vec{e}}_1 + \cos\psi\ \hat{\vec{e}}_2,
\end{align}
their values transform as
\begin{align}
  Q' &= \phantom{-}\cos 2\psi\ Q + \sin 2\psi\ U,\nonumber\\
  U' &= -\sin 2\psi\ Q + \cos 2\psi\ U.
\end{align}
In the \healpix\ polarization convention, the Stokes parameters $Q$ and $U$ of a pixel are always
defined with respect to the $(\hat{\vec{\theta}}, \hat{\vec{\phi}})$ basis of the pixel, and must
be rotated to the $(\hat{\vec{x}}, \hat{\vec{y}})$ basis when projecting onto a tangent plane.
The appropriate angle of rotation can be computed as
\begin{equation}
  \tan\psi = \frac{\hat{\vec{x}}\cdot\hat{\vec{\phi}} - \hat{\vec{y}}\cdot\hat{\vec{\theta}}}{\hat{\vec{x}}\cdot\hat{\vec{\theta}} + \hat{\vec{y}}\cdot\hat{\vec{\phi}}},
\end{equation}
with transformation of Stokes parameters most conveniently implemented as a complex phase rotation
\begin{equation}
  Q' + iU' = e^{-2i\psi}\, (Q + iU).
\end{equation}

A linear-order shift of the average pixel position in the \healpix\ grid breaks the symmetry of the local
Taylor expansion, and there is a unique second-order-accurate nearest-neighbour discretization for
the Laplace operator, as opposed to a one-parameter family on the rectangular grid. Unfortunately, it
turns out to be unconditionally unstable for diffusion-type problems, so we will not discuss it here.
Instead, we will use an approximate first-order stencil based on the isotropic weighting $c_i \equiv
c(\rho_i/h)$, with coefficients normalized by ${\cal L}[\text{const}] = 0$ and ${\cal L}[\rho^2] = 4$,
which leads to
\begin{equation}
\sum\limits_{i>0} c_i = -c_0, \hspace{1em}
\sum\limits_{i>0} c_i \rho_i^2 = 4 h^2.
\end{equation}
In previous studies \citep{planck2013-p09a, planck2014-a19} equal weighting was widely used, but
discretization residuals can be significantly improved at little expense by using Gaussian weighting, i.e.,
\begin{equation}
  c_i = 4 \exp\left[-\frac{1}{\sigma}\frac{\rho_i^2}{h^2}\right]\Big/\sum\limits_{i>0} \frac{\rho_i^2}{h^2}\exp\left[-\frac{1}{\sigma}\frac{\rho_i^2}{h^2}\right],
\end{equation}
where the width $\sigma$ can be tuned.  We chose it to be $\sigma=1.61$, which gives near perfect residual
cancelation at the poles. This is the scalar Laplacian stencil we will adopt in what follows, while
the complex Laplacian stencil for polarization is defined by $\tilde{c}_i = e^{-2i\psi_i} c_i$, as
explained above.

\subsubsection{Multigrid methods}
\label{sec:EB_maps:multigrid}

Inpainting a map $\phi$ can be viewed as a diffusive flow $\partial_t\phi = \nabla^2\phi$ applied
to the masked areas, subject to the Dirichlet boundary conditions provided by the unmasked data. A
first-order forward in time discretization of the flow equation $\phi(t+\delta t) - \phi(t) =
(\delta t/h^2) {\cal L}[\phi]$ updates a pixel according to
\begin{equation}
  \phi_0 \rightarrow (1+\alpha c_0) \phi_0 + \alpha\sum\limits_{i>0}c_i \phi_i,
\end{equation}
using the weighted sum of itself and its neighbours, where $\alpha = \delta t/h^2$, and the largest
time step $\delta t$ that can be taken is determined by Courant-Lewy stability analysis. The scheme
is almost guaranteed to be unstable if the coefficient $(1+\alpha c_0)$ becomes negative, so in
practice the fastest diffusion is often achieved by replacing a pixel by a weighted sum of its neighbours:
\begin{equation}\label{multigrid:diffuse}
  \phi_0 \mapsto \sum\limits_{i>0}c_i \phi_i\Big/\sum\limits_{i>0}c_i.
\end{equation}
$N$ successive iterations will diffuse the solution across roughly $N^{1/2}$ pixels, so running diffusion
flow like this for full convergence to the static solution $\nabla^2\phi=0$ is very expensive on high-resolution maps.
In previous studies \citep{planck2013-p09a, planck2014-a19}, a finite number of steps like Eq.~(\ref{multigrid:diffuse}) was applied to inpaint the masked areas.

Convergence of diffusion flow can be accelerated if one notes that a coarser spatial grid allows bigger time steps, and thus the solution of a linear elliptic boundary value problem on a grid with spacing $h$,
\begin{equation}
  {\cal L}[\phi]_h = \rho_h,
\end{equation}
can be found iteratively from the approximate solution $\tilde{\phi}_h$ by downgrading the residual
\begin{equation}\label{multigrid:residual}
  {\cal R}[\tilde{\phi}]_h = {\cal L}[\tilde{\phi}]_h - \rho_h
\end{equation}
to a coarser grid with spacing $2h$, thus obtaining the coarse-grid correction
\begin{equation}\label{multigrid:correct}
  {\cal L}[\delta\phi]_{2h} = {\cal R}[\tilde{\phi}]_{2h},
\end{equation}
which can be upgraded back to a fine grid with spacing $h$ and used to improve the solution, i.e.,
\begin{equation}\label{multigrid:update}
  \tilde{\phi}_h \mapsto \tilde{\phi}_h + \delta\phi_h,
\end{equation}
as detailed in \citet{Brandt:1977kr}. Recursive application of coarse-grid correction
(Eq.~\ref{multigrid:correct}) interlaced with diffusion steps (Eq.~\ref{multigrid:diffuse}), as
illustrated in Fig.~\ref{fig:purified:multigrid}, achieves convergence to the static
solution in $O(\log N_{\text{side}})$ iterations, most of which are on coarser grids,
and thus very fast. This is the method we use to diffusively inpaint intensity
and polarization maps in this paper.

The algorithm proceeds by first constructing the multigrid structure, which will contain temporary
maps and residuals to be corrected. The inpainting mask is recursively degraded, and Laplacian stencils
are precomputed for pixels to be inpainted at all grid levels. Note that only the strict interior of the
masked region should be inpainted at coarse levels to avoid boundary effects that degrade the convergence rate.
The finest grid is initialized with the map to be inpainted, while the coarse grids will contain corrections
and are initialized to zero. The inpainting is carried out by repeated application of the recursive
``w-stroke,'' as illustrated in Fig.~\ref{fig:purified:multigrid} for an $N_{\text{side}}=128$ example.
The w-stroke takes a small number of diffusion steps (Eq.~\ref{multigrid:diffuse}) on the region to be inpainted,
computes and downgrades the residual (Eq.~\ref{multigrid:residual}), recursively calls itself on a coarse grid
twice to obtain the correction (Eq.~\ref{multigrid:correct}), upgrades the correction and merges it with the solution (Eq.~\ref{multigrid:update}), then takes a small number of diffusion steps (Eq.~\ref{multigrid:diffuse}) again.
The entire pattern is repeated until the desired convergence accuracy is reached. For the masks used in
temperature and polarization analysis in this paper, 18 iterations of the w-stroke are enough to reach the
static solution to double precision. This requires only 288 diffusion steps at full resolution, and
is substantially faster than the 2000 iterations at full resolution that were used in \citet{planck2013-p09a}.
Improved performance of the inpainting routine allows for further applications and testing to be carried out with the same computing resources.

\subsection{Purified inpainting}
\label{sec:EB_maps:purified}

\subsubsection{Method Description}

Here, we demonstrate that the advantages of masking and inpainting can
be combined into a single method.  

The motivation for the approach is to construct a full-sky
polarization map, $\tilde\p$, from a masked one, $\M\cdot\p$, such that it
agrees identically with $\p$ on the unmasked portion of the sky, and
assigns modes either through strict attribution to the $E$ and
$B$ subspaces of the full-sky map, or ``ambiguous'' attribution where
some mode mixing is allowed:
\begin{equation}\label{purified:reconstruction}
  \tilde\p \equiv \a+\e+\b;\hspace{1em}
  \M\cdot\tilde\p = \M\cdot\p.
\end{equation}
Note that the polarization map components $\e$ and $\b$ are not pure in the
sense of being orthogonal to the entire $E$- and $B$-mode linear
spaces on the masked sky, as in \citet{Bunn:2002df}; that
requirement results in a large linear algebra problem, which is
expensive to solve, although efficient methods for that have been
developed recently \citep{Bunn:2016lxi}. Instead, their purpose is to ``purify'' the ambiguous
mode map $\a$, ensuring that most of the pure modes are projected out,
thus minimizing the power leakage between $E$ and $B$ in the
ambiguous mode.

Initially, the entire polarization map $\p$ is assigned to an
ambiguous mode:
\begin{equation}\label{purified:all-ambiguous}
  \a = \p; \hspace{1em}
  \e = 0; \hspace{1em}
  \b = 0.
\end{equation}
Then the method proceeds by peeling off non-ambiguous $E$ and
$B$ modes one by one through the explicit construction of a Krylov
subspace \citep{Krylov1931}
generated by the inverse bi-Laplacian, $\bbK$, acting on the
masked maps. The obvious starting point that contains most of the CMB
power is the $E$-mode projection of the masked polarization map:
\begin{equation}\label{purified:e-mode}
  \w = \E\M\cdot\a.
\end{equation}
To prevent the constructed Krylov subspace basis from becoming
degenerate, we use Lanczos bi-orthogonalization \citep{Lanczos1950}.
To do this we keep two recent
normalized basis vectors, $\u$ and $\v$, initially set to
\begin{equation}\label{purified:krylov:start}
  \beta = \norm{w}{w}^{\frac{1}{2}},\hspace{1em}
  \u = 0,\hspace{1em}
  \v = \frac{\w}{\beta},
\end{equation}
and project them out (from the next $E$-mode generated using the inverse
bi-Laplacian operator, $\bbK$) to obtain a new mode from the previous ones via
\begin{equation}\label{purified:krylov:next}
  \w = \E\bbK\M\cdot\v,\hspace{1em}
  \w \mapsto \w - \beta\,\u,\hspace{1em}
  \w \mapsto \w - \norm{w}{v}\, \v.
\end{equation}
Once the new mode is constructed, it is normalized and the most recent
basis set is updated:
\begin{equation}\label{purified:krylov:update}
  \beta \mapsto \norm{w}{w}^{\frac{1}{2}};\hspace{1em}
  \u \mapsto \v;\hspace{1em}
  \v \mapsto \frac{\w}{\beta}.
\end{equation}
Lanczos bi-orthogonalization guarantees that the dot product of the
constructed basis vectors $\langle\v_i,\v_j\rangle$ forms a
tri-diagonal matrix, and thus avoids any stability issues associated with
the basis vectors becoming nearly linearly dependent. Whenever a new
basis vector, $\v$, is constructed following
Eqs.~(\ref{purified:krylov:start}) or (\ref{purified:krylov:update}), it
is projected out from the ambiguous mode map and assigned to the
$E$-mode map:
\begin{equation}\label{purified:peel-off}
  \alpha = \norm{a}{v};\hspace{1em}
  \a \mapsto \a - \alpha\,\v;\hspace{1em}
  \e \mapsto \e + \alpha\,\v.
\end{equation}
Once a sufficient number of $E$ modes are extracted from the map,
$B$ modes are peeled off next in exactly the same way, via
the $B$-mode projection operator, $\B$, instead of $\E$. The Krylov
subspace construction (Eqs.~\ref{purified:e-mode} through
\ref{purified:krylov:update}) can be restarted several times on the
remaining ambiguous mode map, but it reaches a point of diminishing
returns rather quickly.  The exact number of Krylov modes to peel
depends on a compromise between performance and quality of reconstruction,
and is discussed in the next section.

After the ambiguous modes are purified, they still need to be
reconstructed on the full sky. The correct approach is to inpaint the
masked region by solving the elliptical partial differential equation,
$\nabla^2\a=0$, in the mask interior, subject to fixed values of $\a$ at
the mask boundary, which minimizes the total extra amount of power
introduced to the reconstructed ambiguous-mode map.

Inpainting is a linear
operation, and can be represented by a matrix operator $\F$. A
suitable method for inpainting high-resolution maps is the multi-grid
approach, as discussed above, which has computational costs of
${\cal O}(n \log n)$.  The final reconstructed polarization map is
generated by inpainting the remaining ambiguous modes as
\begin{equation}\label{purified:inpaint}
  \a \mapsto \F\M \cdot \a .
\end{equation}

\subsubsection{Performance considerations}
\label{sec:EB_maps:performance}

Figure~\ref{fig:purified:coupling} presents the coupling matrices
corresponding to the purified inpainting of the polarization data
after application of the common polarization mask at $N_{\rm side}=1024$. The
temperature maps are inpainted following the usual multi-grid
approach.  The low-to-high-$\ell$ mode-mixing is much improved over
the equivalent case with masking alone, while the increase in the
high-to-low-$\ell$ mode-mixing is substantially less than if a simple
inpainting approach were applied. Purified inpainting therefore represents a
good compromise for the $E$- and $B$-mode reconstruction of CMB
polarization maps.

The computational costs of the purified-inpainting method are
dominated by the two round-trip full-resolution spherical transforms
required for the initial projection (Eq.~\ref{purified:e-mode}) of the
$E$ and $B$ modes, and the multi-grid inpainting of the ambiguous
modes. The Krylov subspace construction does eventually produce a
complete basis for the $E$- and $B$-mode subspaces, but is too
expensive to run to completion for high-resolution maps.  An
investigation of the computational cost versus quality of the
reconstruction suggests that a drastic truncation works well. In fact,
we extract only 32 modes in the code used here. The Krylov basis maps
generated by an inverse bi-Laplacian operator generally have very red
spectra, due to the inverse $(\ell-1)\ell(\ell+1)(\ell+2)$ factor in
Eq.~(\ref{purified:bilaplacian}), hence do not require spherical transforms
at full $\ell_{\text{max}}$.

\subsection{Reconstruction residuals and confidence mask}
\label{sec:EB_maps:residuals}

\begin{figure*}[htbp!]
  \centering
  \includegraphics[width=\textwidth]{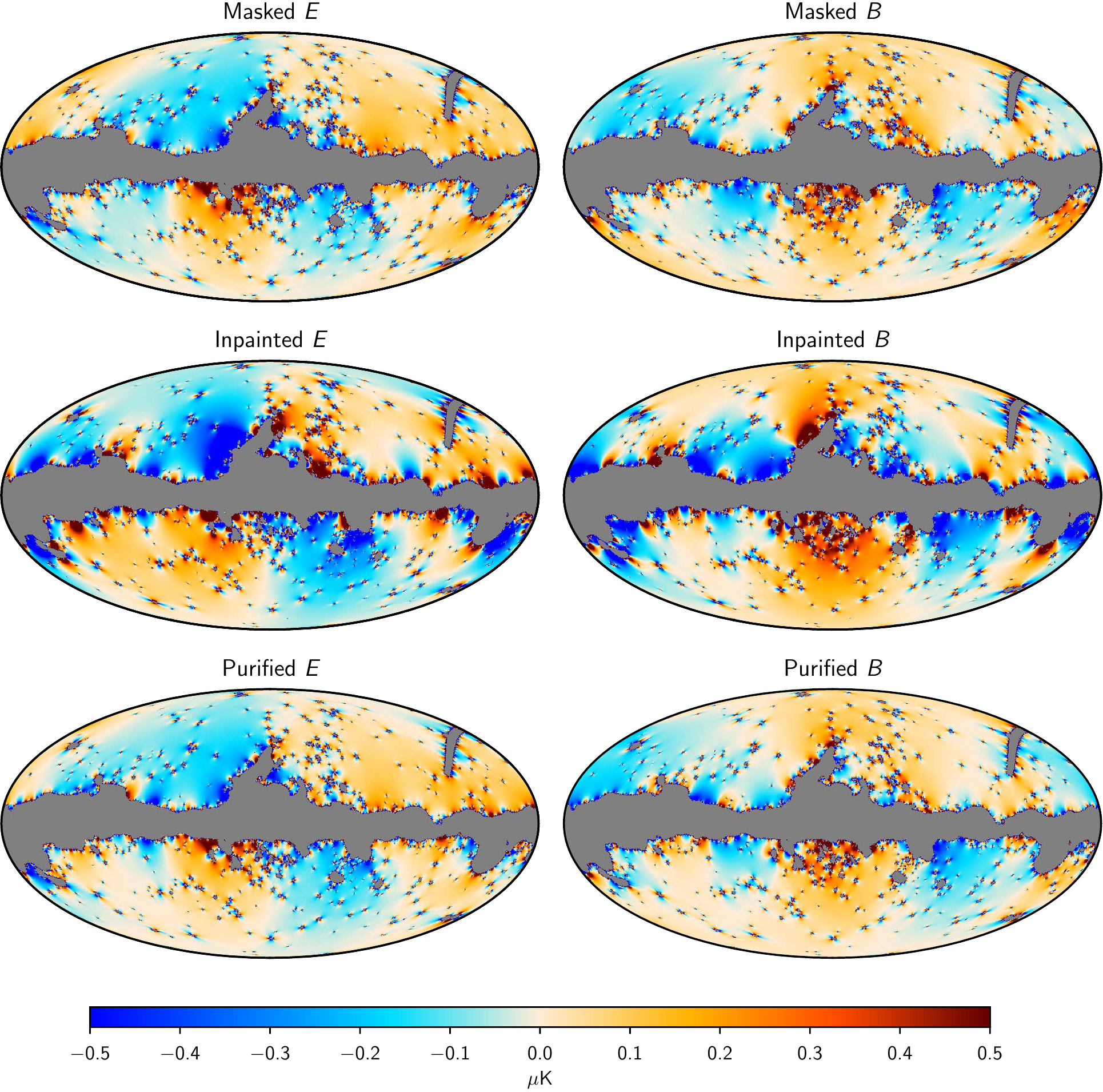}
  \caption{$E$- and $B$-mode reconstruction residuals, shown in the
    left and right columns respectively, for the methods of mask multiplication
    (top), simple inpainting (middle), and purified inpainting
    (bottom). Residuals are shown for a single realization of
    a \smica\ component-separated simulation containing both CMB
    signal and noise at a resolution $N_{\mathrm{side}}=1024$. The
    grey area represents the common polarization mask applied during
    the reconstruction.}
  \label{fig:purified:residuals}
\end{figure*}

\begin{figure*}[htbp!]
  \centering
  \includegraphics[width=\textwidth]{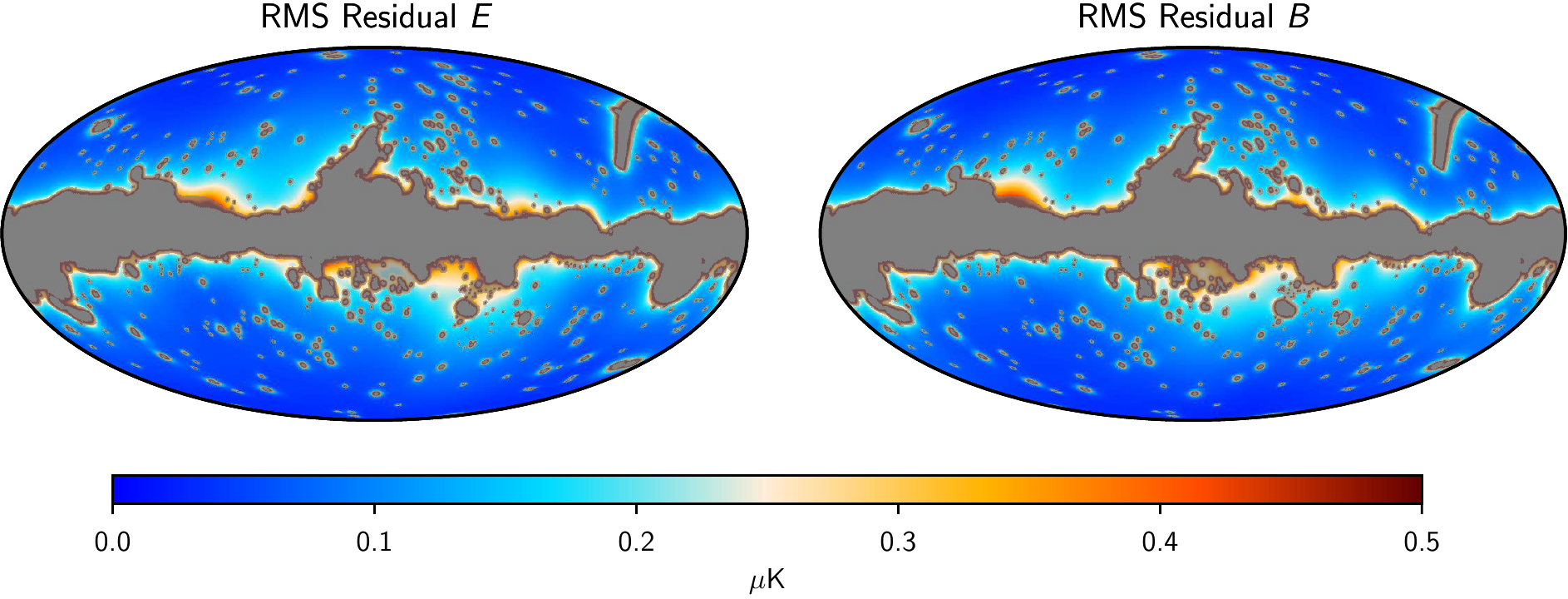}
  \caption{Rms of the combined reconstruction residuals for $E$ and
    $B$ modes (left and right columns, respectively) determined from
    the purified inpainting of the FFP10 simulations for all four
    component-separation methods (i.e., the average of the square of
    each of the four residuals). The grey area represents the common
    polarization mask, while the semi-transparent grey area indicates
    an expansion of the confidence mask, which together admit 65\,\% of
    the sky for further analysis.  The mask increment is determined
    from thresholding the total residual $\delta E^2+\delta B^2$,
    smoothed by an 80\arcmin\ FWHM Gaussian. The rms reconstruction
    accuracy is better than 0.5$\,\mu$K, with the largest deviations
    mainly localized near the mask boundary.}
  \label{fig:purified:confidence}
\end{figure*}

\begin{figure*}[htbp!]
  \centering
  \begin{tabular}{cc}
    \includegraphics[width=\columnwidth]{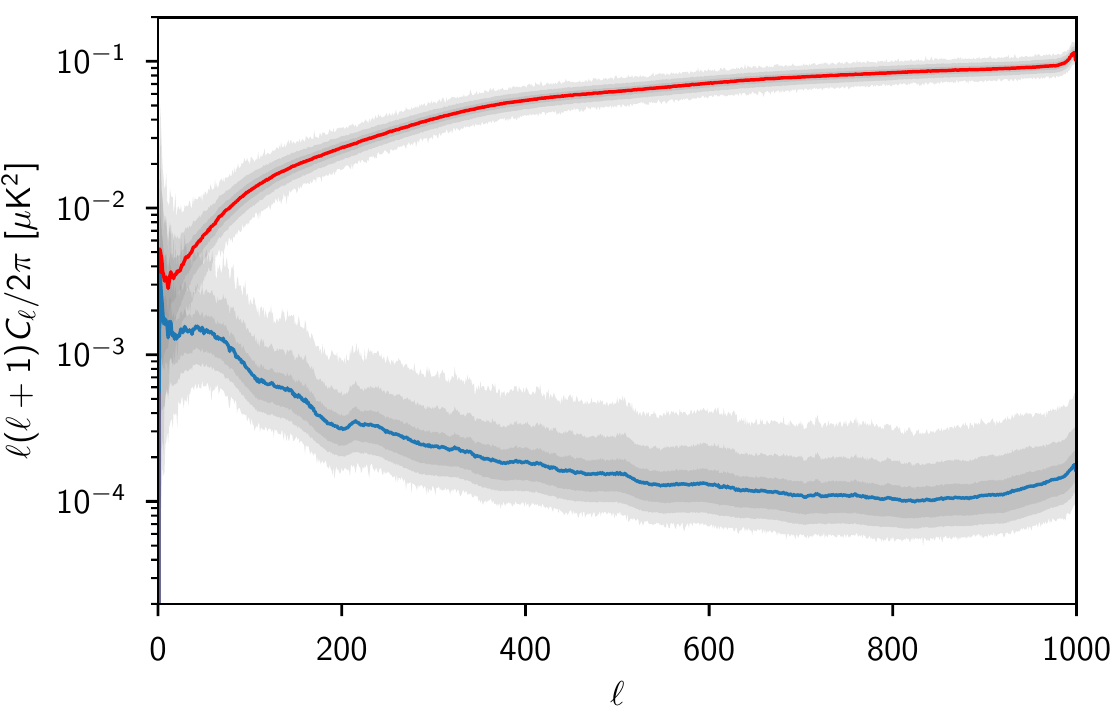} &
    \includegraphics[width=\columnwidth]{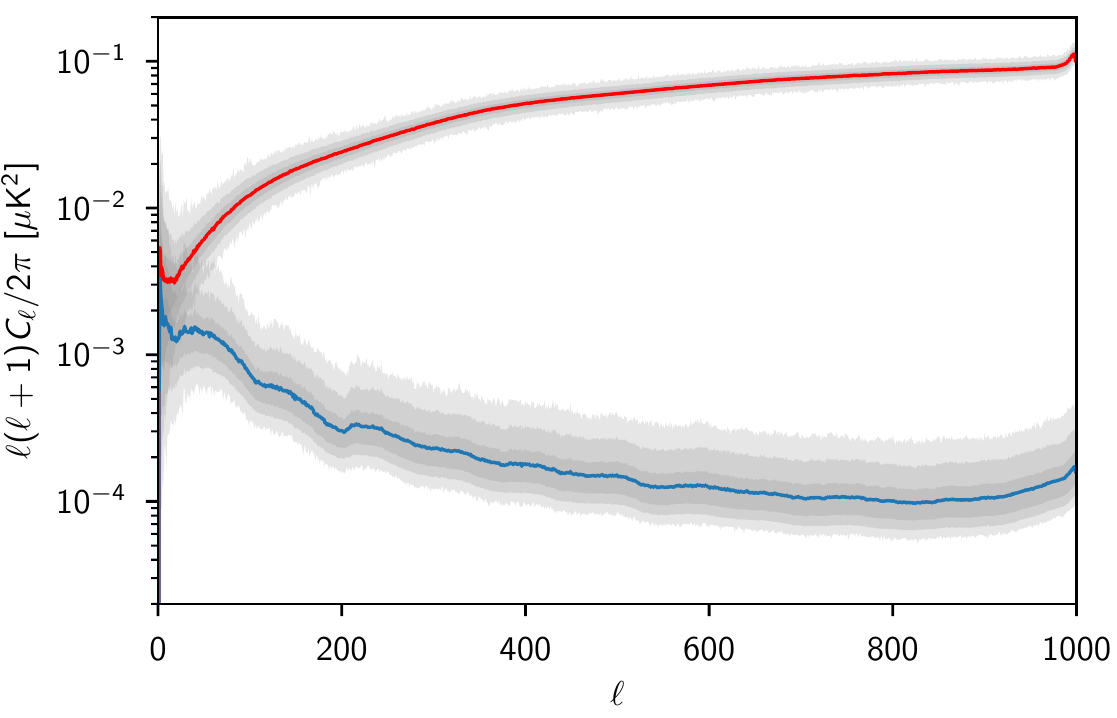} \\
  \end{tabular}
  \caption{Pseudo-$C_\ell$ spectra of residuals for $E$ and
    $B$ modes (left and right columns, respectively) determined from
    the purified inpainting of the FFP10 simulations for all four
    component-separation methods. Solid curves show averaged spectra, with
    red corresponding to the residual masked with the reconstruction mask and
    blue corresponding to the residual masked with the confidence mask.
    Semi-transparent areas filled with grey show 68\,\%, 95\,\%, and
    minimum-to-maximum bounds for individual realizations.
    }
  \label{fig:purified:spectra}
\end{figure*}

The accuracy of reconstruction of $E$- and $B$-mode maps for a given method
can be evaluated from realistic MC simulations of the CMB signal plus
noise, where the true full-sky $E^\ast$ and $B^\ast$ maps are known,
and can be directly compared to the reconstructed ones $\tilde{E}$ and
$\tilde{B}$. Various metrics of performance can be assigned to the
masked residual maps $\delta E \equiv \M\cdot(\tilde{E}-E^\ast)$ and
$\delta B \equiv \M\cdot(\tilde{B}-B^\ast)$.
Figure~\ref{fig:purified:residuals} compares the
residual maps for one \smica\ realization from the FFP10 simulations,
as reconstructed via each of the masking, simple inpainting, and 
purified-inpainting methods.
The latter seems to perform on par or better than other direct reconstruction
methods published so far, and is competitive with the maximum-likelihood
estimators, at substantially lower computational
cost. Apodization and inpainting methods do not perform as well
because of the $E$- and $B$-mode mixing arising from the mask, while
the pure projection method of \citet{2006PhRvD..74h3002S} actually
does much worse as a consequence of the high-$\ell$-to-low-$\ell$
aliasing of the noise power. The purified-inpainting method, presented
here, offers a balance between these two considerations,
and yields an order of magnitude lower residuals for
$N_{\mathrm{side}}=1024$ \Planck\ maps.

\begin{figure}[htbp!]
  \centering
  \includegraphics[width=\columnwidth]{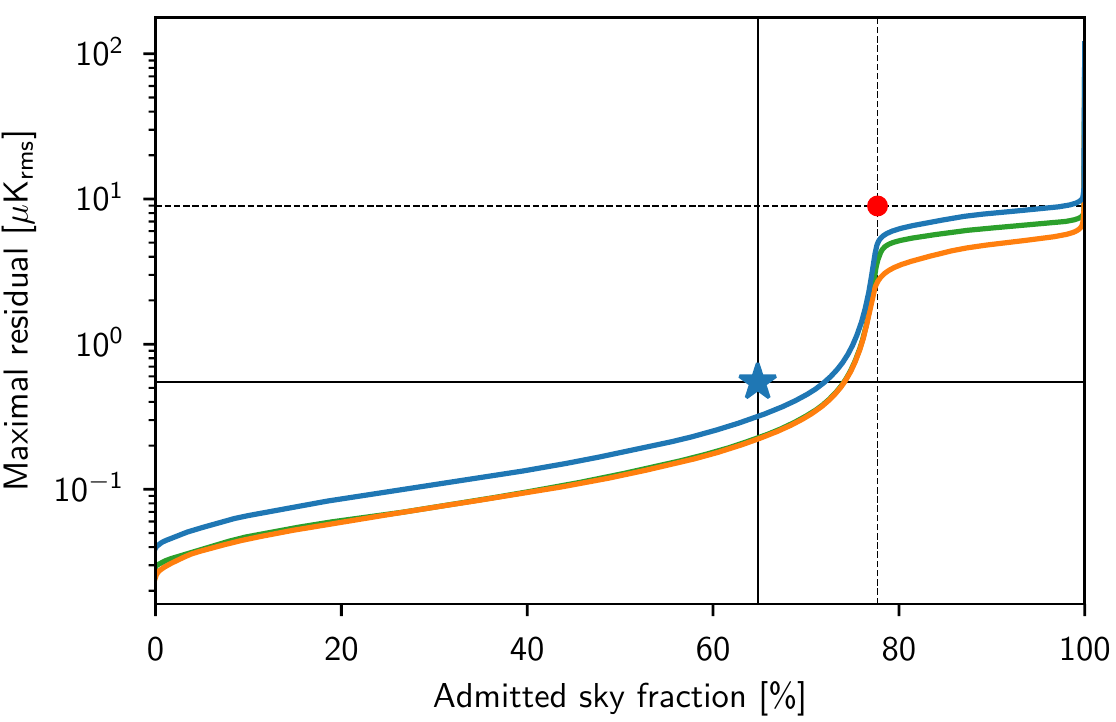}
  \caption{
Maximal rms of reconstruction residuals for $E$ modes (green), $B$ modes
    (orange), and combined $E$ and $B$ modes (blue) determined from purified-inpainting
    simulations as a function of sky fraction.
    The red circle and the thin dashed lines show the reconstruction quality within the
    common polarization mask used in reconstruction (77.7\,\% of the sky
    admitted, maximum rms combined residual of 9.0$\,\mu$K). The blue star and the thin solid
    lines show the reconstruction quality within the confidence mask (64.8\,\% of
    the sky admitted, with a maximum rms combined residual of 0.5$\,\mu$K). }
  \label{fig:purified:tuning}
\end{figure}

Extending the mask beyond that used in the reconstruction
process itself can help to reduce the residuals levels yet further, in
particular given that these residuals tend to be localized near the
mask boundary.
To systematically define the optimal confidence mask, we performed
purified-inpainting reconstructions for the FFP10 CMB-plus-noise
realizations, as propagated through all four \Planck\ component
separation pipelines and subsequently provided at a number of
resolutions, then evaluated the rms signal in the residual maps.
Combined reconstruction residuals for all four component-separation
pipelines are shown in Fig.~\ref{fig:purified:confidence}, along with
the confidence mask that admits 65\,\% of the sky for further analysis,
as derived from the iso-levels of the total rms reconstruction
residual $\delta E^2+\delta B^2$ smoothed by an 80\arcmin\ FWHM
Gaussian beam. Figure~\ref{fig:purified:spectra} shows the pseudo-$C_\ell$
anisotropy power spectra of masked residuals for reconstructed $E$ and $B$ modes
for all four component-separation pipelines.

The choice of the confidence mask threshold is motivated by
Fig.~\ref{fig:purified:tuning}, which shows the maximal rms
reconstruction residuals versus the sky fraction. Note that a slight
reduction in the sky fraction admitted by the confidence mask results
in more than an order of magnitude decrease in the residuals.
Smoothing the residuals in order to simplify the detailed geometry of
the confidence mask results in a slight increase of the maximal
residual, but is nevertheless beneficial for many analyses, given the
dependence of the mode coupling on this structure.  The final
confidence mask is selected, such that the rms reconstruction
residuals are less than $0.5\,\mu$K, significantly below the
cosmological $E$-mode signal, thus allowing sensitive tests to be
undertaken concerning the isotropy and statistical properties of the
\Planck\ polarization data.

\end{appendix}

\raggedright

\end{document}